\documentclass[11pt]{report}

\usepackage[a4paper,twoside,left=35mm,right=25mm,top=40mm,bottom=35mm]{geometry}

\usepackage{subfigure}
\usepackage{masthesis}

\newcommand{\be}{\begin{equation}}
\newcommand{\ee}{\end{equation}}
\renewcommand{\[}{\begin{equation}}
\renewcommand{\]}{\end{equation}}

\newcommand{\A}{\mathcal{A}}

\renewcommand{\O}{\mathcal{O}}
\newcommand{\rar}{\rightarrow} 
\newcommand{\dd}{\mathrm{d}}

\newcommand{\vphi}{\varphi}
\newcommand{\ep}{\varepsilon}
\renewcommand{\a}{\alpha}
\newcommand{\g}{\gamma}

\newcommand{\<}{\langle}
\renewcommand{\>}{\rangle}

\newcommand{\nn}{\nonumber}
\newcommand{\lla}{\langle \! \langle}
\newcommand{\rra}{\rangle \! \rangle}
\renewcommand{\a}{\alpha}
\renewcommand{\b}{\beta}
\renewcommand{\d}{\delta}

\newcommand{\bs}[1]{\boldsymbol{#1}}
\newcommand{\x}{{\bs{x}}}
\newcommand{\y}{{\bs{y}}}
\newcommand{\z}{\zeta}
\newcommand{\p}{\partial}
\newcommand{\q}{{\bs{q}}}

\renewcommand{\t}{\theta}
\newcommand{\tb}{\bar{\theta}}

\DeclareMathOperator{\Li}{Li}

\DeclareMathOperator{\K}{K}

\newcommand{\ino}{i}
\newcommand{\lcb}{{\tt {\char '173}}}
\newcommand{\rcb}{{\tt {\char '175}}}

\begin{document}

  \phdtitle{Shift operators and momentum-space conformal field theory}            {Francesca Caloro}                     {logo}                              {February $2024$}                           {Dr. Paul McFadden}   
  \thispagestyle{empty}
  \cleardoublepage

  \begin{dedication}
\emph{D'autant que je m'int{\'e}resse moins aux math{\'e}matiques,  \\q'aux math{\'e}maticiens, comme en tout autre domaine.}\\~\\ \emph{Simone Weil} \cite{WeilSimone}
\end{dedication}
\thispagestyle{empty}
\cleardoublepage
 
  \begin{acknowledgements}
This thesis represents the end of an incredible journey through public schools. I want to take this opportunity to thank all my teachers from school, and outside school, older and younger than me. This thesis would not exist without the passion for questioning and understanding together with the ability to make choices that I learned from you.

Thanks to my first math teacher Sig.na Anna Licchetta, I felt math was fun and warm. Thank you for telling us of your illness and showing us the dignity of the end.

Thanks to Sig.ra Elena Izzo and Sig.ra Raffaella Cristiani for putting the roots of learning in me. Thanks to Prof. Antonio Marsano and Prof.ssa Martella for helping me speak English and French. Thanks to Prof. Antonio Piri for teaching me art. Thanks to Prof. Luigi Lecci for challenging us with problems in math and physics. Thanks, Prof.ssa Lara De Marco, for the sparkle in your eyes while you spoke about Montale. Thank you, Prof.ssa Daniela Orlando, for making me study physics, even though I probably never stopped thinking mathematically.

Thanks to my professors from Universit\`a del Salento. To Prof. Claudio Corian\`o for his passion for gauge theories, and Prof. Giulio Landolfi and Dr. Mario Angelelli for the fun with complex networks. A special thanks to my master's supervisor Luca Girlanda for all the discussions about physics and life, your kindness, and your time.

Thanks to my professors from the music conservatories in Taranto and Lecce. Thanks to Prof. Fernando Giovinazzi, I learned that there are no first and second classes in music, hence in any field. Thanks to Prof.ssa Luisa Cosi for her enchanting classes of history of music and to Prof.ssa Bianca Maria Dell'Erba for the care, the challenges and the music. 

Finally, thanks to my supervisor Dr. Paul McFadden. The research presented in this thesis would not have been possible without your guidance and dedication. 

The PhD time has been tough sometimes, but full of fun. A special thanks to my friends Ashlin and Devika for sharing the laughs and the tears and, most importantly, very tasty biryani. Thank you, Ashlin, for being there throughout these four years, you have been a safe place. Thank you, Devika, for finding the comedy in the tragedies, and for the cumin. Thanks, Tao, for your genuine friendship and thanks to Aida for coping together with Covid restrictions in Newcastle. Thanks to the whole badminton crew!

Thanks to my friends who stayed through the years, while, unlike symmetric correlators, we changed form. Thanks, Marina, for telling me of your neck pain after a night spent with Shakespeare's sonnets under the pillow. Thank you, Annamaria, for being there and believing in me. Thanks, Isabella, for our walks at Marina di Andrano.

Thanks to my friends from Continuiamo a Rapirci that I met at Londra Scrive. You have lightened and enriched my last two years. 

Thanks to Paola and Michela for growing up together, for the music, the debates, the fun and the love. Grazie mamma e pap\`a per il cibo, i libri, il pianoforte.

\end{acknowledgements}
\thispagestyle{empty}
\cleardoublepage
 
  \begin{abstract}
A momentum-space approach to conformal field theory offers a new perspective on cosmological correlators and better reveals the underlying connections to scattering amplitudes. While correlation functions at up to three points are well understood, the form of higher-point functions is still under active study and few explicit results are available.

A representation for the general $n$-point function of scalar operators was recently proposed in the form of a Feynman integral with the topology of an $(n-1)$-simplex, featuring an arbitrary function of momentum-space cross ratios. In this thesis, we show the graph polynomials for this integral can all be expressed in terms of the first and second minors of the Laplacian matrix for the simplex. Computing the effective resistance between nodes of the corresponding electrical network, an inverse parametrisation is found in terms of the determinant and first minors of the Cayley-Menger matrix. These parametrisations reveal new families of weight-shifting operators expressible as determinants that connect n-point functions in spacetime dimensions differing by two. Furthermore, they enable the validity of the conformal Ward identities to be established directly without recourse to recursion in the number of points.

We then analyse the representation of conformal, and more general, Feynman integrals through a class of multivariable hypergeometric functions proposed by Gelfand, Kapranov \& Zelevinsky.  Among other advantages, this formalism enables the systematic construction of highly non-trivial weight-shifting operators known as ``creation'' operators.  
We discuss these operators from a physics perspective emphasising their close connection to the spectral singularities that arise for special parameter values, and their relationship to the Newton polytope of the integrand. Via these methods we construct novel weight-shifting operators connecting contact Witten diagrams of different operator and spacetime dimensions, as well as exchange diagrams with purely non-derivative vertices.

\end{abstract}
\thispagestyle{empty}
\cleardoublepage
 
\thispagestyle{empty}
\cleardoublepage
  
  \pagenumbering{roman}                                                         
  \tableofcontents

  \clearpage                              \thispagestyle{empty}                   \cleardoublepage                      
  \pagenumbering{arabic}                  \part{Introduction}
  \clearpage{}\begin{chapter}{\label{chap1}Introduction}
\section{Conformal field theory}
Conformal symmetry arises in many different physical contexts. Historically, it was first applied in the 70s to the study of critical phenomena \cite{DiFrancesco} and soon after started to play a major role in string theory \cite{Polchinski}. In the late 90s, conformal field theories (CFTs) were found to be dual to gravitational theories in Anti de Sitter (AdS) space through a paradigm known as holography \cite{Maldacena97, Maldacena99}. 

Among the key observables of quantum field theory, cosmology, and condensed matter are correlation functions of operators. These are scale-invariant at critical points, and surprisingly different physical systems are sometimes found to share the same set of critical exponents. For example, the critical exponent for ferromagnets is the same as for water (liquid-vapor transition) \cite{Elvang:2020lue}. This property is called \emph{universality} and reveals a common underlying conformal symmetry. Polyakov showed that correlators at critical points are indeed invariant under the full conformal group (which also includes special conformal transformations) and opened the path for applications of conformal symmetry in physics and quantum field theory \cite{Polyakov:1970xd}. 
A program to study the implications of conformal symmetry for scalar and tensorial operators in general spacetime dimension was developed in \cite{Mack:1969rr, Parisi:1971zza, FERRARA197277,Mack:1972kq,Fradkin:1994zf, Osborn:1993cr}. The idea was to derive the form of correlators by symmetry principles, and this led to looking for solutions of conformal Ward identities in a spacetime dimension $d>2$ for general scaling dimensions. 
These analyses were carried out in position space, where conformal transformations act directly.

The study of conformal anomalies \cite{Giannotti:2008cv, Armillis:2009pq, Schwimmer:2023nzk} and the application of holography to cosmology \cite{Witten:2001kn, Strominger:2001pn, Antoniadis:2011ib,Maldacena:2011nz, Creminelli:2011mw, Bzowski:2012ih, Mata:2012bx,Kehagias:2012pd, McFadden:2013ria,Ghosh:2014kba, Anninos:2014lwa,Arkani-Hamed:2015bza, Arkani-Hamed:2018kmz, Baumann:2019oyu,Baumann:2020dch,Sleight:2019hfp} motivated the development of momentum-space conformal field theory.
The inflationary epoch is described by an approximately de Sitter spacetime geometry, and the symmetries of this spacetime act on late-time slices as conformal transformations. Therefore, inflationary correlators can equivalently be regarded as CFT correlators.
Moreover, momentum-space CFT also found a central role in the study of renormalisation \cite{apk2016, Bzowski:2017poo, Bzowski:2018fql, Bzowski:2022rlz} and scattering amplitudes, revealing features such as double-copy structure and colour/kinematic duality \cite{Raju:2012zr, Farrow:2018yni,Lipstein:2019mpu, Armstrong:2020woi,Albayrak:2020fyp}. 

The analysis of the implications of conformal symmetry in momentum space started ten years ago. The form of 2- and 3-point functions of scalar and tensorial operators in $d>2$ are strictly constrained by the symmetry and their unique form was found \cite{Bzowski:2013sza, Coriano:2013jba}. Equivalent representations for the 3-point function have been analysed: this can be expressed as an integral over three Bessel functions of the second kind, as a multivariable hypergeometric function Appell $F_4$, or as a one-loop triangle Feynman diagram. Except for some special solutions, evaluating these functions can be difficult. A reduction scheme has been described to construct a class of 3-point functions which are also the building blocks of tensorial correlators \cite{evaluation, apk2016, Bzowski:2017poo}. This reduction is performed via the action of shift operators that connect different solutions with shifted parameters. Moreover, a full understanding of their singularities and renormalisation has also been discussed \cite{Bzowski:2015pba}. Less complete and understood is instead the form of 4- and higher-point correlators for general values of their parameters. A representation for the general $n$-point function of scalar operators was recently proposed in the form of a Feynman integral with the topology of an $(n-1)$-simplex, featuring an arbitrary function of momentum-space cross ratios \cite{Bzowski:2020kfw, Bzowski:2019kwd}. This was shown to be conformally invariant, and a recursive interpretation of its form was given. 

The research presented in this thesis takes its starting point from the simplex representation of $n$-point functions and develops to explore the interplay between integral representations and shift operators.

\section{Shift operators}
Operators that act on a function to shift one or more of its parameters appear in the physical and mathematical literature with different names. Here, we refer to such operators as \emph{shift operators}.

Historically, their relevance in physics was first revealed in quantum mechanics. To solve the Schr\"odinger problem of the harmonic oscillator, Dirac introduced shift operators known as the creation and annihilation operators \cite{Dirac1930-DIRTPO}. 
They were useful both on the computational and physical sides. Indeed, such operators act on an eigenfunction of the Hamiltonian to generate a new eigenstate of the same Hamiltonian, but with shifted eigenvalue. Physically, the action of such shift operators makes an energy quantum $\hslash\omega$ appear or disappear. By knowing these operators, Dirac was able to find the ground state and the full set of solutions of the quantum harmonic oscillator: once the ground state is found, it is sufficient to apply the creation operator to find all the remaining eigenstates. 

The idea of introducing operators that act on a quantum state to shift a quantum number was also applied to describe the physics of the quantum angular momentum \cite{Edmonds:1955fi}. In the case of a three-dimensional angular momentum, it turned out that it was convenient to define the shift operators, 
often referred to as raising/lowering operators. They are complex linear combinations of the quantised spatial $x$ and $y$ components of the angular momentum operator.
In an analogy with the system of the harmonic oscillator, one can show that these operators act on an eigenstate of the $z$ component of the angular momentum to increase or decrease it by an angular momentum quantum $m\hslash$. 
Once the quantisation of the angular momentum is defined, several applications show the utility of such shift operators \cite{Sakurai:2011zz}. Among these, we find the solution of the Schr\"odinger problem of the hydrogen atom \cite{Dirac:1926vy, Pauli:1926qpj}, and the study of hydrogen-like systems in solid state physics, the effect of a magnetic field on the energy levels of an atom, and the Zeeman effect \cite{cohen}.

In quantum field theory (QFT), creation and annihilation operators appear in the solution of dynamical equations. For instance, a scalar field is expressed as a superposition of normal modes whose amplitudes of oscillation are given by operators analogous to the creation and annihilation operators of the quantum harmonic oscillator, since they satisfy the same commutation relations. 
It is interesting to note that according to Pauli's exclusion principle \cite{Pauli:1925nmn}, fermions and bosons are described by different symmetries and this property results in the fact that while bosons' shift operators obey commutation relations, fermions' obey anticommutation relations \cite{Peskin:1995ev}. Therefore shift operators are related to the symmetry of a system.
This relation is extensively used in particle physics as well as in statistical physics.

Along with the development of QFT, Feynman integrals became key objects in various areas of physics. Their accurate evaluation became central to the understanding of physical phenomena. With the development of experiment and theory, the need for precision and accuracy increased. Higher orders in perturbative QFT are required in the computation of observables via Feynman diagrams, whose number and complexity increase with the order in perturbation. For this reason, Feynman integrals continue to be an active research topic needed for example in scattering processes, perturbative quantum chromodynamics, lattice computations, CFT and cosmology \cite{Delto:2023kqv, Tarasov:2022pwt, Bottcher:2023wsr, Devoto:2023rpv, Agarwal:2023suw, Gasparotto:2023roh, Heckelbacher:2022hbq, Caloro:2022zuy, Arkani-Hamed:2023kig}. 

The problem of computing Feynman integrals is often hard to tackle. The main needs are the reduction of tensorial integrals to scalar integrals and the reduction of the latter into a small (finite) number of Feynman integrals known as master integrals. Various techniques have been explored and continue to be discussed \cite{Weinzierl:2022eaz}. Among these we find integration-by-parts (IBP) identities \cite{Chetyrkin:1981qh, Smirnov:2010hn, Grozin:2011mt}, allowing us to express any Feynman integral as a linear combination of master integrals. These identities stem from taking the total derivative of the integrand.  More recently the IBP method has been extended by looking at operators that annihilate the integrand \cite{Bitoun:2017nre}. A complementary method to simplify the computation of Feynman integrals is based on recurrence relations, and the shift operators from which they can be derived \cite{Tarasov:1996br, Tarasov:1997kx, Lee:2010wea}. This method is based on considering the Feynman integral as a function of its parameters, (\emph{i.e.}, the powers of propagators and/or the spacetime dimension) and by acting with appropriate operators on such an integral, one finds a new integral with shifted parameters. This is also useful in tensorial reductions.

Later, when CFT arose in the study of physical phenomena, the methods involving shift operators and recursion relations appeared in this context. Position-space 4-point functions of scalar operators are expressed via the operator product expansion whose terms, known as conformal blocks, depend on the spacetime dimension and the operator dimensions. Dolan and Osborn showed that such conformal blocks satisfy a second-order differential equation and are related to the eigenfunctions of the quadratic Casimir of the conformal group \cite{Dolan:2003hv, Dolan:2011dv}. They define various sets of shift operators that act on the conformal blocks to shift their parameters. These operators were fundamental for the development of numerical bootstrap methods and found applications, for instance, in the study of the three-dimensional Ising model, helping to find bounds on the physical parameters \cite{El-Showk:2012cjh}.

Subsequently, various techniques for computing blocks of operators with spin have been developed. Such operators played an important role, for example, in finding new results on the Regge limit in CFTs \cite{Li:2017lmh, Kulaxizi:2017ixa}, or universal numerical bounds on classes of CFTs \cite{Iliesiu:2015qra, Dymarsky:2017xzb, Dymarsky:2017yzx}. More recently it was also useful for the large-$N$ solution of the SYK model \cite{Polchinski:2016xgd, Murugan:2017eto}. One of the most promising techniques is the method introduced in \cite{Karateev:2017jgd}, based on \emph{weight-shifting} operators. These operators act to increase or decrease the parameters of an operator, for instance, they can shift the spin. Recently, these shift operators found applications in the computation of inflationary correlators \cite{Pimentel2019}. 
Moreover, in momentum-space CFT, a set of shift operators was also introduced to compute 3-point functions of tensorial operators \cite{Bzowski:2015yxv}. In a similar fashion to Feynman integrals, these operators allowed the computation of tensorial correlators requiring only the knowledge of one (master) scalar correlator. 

The search for new shift operators for momentum-space CFT correlators and Feynman integrals that is presented in this thesis was inspired by the works summarised above.
\section{Outline}
The outline of the thesis is as follows. The first part is devoted to illustrating some aspects of the state of the art of conformal field theory.
In Chapter \ref{cha2}, we present the essential features of conformal symmetry in position space and derive the consequent constraints on correlators up to $n$-points. In Chapter \ref{cha3}, we give a broader presentation of conformal symmetry in momentum space, where our research is focused. First, we derive conformal Ward identities in momentum space and construct their solutions up to $n$ points. We give a detailed overview of the properties of 3-point functions by discussing equivalent representations, singularities and shift operators. We then present the simplex representation for $n$-point functions and conclude with an illustration of some special 4-point solutions to the conformal Ward identities, namely the contact and exchange Witten diagrams. All the ingredients necessary to follow the second part of the thesis are then in place. The second part of the thesis is based on the research that appeared in \cite{Caloro:2022zuy} and \cite{Caloro}. Chapter \ref{cha4} focuses on the simplex integral. We derive parametric integral representations for the simplex integral. By using inverse Schwinger parameters, we find that all graph polynomials for this integral can be expressed in terms of the first and second minors of the Laplacian matrix for the simplex. Inspired by the analogy between the simplicial geometry and electrical circuits, we regard the  Schwinger  parameters  as  resistances  in  an  electrical  network and re-parametrise the simplex integral by computing the effective resistance between all vertices  of  the  simplex. This gives a representation in terms of the determinant and first minors of the Cayley-Menger matrix. These parametrisations have various advantages. The diagonal structure of the exponential factor in the integrand allows a Fourier-like correspondence. This reveals new families of shift operators, expressible as determinants, that connect solutions of the conformal Ward identities in spacetime dimension $d$ to new solutions in dimension $d+2$. They are the generalisation to $n$-point of the known 3-point shift operators. Moreover, these novel representations reduce the number of scalar integrals and allow us to verify that the conformal Ward identities are satisfied via direct computation. 
Different integral representations may give complementary perspectives on the same object. Motivated by the description of some 3- and 4-point conformal correlators in terms of hypergeometric functions, in Chapter \ref{cha5} we then move to analyse the representation of conformal, and other more general Feynman integrals through a class of multivariable hypergeometric functions proposed by Ge'lfand, Kapranov \& Zelevinsky known as GKZ functions. A Feynman integral in GKZ form is characterised by a unique denominator and a higher-dimensional space of variables (in the context of Feynman integrals, these are momenta and masses). The strength of this representation is that all the properties of the function can be encoded in a matrix. From this, we can derive a set of partial differential equations satisfied by the integral and the singularities in the parameters (for physical integrals, these are represented by the spacetime dimension and the generalised propagators). These \emph{spectral singularities} are given by an infinite number of hyperplanes parallel to the facets of the Newton polytope associated with the matrix. We discuss how the knowledge of these singularities is the starting point for a systematic construction of non-trivial shift operators known as ‘creation’ operators. For 3-point CFT correlators, these are indeed the inverse operators of the shift operators involved in the reduction scheme, acting on a 3-point function to lower $d$ by two. We derive these shift operators for various Feynman integrals and for special classes of 4-point (and $n$-point) conformal correlators, such as contact Witten diagrams, consisting of integral over multiple Bessel functions. Using this formalism, we also derive novel weight-shifting operators connecting contact and exchange Witten diagrams with different operator dimensions but within the same spacetime $d$. Remarkably, unlike all previous operators \cite{Pimentel2019, Bzowski:2022rlz}, these novel shift operators generate shifted exchange diagrams with purely non-derivative vertices and can be applied for any values of the parameters.
Finally, in Chapter \ref{cha7}, we conclude with a summary and open questions.

\end{chapter}
\clearpage{}
  \part{Conformal field theory}

  \clearpage{}\begin{chapter}{\label{cha2}Conformal field theory in position space}
\allowdisplaybreaks

In this chapter we briefly introduce the main features of conformal symmetry. We define conformal transformations (translations, rotations, dilatations and special conformal transformations) and describe the conformal group by finding the generators of these transformations and their commutation relations. We then analyse the consequences of conformal symmetry on correlation functions of scalar operators: symmetry imposes strong constraints on their form. This material is discussed in many books and reviews from where the content of this chapter is inspired \cite{Rychkov, DiFrancesco:1997nk, Gillioz:2022yze}.
\section{Conformal transformations}
In this section we present conformal transformations and explain their geometrical meaning.

Let us consider a spacetime with metric $g_{\mu\nu}$. A Weyl transformation is a spacetime-dependent rescaling of the metric sending the initial $g_{\mu\nu}$ to the rescaled $g'_{\mu\nu}$ \cite{Wald},
\[
g_{\mu\nu}~\rightarrow g'_{\mu\nu}=\mathrm{e}^{2\sigma(x)} g_{\mu\nu},
\]
where $\sigma(x)$ is a generic function of the coordinates defining the rescaling factor $\mathrm{e}^{2\sigma(x)}$.
The infinitesimal version of this transformation is
\[\label{dsigma}
g_{\mu\nu}~\rightarrow g'_{\mu\nu}=g_{\mu\nu}+\delta g_{\mu\nu},\quad\mathrm{with}\quad \delta g_{\mu\nu}=2\sigma(x)g_{\mu\nu}.
\] 
In general, the effect of a Weyl transformation is a change of the spacetime geometry. If we consider the initial spacetime to be flat, with metric $\eta_{\mu\nu}$, a general Weyl transformation sends this flat metric to $g'_{\mu\nu}$, a curved spacetime metric. 
We look for special Weyl transformations ($\sigma(x)$) that can be undone by a special diffeomorphism so that the metric remains flat. Let us consider the diffeomorphism
\[
x^\mu~\rightarrow ~ x^\mu-\xi^\mu,
\] 
then, the corresponding infinitesimal transformation of the metric reads
\[\label{dxi}
g_{\mu\nu}~\rightarrow g'_{\mu\nu}=g_{\mu\nu}+\delta_\xi g_{\mu\nu},\quad\mathrm{with}\quad \delta_\xi g_{\mu\nu}=2\nabla_{(\mu}\xi_{\nu)},
\] 
where $\nabla$ denotes the covariant derivative.
In order to leave the spacetime metric flat, we require the overall change in the metric -- due to the Weyl transformation and the diffeomorphism -- to vanish. In other words, we require the following condition to be satisfied:
\[
\delta g_{\mu\nu}=\delta_\sigma g_{\mu\nu}+\delta_\xi g_{\mu\nu}=0.
\]
Evaluating \eqref{dsigma} and \eqref{dxi} on a flat metric, this reads
\[\label{conf_kill_1}
2\p_{(\mu}\xi_{\nu)}=-2\sigma\eta_{\mu\nu},
\]
where we now wrote the standard partial derivative 
and $\eta_{\mu\nu}$ denotes the (flat) Euclidean metric $\eta_{\mu\nu}=\mathrm{diag}(1,1,..,1)$. Note that in this thesis we will work in Euclidean signature. By contracting equation \eqref{conf_kill_1} we find the relation between the function $\sigma(x)$ and the vector $\xi^\mu$ 
\[\label{sigma}
\sigma=-\frac{1}{d}\p_\mu\xi^\mu.
\]
Substituting \eqref{sigma} in \eqref{conf_kill_1}, we obtain 
\[\label{conf_kill_2}
\p_{(\mu}\xi_{\nu)}=\frac{1}{d}\p_\rho\xi^\rho\eta_{\mu\nu}.
\]
This is the \emph{conformal Killing equation} and defines
the condition that $\xi^\mu$ must satisfy to generate a diffeomorphism acting on the metric to undo the Weyl transformation.
In the following we show that when $d>2$ the conformal Killing equation \eqref{conf_kill_2} has a finite number of solutions, while when $d=2$ an infinite number of solutions exists. However, in this thesis we are interested in spacetime dimensions $d\geq 3$. To find the general solution of \eqref{conf_kill_2}, we act with the partial derivative $\p_\rho$ on \eqref{conf_kill_1}, giving a second-order differential equation. By taking a linear combination of this equation with Lorentz indexes cyclically permuted, we obtain
\[
\p_\mu\p_\nu\xi_\rho=-\eta_{\mu\rho}\p_\nu\sigma-\eta_{\nu\rho}\p_\mu\sigma+\eta_{\mu\nu}\p_\rho\sigma,
\]
then by contracting it
\[
\p^2\xi_\mu=(d-2)\p_\mu\sigma.
\]
We now act with $\p_\nu$ on this last equation and with $\p^2$ on \eqref{conf_kill_1}. Combining the resulting expressions we have
\[\label{conf_kill_3}
(d-2)\p_\mu\p_\nu\sigma=-(\p^2\sigma)\eta_{\mu\nu},
\]
whose contraction gives
\[
2(d-1)\p^2\sigma=0.
\]
Hence, $\p^2\sigma=0$ for $d>1$. Consequently, from \eqref{conf_kill_3} we deduce that
\[
(d-2)\p_\mu\p_\nu\sigma=0.
\]
Thus, for $d>2$ the following condition holds
\[
\p_\mu\p_\nu\sigma=0,
\]
which amounts to say that $\sigma$ is at most linear in the spacetime coordinates $x^\mu$ and, according to \eqref{sigma}, this means that the Killing vector $\xi^\mu$ is at most quadratic in $x^\mu$:
\[\label{xiansatz}
\xi_\mu=A_\mu+B_{\mu\nu}x^\nu+C_{\mu\nu\rho}x^\nu x^\rho,
\]
with $A_\mu$, $B_{\mu\nu}$ and $C_{\mu\nu\rho}$ some coefficients we are going to find. To this aim, we substitute \eqref{xiansatz} back into equation \eqref{conf_kill_1} and find that $A_\mu\equiv a_\mu$ is an arbitrary constant vector, while $B_{\mu\nu}$ is the sum of an antisymmetric term $\omega_{\mu\nu}$ and a symmetric term proportional to the metric
\[
B_{\mu\nu}=\omega_{\mu\nu}+\lambda\eta_{\mu\nu}.
\]
Finally, taking into account that $C_{\mu\nu\rho}$ is symmetric in the last two indexes, it must be of the form
\[
C_{\mu\nu\rho}=-b_\rho\eta_{\mu\nu}+b_\mu\eta_{\nu\rho}-b_\nu\eta_{\mu\rho},
\]
where $b_\nu$ is an arbitrary constant vector. Hence, the general solution of the conformal Killing equation for $d\geq 3$ is
\[\label{xiConf}
\xi^\mu=a^\mu+\omega^\mu{}_\nu x^\nu + \lambda x^\mu + b^\mu x^2 -2(b_\nu x^\nu)x^\mu.
\]
The infinitesimal change of coordinates generated by this conformal Killing vector $x_i^\mu$ defines four class of transformations:
\begin{enumerate}
\item translations: $\xi^\mu_T=a^\mu$,
\item rotations: $\xi^\mu_R=\omega^\mu{}_\nu x^\nu$,
\item scale transformations (dilatation): $\xi^\mu_D=\lambda x^\mu$,
\item special conformal transformations (SCT): $\xi^\mu_{\mathrm{SCT}}=b^\mu x^2 -2(b_\nu x^\nu)x^\mu$.
\end{enumerate}
This is what we anticipated at the beginning, \emph{i.e.}, that conformal transformations define a group larger than the Poincar\'e one, by including dilatations and special conformal transformations. Let us note that translations and rotations (or Lorentz transformations in the case we are considering the Minkowski metric, instead of the Euclidean one) are isometries, in fact $\sigma(x)=0$ both for $\xi^\mu_T$ and $\xi^\mu_R$. For scale transformations, the metric is rescaled by a constant $\sigma(x)=-\lambda$, independent of the spacetime coordinates. Finally, for SCT we find $\sigma(x)=2b\cdot x$, which corresponds to a spacetime-dependent rescaling. While the first three transformations are intuitive to visualise, the special conformal transformations defined by
\[\label{sctinf}
x^\mu~\rightarrow~x^\mu-\xi^\mu_{\mathrm{SCT}}=x^\mu-b^\mu x^2 +2(b_\nu x^\nu)x^\mu,
\]
are harder to visualise. However, we can see them as a combination of an inversion, a translation by $b^\mu$ and an inversion again:
\[
x^\mu~\rightarrow~\frac{x'^\mu}{x'^2}=\frac{x^\mu}{x^2}-b^\mu,
\]
leading to 
\[\label{sct}
x'^\mu=\frac{x^\mu-b^\mu x^2}{1-2\bs{b}\cdot \x+b^2 x^2}.
\]
This corresponds to the finite version of the special conformal transformations \eqref{sctinf} as one can see by expanding the last expression for an infinitesimal vector $b^\mu$, recovering $x'^\mu=x^\mu-\xi^\mu_{\mathrm{SCT}}$.

To conclude this section, we show that conformal transformations act locally as the composition of a rotation and a scale transformation. To see this, we compute the Jacobian associated to conformal transformations
\begin{align}
\left | \frac{\p x'^\mu}{\p x^\nu}\right |&=(1-\lambda+2b\cdot x)\delta^\mu{}_\nu-\omega^\mu{}_\nu+2(b_\nu x^\mu - b^\mu x_\nu)\notag\\
&\approx (1-\lambda+2b\cdot x)\left(\delta^\mu{}_\nu-\omega^\mu{}_\nu+2(b_\nu x^\mu - b^\mu x_\nu)\right)=e^\sigma R^\mu{}_\nu,
\end{align}
where $R^\mu{}_\nu=\delta^\mu{}_\nu-\omega^\mu{}_\nu+2(b_\nu x^\mu - b^\mu x_\nu)$ is an orthogonal matrix responsible for the rotation, while the factor $(1-\lambda+2b\cdot x)=e^\sigma$ is the local scale transformation. It is now patent why the name \emph{conformal}: conformal transformations preserve angles. 
\section{Conformal group}
Conformal transformations form a group, \emph{i.e.}, given the infinitesimal conformal transformation
\[
x^\mu~\rightarrow~x'^\mu=x^\mu-\xi^\mu_T-\xi^\mu_R-\xi^\mu_D-\xi^\mu_{\mathrm{SCT}},
\]
the identity and inverse elements exist plus the composition of two conformal transformations is still a conformal transformation. Moreover, it is a continuous group since it acts on the spacetime coordinates. This means that we can describe it through its generators and the commutation relations amongst them. 
We start by finding the generator of translations. 
Let us assume $f(x)$ to be an element of the conformal group, and act with an infinitesimal translation described by the parameter $a^\mu$
\[
f(x^\mu)~\rightarrow f(x'^\mu)=f(x^\mu-a^\mu)\approx(1-iP_\mu a^\mu)f(x^\mu),\qquad P_\mu=-i\p_\mu,
\]
we found the expected generator for translations $P_\mu$. The finite version of the transformation is then given by exponentiating the generator
\[
f(x)~\rightarrow~e^{ia_\mu P^\mu}f(x),
\]
therefore the generators are fundamental to define the group. Knowing the four infinitesimal transformations defining the conformal group, we can find the associated generators:
\begin{align}
&P_\mu=-i\p_\mu \quad &\mathrm{(translations)},\\
&M_{\mu\nu}=-i(x_\mu\p_\nu-x_\nu\p_\mu)\quad &\mathrm{(rotations/Lorentz ~transformations)},\\
&D=-ix^\mu\p_\mu \quad &\mathrm{(dilatations)},\\
&K_\mu=-i(x^2\p_\mu-2x_\mu x_\nu \p^\nu)\quad &\mathrm{(special~conformal ~transformations)}.
\end{align}
The algebra of the conformal group is then defined by the following commutation relations
\begin{align}
&[M_{\mu\nu},M_{\rho\sigma}]=-i(\eta_{\mu\rho}M_{\nu\sigma}-\eta_{\mu\sigma}M_{\nu\rho}-\eta_{\nu\rho}M_{\mu\sigma}+\eta_{\nu\sigma}M_{\mu\rho}),\notag\\
&[M_{\mu\nu},P_\rho]=-i(\eta_{\mu\rho}P_\nu-\eta_{\nu\rho}P_\mu),\notag\\
&[D,P_\mu]=iP_\mu,\notag\\
&[P_\mu,K_\nu]=2i(\eta_{\mu\nu}D-M_{\mu\nu}),\notag\\
&[D,K_\mu]=-iK_\mu,\notag\\
&[M_{\mu\nu},K_\rho]=-i(\eta_{\mu\rho}K_\nu-\eta_{\nu\rho}K_\mu),\notag\\
&[P_\mu,P_\nu]=[D,D]=[M_{\mu\nu},D]=[K_\mu,K_\nu]=0.
\end{align}
We can show that the Euclidean conformal algebra in $d$ dimensions is isomorphic to the algebra of the Lorentz group $SO(d+1,1)$. In fact, let us define the operators $J_{ab}=-J_{ba}$, for $a,b=-1,0,1,..,d$:
\begin{align}
&J_{\mu\nu}=M_{\mu\nu},\qquad & J_{-1,0}=D,\\
&J_{-1,\mu}=\frac{1}{2}(P_\mu-K_\mu),\qquad & J_{0,\mu}=\frac{1}{2}(P_\mu+K_\mu).
\end{align}
They satisfy the commutation relations of the $SO(d+1,1)$ algebra:
\[
[J_{ab},J_{cd}]=-i(\eta_{ac}M_{bd}-\eta_{ad}M_{bc}-\eta_{bc}M_{ad}+\eta_{bd}M_{ac}),
\]
where $\eta_{ab}=(-1,..,1,1,1)$. In the same way one can show that the Lorentzian conformal $d-$dimensional group is isomorphic to $SO(d,2)$. Finally, it is worth noticing that the commutation relations above show that while $M_{\mu\nu}$ and $P_\mu$ form a group, which is the Poincar\'e group, the generators $M_{\mu\nu}$, $P_\mu$ and $D$ also form a group. This implies that if we enhanced the Poincar\'e group only by introducing the scale transformation, we would not obtain the full conformal group. 
\section{Conformal transformations for operators}
In the first section we defined the action of conformal transformations on the spacetime coordinates and the metric. However, in a conformal field theory, the fields also transform. To find their transformation 
we combine, as before, a general Weyl transformation with the diffeomorphism found in section 1 and we find that, while we require the metric to stay flat, other fields do not transform trivially. According to Weyl transformations, if the metric transforms as $g_{\mu\nu}\rightarrow g'_{\mu\nu}=\mathrm{e}^{2\sigma(x)} g_{\mu\nu}$, then a Weyl transformation of a scalar field $\O$ is
\[
\O~\rightarrow~\O'=\mathrm{e}^{-\Delta\sigma}\O,
\]
where $\Delta$ is the Weyl weight. 
Note that for scalar fields the Weyl weight coincides with the scaling dimension of the operator. For instance in a Weyl-invariant free scalar field theory described by the action
\[
S=-\frac{1}{2}\int \mathrm{d}^d x\sqrt{-g}\left( g_{\mu\nu}\p^\mu\O\p^\nu\O+\zeta R \O^2\right),
\]
one finds that the Weyl weight is the same as the canonical scaling dimension $\Delta$, \emph{i.e.}, $\Delta=\frac{d}{2}-1$. Moreover, conformal invariance also requires $\zeta=(d-2)/(4(d-1))$. 
In the following discussion, however, we will not need any Lagrangian to study conformal field theory.\\To determine the infinitesimal conformal transformation of scalar fields we compose the infinitesimal Weyl transformation 
\[
\delta_\sigma\O=-\Delta \sigma \O
\]
with the transformation of the scalar field due to the diffeomorphism
\[
\delta_\xi\O=\xi^\mu \p_\mu \O,
\]
giving 
\[\label{infGeneral}
\delta\O=\delta_\sigma\O+\delta_\xi\O=\left[\xi^\mu\p_\mu+\frac{\Delta}{d}(\p_\mu\xi^\mu)\right]\O,
\]
where we used \eqref{sigma} to express the Weyl parameter $\sigma$ in terms of the vector $\xi^\mu$. To find how the field $\O$ transforms under purely dilatations or special conformal transformations, we consider $\xi^\mu=\xi^\mu_D=\lambda x^\mu$ and $\xi^\mu=\xi^\mu_{\mathrm{SCT}}=b^\mu x^2 -2(b_\nu x^\nu)x^\mu$ in the general relation \eqref{infGeneral}:
\begin{align}
&\delta_D\O(x)=\lambda (x^\mu\p_\mu+\Delta)\O(x),\label{infD}\\
&\delta_{\mathrm{SCT}}\O(x)=b_\nu\left[(x^2\eta^{\mu\nu}-2x^\mu x^\nu)\p_\mu-2\Delta x^\nu\right]\O(x).\label{infSCT}
\end{align}
The finite conformal transformation for scalar operators is
\[\label{CTscalar}
\O(x)~\rightarrow~\O'(x')=\left| \frac{\p x'}{\p x} \right|^{-\frac{\Delta}{d}}\O(x),
\]
which we can see as a rescaling of the field $\O$ by a power $-\Delta$ of the length rescaling factor $|\p x'/ \p x|^{1/d}=e^\sigma$. In fact $|\p x'/ \p x|$ is a local hypervolume in $d$ dimensions, consequently $|\p x'/ \p x|^{1/d}$ is a length rescale. This transformation rule defines a so-called \emph{primary} operator.

\section{Conformal Ward Identities}
In this section we analyse the consequences of conformal symmetry. We will show that conformal symmetry imposes constraints on the observables allowing a non-perturbative approach to CFT.

In field theory, invariance implies the following equivalence between correlators \cite{Peskin:1995ev}
\[\label{PSinv1}
\langle \mathcal{O}_1(\x_1)\mathcal{O}_2(\x_2)\cdots \mathcal{O}_n(\x_n)\rangle = \langle \mathcal{O}'_1(\x_1)\mathcal{O}'_1(\x_2)\cdots \mathcal{O'}_n(\x_n)\rangle,
\]  
where $\mathcal{O}_i(\x_i)$ is a scalar operator with scaling dimension $\Delta_i$ and $\mathcal{O'}_i(\x_i)$ is the transformed operator. One can show that the above equation holds by considering the path-integral formalism for correlators. Let $S_{\mathrm{CI}}$ be a conformally invariant action, then
\begin{align}
\langle \mathcal{O}_1(\x_1)\mathcal{O}_2(\x_2)\cdots \mathcal{O}_n(\x_n)\rangle &= \int \left[\mathcal{D}\mathcal{O}\right]\mathcal{O}_1(\x_1)\mathcal{O}_2(\x_2)\cdots \mathcal{O}_n(\x_n)\mathrm{e}^{-S_{\mathrm{CI}}[\mathcal{O}]}\nn \\
&=\int \left[\mathcal{D}\mathcal{O'}\right]\mathcal{O'}_1(\x_1)\mathcal{O'}_2(\x_2)\cdots \mathcal{O'}_n(\x_n)\mathrm{e}^{-S_{\mathrm{CI}}[\mathcal{O'}]}\nn \\
&=\int \left[\mathcal{D}\mathcal{O}\right]\mathcal{O'}_1(\x_1)\mathcal{O'}_2(\x_2)\cdots \mathcal{O'}_n(\x_n)\mathrm{e}^{-S_{\mathrm{CI}}[\mathcal{O}]}\nn \\
&=\langle \mathcal{O}'_1(\x_1)\mathcal{O}'_2(\x_2)\cdots \mathcal{O'}_n(\x_n)\rangle,
\end{align}
where in the second line we renamed $\mathcal{O}\rightarrow \mathcal{O}'$, while in the third line we considered conformal invariance both of the action and the functional measure $\left[\mathcal{D}\mathcal{O}\right]$. When the latter is not invariant, however, there will be an anomaly corresponding to the symmetry breaking and equation \eqref{PSinv1} will be modified with an additional term. This is related to renormalisation which we will discuss later in Chapter \ref{cha3}. In the following we assume the symmetry is not broken. At the infinitesimal level, equation \eqref{PSinv1} reads
\begin{align}\label{cwi0}
0&=\d\langle \mathcal{O}_1(\x_1)\mathcal{O}_2(\x_2)\cdots \mathcal{O}_n(\x_n)\rangle \nn \\
&=\sum_{i=1}^n\langle \mathcal{O}_1(\x_1)\mathcal{O}_2(\x_2)\cdots \d\mathcal{O}_i(\x_i)\cdots \mathcal{O}_n(\x_n)\rangle.
\end{align}
This equation amounts to a set of differential equations, known as conformal Ward identities (CWIs). To obtain their expressions, we express $\d\mathcal{O}_i(\x_i)=\O(\x_i')-\O(\x)$ in \eqref{cwi0} using  \eqref{infGeneral}. For translations, $\delta\O(\x)=a^\mu\p_\mu\O(\x)$ and the Ward identity reads
\[\label{translation}
\sum_{i=1}^n\frac{\p}{\p x_i^\mu}\langle \mathcal{O}_1(\x_1)\mathcal{O}_2(\x_2)\cdots \mathcal{O}_n(\x_n)\rangle =0,
\]
while rotation Ward identity is
\[\label{rotation}
\sum_{i=1}^n\left(x_i^\mu\p_i^\nu-x_i^\nu\p_i^\mu\right)\langle \mathcal{O}_1(\x_1)\mathcal{O}_2(\x_2)\cdots \mathcal{O}_n(\x_n)\rangle =0.
\]
These two transformations require the scalar $n$-point correlator to be a function of the coordinate separations 
\[\label{xij}
x_{ij}=|\bs{x}_i-\bs{x}_j|,\quad i,j=1,..,n.
\]
Less trivial are the constraints imposed by dilatation and special conformal Ward identities which read respectively 
\begin{align}
&\sum_{i=1}^n\left(x_i^\mu\frac{\p}{\p x_i^\mu}+\Delta_i\right)\langle \mathcal{O}_1(\x_1)\mathcal{O}_2(\x_2)\cdots \mathcal{O}_n(\x_n)\rangle =0,\label{DWIpos}\\
&\sum_{i=1}^n\left\{\left(x_i^2\eta^{\mu\nu}-2x_i^\mu x_i^\nu\right)\frac{\p}{\p x_i^\mu}-2\Delta_i x_i^\nu\right\}\langle\mathcal{O}_1(\x_1)\mathcal{O}_2(\x_2)\cdots \mathcal{O}_n(\x_n)\rangle =0,\label{SCWIpos}
\end{align}
where we used equations \eqref{infD} and \eqref{infSCT}. The dilatation Ward identity (DWI) implies that the correlator has to be a homogeneous function of the positions of degree $-\Delta_t$, with $\Delta_t=\sum_{i=1}^n\Delta_i$. As it is now evident, each Ward identity constrains the shape of correlators.
Finally, let us mention that conformal symmetry also implies the following equivalence between correlators
\[\label{cwi2}
\langle\mathcal{O}_1(\x'_1)\cdots \mathcal{O}_n(\x'_n)\rangle=\left |\frac{\p \x'_1}{\p \x_1}\right |^{-\Delta_1/d}\cdots\left |\frac{\p \x'_n}{\p \x_n}\right |^{-\Delta_n/d}\langle\mathcal{O}_1(\x_1)\cdots \mathcal{O}_n(\x_n)\rangle,
\]
which directly stems from equation \eqref{CTscalar}. This is how the correlator transforms under a finite conformal transformation.\\In the following we will list the solutions for 2-, 3-, 4- and $n-$point scalar correlators obtained by solving these constraints.
\section{Position-space conformal correlators}
In this section we present the solutions of position-space conformal Ward identities. We will show that up to 3-point functions, the solution is unique. 

First, let us note that the 1-point function vanishes
\[
\langle \mathcal{O}_1(\x_1)\rangle=0.
\]
In fact, translations and rotations require it to be a constant, $\langle \mathcal{O}_1(\x_1)\rangle=$const. A non-vanishing constant, however, would violate scaling invariance.
To find solutions for $n\geq 2$ we need instead the full set of CWIs as we show in the following sections.
\subsection{2-point function}
Translation and rotation symmetries imply that the 2-point correlator is a function of $x_{12}$, while dilatations require the correlator to be a homogeneous function in the positions with degree $-\Delta_t$. Hence the 2-point correlator must be of the form
\[
\langle\mathcal{O}_1(\x_1)\mathcal{O}_2(\x_2)\rangle \propto x_{12}^{-\Delta_1-\Delta_2}.
\]
Finally, the SCWI imposes a further constraint on the scaling dimension. By acting with the SCWI on the 2-point function above, one finds that a non-vanishing solution exists if and only if $\Delta_1=\Delta_2=\Delta$. Therefore the general 2-point scalar function is
\begin{align}\label{2ptPos}
\langle\mathcal{O}_1(\x_1)\mathcal{O}_2(\x_2)\rangle =\begin{cases} C_{12}x_{12}^{-2\Delta},\quad &\Delta_1=\Delta_2=\Delta \\
0 ,\quad &\Delta_1\neq\Delta_2
\end{cases},
\end{align}
where $C_{12}$ is a normalisation constant, it can be set to one by normalising the scalar operators.
\subsection{3-point function}
The solution of 3-point function CWIs is also unique. As above, translation and rotation invariance requires the 3-point scalar correlator to be a function of $x_{ij}$, with $i\neq j=1,2,3$
\[
\langle\mathcal{O}_1(\x_1)\mathcal{O}_2(\x_2)\mathcal{O}_3(\x_3)\rangle =f(x_{12},x_{13},x_{23}).
\]
The dilatation Ward identity specifies the function to be
\[
\langle\mathcal{O}_1(\x_1)\mathcal{O}_2(\x_2)\mathcal{O}_3(\x_3)\rangle\propto x_{12}^{2\a_{12}} x_{13}^{2\a_{13}} x_{23}^{2\a_{23}},
\] 
where $\a_{ij}$ are some constants that satisfy
\[
\sum_{1\leq i<j\leq 3}\a_{ij}=-\Delta_t.
\]
Finally, the SCWI fixes uniquely the values of the parameters $\a_{ij}$, since the terms of the sum in equation \eqref{SCWIpos} are three, as the number of the unknown parameters. Instead of substituting the ansatz in the SCWI, we use equation \eqref{cwi2} to find their values. Squaring equation \eqref{sct}, one can show that
\[\label{sctsq}
x'^2_{ij}=\frac{x^2_{ij}}{\g_i\g_j},
\] 
where we defined $\g_i=1-2\bs{b}\cdot \x_i+b^2 x_i^2$. And taking into account that for SCTs
\[
\left |\frac{\p \x'_i}{\p \x_i}\right |^{1/d}=\g_i^{-1},
\] 
using equation \eqref{cwi2} we find
\[
\Delta_i=-\sum_{j=1}^3 \a_{ij},\quad i=1,2,3,
\]
which fixes $\a_{ij}$ to 
\[\label{aij3pt}
2\a_{12}=2\Delta_3-\Delta_t,
\]
along with cyclic permutations. Hence the general solution of 3-point scalar CWIs is unique and it reads
\[\label{3pt-pos}
\langle\mathcal{O}_1(\x_1)\mathcal{O}_2(\x_2)\mathcal{O}_3(\x_3)\rangle=\frac{C_{123}}{x_{12}^{\Delta_1+\Delta_2-\Delta_3} x_{13}^{\Delta_1-\Delta_2+\Delta_3} x_{23}^{-\Delta_1+\Delta_2+\Delta_3}}.
\]
Note that while the constant $C_{12}$ could be set to one, this is not possible with the constant $C_{123}$. In fact, the latter is related to physical properties and is called ``OPE constant" or ``structure constant".
\subsection{4-point function} 
While symmetry fixes 2- and 3-point functions completely up to constants, 4-point and higher-point functions are not uniquely fixed. However, conformal symmetry imposes strong constraints on their form. Here we will analyse the solution of 4-point CWIs and in the next section we will generalise the result to $n$-point functions. 

Translation, rotation and dilatation invariance constrain the form of the solution to be
\[
\langle\mathcal{O}_1(\x_1)\mathcal{O}_2(\x_2)\mathcal{O}_3(\x_3)\mathcal{O}_4(\x_4)\rangle\propto\prod_{1\leq i<j\leq 4}x_{ij}^{2\a_{ij}},
\]
where 
\[
\sum_{1\leq i<j\leq 4}2\a_{ij}=-\Delta_t,
\]
and without loss of generality we assume $\a_{ij}=\a_{ji}$ and $\a_{ii}=0$.

Note that in this case the number of coordinate separations $x_{ij}$ is larger than the number $n$ of constraints following from the SCWI. To be more specific, there are $n(n-1)/2$ coordinate separations and $n$ constraints from the SCWI. This implies that there are $n(n-3)/2$ degree of freedom in the general solution. As showed earlier, under special conformal transformations, equation \eqref{sctsq} holds, therefore
\[\label{par}
\Delta_i=-\sum_{j=1}^4 \a_{ij},\quad i=1,..,4.
\] 
Note that, unlike for $n=3$, this condition does not fully fix the parameters $\a_{ij}$. Moreover, due to equation \eqref{sctsq}, 
4-point functions admit two simple conformal invariants, the so called \emph{conformal cross ratios}: 
\[\label{crosspos}
u=\frac{x_{12}^2x_{34}^2}{x_{13}^2x_{24}^2},\qquad v=\frac{x_{13}^2x_{24}^2}{x_{14}^2x_{23}^2}.
\]
Therefore, the general 4-point function also depends on an arbitrary function $f$ of cross ratios:
\[\label{4pt-pos}
\langle\mathcal{O}_1(\x_1)\mathcal{O}_2(\x_2)\mathcal{O}_3(\x_3)\mathcal{O}_4(\x_4)\rangle=f(u,v)\prod_{1\leq i<j\leq 4}x_{ij}^{2\a_{ij}},
\]
where the parameters $\a_{ij}$ are related to the scaling dimension as in equation \eqref{par}. In the following section we generalise this result to $n$-point and discuss the dependence of the number of independent cross ratios on the spacetime dimension $d$ and the number of points $n$.

\subsection[{\texorpdfstring{$n$-point function}{n-point function}}]{\texorpdfstring{\boldmath{$n$}-point function}{n-point function}}

The result given in the previous section can be generalised to $n$ points. The general solution is the following conformally invariant $n$-point function of scalar operators $\mathcal{O}_1,...,\mathcal{O}_n$ with scaling dimensions $\Delta_1,...,\Delta_n$:
\[\label{n-pt}
\langle\mathcal{O}_1(\x_1)\cdots \mathcal{O}_n(\x_n)\rangle=\prod_{1\leq i<j\leq n}x_{ij}^{2\a_{ij}}f(\bs{u}),
\]
where the parameters $\a_{ij}$ are related to the scaling dimensions by the formula implied by special conformal invariance
\[
\Delta_i=-\sum_{j=1}^n\a_{ij},\quad i=1,2,..,n,
\]
and $f$ is an arbitrary function of $N_{d,n}$ independent cross ratios
\[
u_{pqrs}=\frac{x_{pr}^2x_{qs}^2}{x_{pq}^2x_{rs}^2},
\]
where $p,q,r,s=1,2,..,n$. We denote the set of all independent cross ratios with the symbol $\bs{u}$.
As discussed in \cite{Kravchuk:2016qvl, Simmons-Duffin:2016gjk, Hogervorst:2013sma}, the number of independent cross ratios $N_{d,n}$ depends on the number of points $n$ and the spacetime dimension $d$ 
\begin{align}
&N_{d,n}=n(n-3)/2,\qquad & n\leq d+2,\nn\\
&N_{d,n}=nd-(d+2)(d+1)/2 ,\qquad &  n> d+2.
\end{align}
To understand this counting, let us consider $n$ points $\x_1,...,\x_n$ in a $d-$dimensional spacetime. Using conformal transformations some of these $n$ points can be fixed in the spacetime. For example, $\x_1$ can be sent to infinity by using a special conformal transformation, while using translations $\x_2$ can be fixed at the origin. Then $\x_3$ can be set at $(1,0,...,0)$ by performing a rotation together with a dilatation (which fixes the non-zero coordinate to be equal to one):
\[
\x_3=(1,0,...,0), \]
where the vector contains $d-1$ zero components. For the other points, we can use the remaining rotations if available, depending on the spacetime dimension $d$. By doing a rotation in a $(d-1)-$dimensional spacetime (so that $\x_3$ remains fixed) we can move $\x_4$ to lie in the plane spanned by the first two axes, \emph{i.e.},
\[
\x_4=(X_1^{(4)},X_2^{(4)},0,...,0), \]
which has $m=2$ degrees of freedom. We iterate this procedure by performing rotations in successively lower-dimensional spaces, giving for instance
\[
\x_5=(X_1^{(5)},X_2^{(5)},X_3^{(5)},0,...,0), \]
that has $m=3$ degrees of freedom. The $n$th point will be 
\[
\x_n=(X_1^{(n)},X_2^{(n)},...,X_{n-2}^{(n)},0,...,0). \]
Summing all the degrees of freedom $m$ for each point, we find
\[\label{Nd1}
N_{d,n}=\sum_{m=2}^{n-2}m=\frac{1}{2}n(n-3).
\]
Note that we implicitly assumed $d\geq n-2$ so far. When $n=4$, for $d\geq 2$ (and in this thesis we are considering $d\geq 3$), this assumption holds. Hence this counting of cross ratios holds and gives, as expected, $N_{d,n}=2$.
On the other hand, if $d<n-2$ (or equivalently $n>d+2$), then all $\x_k$, with $k\geq d+2$, will have $d$ free parameters since there are no rotational degrees of freedom left to fix them. So we have
\[
\x_k=(X_1,X_2,...,X_{d}),\quad k= d+3,..,n. 
\]
Hence, the counting of the degrees of freedom becomes
\[\label{Nd2}
N_{d,n}=\Big(\sum_{m=2}^{d}m\Big)+d\left(n-(d+2)\right)=nd-\frac{1}{2}(d+2)(d+1).
\]
Note that the two values for $N_{d,n}$ in equations \eqref{Nd1} and \eqref{Nd2} coincide when $n=d+1$ or $n=d+2$.

\end{chapter}
\clearpage{} 
  \clearpage{}\begin{chapter}{\label{cha3}Conformal field theory in momentum space}
\allowdisplaybreaks
\section{Introduction}

In this chapter we give an overview of the main results in momentum-space CFT. As a counterpart to the previous chapter, we first derive CWIs in momentum space. Then, we solve them from 2- to general $n$-point. We show that up to three points the symmetry fixes uniquely the form of correlators and discuss their singularities and renormalisation. In particular, we derive the 3-point function in the form of the \emph{triple-$K$ integral}, an integral of three Bessel functions of the second kind and find equivalent representations. Then, we introduce shift operators acting on the 3-point function to shift the parameters. This helps the evaluation of special 3-point functions which would be otherwise difficult. While 3-point functions are well understood, the knowledge of higher-point functions is less complete. We then present the general $n$-point function recently found as a Feynman integral over a $(n-1)$-simplex, featuring an arbitrary function of momentum-space cross ratios and conclude with a summary and open questions which will be addressed in the second part of the thesis.
\section{Conformal Ward identities}
In Chapter \ref{cha2} we presented the scalar conformal Ward identities in position space. To explore the implications of conformal symmetry in momentum space we start with deriving the
corresponding momentum-space conformal Ward identities. We obtain a set of differential equations that must be satisfied by conformal correlators. This means that, as in position space, CWIs constrain the form of conformal correlators. Moreover, the theory of differential equations and multivariable hypergeometric functions reveals a description of 3-point functions and certain special 4-point functions \cite{Coriano:2020ees, Coriano:2019sth}
in terms of known hypergeometric functions such as Appell $F_4$ or Lauricella $F_C$. In Chapter \ref{cha5} we will also describe some of these solutions as generalised hypergeometric functions introduced by Gel’fand, Kapranov, and Zelevinsky (GKZ systems).

The momentum-space dilatation and special conformal Ward identities read
\begin{align}
&0=D\lla\O(\bs{p}_1)...\O(\bs{p}_n)\rra,\label{DWI0}\\
&0=\mathcal{K}^\mu\lla\O(\bs{p}_1)...\O(\bs{p}_n)\rra,\label{scWI}
\end{align}
where double brackets denote the \emph{reduced} correlators related to the standard correlator by pulling out the delta function of momentum-conservation:
\[
\langle\O_1(\bs{p}_1)\cdots\O_n(\bs{p}_n)\rangle=(2\pi)^d\delta(\bs{p}_1+\dots+\bs{p}_n)\lla\O_1(\bs{p}_1)\cdots\O_n(\bs{p}_n)\rra,
\]
while $D$ and $\mathcal{K}^\mu$ respectively are the dilatation and the special conformal operators:
\begin{align}\label{CWIop}
&D=-(n-1)d+\Delta_t-\sum_{j=1}^{n-1}p_j^\mu\frac{\p}{\p p_j^\mu},\nn\\
&\mathcal{K}^\mu=\sum_{j=1}^{n-1}\mathcal{K}_j^\mu,
\end{align}
with
\[\label{Kjred}
\mathcal{K}_j^\mu=2(\Delta_j-d)\frac{\p}{\p p_{j\mu}}-2p_j^\nu\frac{\p}{\p p_j^\nu}\frac{\p}{\p p_{j\mu}}+p_j^\mu\frac{\p}{\p p_j^\nu}\frac{\p}{\p p_{j\nu}},
\]
and $\Delta_t=\sum_{j=1}^n \Delta_j$. To derive the momentum-space CWIs above, we consider the inverse Fourier transform of position-space correlators
\[
\langle \O_1(\bs{x}_1)\cdots\O_n(\bs{x}_n)\rangle= \left(\prod_{j=1}^n \int \frac{\dd^d\bs{p}_j}{(2\pi)^d}\mathrm{e}^{i\bs{p}_j\cdot \x_j}\right)\langle\O_1(\bs{p}_1)\cdots\O_n(\bs{p}_n)\rangle.
\]
Translational invariance corresponds to pulling out a momentum-conserving delta function. To see this, we take the inverse Fourier transform of 
\[
\langle \O_1(\bs{x}_1)\cdots\O_n(\bs{x}_n)\rangle=\langle \O_1(\bs{x}_1-\bs{x}_n)\cdots\O_n(\bs{0})\rangle,
\]
leading to
\[\label{invFou}
\langle \O_1(\bs{x}_1)\cdots\O_n(\bs{x}_n)\rangle = \left(\prod_{j=1}^{n-1} \int \frac{\dd^d\bs{p}_j}{(2\pi)^d}\mathrm{e}^{i\bs{p}_j\cdot \x_{jn}}\right)\lla\O_1(\bs{p}_1)\cdots\O_n(\bs{p}_n)\rra,
\]
with $\x_{jn}=\x_j-\x_n$.
We act with the position-space dilatation and special conformal operators, \eqref{DWIpos} and \eqref{SCWIpos}, on the right-hand side of \eqref{invFou}. This amounts to considering the action only on the exponential factor $\exp\left(\sum_{j=1}^{n-1}\bs{p}_j\cdot \x_{jn}\right)$. For this purpose, we re-write the position-space CWIs by eliminating the derivative with respect to $x_n^\mu$ via the translational Ward identity \eqref{translation},
\[
\frac{\p}{\p x_n^\mu}~\rar~-\sum_{j=1}^{n-1}\frac{\p}{\p x_j^\mu}.
\]
Thus, the rotation and dilatation Ward identities in position space read
\begin{align}
&0=\sum_{j=1}^{n-1}\left(x_{jn}^\mu\frac{\p}{\p_j^\nu}-x_{jn}^\nu\frac{\p}{\p_j^\mu}\right)\langle \mathcal{O}_1(\x_1)\mathcal{O}_2(\x_2)\cdots \mathcal{O}_n(\x_n)\rangle ,\label{rot2}\\
&0=\left(\Delta_t+\sum_{j=1}^{n-1}x_{jn}^\mu\frac{\p}{\p x_{j}^\mu}\right)\langle \mathcal{O}_1(\x_1)\mathcal{O}_2(\x_2)\cdots \mathcal{O}_n(\x_n)\rangle \label{dil2}.
\end{align}
Rearranging the SCWI we have
\begin{align}\label{scwi2}
0=&\sum_{j=1}^{n-1}\left\{\left(x_{jn}^2\eta^{\mu\nu}-2x_{jn}^\mu x_{jn}^\nu\right)\frac{\p}{\p x_j^\mu}-2\Delta_j x_{jn}^\nu\right\}\langle\mathcal{O}_1(\x_1)\mathcal{O}_2(\x_2)\cdots \mathcal{O}_n(\x_n)\rangle \nn\\
&+2x_n^\nu\sum_{j=1}^{n-1}\left(x_{jn}^\mu\frac{\p}{\p_j^\nu}-x_{jn}^\nu\frac{\p}{\p_j^\mu}\right)\langle \mathcal{O}_1(\x_1)\mathcal{O}_2(\x_2)\cdots \mathcal{O}_n(\x_n)\rangle\nn\\
&-2x_n^\mu\left(\Delta_t+\sum_{j=1}^{n-1}x_{jn}^\mu\frac{\p}{\p x_{j}^\mu}\right)\langle \mathcal{O}_1(\x_1)\mathcal{O}_2(\x_2)\cdots \mathcal{O}_n(\x_n)\rangle,
\end{align}
where the last two lines vanish upon \eqref{rot2} and \eqref{dil2} respectively. After having expressed the position-space CWIs only in terms of $\bs{x}_{jn}$ and $\p/\p x_j^\mu$, with $j=1,...,n-1$, the standard Fourier correspondence holds
\[
x_{jn}^\mu\rar -i\frac{\p}{\p p_j^\mu},\qquad \frac{\p}{\p x_j^\mu}\rar i p_j^\mu, \qquad j=1,..,n-1.
\]
To find the dilatation and special conformal Ward identities in momentum space, we then act with the operators in \eqref{dil2} and \eqref{scwi2} on the right-hand side of \eqref{invFou} and integrate by parts with respect to the momenta. This leads to the CWIs \eqref{DWI0} and \eqref{scWI}.

In the following section we will solve 2- and 3-point CWIs directly and find a representation for the general $n$-point function that solves the CWIs. 
Before moving to the next sections, let us note that we can obtain a set of scalar SCWIs. In fact, in momentum space we have the advantage of decomposing the operator $\mathcal{K}^\mu$ into a basis of $n-1$ independent vectors $p_j^\mu$, with $j=1,..,n-1$:
\[\label{dec}
\mathcal{K}^\mu=p_1^\mu\mathcal{K}_1+\dots+p_{n-1}^\mu\mathcal{K}_{n-1}.
\]
Hence, the SCWI \eqref{scWI} is equivalent to $n-1$ scalar equations
\[\label{SCWIscal}
\mathcal{K}_j\lla\O(\bs{p}_1)...\O(\bs{p}_n)\rra=0,\qquad j=1,...,n-1.
\]
\section{2-point function}\label{cha3sec2pt}
Conformal invariance implies that the 1-point function vanishes, hence the first non trivial correlator is the 2-point function. In this section we solve momentum-space CWIs at two points. The complexity of the set of differential equations to be solved and their solutions increases with the number of points $n$. 

Translational invariance corresponds to momentum conservation, hence the 2-point correlator depends only on $\bs{p}\equiv\bs{p}_1=-\bs{p}_2$. Therefore the expansion in \eqref{dec} contains only one term. Moreover, rotational invariance implies that the scalar correlator only depends on scalar quantities, which in this case is the magnitude $p$ of the momentum $\bs{p}$. Using the chain rule
\[
\frac{\dd}{\dd p_\mu}=\frac{p^\mu}{p}\frac{\dd}{\dd p},
\]
from \eqref{DWI0} and \eqref{SCWIscal}, we obtain the 2-point DWI and SCWI 
\begin{align}
&0=\left(d-\Delta_1-\Delta_2+p\frac{\dd}{\dd p}\right)\lla\O(\bs{p})\O(-\bs{p})\rra,\label{2ptDWI}\\
&0=\left(\frac{\dd^2}{\dd p^2}+\frac{d-2\Delta_1+1}{p}\frac{\dd}{\dd p}\right)\lla\O(\bs{p})\O(-\bs{p})\rra.\label{2ptSCWI}
\end{align}
We start with solving the SCWI. 
The general solution of \eqref{2ptSCWI} reads
\[
\lla\O(\bs{p})\O(-\bs{p})\rra=c_0p^{2\Delta_1-d}+c_1,
\]
where $c_0$ and $c_1$ are integration constants. We then plug this expression into the dilatation Ward identity \eqref{2ptDWI}
and find
\[
\Delta_1=\Delta_2\equiv\Delta,\quad c_1=0,
\]
as expected from the 2-point position-space solution. So the general 2-point conformal correlator is
\[
\lla\O(\bs{p})\O(-\bs{p})\rra=c_0p^{2\Delta-d}.
\]

We could have found this solution by Fourier transforming the known position-space solution. Setting $C_{12}=1$ in \eqref{2ptPos}, we find the correspondent solution in momentum space
\[\label{2ptFT}
\lla\O(\bs{p})\O(-\bs{p})\rra=\int \frac{\dd ^d\x}{(2\pi)^d}\mathrm{e}^{-i\bs{p}\cdot \x} x^{-2\Delta}=\frac{2^{d-2\Delta}\pi^{d/2}\Gamma\left(\frac{d}{2}-\Delta\right)}{\Gamma(\Delta)}p^{2\Delta-d},
\]
where we wrote $x^{-2\Delta}$ using the Schwinger parametrisation \cite{Weinzierl:2022eaz}
\[\label{schwinger}
\frac{1}{A^\nu}=\frac{1}{\Gamma(\nu)}\int_0^\infty \dd \lambda \lambda^{\nu-1}\mathrm{e}^{-\lambda A},
\]
and performed the resulting Gaussian integral.
\section{3-point function}\label{cha3sec3pt}
In this section we present the 3-point function. First we solve the 3-point conformal Ward identities by separation of variables, giving a scalar representation of the solution. Then, we discuss the uniqueness of this solution by looking at its asymptotic behaviour and the singularities arising from collinear configurations of the momenta. This leads to a unique 3-point function known as the \emph{triple-$K$ integral}. We then present equivalent representations in terms of the generalised hypergeometric function Appell $F_4$ and of a 1-loop triangle Feynman diagram.

Let us first write the 3-point CWIs in terms of scalar invariants. Poincar\'e invariance implies that 3-point correlators depend on the scalars formed with $\bs{p}_1$ and $\bs{p}_2$: we choose the momenta magnitudes $p_j$, with $j=1,2,3$. Let us start with the dilatation Ward identity. Taking into account that $\bs{p}_3=-\bs{p}_1-\bs{p}_2$, we find the chain rules
\begin{align}\label{chain3}
&\frac{\p}{\p p_1^\mu}=\frac{p_1^\mu}{p_1}\frac{\p}{\p p_1}+\frac{p_1^\mu+p_2^\mu}{p_3}\frac{\p}{\p p_3},\nn\\
&\frac{\p}{\p p_2^\mu}=\frac{p_2^\mu}{p_2}\frac{\p}{\p p_2}+\frac{p_1^\mu+p_2^\mu}{p_3}\frac{\p}{\p p_3}.
\end{align}
Hence, we can write the dilatation Ward identity in terms of scalar variables:
\[\label{3ptDWIsc}
0=D\lla\O_1(\bs{p}_1)\O_2\bs{p}_2\O_3\bs{p}_3)\rra=\left(2d-\Delta_t+\sum_{j=1}^3p_j\frac{\p}{\p p_j}\right)\lla\O_1(\bs{p}_1)\O_2\bs{p}_2\O_3\bs{p}_3)\rra.
\] 
As noted earlier, this equation tells us that its solution is a homogeneous function of degree $\Delta_t-2d$. This means that we can write the solution in the following way
\[\label{3pt_ansatz1}
\lla\O_1(\bs{p}_1)\O_2\bs{p}_2\O_3\bs{p}_3)\rra=p_3^{\Delta_t-2d}F\left(\frac{p_1}{p_3},\frac{p_2}{p_3}\right),
\]
where here $F$ is a general function. To determine its explicit expression we need the special conformal Ward identity
\[\label{3ptSCWIvec}
\mathcal{K}^\mu\lla\O_1(\bs{p}_1)\O_2\bs{p}_2\O_3\bs{p}_3)\rra=0,
\]
where 
\[
\mathcal{K}^\mu=p_1^\mu\mathcal{K}_1+p_2^\mu\mathcal{K}_2.
\]
Therefore the SCWI \eqref{3ptSCWIvec} is equivalent to the system formed by the following two scalar equations
\[\label{3ptSCWIsc}
\mathcal{K}_j\lla\O_1(\bs{p}_1)\O_2(\bs{p}_2)\O_3(\bs{p}_3)\rra=0,\quad j=1,2.
\]
Using the chain rule \eqref{chain3} we find the operators $\mathcal{K}_j$. The explicit expression of $\mathcal{K}_1$ is
\begin{align}
\mathcal{K}_1=&\frac{\p^2}{\p p_1^2}+\frac{\p^2}{\p p_3^2}+\frac{2p_1}{p_3}\frac{\p^2}{\p p_1 \p p_3}+\frac{2p_2}{p_3}\frac{\p^2}{\p p_2 \p p_3}-\frac{2\Delta_1-d-1}{p_1}\frac{\p}{\p p_1}\nn\\&-\frac{2\Delta_1+2\Delta_2-3d-1}{p_3}\frac{\p}{\p p_3},
\end{align}
and $\mathcal{K}_2$ can be obtained by permuting $1\leftrightarrow 2$ indices. Since these operators contain mixed derivatives, we eliminate the mixed terms by invoking the DWI \eqref{3ptDWIsc}. We define
\[
\K_{13}=\mathcal{K}_1-\frac{2}{p_3}\frac{\p}{\p p_3}D,\qquad \K_{23}=\mathcal{K}_2-\frac{2}{p_3}\frac{\p}{\p p_3}D,
\]
hence the set of equations \eqref{3ptDWIsc} amounts to the following simpler version
\[\label{3ptSCWIsep}
\mathcal{\K}_{j3}\lla\O_1(\bs{p}_1)\O_2(\bs{p}_2)\O_3(\bs{p}_3)\rra=0,\quad j=1,2,
\]
where
\begin{align}\label{defK}
&\K_{ij}=\K_i-\K_j,\nn\\
&\K_i\equiv \frac{\p^2}{\p p_i^2}-\frac{2\Delta_i-d-1}{p_i}\frac{\p}{\p_i},\qquad i\neq j=1,2,3.
\end{align}
Note that the operator $\K_i$ appeared already in the 2-point special conformal Ward identity in equation \eqref{2ptSCWI}.

In the following we find the general solution of the CWIs that -- to summarise -- we wrote as the following set of equations
\begin{align}\label{cwisum}
&D\lla\O_1(\bs{p}_1)\O_2(\bs{p}_2)\O_3(\bs{p}_3)\rra=0,\nn\\
&\K_{13}\lla\O_1(\bs{p}_1)\O_2(\bs{p}_2)\O_3(\bs{p}_3)\rra=0,\nn\\
&\K_{23}\lla\O_1(\bs{p}_1)\O_2(\bs{p}_2)\O_3(\bs{p}_3)\rra=0.
\end{align}
Before showing how to solve these, we give the general solution known as the \emph{triple-K integral}:
\[\label{tripleK}
\lla\O_1(\bs{p}_1)\O_2(\bs{p}_2)\O_3(\bs{p}_3)\rra=c_{123}\int_0^{\infty} \dd x \: x^\a\prod_{j=1}^3p_j^{\b_j}K_{\b_j}(p_jx),
\]
where $c_{123}$ is an integration constant and $K_{\b_j}$ is the modified Bessel function of the second kind, also known as Bessel-K function \cite{abramowitz1972handbook}. The parameters $\a$ and $\b_j$ are related to the physical spacetime dimension and the scaling dimensions as follows:
\[\label{bjdef}
\a=\frac{d}{2}-1,\qquad \qquad \b_j=\Delta_j-\frac{d}{2},\qquad j=1,2,3.
\]
In the following we will denote the triple-$K$ integral \eqref{tripleK} with $I_{\a\{\b_1\b_2\b_3\}}$.

Let us present the derivation of the general solution. We denote the general solution as $f$. Taking into account the definitions in \eqref{defK}, we can write the last two equations of \eqref{cwisum} as
\[
\K_1f=\K_2f=\K_3f.
\]
Therefore we use separation of variables
\[
\lla\O_1(\bs{p}_1)\O_2(\bs{p}_2)\O_3(\bs{p}_3)\rra=f(p_1)f(p_2)f(p_3),
\]
leading to
\[
\frac{\K_1f_1}{f_1}=\frac{\K_2f_2}{f_2}=\frac{\K_3f_3}{f_3}=x^2,
\]
where $x^2$ is a constant. Then, we have to solve the equation
\[\label{Keq}
\K_if_i=\left(\frac{\p^2}{\p p_i}+\frac{1-2\b_i}{p_i}\frac{\p}{\p p_i}\right)f_i=x^2f_i,
\]
which is equivalent to the modified Bessel's equation,
\[
\left[p_i^2\p_i^2+p_i\p_i-(p_i^2x^2+\b^2)\right]u_i=0,
\]
with $f_i=p_i^{\b_i}u_i$ and $\p_i=\p/\p p_i$. Hence, the general solution of \eqref{Keq} is a linear combination of Bessel-K and Bessel-I functions:
\[\label{sol_mod_bes}
f_i(p_i)=p_i^{\b_i}\left[a_K K_{\b_i}(p_i x)+a_I I_{\b_i}(p_i x)\right],
\]
and $f=\prod_i f_i(p_i)$ solves the SCWIs. By linearity, indeed, any integral $\int \dd x\rho(x) f$ over a spectral function $\rho(x)$ solves the CWIs. The dilatation Ward identity fixes the form of this function. Noting that
\[
\sum_{i=1}^3p_i\p_i f(p_1x,p_2x,p_3x)=x\p_x f(p_1x,p_2x,p_3x),
\]
we have
\[
\int_0^\infty \dd x \rho(x) x\p_x f(p_1x,p_2x,p_3x)=(\b_t-\a-1)\int_0^\infty\dd x \rho(x)f(p_1x,p_2x,p_3x),
\]
where we defined $\b_t=\b_1+\b_2+\b_3$. Integrating by parts we find $\rho(x)=x^{\a-\b_t}$. Therefore we conclude that if $f(p_1,p_2,p_3)$ is the solution of the SCWIs, then 
\[\label{3pt_roughsol}
I=\int_0^\infty \dd x x^{\a-\b_t} f(p_1x,p_2x,p_3x)
\]
solves the DWI. The solution we found is not exactly the triple-K integral in \eqref{tripleK}, since the function $f$ involves also Bessel-I functions. A discussion based on physical properties of the solution restricts $I$ to the final form of a triple-$K$ integral, as we explain in the following section.
\subsection{Collinear singularities}
For \eqref{3pt_roughsol} to converge, at least one of the $f_i(p_i)$ needs to be a Bessel-K function. Let us assume $f_3\sim K_{\b_3}$, then the conformal Ward identities admit four independent solutions that schematically we can refer to as $IIK,~KIK,~IKK,~KKK$. In this section we show that only one of them is physically acceptable, leading to the triple-K integral \eqref{tripleK}.
This conclusion agrees with the result in position-space, where the 3-point function is unique. 

To see this, consider the behaviour of modified Bessel functions when their argument is large, 
\[\label{BesselLarge}
I_\nu(x)=\frac{1}{\sqrt{2\pi}}\frac{\mathrm{e}^x}{\sqrt{x}}+\dots,\qquad K_\nu(x)=\sqrt{\frac{\pi}{2}}\frac{\mathrm{e}^{-x}}{\sqrt{x}}+\dots,\qquad x\rightarrow \infty.
\]
Since the Bessel-I function diverges for large $x$, as we anticipated, the solution must have at least one Bessel-K function in the integrand.
Now, let us consider the case where the integrand is of the form $IIK$. According to the asymptotic behaviours \eqref{BesselLarge}, in order for the integral to converge, the following condition must hold
\[
p_3\geq p_1+p_2.
\]
This violates the triangle inequality ($p_3\leq p_1+p_2$), therefore we exclude solutions of the form $IIK$. If we consider the integrand with only one Bessel-I ($IKK$), we find that this is also forbidden since it is singular for collinear momentum configurations. In fact, at large $x$ the integral would approximately be
\[
I\sim \int \dd x x^{\a-\frac{3}{2}}\mathrm{e}^{(p_1-p_2-p_3)x}.
\]
While the triangle inequality is respected, if the momenta are collinear, \ie ~$p_1=p_2+p_3$, the integral diverges when $\a\geq 1/2$, hence $d\geq 3$. Therefore we find that only a unique 3-point function exists, given by the triple-$K$ integral
\[
I_{\a\{\b_1\b_2\b_3\}}=c_{123}\int_0^{\infty} \dd x x^\a\prod_{j=1}^3p_j^{\b_j}K_{\b_j}(p_jx).
\]
In the following, we present some explicit examples of triple-$K$ integrals, for special values of the parameters.
\subsection{Examples}
In this section we consider two examples. When the $\b_i$ are half-integer, the integral is given by elementary functions, while for integer $\b_i$, the triple-$K$ can be expressed in terms of the Bloch-Wigner function.
\begin{enumerate}
\item $d=4,~\Delta_1=\Delta_2=\Delta_3=5/2$.\\In this case all $\b_i$ are half integers:
\[
\a=\frac{d}{2}-1=1,\qquad \b_i=\Delta_i-\frac{d}{2}=\frac{1}{2},
\]
and the 3-point correlator reads 
\begin{equation}
\langle\!\langle O_1(\bs{p}_1)O_2(\bs{p}_2)O_3(\bs{p}_3)\rangle\!\rangle=c(p_1p_2p_3)^{\frac{1}{2}}\int_0^\infty \dd x \: x K_{\frac{1}{2}}(p_1x)K_{\frac{1}{2}}(p_2x)K_{\frac{1}{2}}(p_3x).
\end{equation}
The Bessel-K function in the integrand is 
\begin{equation}
K_{\frac{1}{2}}(x)=\sqrt{\frac{\pi}{2}}\frac{\mathrm{e}^{-x}}{x^{1/2}},
\end{equation}
and the integral is convergent giving
\begin{equation}
\langle\!\langle O_1(\bs{p}_1)O_2(\bs{p}_2)O_3(\bs{p}_3)\rangle\!\rangle=\frac{c\pi^2}{2^{\frac{3}{2}}}\frac{1}{\sqrt{p_1+p_2+p_3}}.
\end{equation}
\item $d=4,~\Delta_1=\Delta_2=\Delta_3=2$.\\In this case all $\b_i$ are integers
\[
\a=\frac{d}{2}-1=1,\qquad \b_i=\Delta_i-\frac{d}{2}=0,
\]
and we denote the 3-point function as 
\begin{equation}
\langle\!\langle O_1(\bs{p}_1)O_2(\bs{p}_2)O_3(\bs{p}_3)\rangle\!\rangle=cI_{1\{000\}},
\end{equation}
where the integral $I_{1\{000\}}$ is a known integral in literature \cite{Davydychev1992,tHooft1978} and is expressed in terms of the dilogarithm function Li$_2$ \cite{Zagier}. We show the explicit computation of $I_{1\{000\}}$ in appendix \ref{I1000comp} and quote here the final result:
\[\label{I1000z}
I_{1\{000\}}=\frac{1}{2\sqrt{-J^2}}\left[\mathrm{Li}_2(\bar{z})-\mathrm{Li}_2(z)-\frac{1}{2}\ln(z\bar{z})\ln\left(\frac{1-z}{1-\bar{z}}\right)\right],
\]
where the $z$-variables are related to the momenta magnitudes by
\[\label{z_var_def}
z\bar{z}=\frac{p_1^2}{p_3^2},\qquad(1-z)(1-\bar{z})=\frac{p_2^2}{p_3^2},
\]
or equivalently
\[
z=\frac{1}{2p_3^2}\big(p_1^2 - p_2^2 + p_3^2 +\sqrt{-J^2}\big), \qquad \bar{z}=\frac{1}{2p_3^2}\big(p_1^2 - p_2^2 + p_3^2 -\sqrt{-J^2}\big),
\]
where we defined
\begin{align}
J^2&=(p_1+p_2-p_3)(p_1-p_2+p_3)(-p_1+p_2+p_3)(p_1+p_2+p_3)\nn\\
&=-(z-\bar{z})^2p_3^4.
\end{align}
Note that $J^2$ is related to the Gram determinant of $\bs{p}_1$ and $\bs{p}_2$:
\[
J^2=4G(\bs{p}_1,\bs{p}_2),
\]
and hence $\sqrt{J^2}$ is proportional to the area of the triangle $A_{\text{T}}(p_1,p_2,p_3)$ whose sides are $p_1,p_2,p_3$.
The solution $I_{1\{000\}}$ is related to the Bloch-Wigner function $D(z)$ in the following way
\[
I_{1\{000\}}=\frac{D(z)}{\sqrt{-J^2}}=\frac{\text{Vol}(\Delta)}{4A_{\text{T}}(p_1,p_2,p_3)}.
\]

Finally, let us note that the integral $I_{1\{000\}}$ is an example of a \emph{master integral}: we can obtain all triple-$K$ integrals with integral $\beta_i$ from it by acting with shift operators, as we will explain later in the thesis.

\end{enumerate}
Below we summarise the dependence of the triple-$K$ integral on $\beta_i$: 
\begin{itemize}
\item all $\beta_i$ half-integral~~$\Rightarrow$~~$I_{\alpha\{\beta_1\beta_2\beta_3\}}$ in terms of elementary functions of momentum magnitudes
\item all $\beta_i$ integral~~$\Rightarrow$~~$I_{\alpha\{\beta_1\beta_2\beta_3\}}$ in terms of dilogarithms.
\end{itemize}
\subsection{The 3-point function as a hypergeometric system}
In this section we show an alternative way of solving the conformal Ward identities with the aim of stressing the connection between CWIs and hypergeometric systems. It has been shown independently in \cite{Coriano:2013jba} and \cite{apk2014} that the solution of 3-point CWIs can be expressed in terms of the generalised hypergeometric function of two variables Appell $F_4$ \cite{appell1926fonctions, H_Exton_1995}. In fact, the system of dilatation and special conformal Ward identities is equivalent to the set of differential equations defining Appell $F_4$ function:
\begin{align}\label{AppEq1def}
0=\biggl[& \xi(1-\xi)\frac{\partial^2}{\partial\xi^2}-\eta^2\frac{\partial^2}{\partial\eta^2}-2\xi\eta\frac{\partial^2}{\partial\xi\partial\eta} \notag\\&+\bigl(\gamma-(\alpha+\beta+1)\xi\bigr)\frac{\partial}{\partial\xi}-(\alpha+\beta+1)\eta\frac{\partial}{\partial\eta}-\alpha\beta\biggr]F(\xi,\eta),
\end{align}
\begin{align}\label{AppEq2def}
0=\biggl[& \eta(1-\eta)\frac{\partial^2}{\partial\eta^2}-\xi^2\frac{\partial^2}{\partial\xi^2}-2\xi\eta\frac{\partial^2}{\partial\xi\partial\eta} \notag\\&+\bigl(\gamma'-(\alpha+\beta+1)\eta\bigr)\frac{\partial}{\partial\eta}-(\alpha+\beta+1)\xi\frac{\partial}{\partial\xi}-\alpha\beta\biggr]F(\xi,\eta).
\end{align}
where
\begin{equation}
\xi=\frac{p_1^2}{p_3^2},~~~~~~\eta=\frac{p_2^2}{p_3^2},
\end{equation}
and $\a,\b,\g,\g'$ are some linear combinations of the spacetime $d$ and the scaling dimensions $\Delta_i$. Amongst the well known properties of the Appell $F_4$ system, let us note that it admits four linearly independent solutions:
\begin{align}\label{indsol}
&F_4(\alpha,\beta;\gamma,\gamma';\xi,\eta), \nn\\
&\xi^{1-\gamma}F_4(\alpha+1-\gamma,\beta+1-\gamma;2-\gamma,\gamma';\xi,\eta),\nn\\
&\eta^{1-\gamma'}F_4(\alpha+1-\gamma',\beta+1-\gamma';\gamma,2-\gamma';\xi,\eta),\nn\\
&\xi^{1-\gamma}\eta^{1-\gamma'}F_4(\alpha+2-\gamma-\gamma',\beta+2-\gamma-\gamma';2-\gamma,2-\gamma';\xi\,\eta).
\end{align}
In the following we briefly explain how to show the equivalence between the Appell $F_4$ system and the conformal Ward identities. First, we recall that the dilatation Ward identity allows us to write the following ansatz for the general solution:
\[\label{3pt_ansatz2}
\lla O(p_1)O(p_2)O(p_3)\rra=p_3^{\Delta_t-2d}\left(\frac{p_1^2}{p_3^2}\right)^\mu\left(\frac{p_2^2}{p_3^2}\right)^\lambda F\left(\frac{p_1^2}{p_3^2},\frac{p_2^2}{p_3^2}\right),
\]
with $\lambda$ and $\mu$ arbitrary parameters.
We also use the dilatation Ward identity \eqref{3ptDWIsc} to eliminate the derivatives with respect to $p_3$ appearing in the special conformal Ward identities:
\[
\frac{\p}{\p p_3}\rar \frac{1}{p_3}\left(\Delta_t-2d-p_1\frac{\p}{\p p_1}-p_2\frac{\p}{\p p_2}\right).
\]
Then, by using the chain rule to rewrite the special conformal Ward identities in terms of $\xi$, $\eta$ and their derivatives, we find that these equations coincide with those defining the Appell $F_4$ system, when
\[
\mu=\frac{1}{2}\left(\Delta_1-\frac{d}{2}\right)(\epsilon_1+1),\qquad \lambda=\frac{1}{2}\left(\Delta_2-\frac{d}{2}\right)(\epsilon_2+1),
\]
with $\epsilon_1,\epsilon_2\in\{-1,+1\}$. And
\begin{align}\label{parchoices}
&\alpha=\frac{1}{2}\left[\epsilon_1\left(\Delta_1-\frac{d}{2}\right)+\epsilon_2\left(\Delta_2-\frac{d}{2}\right)+\Delta_3\right],\notag \\ &\beta=\alpha-\left(\Delta_3-\frac{d}{2}\right),\notag\\&\gamma=1+\epsilon_1\left(\Delta_1-\frac{d}{2}\right),\notag\\&\gamma'=1+\epsilon_2\left(\Delta_2-\frac{d}{2}\right).
\end{align}
As showed in the previous section, the 3-point function is unique. Here, again, to avoid collinear singularities, only a particular linear combination of the four linearly independent solutions \eqref{indsol} is acceptable. In fact, Appell $F_4$ has the integral representation \cite{Prudnikov}
\begin{align}
F_4 \left( \alpha, \beta; \gamma, \gamma'; \frac{p_1^2}{p_3^2}, \frac{p_2^2}{p_3^2} \right) & = \frac{\Gamma(\gamma) \Gamma(\gamma')}{2^{\alpha+\beta-\gamma-\gamma'} \Gamma(\alpha) \Gamma(\beta)} \cdot \frac{p_3^{\alpha+\beta}}{p_1^{\gamma-1} p_2^{\gamma'-1}} \times \nn \\
& \times \int_0^\infty \dd x x^{\alpha + \beta - \gamma - \gamma'+ 1} I_{\gamma - 1}(p_1 x) I_{\gamma' - 1}(p_2 x) K_{\beta - \alpha}(p_3 x), \label{e:F4toIIK}
\end{align}
and taking into account \eqref{parchoices}, the four solutions read
\begin{equation} \label{e:IIK}
p_1^{\Delta_1 - \frac{d}{2}} p_2^{\Delta_2 - \frac{d}{2}} p_3^{\Delta_3 - \frac{d}{2}} \int_0^\infty \dd x \: x^{\frac{d}{2} - 1} I_{\pm(\Delta_1 - \frac{d}{2})} (p_1 x) I_{\pm(\Delta_2 - \frac{d}{2})} (p_2 x) K_{\Delta_3 - \frac{d}{2}} (p_3 x).
\end{equation}
We explained earlier that `$IIK$' integrals have singularities for collinear configurations of the momenta. However, since
\begin{equation}\label{KlcI}
K_\nu(x) = \frac{\pi}{2 \sin ( \nu \pi)} \left[ I_\nu(x) - I_{-\nu}(x) \right],
\end{equation}
we recover the triple-$K$ solution by taking the following linear combination of the integrals in \eqref{e:IIK}:
\begin{align}
& \int_0^\infty \dd x \: x^{\alpha - 1} K_{\b_1}(p_1 x) K_{\b_2}(p_2 x) K_{\b_3}(p_3 x)  \nn\\
& \qquad = \frac{2^{\alpha - 4}}{c^\alpha} \left[ A(\b_1, \b_2) + A(\b_1, -\b_2) + A(-\b_1, \b_2) + A(-\b_1, -\b_2) \right], \label{e:KKK}
\end{align}
where
\begin{align}
A(\b_1, \b_2) & = \left( \frac{p_1}{p_3} \right)^{\b_1} \left( \frac{p_2}{p_3} \right)^{\b_2} \Gamma \left( \frac{\alpha + \b_1 + \b_2 - \b_3}{2} \right) \Gamma \left( \frac{\alpha + \b_1 + \b_2 + \b_3}{2} \right) \Gamma(-\b_1) \Gamma(-\b_2)  \nn\\
& \qquad \times F_4 \left( \frac{\alpha + \b_1 + \b_2 - \b_3}{2}, \frac{\alpha + \b_1 + \b_2 + \b_3}{2}; \b_1 + 1, \b_2 + 1; \frac{p_1^2}{p_3^2}, \frac{p_2^2}{p_3^2} \right).
\end{align}

\subsection[{\texorpdfstring{The triple-$K$ integral as a triangle Feynman integral}{The triple-K integral as a triangle Feynman integral}}]{\texorpdfstring{The triple-\boldmath{$K$} integral as a triangle Feynman integral}{The triple-K integral as a triangle Feynman integral}}
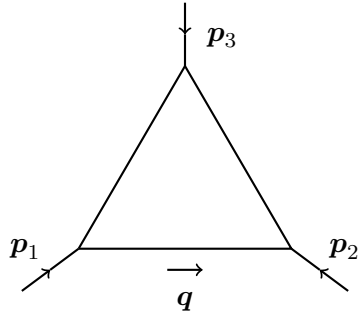
\begin{figure}
\centering
\begin{tikzpicture}[scale=1.4]
   \draw[thick] (-1,0)--(1,0) -- (0,1.73); 
   \draw[,thick] (0,1.73) -- (0,1.73+0.3);
   \draw[<-,thick] (0,1.73+0.3)--   (0,1.73+0.6) ;
   \draw[thick] (0,1.73) -- (-1,0);
  \draw[thick] (-1,0)--(-1.2666,-0.2);
  \draw[<-,thick](-1.2666,-0.2)-- (-4/3-0.2,-0.4) ;
   \draw[thick] (1,0) -- (1.2666,-0.2);
   \draw[<-,thick] (1.2666,-0.2)--(4/3+0.2,-0.4) ;
   
   \draw[->,thick] (-1/6,-1/5)--(1/6,-1/5) ;
   
   \node[text width=0.5cm, text centered ] at (-1.5,0) {$\bs{p}_1$};
   \node[text width=0.5cm, text centered ] at (1.5,0) {$\bs{p}_2$};
   \node[text width=0.5cm, text centered ] at (1/4+0.1,2) {$\bs{p}_3$};
   \node[text width=0.5cm, text centered ] at (0,-1/2) {$\bs{q}$};

\end{tikzpicture}
\caption{The 1-loop massless triangle graph \eqref{triangle}.}
\label{fig:tri}
\end{figure}
In this section we introduce a further representation of the 3-point function: we show the equivalence between the triple-$K$ integral and the 1-loop triangle Feynman integral (see fig. \ref{fig:tri}
\[\label{triangle}
I_{d\{\a_{12},\a_{13},\a_{23}\}}=\int \frac{\dd ^d\bs{q}}{(2\pi)^d}\frac{1}{|\bs{q}|^{2\a_{12}+d}|\bs{q}-\bs{p}_1|^{2\a_{13}+d}|\bs{q}+\bs{p}_2|^{2\a_{23}+d}}.
\]
The relation between these two representations is
\[\label{tri-to-3K}
I_{\a\{\b_1\b_2\b_3\}}=C_T I_{d\{\a_{12},\a_{13},\a_{23}\}}
\]
with 
\[\label{Ct}
C_T=c_{123}2^{\frac{3}{2}d-4}\pi^{d/2}\Gamma\left(\frac{\Delta_t-d}{2}\right)\prod_{1\leq j<k\leq 3}\Gamma\left(\a_{jk}+\frac{d}{2}\right).
\]

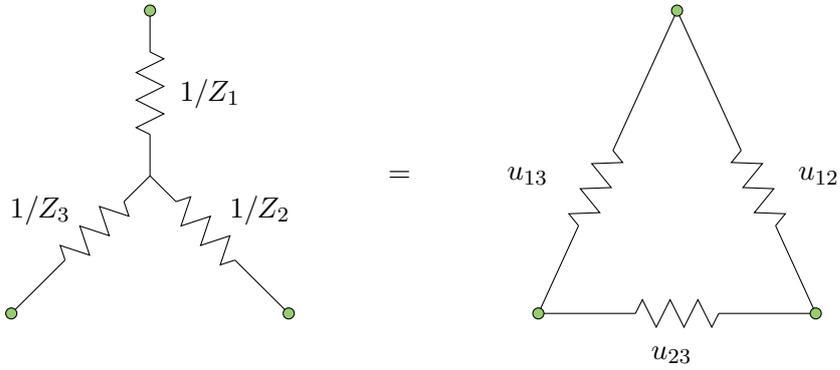
\begin{figure}
\centering
\hspace{2mm}\begin{tikzpicture}[scale=3.65]
\def \start{0.106066};
\def \del{0.0176777};
\def \var{0.035355};
\def \disp{-1.9};

\draw (0,0) -- (0.35,0);
\draw (0.65,0) -- (1,0);
\draw (0,0) -- (0.144831,0.318628);
\draw (0.268972,0.591738) -- (0.5,1.1);
\draw (1,0) --  (1-0.144831,0.318628);
\draw  (1-0.268972,0.591738) -- (1-0.5,1.1);

\draw (0.144831,0.318628) -- (0.109658,0.362078) --(0.221385,0.366216) --
(0.151038,0.453114) -- (0.262765,0.457252)--(0.192418,0.544151)--
(0.304145,0.548289)--(0.268972,0.591738);

\draw (1-0.144831,0.318628) -- (1-0.109658,0.362078) --(1-0.221385,0.366216) --
(1-0.151038,0.453114) -- (1-0.262765,0.457252)--(1-0.192418,0.544151)--
(1-0.304145,0.548289)--(1-0.268972,0.591738);

\draw (0.35,0) -- (0.375,0.05) -- (0.05+0.375,-0.05) -- (0.05+0.425,0.05) 
-- (0.05+0.475,-0.05) -- (0.05+0.525,0.05) -- (0.05+0.575,-0.05) -- (0.05+0.6,0);

\draw[black,fill=YellowGreen] (0,0) circle [radius=0.02];
\draw[black,fill=YellowGreen] (1,0) circle [radius=0.02];
\draw[black,fill=YellowGreen] (1/2,1.1) circle [radius=0.02];

\node[right] at (0.37,-0.15) {$u_{23}$};
\node[right] at (-0.15,0.5) {$u_{13}$};
\node[right] at (0.55+0.35,0.5) {$u_{12}$};

\node at (-0.5,0.5) {$=$};

\draw (\disp+0,0) -- (\disp+0.3-\start,0.3-\start);
\draw (\disp+0.3+\start,0.3+\start) -- (\disp+0.5,0.5); 
\draw (\disp+0.5,0.5) -- (\disp+0.7-\start,0.3+\start);
\draw (\disp+0.7+\start,0.3-\start) -- (\disp+1,0);
\draw (\disp+0.5,0.5) -- (\disp+0.5,0.65);
\draw (\disp+0.5,0.95) -- (\disp+0.5,1.1);

\draw (\disp+0.5,0.65) -- (\disp+0.5-0.05,0.675) -- (\disp+0.5+0.05,0.725) -- (\disp+0.5-0.05,0.775) -- (\disp+0.5+0.05,0.825) -- (\disp+0.5-0.05,0.875) -- (\disp+0.5+0.05,0.925) -- (\disp+0.5,0.95);

\draw (\disp+0.3-\start,0.3-\start) -- (\disp+0.3-\start+\del-\var,0.3-\start+\del+\var) -- (\disp+0.3-\start+3*\del+\var,0.3-\start+3*\del-\var) -- (\disp+0.3-\start+5*\del-\var,0.3-\start+5*\del+\var) -- (\disp+0.3-\start+7*\del+\var,0.3-\start+7*\del-\var) -- (\disp+0.3-\start+9*\del-\var,0.3-\start+9*\del+\var) -- (\disp+0.3-\start+11*\del+\var,0.3-\start+11*\del-\var) -- (\disp+0.3+\start,0.3+\start);
\draw (\disp+0.7-\start,0.3+\start) 
  -- (\disp+0.7-\start+\del+\var, 0.3+\start-\del+\var) 
  -- (\disp+0.7-\start+3*\del-\var, 0.3+\start-3*\del-\var) 
  -- (\disp+0.7-\start+5*\del+\var, 0.3+\start-5*\del+\var) 
  -- (\disp+0.7-\start+7*\del-\var, 0.3+\start-7*\del-\var) 
  -- (\disp+0.7-\start+9*\del+\var, 0.3+\start-9*\del+\var) 
  -- (\disp+0.7-\start+11*\del-\var, 0.3+\start-11*\del-\var) 
  -- (\disp+0.7+\start, 0.3-\start);

\draw[black,fill=YellowGreen] (0+\disp,0) circle [radius=0.02];
\draw[black,fill=YellowGreen] (1+\disp,0) circle [radius=0.02];
\draw[black,fill=YellowGreen] (1/2+\disp,1.1) circle [radius=0.02];

\node[left] at (\disp+0.25,0.375) {$1/Z_3$};
\node[right] at (\disp+0.75,0.375) {$1/Z_2$};
\node[right] at (\disp+0.57,0.8) {$1/Z_1$};
\end{tikzpicture}

\caption{Equivalent electrical networks of resistors under star-mesh duality, where the resistances are related as given in \eqref{starmesh}.   
The external currents flowing into the corresponding dotted nodes and the overall power dissipation are equal. 
 \label{fig:4}}
\end{figure}

To show that equation \eqref{tri-to-3K} holds, here we will manipulate both integrals. The reader can find how to derive the triangle integral starting from the triple-K (and vice versa) in \cite{Bzowski:2020kfw} and \cite{Bzowski:2013sza}. 
The equivalence between these representations is the star-mesh equivalence of electrical circuits in disguise. To see this, we will Schwinger parametrise both representations and perform a star-mesh transformation between the respective Schwinger parameters $u_{ij}$ and $Z_j$. The triangle integral can be thought as the `mesh' circuit with resistances $u_{ij}$, while the triple$-K$ integral corresponds to the `star' circuit with resistances $1/Z_j$ (see fig. \ref{fig:4}) with the external momenta $\bs{p}_j$ the ingoing currents.  
Later in this thesis we will see how the connection between electrical networks and integral representations of conformal correlation functions can be used to find new scalar representations of the general $n$-point function.

Let us consider the triple-K integral \eqref{tripleK}. We re-express the Bessel-K functions using the integral formula \cite{abramowitz1972handbook}
\[
K_\nu(z)=\frac{1}{2}\left(\frac{1}{2}z\right)^\nu\int_0^\infty \dd t \: t^{-\nu-1}\exp\left(-t-\frac{z^2}{4t}\right),
\]
giving
\[
p_j^{\b_j}K_{\b_j}(p_jx)=\frac{1}{2}\left(\frac{x}{2}\right)^{\b_j}\int_0^\infty \dd Z_j \: Z_j^{\b_j-1}\exp\left(-\frac{p_j^2}{Z_j}-x^2\frac{Z_j}{4}\right),
\]
where $Z_j=p_j^2/t$. We set $z=x^2/4$ and perform the integral over $z$ using \eqref{schwinger}, leading to
\[\label{3KtoTri}
I_{\a\{\b_1\b_2\b_3\}}=c_{123}2^{\a-1}\Gamma\left(\frac{\Delta_t-d}{2}\right)\left(\prod_{j=1}^3\int_0^\infty \dd Z_j \: Z_j^{\b_j-1}\right)Z_t^{\frac{d-\Delta_t}{2}}\exp\left(-\sum_j\frac{p_j^2}{Z_j}\right),
\]
where we defined $Z_t=Z_1+Z_2+Z_3$. The $Z_j$ variables can be interpreted as the conductivities of the `star' network and the argument of the exponential can be seen as the power dissipated in the same circuit. 

Let us now work on the triangle Feynman integral \eqref{triangle}. We Schwinger-parametrise the factors in the denominator by using equation \eqref{schwinger}
\[
I_{d\{\a_{12}\a_{13}\a_{23}\}}=C\left(\prod_{1\leq j<k \leq 3}\int_0^\infty \dd u_{jk} \: u_{jk}^{\a_{jk}+\frac{d}{2}-1}\right)U^{-\frac{d}{2}}\exp\left({-\frac{F}{U}}\right),
\]
with 
\[
C=(4\pi)^{-\frac{d}{2}}\prod_{1\leq j<k \leq 3}\Gamma\left(\a_{jk}+\frac{d}{2}\right),
\]
and $U$ and $F$ the Symanzik polynomials
\[
U=u_{12}+u_{13}+u_{23},\qquad F=p_1^2u_{12}u_{13}+p_2^2u_{12}u_{23}+p_3^2u_{13}u_{23}.
\]
Finally, we diagonalise the exponential by performing the `star-mesh' change of variables. In fact, the Schwinger parameters $u_{jk}$ can be interpreted as the resistances of the triangle (or mesh) circuit which are related to the conductivities $Z_i$ of the star circuit via
\[\label{starmesh}
\frac{u_{ik}u_{jk}}{u_t}=\frac{1}{Z_k}, \qquad i,j,k=1,2,3,
\]
giving
\[\label{Trito3K}
I_{d\{\a_{12}\a_{13}\a_{23}\}}=C\left(\prod_{j=1}^3\int_0^\infty \dd Z_j \: Z_j^{(\a_{ik}-\a_t-\frac{d}{2})-1}\right)Z_t^{\a_t+\frac{d}{2}}\exp\left(-\sum_j\frac{p_j^2}{Z_j}\right),
\]
with $\a_t=\a_{12}+\a_{13}+\a_{23}$. It is evident that this integral is the same as in equation \eqref{3KtoTri} when
\[
\a_{jk}=-\b_j-\b_k+\frac{1}{2}\left(\b_t-\frac{d}{2}\right),
\]
or equivalently
\[\label{aijtobi}
\b_i=-\frac{d}{2}-\sum_{j\neq i}\a_{ij}.
\]
Hence, we showed the equivalence in \eqref{tri-to-3K}.

\section{Singularities and renormalisation}
So far we have presented 2- and 3-point functions. Before moving to higher-point functions we discuss when divergences arise and briefly explain how to renormalise the solutions in these cases. We will see that renormalisation is necessary when the solutions are analytic functions of the squared momenta $p_i^2$, since this corresponds to local solutions in position-space. 
\subsection{2-point function}
In section \ref{cha3sec2pt} we found the solution of the 2-point CWIs, but we have not discussed its domain of existence.
Let us recall the Fourier transformed 2-point solution \eqref{2ptFT}:
\[
\lla\O(\bs{p})\O(-\bs{p})\rra=c_0p^{2\left(\Delta-\frac{d}{2}\right)},\qquad c_0=c\frac{2^{d-2\Delta}\pi^{d/2}\Gamma\left(\frac{d}{2}-\Delta\right)}{\Gamma(\Delta)}.
\]
For finite $c_0$ and generic $\Delta$ this solves the CWIs. However, for values of $d$ and $\Delta$ such that
\[\label{sing2pt}
\frac{d}{2}-\Delta=-n,\qquad n \in \mathbb{Z}^+,
\]
the 2-point function is divergent and needs to be regulated. One way is to dimensionally regulate the correlator by shifting $d$ and $\Delta$ as follows
\[
\frac{d}{2}-\Delta=-n-\epsilon,\qquad \epsilon\ll 1.
\]
Then, expanding in $\epsilon$, the regulated correlator reads
\[\label{pole2pt}
\lla\O(\bs{p})\O(-\bs{p})\rra_{\mathrm{reg}}=\frac{c_0^{(-1)}}{\epsilon}p^{2n}+c_0^{(-1)}p^{2n}\ln p^2+ c_0^{(0)}p^{2n}+O(\epsilon),
\]
which has a pole in $\epsilon$ in the limit $\epsilon\rar 0$. 
Can we cancel this singularity? We cannot rescale $c_0\rar \epsilon c_0$ since when the condition \eqref{sing2pt} holds, the 2-point function is an analytic function of $p^2$:
\[
\lla\O(\bs{p})\O(-\bs{p})\rra=c_0p^{2n}.
\]
This corresponds to a \emph{local} solution, which means it has support only at $x^2=0$ as one can see from its Fourier transform: this is given by $2n$ derivatives acting on a delta function
\[\label{local2pt}
\langle\O(\x)\O(0)\rangle=c_0(-\square)^n\delta(\x).
\]
This is not a physically acceptable solution and, as we showed in Chapter \ref{cha2}, all position-space correlators are non-local. Moreover, when condition \eqref{sing2pt} holds, there is a new local term in the action of the form
\[
\phi\square^n\phi,
\]
where $\phi$ is the source of the operator $\O$. With an appropriate choice of the coefficient, this can be treated as a counterterm that cancels the divergence of the correlator.

The contribution to the momentum-space correlator of this counterterm is
\[\label{2ptct}
\lla\O(\bs{p})\O(-\bs{p})\rra_{\mathrm{ct}}=\tilde{c}(\epsilon)p^{2n}\mu^{2\epsilon}=\frac{\tilde{c}^{(-1)}}{\epsilon}p^{2n}+\tilde{c}^{(-1)}p^{2n}\ln \mu^2+ \tilde{c}^{(0)}p^{2n}+O(\epsilon),
\]
where the renormalisation group (RG) scale $\mu$ appeared on dimensional grounds. If we then sum the two contribution to the correlator, \eqref{pole2pt} and \eqref{2ptct}, we can cancel the divergence by setting $\tilde{c}^{(-1)}=-c_0^{(-1)}$ and take the limit $\epsilon\rar 0$, leading to the renormalised 2-point function:
\[
\lla\O(\bs{p})\O(-\bs{p})\rra_{\mathrm{ren}}=p^{2n}\left(c_0^{(-1)}\ln\frac{p^2}{\mu^2}+c'_0\right).
\]
This solution depends on the renormalisation scale $\mu$ which breaks conformal symmetry resulting in a conformal anomaly. This has been studied in more detail in \cite{Bzowski:2015pba} and we will not need it in this thesis.
\subsection{3-point function}
In section \ref{cha3sec3pt} we found the general solution of the 3-point CWIs and discussed its uniqueness due to the absence of collinear singularities, leading to the triple-$K$ integral \eqref{tripleK}. This converges at large $x$, according to the expansion in \eqref{BesselLarge}. However, there could be divergences in the limit of $x\rightarrow 0$. The condition for this to happen is
\[\label{singularities}
\a+1\pm \b_1 \pm \b_2 \pm \b_3 = -2n,\quad n \in \mathbb{Z}^+.
\]
To see this, we consider the series expansion of the Bessel-K function around $x=0$. Let $\b_j$ be a non-integer number. Then, taking into account the series expansion of the Bessel-I, $I_\b=\sum_{n=0}^\infty a_I x^{\b+2n}$, and equation \eqref{KlcI}, the expansion of Bessel-K is
\[
K_{\b_j}(x)=x^{\b_j}\sum_{n_j=0}^\infty a_+x^{2n_j}+x^{-\b_j}\sum_{n_j=0}^\infty a_-x^{2n_j}.
\]
We use this equation to expand the integrand of the triple-$K$ \eqref{tripleK}, leading to integrals of the form
\[\label{etaint}
\int_0^\infty \dd x \: x^\eta=\frac{x^{\eta+1}}{\eta+1},\qquad \eta=\a\pm \b_1\pm \b_2 \pm \b_3+2n_t,
\]
with $n_t=n_1+n_2+n_3$. This integral converges at $x=0$ when
\[
\a\pm \b_1 \pm \b_2 \pm \b_3 > -1.
\]
However, if we consider the integral as a function of its parameters while the momenta are fixed, we can perform an analytic continuation by considering the complex $\eta$-plane. Here the integral in \eqref{etaint} is well defined unless $\eta=-1$, where it has a pole. This gives the singularity condition in \eqref{singularities}. For integer $\b_j$, the series expansion of Bessel-K functions around $x=0$ is
\[
K_{\b_j}(x)=x^{\b_j}\sum_{n_j=0}^\infty a_+x^{2n_j}+x^{-\b_j}\sum_{n_j=0}^\infty a_-x^{2n_j}+c_I\log(x)I_{\b_j}(x).
\]
Then the corresponding expansion of the integrand in the triple-$K$ contains also logarithms, 
\[\label{etaint2}
\int_0^\infty \dd x \: x^\eta\log(x)=-\frac{\mathrm{const_1}}{(\eta+1)^2}+\frac{\mathrm{const_2}}{(\eta+1)}.
\]
This, however, does not change the loci of singularities $\eta=-1$, it may only increase the order of the pole.

When equation \eqref{singularities} holds, we need to regulate the triple-$K$ integral. We dimensional regulate it by shifting the parameters as follows
\[
\a\rar \tilde{\a}=\a+u\epsilon,\qquad \b_i\rar \tilde{\b}_i=\b_i+v\epsilon,\quad i=1,2,3.
\]
Note that this leaves the form of the triple-$K$ unchanged, but the parameters are now $\tilde{\a}$ and $\tilde{\b}_i$ and the constant $c_{123}$ depends on the regularisation parameters $\epsilon,u$ and $v$. When the singularity condition \eqref{singularities} holds, we find that the regulated solution is singular when we take the limit $\epsilon\rar 0$. Depending on the type of singularity, this is canceled either by renormalisation or by an appropriate choice of the constant $c_{123}$.

Four different singularity conditions arise from \eqref{singularities}, depending on the relative signs of the $\b_i$. These are the $(---)$, $(--+)$, $(-++)$, $(+++)$ conditions. The first two singularities, \ie ~$(---)$ and $(--+)$, correspond respectively to ultralocal and semilocal solutions. By ultralocal we mean that the solution only has support on configurations where three positions collapse to one single point. By semilocal, we refer to solutions that have support where two of the three positions coincide at one point. In these cases, counterterms exist and divergences are canceled via renormalisation. The latter leads to a conformal anomaly for the condition $(---)$, while the condition $(--+)$ corresponds to beta functions\footnote{Note, this is not in contradiction with conformal symmetry, since this type of beta functions are associated to couplings of composite operators which are not couplings appearing in the Lagrangian of fundamental fields.}.

On the other hand, the other two conditions with mostly `+' signs, correspond to non-local solutions. Consequently, counterterms do not exist and the singularities are just singularities of the triple-$K$ integral that can be cured by choosing the constant $c_{123}$ to be proportional to $\epsilon$. \\Finally, note that more than one condition can be satisfied simultaneously, resulting in higher-order singularities.

In the next section we illustrate some examples where the condition \eqref{singularities} holds.

\subsection{Examples}
\begin{enumerate}
\item $(+++)$ condition: $d=3$, $\Delta_1=\Delta_2=\Delta_3=1$ ($\a=1/2$ and $\b_i=-1/2$, with $i=1,2,3$).\\
Let us consider the following regularisation scheme
\[\label{regscheme}
d\rar d+2\epsilon,\qquad \Delta_i\rar \Delta_i+\epsilon,
\]
so that the indices $\b_i$ \eqref{bjdef} of Bessel-K functions don't change. Then, the triple-$K$ integral reads
\begin{equation}\label{ex1}
\langle\!\langle O_1(\bs{p}_1)O_2(\bs{p}_2)O_3(\bs{p}_3)\rangle\!\rangle=c(p_1p_2p_3)^{-\frac{1}{2}}\int_0^\infty \dd x \: x^{\frac{1}{2}+\epsilon} K_{\frac{1}{2}}(p_1x)K_{\frac{1}{2}}(p_2x)K_{\frac{1}{2}}(p_3x),
\end{equation}
where the Bessel-K are elementary functions,  
\begin{equation}
K_{\frac{1}{2}}(x)=\sqrt{\frac{\pi}{2}}\frac{\mathrm{e}^{-x}}{x^{1/2}}.
\end{equation}
Then, the integral evaluates to
\begin{align}\label{expansion1}
\lla \O_1(\bs{p}_1)\O_2(\bs{p}_2)\O_3(\bs{p}_3)\rra &=\frac{c_{123}}{p_1p_2p_3}\left(\frac{\pi}{2}\right)^{3/2}\int_0^\infty \dd x \: x^{-1+\epsilon}\mathrm{e}^{-(p_1+p_2+p_3)x}\notag\\
&=\frac{c_{123}}{p_1p_2p_3}\left(\frac{\pi}{2}\right)^{3/2}(p_1+p_2+p_3)^{-\epsilon}\Gamma(\epsilon)\notag\\
&=\frac{c_{123}}{p_1p_2p_3}\left(\frac{\pi}{2}\right)^{3/2}\left [\frac{1}{\epsilon}-\log(p_1+p_2+p_3)-\gamma_E+O(\epsilon)\right],
\end{align}
where in the last equality we have expanded in $\epsilon$, using 
\[
\Gamma(\epsilon)=\frac{1}{\epsilon}-\gamma_E+O(\epsilon),
\]
with $\gamma_E$ the Euler-Mascheroni constant. This 3-point function is divergent for $\epsilon\rar 0$. However, the $(+++)$ condition does not admit any counterterm. This is because the leading term of \eqref{expansion1} is a non-analytic function of the squared momentum magnitudes and hence this is a non-local, physically acceptable, solution. We then eliminate the divergence by choosing $c_{123}= C_{123}\epsilon$. Thus,
\[
\lla \O_1(\bs{p}_1)\O_2(\bs{p}_2)\O_3(\bs{p}_3)\rra=\left(\frac{\pi}{2}\right)^{3/2}\frac{C_{123}}{p_1p_2p_3}.
\]
\item $(---)$ condition: $d=3$, $\Delta_1=\Delta_2=\Delta_3=2$ ($\a=1/2$ and $\b_i=1/2$, with $i=1,2,3$).
\\Using the regularisation scheme defined in \eqref{regscheme}, this 3-point function reads
\[
\lla \O_1(\bs{p}_1)\O_2(\bs{p}_2)\O_3(\bs{p}_3)\rra=-c_{123}(p_1p_2p_3)^{\frac{1}{2}}\int_0^\infty \dd x \: x^{\frac{1}{2}+\epsilon} K_{\frac{1}{2}}(p_1x)K_{\frac{1}{2}}(p_2x)K_{\frac{1}{2}}(p_3x),
\]
where the overall minus sign is for convenience. This integral is the same as the one in \eqref{ex1}, so here we have
\[\label{expansion2}
\lla \O_1(\bs{p}_1)\O_2(\bs{p}_2)\O_3(\bs{p}_3)\rra =c_{123}\left(\frac{\pi}{2}\right)^{3/2}\left [-\frac{1}{\epsilon}+\log(p_1+p_2+p_3)+\gamma_E+O(\epsilon)\right].
\]
However, the leading term in \eqref{expansion2} is a constant and the divergence for $\epsilon\rar 0$ is ultralocal. This divergence is canceled by the counterterm
 \[
S_{ct}=a(\epsilon)\int \dd^{3+2\epsilon}\bs{x}\phi^3\mu^{-\epsilon},
\]
where $\mu$ is the renormalisation scale and $\phi$ is the source field of the operators $\O$. To cancel the divergence, we choose
\[
a(\epsilon)=\frac{1}{6}c_{123}\left(\frac{\pi}{2}\right)^{3/2}\left(\frac{1}{\epsilon}+a_0\right),
\]
where $a_0$ is an arbitrary constant dependent on the regularisation scheme. Then, the renormalised 3-point function is
\[
\lla \O_1(\bs{p}_1)\O_2(\bs{p}_2)\O_3(\bs{p}_3)\rra_{\mathrm{ren}} =c_{123}\left(\frac{\pi}{2}\right)^{3/2}\left[\log\left(\frac{p_1+p_2+p_3}{\mu}\right)+c_1\right],
\]
where $c_1=c_0+\gamma_E$. Therefore we find, as expected for the $(---)$ condition, that the correlator depends on the RG scale $\mu$. Hence the conformal symmetry is broken and a conformal anomaly $A$ exists:
\[
A=\int\dd^d \bs{x} \mathcal{A}_{222}(\square^2\phi)^3,
\]
where we defined
\[
\mathcal{A}_{222}=\mu\frac{\mu}{\p \mu}\lla \O_1(\bs{p}_1)\O_2(\bs{p}_2)\O_3(\bs{p}_3)\rra_{\mathrm{ren}}=-c_{123}\left(\frac{\pi}{2}\right)^{3/2}.
\]
\end{enumerate}
\section{Shift operators}
We conclude the analysis on 3-point functions by presenting their shift operators. These are operators that act on a 3-point function to shift the parameters $d$ and $\Delta_i$ ($i=1,2,3$), or equivalently $\a$ and $\b_i$. In other words, they connect two solutions of the CWIs with shifted parameters. Earlier in this chapter we discussed the form of 3-point functions depending on the values of $\b_i$. For half-integer $\b_i$ the triple-$K$ integral can easily be computed in terms of elementary functions. For integer $\b_i$ the 3-point function is expressible in terms of the dilogarithm function, however, the computation is cumbersome. We showed the evaluation of the master integral $I_{1\{000\}}$ ($\a=1$ and $\b_i=0$) in appendix \ref{I1000comp}. For larger integer values of $\b_i$ the computation is more complicated. One strategy to obtain this class of triple-$K$ is to generate them by acting on the master integral with a shift operator. This leads to the reduction scheme for the evaluation of 3-point functions discussed in \cite{evaluation}.

In this section we show that two types of such operators exist. The first family of shift operators acts to shift both the spacetime dimension $d$ up by two (hence $\a$ up by one) and one $\b_i$ up or down by one. We derive their form using the properties of Bessel-$K$ functions, as shown in \cite{Bzowski:2013sza, Bzowski:2015yxv}. In Chapter \ref{cha4}, we will find the general form of these operators acting on $n-$point functions, using a new scalar representation of the $n-$point functions. The second set of operators act to shift two of the $\b_i$ up or down while leaving the spacetime dimension invariant and will be further discussed in Chapter \ref{cha5}. We show that their expression can be understood easily in position space. Finally, we derive some recursion relations for triple-$K$ integrals.

\subsection[{\texorpdfstring{Operators shifting $d$}{Operators shifting d}}]{\texorpdfstring{Operators shifting \boldmath{$d$}}{Operators shifting d}}
The 3-point $d-$shifting operators are
\[\label{LRdef00}
\mathcal{L}_i = -\frac{1}{p_i}\frac{\partial}{\partial p_i}, \qquad
\mathcal{R}_i = 2\beta_i - p_i\frac{\partial}{\partial p_i},\qquad \beta_i=\Delta_i-\frac{d}{2}.
\]
They act on the 3-point function by sending
\begin{align}\label{LRaction0}
\mathcal{L}_i: \quad\beta_i\rightarrow \beta_i -1,\quad \a\rightarrow \a+1, \qquad
\mathcal{R}_i:\quad \beta_i\rightarrow\beta_i+1,\quad \a\rightarrow \a+1,
\end{align}
or equivalently, 
\begin{align}
&\mathcal{L}_1:\quad  (d,\Delta_1,\Delta_2,\Delta_3)\rightarrow (d+2,\Delta_1,\Delta_2+1,\Delta_3+1), \\
&\mathcal{R}_1:\quad (d,\Delta_1,\Delta_2,\Delta_3)\rightarrow (d+2,\Delta_1+2,\Delta_2+1,\Delta_3+1), 
\end{align}
and similarly under any permutation in the set $\{1,2,3\}$. To understand their expressions and actions, consider the following properties of Bessel functions
\begin{align}
&K_\nu=K_{-\nu},\label{parity-bess}\\&\frac{\p}{\p p}\left[p^\nu K_\nu(px)\right]=-xp^\nu K_{\nu-1}(px).\label{prop-bess-der}
\end{align}
The first property \eqref{parity-bess}, together with the definition of the triple-$K$ integral \eqref{tripleK}, imply that
\[
I_{\a\{-\b_1\b_2\b_3\}}=p_1^{-2\b_1}I_{\a\{\b_1\b_2\b_3\}},
\]
hence the operator $p_i^{-2\b_i}$ sends $\b_i\rar -\b_i$. This is effectively a \emph{shadow transformation}, sending $\Delta_i\rar d-\Delta_i$. From the property \eqref{prop-bess-der}, we see that $\mathcal{L}_1$ acts on the triple-$K$ integral as 
\[\label{Lact}
\mathcal{L}_1I_{\a\{\b_1\b_2\b_3\}}=I_{\a+1\{\b_1-1,\b_2\b_3\}}.
\]
This can be seen by direct computation
\begin{align}
\mathcal{L}_1I_{\a\{\b_1\b_2\b_3\}}&=c_{123}\int_0^{\infty} \dd x \: x^\a \left[-\frac{1}{p_1}\frac{\p}{\p p_1}\left(p_1^{\b_1}K_{\b_1}(p_1 x)\right)\right]p_2^{\b_2}p_3^{\b_3}K_{\b_2}(p_2 x)K_{\b_3}(p_3 x)\nn\\
&=c_{123}\int_0^{\infty} \dd x \: x^\a \left(x \: p_1^{\b_1-1}K_{\b_1-1}(p_1 x)\right)p_2^{\b_2}p_3^{\b_3}K_{\b_2}(p_2 x)K_{\b_3}(p_3 x)\nn\\
&=c_{123}\int_0^{\infty} \dd x \: x^{\a+1}p_1^{\b_1-1}p_2^{\b_2}p_3^{\b_3}K_{\b_1-1}(p_1 x)K_{\b_2}(p_2 x)K_{\b_3}(p_3 x).
\end{align}
Combining $\mathcal{L}_1$ and the shadow transform we obtain the operator $\mathcal{R}_1$
\[
\mathcal{R}_1=p_1^{2\beta_1+2}\,\mathcal{L}_1 \,p_1^{-2\beta_1},
\]
which acts on the triple-$K$ as
\[\label{Ract}
\mathcal{R}_1I_{\a\{\b_1\b_2\b_3\}}=I_{\a+1\{\b_1+1,\b_2\b_3\}},
\]
since
\begin{align}
(\b_i,\a) \xrightarrow{p_i^{-2\b_i}} (-\b_i,\a)
\xrightarrow{\mathcal{L}_i} (-\b_i-1, \a+1)
\xrightarrow{p_i^{-2(-\b_i-1)}=p_i^{2(\b_i+1)}} (\b_i+1, \a+1).
\end{align}
It is interesting to note that the combination $\mathcal{L}_i\mathcal{R}_i$ amounts to the special conformal operator we introduced in \eqref{defK},
\[
\mathcal{L}_i\mathcal{R}_i=\K_i=\frac{\p^2}{\p p_i^2}+\frac{1-2\b_i}{\p}\frac{\p}{\p p_i}.
\]
This operator acts on a triple-$K$ integral to shift $\a$ up by two, since $\mathcal{L}_i$ and $\mathcal{R}_i$ shift $\b_i$ in opposite directions but both increase $\a$ by one:
\[\label{Kshift}
\K_iI_{\a\{\b_1\b_2\b_3\}}=I_{\a+2\{\b_1\b_2\b_3\}},\qquad i=1,2,3.
\]
\subsubsection[{\texorpdfstring{Triple-$K$ recursion relations}{Triple-K recursion relations}}]{\texorpdfstring{Triple-\boldmath{$K$} recursion relations}{Triple-K recursion relations}}
Using the actions of the above operators above and the CWIs we can derive some additional recursion relations. From the dilatation Ward identity \eqref{3ptDWIsc}, expressing the operators $p_i\p_{p_i}$ in terms of $\mathcal{L} _i$ and $\mathcal{R}_i$, we have
\begin{align}
&(\a+1-\b_t)I_{\a\{\b_1\b_2\b_3\}}=\left(p_1^2\mathcal{L}_1+p_2^2\mathcal{L}_2+p_3^2\mathcal{L}_3\right)I_{\a\{\b_1\b_2\b_3\}},\label{rec1}\\
&(\a+1+\b_t)I_{\a\{\b_1\b_2\b_3\}}=\left(\mathcal{R}_1+\mathcal{R}_2+\mathcal{R}_3\right)I_{\a\{\b_1\b_2\b_3\}},\label{rec2}
\end{align}
and considering the action \eqref{Lact} of the operators $\mathcal{L}_i$ and \eqref{Ract} of $\mathcal{R}_i$, we obtain
\begin{align}
&(\a+1-\b_t)I_{\a\{\b_1\b_2\b_3\}}=p_1^2I_{\a+1\{\b_1-1,\b_2,\b_3\}}+p_2^2I_{\a+1\{\b_1,\b_2-1,\b_3\}}+p_3^2I_{\a+1\{\b_1,\b_2,\b_3-1\}}\nn\\
&(\a+1+\b_t)I_{\a\{\b_1\b_2\b_3\}}=I_{\a+1\{\b_1+1,\b_2,\b_3\}}+I_{\a+1\{\b_1,\b_2+1,\b_3\}}+I_{\a+1\{\b_1,\b_2,\b_3+1\}}.
\end{align}
Taking into account that 
\begin{align}
&\mathcal{L}_1I_{\a\{\b_1\b_2\b_3\}}=\mathcal{R}_2\mathcal{R}_3I_{\a-1\{\b_1-1\b_2-1\b_3-1\}},\nn\\
&\mathcal{R}_1I_{\a\{\b_1\b_2\b_3\}}=\mathcal{L}_2\mathcal{L}_3I_{\a-1\{\b_1+1,\b_2+1,\b_3+1\}},
\end{align}
equations \eqref{rec1} and \eqref{rec2} are equivalent to the following identities 
\begin{align}
&(\a+1-\b_t)I_{\a\{\b_1\b_2\b_3\}}=\left[p_1^2\mathcal{R}_2\mathcal{R}_3+p_2^2\mathcal{R}_1\mathcal{R}_3+p_3^2\mathcal{R}_1\mathcal{R}_2\right]I_{\a-1\{\b_1-1\b_2-1\b_3-1\}},\\
&(\a+1+\b_t)I_{\a\{\b_1\b_2\b_3\}}=\left(\mathcal{L}_1\mathcal{L}_2+\mathcal{L}_1\mathcal{L}_3+\mathcal{L}_2\mathcal{L}_3\right)I_{\a-1\{\b_1+1,\b_2+1,\b_3+1\}}.
\end{align}
\subsection[{\texorpdfstring{Operators preserving $d$}{Operators preserving d}}]{\texorpdfstring{Operators preserving \boldmath{$d$}}{Operators preserving d}}\label{wshiftop}
In this section we discuss shift operators acting on solutions of the CWIs to shift the scaling dimensions while preserving the spacetime dimension. The best known operators in literature \cite{Baumann:2019oyu, Karateev:2017jgd} are those acting on the CWI solutions to shift two scaling dimensions $\Delta_i$ up or down by one unit. They are denoted by $\mathcal{W}_{ij}^{\sigma_i\sigma_j}$, with $\sigma_i=\pm 1$, and their action is to send
\[
d\rar d\qquad\Delta_i\rar \Delta_i+\sigma_i,\qquad \Delta_j\rar \Delta_j+\sigma_j,
\]
or equivalently,
\[
\a\rar \a,\qquad \b_i\rar \b_i+\sigma_i,\qquad \b_j\rar \b_j+\sigma_j.
\]
One might wonder why the shift operators introduced until now act either to shift two parameters by one unit or act to shift one parameter by two units (see \eqref{Kshift}). 
This is not a trivial question and we will address it in Chapter \ref{cha5}.

In the following we introduce the expressions of $\mathcal{W}$-operators and derive their action on the 3-point functions. We will consider $\mathcal{W}_{12}^{\pm\pm}$, since the expressions for general $i,j$ can be obtained by permutation. According to recent understanding of these operators \cite{Bzowski:2022rlz} in momentum space, their expressions read
\begin{align}
&\mathcal{W}_{12}^{--}=\frac{1}{2}\biggl(\frac{\p}{\p p_1^{\mu}}-\frac{\p}{\p p_2^{\mu}}\biggr)\biggl(\frac{\p}{\p p_{1,\mu}}-\frac{\p}{\p p_{2,\mu}}\biggr),\label{Wmmdef}\\
&\mathcal{W}_{12}^{+-}=p_1^{2(\b_1+1)}\mathcal{W}_{12}^{--}p_1^{-2\b_1},\label{Wpmdef}\\
&\mathcal{W}_{12}^{-+}=p_2^{2(\b_2+1)}\mathcal{W}_{12}^{--}p_2^{-2\b_2},\label{Wmpdef}\\
&\mathcal{W}_{12}^{++}=p_1^{2(\b_1+1)}p_2^{2(\b_2+1)}\mathcal{W}_{12}^{--}p_1^{-2\b_1}p_2^{-2\b_2}.\label{Wppdef}
\end{align}
We explain below the expression of $\mathcal{W}_{12}^{--}$, while the remaining ones are obtained by shadow transforming $\mathcal{W}_{12}^{--}$. 
\subsubsection{Lowering operator $\mathcal{W}_{12}^{--}$}
The easiest way to derive equation \eqref{Wmmdef}, is to start from position space. In fact given the form of the general $n$-point solution \eqref{n-pt}, it is intuitive to find the position-space expression of the lowering operator $\mathcal{W}_{12}^{--}$. As we showed in Chapter \ref{cha2}, the general position-space $n$-point function can be written as
\[\label{nptbis}
\phi_n=\langle O(\bs{x}_1)...O(\bs{x}_n)\rangle=\prod_{1\leq i<j\leq n}x_{ij}^{2\a_{ij}}f(\bs{u}),
\]
and the parameters $\a_{ij}$ are related to the scaling dimensions $\Delta_i$ via 
\[
\Delta_i=-\sum_{j=1}^n\a_{ij},\quad i=1,2,..,n
\]
where $\a_{ij}=\a_{ji}$ and $\a_{ii}=0$. It is straightforward to understand that $x_{ij}^2$ is an operator shifting $\Delta_i$ and $\Delta_j$ down by one. If we multiply \eqref{nptbis} by $x_{ij}^2$, for some specific choice of $i$ and $j$, this serves to shift $\a_{ij}\rightarrow \a_{ij}+1$ and hence $\Delta_i\rightarrow \Delta_i-1$ and $\Delta_j\rightarrow \Delta_j-1$. Multiplying by $x_{ij}^2$ thus acts as a lowering operator generating a new solution of the $n$-point conformal Ward identities in which the dimensions $\Delta_i$ and $\Delta_j$ are reduced by one while preserving $f(\bs{u})$ and the spacetime dimension $d$. 

To find the corresponding expression in momentum space we perform a Fourier transform. The Fourier transformed $n$-point function is
\[
\Phi_n=\mathcal{F}[\phi_n]=\int \text{d}\bs{x}_1...\text{d}\bs{x}_n\: \mathrm{e}^{-i\sum_{j=1}^n\bs{x}_j\cdot \bs{p}_j} \phi_n(\bs{x}_1,..,\bs{x}_n),
\]
where, taking into account momentum conservation, 
\[
\sum_{j=1}^n\bs{x}_j\cdot \bs{p}_j=\sum_{j=1}^{n-1}\bs{p}_j\bs{x}_j-\left(\sum_{j=1}^{n-1}\bs{p}_j\right)\bs{x}_n=\sum_{j=1}^{n-1}\bs{p}_j\bs{x}_{jn}.
\]
Therefore, pulling down a factor of $\bs{x}_{12}$ is equivalent to acting on the Fourier transformed $n-$point function with the difference of derivatives with respect to $p_1^\mu$ and $p_2^\mu$:
\[
\mathcal{F}\left[\bs{x}_{12}\phi_4\right]=i\left(\frac{\p}{\p p_1^{\mu}}-\frac{\p}{\p p_2^{\mu}}\right)\Phi_4,
\]
hence 
\[
\mathcal{W}_{12}^{--}=\mathcal{F}\left[-\frac{1}{2}x_{12}^2\right]=\frac{1}{2}\biggl(\frac{\p}{\p p_1^{\mu}}-\frac{\p}{\p p_2^{\mu}}\biggr)\biggl(\frac{\p}{\p p_{1,\mu}}-\frac{\p}{\p p_{2,\mu}}\biggr),
\]
where the factor $-1/2$ is purely conventional \cite{Baumann:2019oyu}. This is valid at $n$-point. However, here we focus on the 3-point function. Using the chain rule \eqref{chain3} we obtain the expression of the 3-point $\mathcal{W}_{12}^{--}$ in terms of Mandelstam variables:
\[\label{W-3pt}
\mathcal{W}_{12}^{--}=\frac{1}{2}\biggl[\p_1^2+\p_2^2+\frac{d-1}{p_1}\p_1+\frac{d-1}{p_2}\p_2+\frac{p_1^2+p_2^2-p_3^2}{p_1p_2}\p_1\p_2\biggr].
\]
Let us now derive the action on the 3-point function.
By construction, the lowering operator acts to generate a 3-point function with $\Delta_1$ and $\Delta_2$ lowered by one. 
To show this, we use the triangle representation \eqref{triangle}. Using the equivalence between the 1-loop triangle integral and the triple-$K$ representation, we then derive the action on the triple-$K$ integral. To simplify the computation, we write the triangle integral by re-parametrising the loop momentum:
\[\label{triangleW}
I_{d\{\a_{12},\a_{13},\a_{23}\}}=\int \frac{\text{d}^d\bs{q}}{(2\pi)^d}\frac{1}{|\bs{p}_1+\bs{p}_2+\bs{q}|^{2\a_{23}+d}|\bs{q}|^{2\a_{13}+d}|\bs{p}_1+\bs{q}|^{2\a_{12}+d}},
\]
and use the following identities
\begin{align}
&\biggl(\frac{\p}{\p p_1^{\mu}}-\frac{\p}{\p p_2^{\mu}}\biggr)f(\bs{p}_1+\bs{p}_2)=0,\\
&\biggl(\frac{\p}{\p p_1^{\mu}}-\frac{\p}{\p p_2^{\mu}}\biggr)f(\bs{p}_1)=\frac{\p}{\p p_1^{\mu}}f(\bs{p}_1),
\end{align}
where $f$ is a generic function. Then, we only need to compute
\[
\delta^{\mu\nu}\frac{\p}{\p p_1^{\mu}}\frac{\p}{\p p_1^{\nu}}\frac{1}{|\bs{p}_1+\bs{q}|^{2\a_{12}+d}}=\frac{2\left(\a_{12}+\frac{d}{2}\right)(\a_{12}+1)}{|\bs{p}_1+\bs{q}|^{2(\a_{12}+1)+d}}.
\]
Hence the action of $\mathcal{W}_{12}^{--}$ on the triangle integral is
\[\label{WonTri}
\mathcal{W}_{12}^{--}I_{d\{\a_{12},\a_{13},\a_{23}\}}=2\left(\a_{12}+\frac{d}{2}\right)(\a_{12}+1)I_{d\{\a_{12}+1,\a_{13},\a_{23}\}}.
\]
Note that, according to \eqref{aijtobi}, sending $\a_{12}\rar\a_{12}+1$ is equivalent to send $\b_1\rar\b_1-1$ and $\b_2\rar\b_2-1$. We derive the action on the triple$-K$ by combining \eqref{WonTri} with the triangle/triple-$K$ equivalence \eqref{tri-to-3K}
\begin{align}\label{W12on3pt}
\mathcal{W}_{12}^{--}I_{\a\{\b_1\b_2\b_3\}}&=2\left(\a_{12}+\frac{d}{2}\right)(\a_{12}+1)\frac{C_T}{C_T|_{\a_{12}+1}}I_{\a\{\b_1-1,\b_2-1,\b_3\}}\nn\\
&=\frac{1}{2}\big[\beta_3^2 - (\alpha-1+\beta_1+\beta_2)^2\big] I_{\alpha,\{\beta_1 - 1,\beta_2-1,\beta_3\}},
\end{align}
where $C_T$ is given in \eqref{Ct}. Let us anticipate here that in the second part of this thesis, following the work \cite{Caloro}, we explain how to derive the factor involving the parameters $\a$ and $\b_i$ by knowing the singularities of the triple-$K$. 
\subsubsection{Lowering-Raising $\mathcal{W}_{12}^{-+}$ and raising operator $\mathcal{W}_{12}^{++}$}
The momentum-space expressions of $\mathcal{W}_{12}^{-+}$ and $\mathcal{W}_{12}^{++}$ can be derived using the shadow transform as in (\ref{Wpmdef}-\ref{Wppdef}). In fact, let us show that these definitions shift the parameters as we want, \ie ~ $\b_1\rar\b_1-1,\b_2\rar\b_2+1$ and $\b_1\rar\b_1+1,\b_2\rar\b_2+1$ respectively :
\begin{align}
\mathcal{W}_{12}^{-+}:\quad(\b_1,\b_2,\b_3) \xrightarrow{p_2^{-2\b_2}} (\b_1,-\b_2,\b_3)
&\xrightarrow{\mathcal{W}_{12}^{--}} (\b_1-1,-\b_2-1,\b_3 )\nn\\&
\xrightarrow{p_2^{2(\b_2+1)}} (\b_1-1,\b_2+1,\b_3 ),
\end{align}
\begin{align}
\mathcal{W}_{12}^{++}:\quad(\b_1,\b_2,\b_3) \xrightarrow{p_1^{-2\b_1}p_2^{-2\b_2}} (-\b_1,&-\b_2,\b_3)
\xrightarrow{\mathcal{W}_{12}^{--}} (-\b_1-1,-\b_2-1,\b_3 )\nn\\&
\xrightarrow{p_1^{2(\b_1+1)}p_2^{2(\b_2+1)}} (\b_1+1,\b_2+1,\b_3 ).
\end{align}
By multiplying out the right-hand sides in \eqref{Wpmdef}-\eqref{Wppdef}, we obtain their explicit expressions 
\begin{align}
\mathcal{W}^{-+}_{12}&=p_2^2 \mathcal{W}_{12}^{--} +2\beta_2\Big(\beta_2+1-\frac{d}{2}+p_2^\mu\partial_{12\mu}\Big)\\
\mathcal{W}^{+-}_{12}&=p_1^2 \mathcal{W}_{12}^{--} +2\beta_1\Big(\beta_1+1-\frac{d}{2}-p_1^\mu\partial_{12\mu}\Big),\\
\mathcal{W}_{12}^{++} 
&=p_1^2p_2^2\mathcal{W}_{12}^{--}+2\beta_1\beta_2(p_1^2+p_2^2-p_3^2)\nn\\[0ex]&\qquad
+2\beta_1 p_2^2 \Big(\beta_1+1-\frac{d}{2}-p_1^\mu\partial_{12\mu}\Big)
+2\beta_2 p_1^2 \Big(\beta_2+1-\frac{d}{2}+p_2^\mu\partial_{12\mu}\Big),
\end{align}
where $\p_{12\mu}$ denotes the difference $\p/\p p_1^\mu-\p/\p p_2^\mu$. \subsection{Intertwining relations}
In this section we conclude the discussion on shift operators presenting an algebraic method to verify the action of a shift operator. Let $\mathcal{I}_{\{d,\Delta_i\}}$ be the solution of the CWIs \eqref{DWI0}-\eqref{scWI} and $X_{i}$ a generic shift operator acting on the solution to shift $d\rightarrow d'$ and $\Delta \rightarrow \Delta_i'$, for some $i$. Then $X_{i}$ has to satisfy the following intertwining relation
\[\label{intert1}
\mathcal{K}^\mu[d',\Delta_i']X_{i}-X_{i}\mathcal{K}^\mu[d,\Delta_i]=\hat{O}_1\mathcal{K}^\mu[d,\Delta_i]+\hat{O}_2^\mu D[d,\Delta_i],
\]
where $\mathcal{K}^\mu$ is the special conformal operator in \eqref{CWIop} and $\hat{O}_j$ are some differential operators such that the homogeneity of the equation holds. Note that the right-hand side is an operator that annihilates the solution $\mathcal{I}_{\{d,\Delta_i\}}$. 
While the left-hand side holds, since
\[
0=\mathcal{K}^\mu[d',\Delta_i']\mathcal{I}_{\{d',\Delta_i'\}}=\mathcal{K}^\mu[d',\Delta_i']X_{i}\mathcal{I}_{\{d,\Delta_i\}}=X_{i}\mathcal{K}^\mu[d,\Delta_i]\mathcal{I}_{\{d,\Delta_i\}}.
\]
We can also consider the CWIs in Mandelstam variables, leading to the analogous relations:
\[\label{intert2}
\K_{ij}[d',\Delta_i']X_{i}-X_{i}\K_{ij}[d,\Delta_i]=\hat{O}_1\K_{ij}[d,\Delta_i]+\hat{O}_2D[d,\Delta_i]\quad i,j=1,2,3.
\]
By direct computation we verified that the $d$-shifting operators $\mathcal{L}_i$ and $\mathcal{R}_i$ satisfy the intertwining equation \eqref{intert2} with the right-hand side equal to zero, \ie
\[
\K_{12}[d+2,\Delta_1,\Delta_2+1,\Delta_3+1]\mathcal{L}_i=\mathcal{L}_{i}\K_{12}[d,\Delta_1,\Delta_2,\Delta_3].
\]
The weight-shifting operator $\mathcal{W}_{12}$, instead, satisfies \eqref{intert2} with a non zero right-hand side of \eqref{intert2} different from zero. To be precise, its intertwining relation is
\[
\K_{12}[d,\Delta_1-1,\Delta_2-1,\Delta_3]\mathcal{W}_{12}^{--}-\mathcal{W}_{12}^{--}\K_{12}[d,\Delta_1,\Delta_2,\Delta_3]=\left(\frac{1}{p_1}\p_1+\frac{1}{p_2}\p_2\right)\K_{12}.
\]
\section{Higher-point functions}
In Chapter \ref{cha2} we derived the general solution of the scalar 4- and $n$-point CWIs in position space. While 3-point functions are uniquely fixed by conformal symmetry, 4- and higher-point functions are less tightly constrained and depend on an arbitrary function of the cross ratios. In this section we review recent work on general $n$-point correlation functions in momentum space \cite{Bzowski:2019kwd, Bzowski:2020kfw}. The representation found has the form of a Feynman integral with the topology of an $(n-1)$-simplex, featuring an arbitrary function of momentum-space cross ratios. This will also be developed further in the following chapter.

\subsection{4-point function}
We first discuss 4-point correlation functions of scalar operators and then we generalise the results to $n$-point functions. The general 4-point function in momentum space has been shown to be expressible as a 3-loop Feynman integral, where the (massless) scalar propagators are raised to generalised powers $\a_{ij}+d/2$:
\begin{align} \label{4ptsimplex}
\langle \O_1(\bs{p}_1)\O_2(\bs{p}_2)\O_3(\bs{p}_3)\O_4(\bs{p}_4)\rangle=\left(\int \prod_{i=1}^3\frac{\mathrm{d}^d \bs{q}_{i}}{(2 \pi)^d} \right) \frac{\hat{f}(\hat{u},\hat{v})}{\text{Den}(\bs{q}_{j}, \bs{p}_k)}(2\pi)^d\delta\Big(\sum_{j=1}^4\bs{p}_j\Big),
\end{align}
where the denominator is 
\begin{align} \label{Den3}
 \text{Den}(\bs{q}_{j}, \bs{p}_k) = &q_{3}^{2\a_{12}+d} q_{2}^{2 \a_{13}+d}  q_{1}^{2 \a_{23}+d}  |\bs{p}_1 + \bs{q}_{2} - \bs{q}_{3}|^{2 \a_{14}+d} 
\nn\\
& \qquad \times
 |\bs{p}_2 + \bs{q}_{3} - \bs{q}_{1}|^{2 \a_{24}+d}  |\bs{p}_3 + \bs{q}_{1} - \bs{q}_{2}|^{2 \a_{34}+d}
\end{align}
and $\hat{f}$ is an arbitrary function that depends on the two dependent momentum-space cross ratios 
\[\label{momcross}
\hat{u}=\frac{q_2^2|\bs{p}_2+\bs{q}_3-\bs{q}_1|^2}{q_3^2|\bs{p}_3+\bs{q}_1-\bs{q}_2|^2},\qquad \hat{v}=\frac{q_1^2|\bs{p}_1+\bs{q}_2-\bs{q}_3|^2}{q_2^2|\bs{p}_2+\bs{q}_3-\bs{q}_1|^2}.
\]
They play a role analogous to the position-space cross ratios \eqref{crosspos}, however they depend on integration variables $\bs{q}_j$, and so are not independent conformal invariants in their own right.
In the following, we show the representation \eqref{4ptsimplex} is conformally invariant. First, we Fourier transform the position-space 4-point function \eqref{4pt-pos}. Then, we briefly discuss conformal invariance from a purely momentum-space perspective.

We already computed the Fourier transform of the 2-point function and this will be useful in our discussion. As a warm up example, we first Fourier transform the position-space 3-point function \eqref{3pt-pos} that we denote with $\phi_3$ here
\[
\phi_3=x_{12}^{2\a_{12}}x_{13}^{2\a_{13}}x_{23}^{2\a_{23}}.
\]
To Fourier transform $\phi_3$, we use the convolution theorem as follows
\begin{align}\label{3ptFT1}
\mathcal{F}[\phi_3]&=c_{123}\mathcal{F}\left[x_{12}^{2\a_{12}}\right]\ast\mathcal{F}\left[x_{13}^{2\a_{13}}x_{23}^{2\a_{23}}\right]\nn\\
&=c_{123}\int \frac{\dd ^d\bs{q}_1}{(2\pi)^d}\frac{\dd ^d\bs{q}_2}{(2\pi)^d}\mathcal{F}\left[x_{12}^{2\a_{12}}\right](\bs{p}_j-\bs{q}_j)\mathcal{F}\left[x_{13}^{2\a_{13}}x_{23}^{2\a_{23}}\right](\bs{p}_j),\quad j=1,2,3.
\end{align}
Using equation \eqref{2ptFT} and taking into account that the Fourier transform in the integrand also depends on $\bs{p}_3$, we have
\[\label{3ptFT12}
\mathcal{F}\left[x_{12}^{2\a_{12}}\right](\bs{p}_1,\bs{p}_2,\bs{p}_3)=(2\pi)^d\delta(\bs{p}_1+\bs{p}_2)\delta(\bs{p}_3)\frac{C_{12}}{p_1^{2\a_{12}+d}},
\]
and
\[\label{3ptFT1n}
\mathcal{F}\left[x_{13}^{2\a_{13}}x_{23}^{2\a_{23}}\right](\bs{p}_1,\bs{p}_2,\bs{p}_3)=(2\pi)^d\delta(\bs{p}_1+\bs{p}_2+\bs{p}_3)\frac{C_{13}C_{23}}{p_1^{2\a_{13}+d}p_2^{2\a_{13}+d}},
\]
where $\mathcal{F}[x_{ij}^{2\a_{ij}}...]$ denotes the Fourier transform over $\bs{p}_1$, $\bs{p}_2$ and $\bs{p}_3$. The parameters $\a_{ij}$ are given in \eqref{aij3pt} and
\[\label{Cij}
C_{ij}=\frac{\pi^{d/2}2^{2\a_{ij}+d}}{\Gamma(-\a_{ij})}\Gamma\left(\frac{d}{2}+\a_{ij}\right).
\]
Then, substituting \eqref{3ptFT12} and \eqref{3ptFT1n} in \eqref{3ptFT1}, we find
\[
\mathcal{F}[\phi_3]=c_{123}C_{12}C_{13}C_{23}\int \frac{\dd ^d\bs{q}_1}{(2\pi)^d}\frac{\dd ^d\bs{q}_2}{(2\pi)^d}\frac{(2\pi)^d\delta(\bs{p}_3+\bs{q}_1+\bs{q}_2)}{q_1^{2\a_{13}+d}q_2^{2\a_{23}+d}}\frac{(2\pi)^d\delta(\bs{p}_1+\bs{p}_2+\bs{p}_3)}{|\bs{p}_1-\bs{q}_1|^{2\a_{12}+d}}.
\]
Integrating over $\bs{q}_2$ and sending $\bs{q}_1\rightarrow -\bs{q}_1+\bs{p}_1$ we recover the 1-loop triangle integral \eqref{triangle}:
\[\label{3ptFT}
\mathcal{F}[\phi_3]=c_{123}C_{12}C_{13}C_{23}\int \frac{\text{d}^d\bs{q}}{(2\pi)^d}\frac{(2\pi)^d\delta(\bs{p}_1+\bs{p}_2+\bs{p}_3)}{|\bs{q}|^{2\a_{12}+d}|\bs{q}-\bs{p}_1|^{2\a_{13}+d}|\bs{q}+\bs{p}_2|^{2\a_{23}+d}}.
\]
Now, we move to 4-point functions. To find the simplex representation of 4-point functions  \eqref{4ptsimplex}, we Fourier transform the position-space 4-point function in identical fashion to the 3-point function. Note that in this case we also have to consider the arbitrary function. First, let us assume a monomial position-space arbitrary function $f(u,v)=u^\a v^\b$ in equation \eqref{4pt-pos}, then the position-space 4-point function is just a product of powers of $x_{ij}^2$:
\begin{align}\label{4ptab}
\phi_4(\a,\b)&=x_{12}^{2(\a_{12}+\a)}x_{13}^{2(\a_{13}-\a+\b)}x_{14}^{2(\a_{14}-\b)}x_{23}^{2(\a_{23}-\b)}x_{24}^{2(\a_{24}-\a+\b)}x_{34}^{2(\a_{34}+\a)}\nn\\&=\prod_{1\leq i< j\leq n} x_{ij}^{2\g_{ij}},
\end{align}
where
\begin{align}\label{gij}
&\g_{12}=\a_{12}+\a,& \quad & \g_{13}=\a_{13}-\a+\b,& \quad & \g_{14}=\a_{14}-\b,\nn \\
&\g_{23}=\a_{23}-\b,& \quad & \g_{24}=\a_{24}-\a+\b, & \quad & \g_{34}=\a_{34}+\a.
\end{align}
Then, we use the convolution theorem by grouping the powers of $x_{ij}$ in a way that a recursive approach can be used:
\begin{align}
\mathcal{F}[\phi_4(\a,\b)]&=\mathcal{F}\left[x_{14}^{2\g_{14}}x_{24}^{2\g_{24}}x_{34}^{2\g_{34}}\right]\ast\mathcal{F}\left[x_{12}^{2\g_{12}}x_{13}^{2\g_{13}}x_{23}^{2\g_{23}}\right]\nn\\
&=\mathcal{F}\left[x_{14}^{2\g_{14}}x_{24}^{2\g_{24}}x_{34}^{2\g_{34}}\right]\ast\mathcal{F}[\phi_3](2\pi)^d\delta(\bs{p_4}),
\end{align}
where the second factor on the right-hand side is given by the 3-point Fourier transform \eqref{3ptFT}. This makes the the recursive structure evident and will be useful for the generalisation to $n$-point functions. Taking into account that
\[
\mathcal{F}\left[x_{14}^{2\g_{14}}x_{24}^{2\g_{24}}x_{34}^{2\g_{34}}\right]=\frac{(2\pi)^d\delta\left(\sum_{j=1}^4\bs{p}_j\right)\prod_{i=1}^4C_{i4}}{p_1^{2\g_{14}+d}p_2^{2\g_{24}+d}p_3^{2\g_{34}+d}},
\]
using the result \eqref{3ptFT} and applying the convolution theorem we find
\begin{align} \label{4ptFT}
\mathcal{F}[\phi_4(\a,\b)]=\int \frac{\mathrm{d}^d \bs{q}_{1}}{(2 \pi)^d} \frac{\mathrm{d}^d \bs{q}_{2}}{(2 \pi)^d} \frac{\mathrm{d}^d \bs{q}_{3}}{(2 \pi)^d} \frac{\prod_{1\leq i< j\leq 4}C_{ij}}{\text{Den}_3^{\a\b}(\bs{q}_{j}, \bs{p}_k)}(2\pi)^d\delta\Big(\sum_{j=1}^4\bs{p}_j\Big),
\end{align}
with
\begin{align} \label{Den3ab}
 \text{Den}_3^{\a\b}(\bs{q}_{j}, \bs{p}_k) = &q_{3}^{2\g_{12}+d} q_{2}^{2 \g_{13}+d}  q_{1}^{2 \g_{23}+d}  |\bs{p}_1 + \bs{q}_{2} - \bs{q}_{3}|^{2 \g_{14}+d} 
\nn\\
& \qquad \times
 |\bs{p}_2 + \bs{q}_{3} - \bs{q}_{1}|^{2 \g_{24}+d}  |\bs{p}_3 + \bs{q}_{1} - \bs{q}_{2}|^{2 \g_{34}+d},
\end{align}
where the $\g_{ij}$ depend on $\a$ and $\b$ as in \eqref{gij}. Note that this is the right-hand side of \eqref{4ptsimplex} with the momentum-space arbitrary function
\[\label{fhatpowers}
\hat{f}(\hat{u},\hat{v})=\prod_{1\leq i< j\leq 4}C_{ij}\hat{u}^\a \hat{v}^\b.
\]
So far we have shown that a 3-simplex integral \eqref{4ptsimplex} with $\hat{f}$ in \eqref{fhatpowers} is a solution of CWIs. Next, we want to show that this is also valid for any arbitrary function $\hat{f}(\hat{u},\hat{v})$. To see this, we take into account that \eqref{4ptFT} is a solution of CWIs and use the inverse Mellin transform to express a general $\hat{f}$:
\[\label{mellin}
\hat{f}(\hat{u},\hat{v})=\frac{1}{(2\pi i)^2}\int_{a-i\infty}^{a+i\infty}\int_{b-i\infty}^{b+i\infty}\dd \a \:\dd \b \:\rho(\a,\b) \hat{u}^\a\hat{v}^\b,
\]
for some appropriate choice of integration contour specified by $a$ and $b$. Since equation \eqref{4ptsimplex} can be written as
\begin{align} 
\langle \O_1(\bs{p}_1)\O_2(\bs{p}_2)\O_3(\bs{p}_3)\O_4(\bs{p}_4)\rangle=\frac{1}{(2\pi i)^2}\int_{a-i\infty}^{a+i\infty}\int_{b-i\infty}^{b+i\infty}\dd \a \: \dd \b \: \rho(\a,\b)W_{\a,\b},
\end{align}
where we defined
\[
W_{\a,\b}=\frac{1}{\prod_{1\leq i< j\leq 4}C_{ij}}\mathcal{F}[\phi_4(\a,\b)],
\]
we showed that \eqref{4ptsimplex} is the general conformal 4-point function in momentum space. Note that the solution \eqref{4ptsimplex} and the Fourier transform \eqref{4ptFT} differs by a factor consisting in the product of the constants $C_{ij}$ defined in \eqref{Cij}.
\subsubsection{Conformal invariance via total derivative}
In this section we discuss the solution \eqref{4ptsimplex} from the point of view of momentum space only. We will briefly discuss why \eqref{4ptsimplex} is a solution of the momentum-space CWIs \eqref{DWI0}, \eqref{scWI}.

The DWI for the reduced $n$-point conformal correlators given in \eqref{DWI0} constrains their expressions to scale as $\Delta_t-(n-1)d$. For $n=4$, it means that the conformal 4-point function must scale as $\Delta_t-3d$. By power counting, we find that this is indeed the scaling of the simplex representation \eqref{4ptsimplex} of 4-point functions. In fact, the three integrals contribute with dimension $3d$, and each propagator scales as $-2\a_{ij}-d$, with $1\leq i<j\leq 4$. Hence the scaling dimension of \eqref{4ptsimplex} is
\[
-2\sum_{1\leq i<j\leq 4}\a_{ij}-6d+3d=-\sum_{i,j=1}^4\a_{ij}-3d=\Delta_t-3d,
\]
where in the last equality we used the equation \eqref{par} relating $\a_{ij}$ and the scaling dimensions of scalar operators. 

Let us now move to the special conformal Ward identity. First, we recall that this is given in \eqref{CWIop}, and in this case $\mathcal{K}^\mu=\mathcal{K}_1^\mu+\mathcal{K}_2^\mu$, with
\[\label{Kj}
\mathcal{K}_j^\mu=2(\Delta_j-d)\frac{\p}{\p p_{j\mu}}-2p_j^\nu\frac{\p}{\p p_j^\nu}\frac{\p}{\p p_{j\mu}}+p_j^\mu\frac{\p}{\p p_j^\nu}\frac{\p}{\p p_{j\nu}}.
\]
To prove that \eqref{4ptsimplex} satisfies the 4-point special conformal Ward identity we have to show that the action of the SCWI operator $\mathcal{K}^\mu$ on the integrand of the simplex representation corresponds to a total derivative. Taking into account that $\mathcal{K}^\mu$ is a second-order differential operator, we expect its action on the integrand of \eqref{4ptsimplex} to be of the form
\[
\mathcal{K}^\mu\left(\frac{\hat{f}(\hat{u},\hat{v})}{\text{Den}(\bs{q}_{j}, \bs{p}_k)}\right)=\sum_{j=1}^2\frac{\p}{\p q_j^k}\left(A_j^{k\mu}\hat{f}+B_j^{k\mu}\frac{\p \hat{f}}{\p \hat{u}}+C_j^{k\mu}\frac{\p \hat{f}}{\p \hat{v}}\right),
\]
for some coefficients $A_j^{k\mu},B_j^{k\mu}$ and $C_j^{k\mu}$ that are \emph{independent} of the arbitrary function $\hat{f}$. Their explicit computation and expressions can be found in \cite{Bzowski:2019kwd} and \cite{Bzowski:2020kfw}. In the following chapter we will prove the validity of the $n$-point CWIs using a new scalar parametrisation of the simplex integral. Hence, we will omit further details here of the explicit computation of these coefficients.

Before showing some special solutions of 4-point CWIs, we first extend the simplex representation to the $n$-point functions. This is, indeed, a generalisation of what we analysed in this section.

\subsection[{\texorpdfstring{$n$-point function: the simplex representation}{n-point function: the simplex representation}}]{\texorpdfstring{\boldmath{$n$}-point function: the simplex representation}{n-point function: the simplex representation}}
In this section we present the conformal $n$-point function of scalar operators known as \emph{simplex integral}. This is a Feynman integral with the topology of an $(n-1)$-simplex, featuring an arbitrary function of momentum-space cross ratios. This generalises the $4$-point solution we introduced in the previous section and reads
\begin{align} \label{simplex0}
&\< \O_1(\bs{p}_1) \ldots \O_n(\bs{p}_n) \> =
 \prod_{1 \leq i < j \leq n} \int \frac{\dd ^d \bs{q}_{ij}}{(2 \pi)^d} \frac{\hat{f}(\hat{\bs{u}})}{q_{ij}^{2 \alpha_{ij} + d}}  \prod_{k=1}^n (2\pi)^d \delta \Big( \bs{p}_k + \sum_{l=1}^n \bs{q}_{lk} \Big),
\end{align}
where $\bs{q}_{ij}$ is the momentum running in the oriented edge from vertex $i$ to $j$ so that $\bs{q}_{ij} = -\bs{q}_{ji}$  and $\bs{q}_{jj} = 0$. The parameters $\a_{ij}$ are the same appearing in the position-space solution and satisfy the condition \eqref{par}. We denoted the set of independent cross ratios with $\bs{\hat{u}}$, collecting the independent \emph{momentum-space cross ratios}:
\[ \label{conf_ratio_q0}
\hat{u}_{[pqrs]} = \frac{q_{pq}^2 q_{rs}^2}{q_{pr}^2 q_{qs}^2}.
\]
Since the simplex integral is related to the position-space $n$-point solution through a Fourier transform, their number follows from the position-space argument explained at the end of chapter \ref{cha2}, \ie ~ there are $n(n-3)/2$ for $n\leq d+2$ and $nd-(d+2)(d+1)/2$ when $n>d+2$. 
Moreover, as noted for the 4-point functions, momentum-space cross ratios depend on the integration variables.

Note that the integrand in \eqref{simplex0} contains a product of $n$ delta functions. To define the \emph{reduced simplex integral} (denoted with the double-brackets), we set aside the delta function corresponding to momentum conservation
\[
\< \O_1(\bs{p}_1) \ldots \O_n(\bs{p}_n) \> = (2 \pi)^d \delta \Big( \sum_{i=1}^n \bs{p}_i \Big) \lla \O_1(\bs{p}_1) \ldots \O_n(\bs{p}_n) \rra,
\]
so that the reduced simplex integral depends only on $n-1$ independent external momenta. Then, we are left with $n-1$ delta functions. We choose the loop-parametrisation of the simplex by integrating over the variables $\bs{q}_{in}$ for $i=1,2,\ldots, n-1$. We obtain the following reduced simplex integral
\begin{align} \label{simplex_red}
\lla \O_1(\bs{p}_1) \ldots \O_n(\bs{p}_n) \rra = \prod_{1 \leq i < j \leq n-1} \int \frac{\dd^d \bs{q}_{ij}}{(2 \pi)^d} \frac{\hat{f}(\hat{\bs{u}})}{\text{Den}_{n}(\bs{\alpha})}
\end{align}
where the denominator reads
\begin{align}
\text{Den}_{n}(\bs{\alpha}) & = \prod_{1 \leq i < j \leq n-1} q_{ij}^{2 \alpha_{ij} + d} \times \prod_{m=1}^{n-1} |\bs{l}_m - \bs{p}_m|^{2 \alpha_{mn} + d} \label{Den}
\end{align}
and $\bs{l}_m$ depends only on the remaining internal momenta,
\begin{align}
\bs{l}_m =- \bs{q}_{mn} + \bs{p}_m = \sum_{j=1}^{n-1} \bs{q}_{mj} = - \sum_{j=1}^{m-1} \bs{q}_{jm} + \sum_{j=m+1}^{n-1} \bs{q}_{mj}.
\end{align}
The integral \eqref{simplex_red} displays $(n-1)(n-2)/2$ $d$-dimensional integrals. In the following chapter we describe instead new purely scalar parametrisations for the simplex which, amongst other advantages, feature fewer integrals.

\subsubsection{Conformal invariance via Fourier transform}
Here we want to understand the simplex integral \eqref{simplex0} from Fourier transforming the position-space $n$-point function \eqref{nptbis}. The idea is to use the recursive structure already illustrated in the previous section. A more direct proof of conformal invariance for the simplex integral will be given in the next chapter. 

We follow the same strategy used in the previous section, \ie ~we first consider a monomial position-space arbitrary function $f$ and derive the Fourier transform of the associated $n$-point function. We will see that this corresponds to the simplex integral with a monomial momentum-space arbitrary function. To show that for any generic arbitrary function the simplex is conformally invariant, we then take an inverse Mellin transform.\\Let $F_n$ be the position-space $n$-point correlator \eqref{nptbis} with a monomial arbitrary function. This is expressed as a product of powers of the independent cross ratios 
\[\label{Fn}
F_n(\bs{\alpha}; \bs{x}_1,\ldots,\bs{x}_n) =\prod_{1\leq i<j\leq n}x_{ij}^{2\a_{ij}},
\]
where the parameters $\a_{ij}$ satisfy \eqref{par}. We want to show that its Fourier transform is
\begin{align}\label{mesh}
\mathcal{F}_{(n)}[F_n]&=
\prod_{1 \leq i < j \leq n} C_{ij} \int \frac{\dd^d \bs{q}_{ij}}{(2 \pi)^d} \frac{1}{q_{ij}^{2 \alpha_{ij} + d}}  \prod_{k=1}^n (2\pi)^d \delta \Big( \bs{p}_k + \sum_{l=1}^n \bs{q}_{lk} \Big),
\end{align}
where $C_{ij}$ is given in \eqref{Cij}. This integral is referred to as the \emph{mesh integral} in \cite{Bzowski:2020lip}, corresponding to a generalised Feynman integral with $n$ points and $n(n-1)/2$ generalised (scalar, massless) propagators with every pair of points connected. Note that with `generalised' we intend that the powers $\a_{ij}$ are not necessarily equal to one. This integral also corresponds to the simplex integral \eqref{simplex0} when $\hat{f}=\prod_{1\leq i <j\leq n}C_{ij}$. First, let us note that \eqref{Fn} is satisfied for $n=2$:
\begin{align}
\mathcal{F}_{(2)}[x_{12}^{2\a_{12}}](\alpha_{12}; \bs{p}_1, \bs{p}_2) & = C_{12} \int \frac{\dd^d \bs{q}_{12}}{(2 \pi)^d} \frac{1}{q_{12}^{2 \alpha_{12} + d}} (2 \pi)^d \delta(\bs{p}_1 - \bs{q}_{12}) (2 \pi)^d \delta(\bs{p}_2 + \bs{q}_{12}) \nn\\
& = (2 \pi)^d \delta(\bs{p}_1 + \bs{p}_2) \frac{C_{12}}{p_1^{2 \alpha_{12} + d}}.
\end{align}
This result is in agreement with the Fourier transform of the 2-point function \eqref{2ptFT}. Before we perform the Fourier transform of \eqref{Fn} and make the recursive structure manifest, let us introduce the following notation 
\begin{align}
&\mathcal{F}_{(n)}[F_n]=\int \dd ^d \bs{x}_1\ldots \dd ^d\bs{x}_n\mathrm{e}^{-i\sum_{j=1}^n\bs{x}_j\cdot \bs{p}_j} F_n(\bs{\alpha}; \bs{x}_1,\ldots,\bs{x}_n) \nn\\
&\mathcal{F}_{(n)}[F_{n-1}]=\int \dd ^d \bs{x}_1\ldots \dd ^d\bs{x}_n\mathrm{e}^{-i\sum_{j=1}^n\bs{x}_j\cdot \bs{p}_j} F_{n-1}(\bs{\alpha}; \bs{x}_1,\ldots,\bs{x}_{n-1}) ,
\end{align}
and one can show that
\[
\mathcal{F}_{(n)}[F_{n-1}]=\mathcal{F}_{(n-1)}[F_{n-1}](2\pi)^d\delta(\bs{p}_n).
\]
We are now ready to prove equation \eqref{mesh} by induction. The key point is the recursive structure in $n$ of $\mathcal{F}_{(n)}[F_n]$. We already showed this property for $n\leq 4$ at the beginning of this section. We discuss it more systematically here, for any $n$. Let us consider the position-space function $F_n$ and factorise it to display a recursive structure (see figure \ref{fig:recur}):
\[\label{Fnpos}
F_n=x_{1n}^{2\a_{1n}}x_{2n}^{2\a_{2n}}\ldots x_{n-1,n}^{2\a_{n-1,n}}\times F_{n-1}.
\]

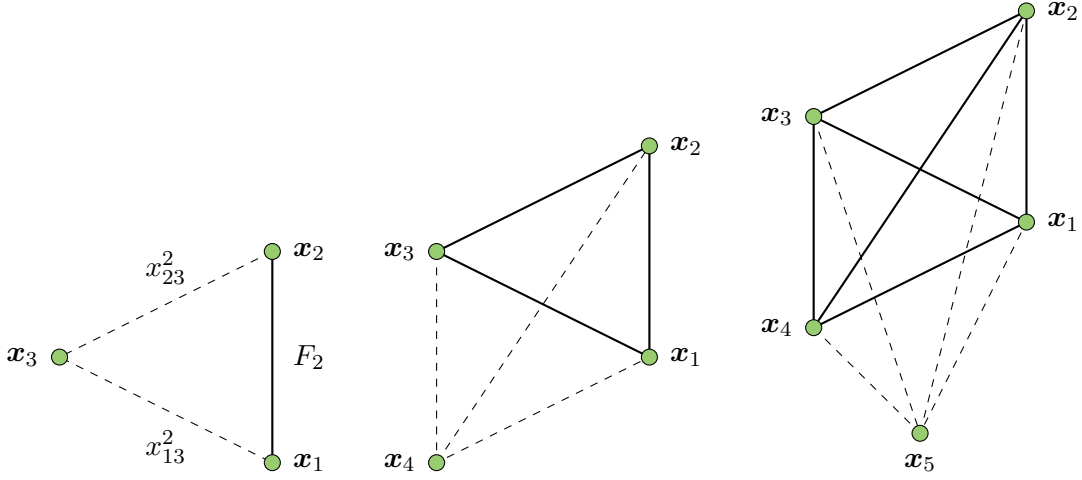
\begin{figure}[t]
\centering
\def\r{6pt}
\tikzset{d/.style={draw,circle,fill=YellowGreen,minimum size=\r,inner sep=0pt, anchor=center}}
\begin{tikzpicture}[scale=1.4]
   \draw[thick] (2,-1)--(2,1); 
   \draw[dashed] (2,1)-- (0,0) -- (2,-1);

   \node[right ] at (2.1,-1) {$\x_1$};
   \node[right ] at (2.1,1) {$\x_2$};
   \node[left ] at (-0.1,0) {$\x_3$};

   \node[above] at (1,0.6) {$x_{23}^2$};
      \node[below] at (1,-0.6) {$x_{13}^2$};
         \node[right] at (2.1,0) {$F_2$};
         
    \node[d] at (2,-1) {};
   \node[d ] at (2,1) {};
   \node[d] at (0,0) {};

\end{tikzpicture}
\quad
\begin{tikzpicture}[scale=1.4]
   \draw[thick] (2,-1)--(2,1); 
   \draw[thick] (2,1)-- (0,0) -- (2,-1);
   \draw[dashed] (0,0)-- (0,-2) -- (2,-1);
   \draw[dashed] (0,-2)--(2,1);
   
   \node[right ] at (2.1,-1) {$\x_1$};
   \node[right ] at (2.1,1) {$\x_2$};
   \node[left ] at (-0.1,0) {$\x_3$};
   \node[left ] at (-0.1,-2) {$\x_4$};
   
   \node[d ] at (2,-1) {};
   \node[d ] at (2,1) {};
   \node[d ] at (0,0) {};
   \node[d ] at (0,-2) {};

\end{tikzpicture}
\quad
\begin{tikzpicture}[scale=1.4]
   \draw[thick] (2,-1)--(2,1); 
   \draw[thick] (2,1)-- (0,0) -- (2,-1);
   \draw[thick] (0,0)-- (0,-2) -- (2,-1);
   \draw[thick] (0,-2)--(2,1);
   \draw[dashed] (0,-2)--(1,-3)--(2,-1);
   \draw[dashed] (0,0)--(1,-3)--(2,1);
   
   \node[right ] at (2.1,-1) {$\x_1$};
   \node[right ] at (2.1,1) {$\x_2$};
   \node[left ] at (-0.1,0) {$\x_3$};
   \node[left ] at (-0.1,-2) {$\x_4$};
   \node[below ] at (1,-3.1) {$\x_5$};
   
     \node[d ] at (2,-1) {};
   \node[d ] at (2,1) {};
   \node[d ] at (0,0) {};
   \node[d ] at (0,-2) {};
   \node[d ] at (1,-3) {};

\end{tikzpicture}

\caption{Illustration of the recursive structure of $F_n$ \eqref{Fnpos}. The continuous lines denote $F_{n-1}$, and the dashed lines correspond to $x_{jn}^2$ ($j=1,\ldots n-1$).}
\label{fig:recur}
\end{figure}

Then, we Fourier transform it by applying the convolution theorem:
\begin{align}\label{induc}
&\mathcal{F}_{(n)}[F_n]=\mathcal{F}_{(n)}[x_{1n}^{2\a_{1n}}x_{2n}^{2\a_{2n}}\ldots x_{n-1,n}^{2\a_{n-1,n}}]\ast \mathcal{F}_{(n)}[F_{n-1}]\nn\\
& = \left[ \frac{(2 \pi)^d \delta \left( \sum_{j=1}^n \bs{p}_j \right)\prod_{i=1}^{n-1}C_{in}}{p_1^{2 \alpha_{1n} +d} p_2^{2 \alpha_{2n} +d} \ldots p_{n-1}^{2 \alpha_{n-1,n} +d}} \right] \ast \left[ \mathcal{F}_{(n-1)}[F_{n-1}](2\pi)^d\delta(\bs{p}_n) \, (2 \pi)^d \delta(\bs{p}_n) \right] \nn\\&=
\prod_{i = 1}^{n-1} C_{in} \int \frac{\dd^d \bs{q}_{i}}{(2 \pi)^d} \frac{\mathcal{F}_{(n-1)}[F_{n-1}](\bs{\alpha}; \bs{p}_1 - \bs{q}_1, \ldots, \bs{p}_{n-1} - \bs{q}_{n-1})}{q_1^{2 \alpha_{1n} + d} q_2^{2 \alpha_{2n} + d} \ldots q_{n-1}^{2 \alpha_{n-1,n} + d}} (2 \pi)^d \delta \Big( \bs{p}_n + \sum_{j=1}^{n-1} \bs{q}_j \Big)
\end{align} 
Now, first we rename $\bs{q}_j\rightarrow \bs{q}_{jn}$ ($j=1,...,n-1$). Then, taking into account that equation \eqref{mesh} holds for $n=2$, equation \eqref{induc} implies that \eqref{mesh} holds for all $n\geq 2$ by induction. This proves that the simplex integral \eqref{simplex0} is conformally invariant when $\hat{f}$ is a monomial in the momentum-space cross ratios. In fact, this is equivalent to shifting the $\a_{ij}$ by some amount and re-defining $\hat{f}=1$, as we showed for $n=4$. In other words, the simplex integral with monomial $\hat{f}$ satisfies the CWIs just as the simplex with $\hat{f}=1$.

Finally, to prove that the simplex integral is conformally invariant for any arbitrary function, we express $\hat{f}$ as an inverse Mellin transform in identical fashion we did in equation \eqref{mellin} for $n=4$. 
\subsubsection{Conformal invariance via CWIs}
To directly prove that the simplex integral \eqref{simplex0} satisfies the CWIs we proceed in an analogous way of $n=4$. First, to show the scale invariance of \eqref{simplex_red}, we need to verify that it scales as $\Delta_t-(n-1)d$. Each integration increases the dimension by $d$, while each propagator decreases the dimension by $2\a_{ij}+d$. Taking into account that the number of integrals is $(n-1)(n-2)/2$ and the number of propagators appearing in the denominator \eqref{Den} is $(n-1)(n-2)/2+(n-1)$, the total scaling is
\[
\frac{1}{2}(n-1)(n-2)d-\left[\frac{1}{2}(n-1)(n-2)-(n-1)\right]d-\sum_{1\leq i<j \leq n}2\a_{ij}=\Delta_t-(n-1)d,
\]
where we used \eqref{par}. Hence, the DWI \eqref{DWI0} is satisfied. 

To prove that the simplex solves the SCWIs, one strategy is to show that the SCWI operator $\mathcal{K}^\mu=\sum_{j=1}^{n-1}\mathcal{K}_j^\mu$ acting on the integrand of \eqref{simplex_red} is equivalent to a sum of total derivatives with respect to $\bs{q}_{ij}$. This has been shown in \cite{Bzowski:2020lip}, here we cite the result:
\begin{align} \label{to_show_simp_k}
& \mathcal{K}^{\kappa}(\bs{\Delta}) \left[ \frac{\hat{f}(\hat{\bs{u}})}{\text{Den}_n(\bs{\alpha})} \right] = \sum_{\substack{i,j = 1\\i \neq j}}^{n-1} \frac{\partial}{\partial q_{ij}^\mu} \left[ \Gamma_{ij}^{\kappa \mu}(\bs{\alpha}) \hat{f}(\hat{\bs{u}}) + \sum_{I \in \mathcal{U}} \Gamma^{\kappa \mu}_{ij, I}(\bs{\alpha}) \frac{\partial \hat{f}(\hat{\bs{u}})}{\partial \hat{u}_I}  \right],
\end{align}
with
\begin{align}
\Gamma^{\kappa \mu}_{ij}(\bs{\alpha}) & = (2 \alpha_{in} + d) \times \frac{A^{\kappa \mu}_{ij}}{\text{Den}_n(\bs{\alpha})}, \label{G0} \\
\Gamma^{\kappa \mu}_{ij, [pqrs]}(\bs{\alpha}) & = 2 (\delta_{ip} \delta_{rn} + \delta_{iq} \delta_{sn} - \delta_{ip} \delta_{qn} - \delta_{ir} \delta_{sn} ) \times \frac{A^{\kappa \mu}_{ij} \hat{u}_{[pqrs]}}{\text{Den}_n(\bs{\alpha})}, \label{GI}
\end{align}
and
\begin{align} \label{A_def}
& A_{ij}^{\kappa \mu} = (\delta^{\kappa \mu} \delta_{\alpha \beta} + \delta^\kappa_\beta \delta^\mu_\alpha - \delta^\kappa_\alpha \delta^\mu_\beta ) \frac{q_{ij}^{\alpha} (\bs{l}_i - \bs{p}_i)^\beta}{(\bs{l}_i - \bs{p}_i)^2}.
\end{align}
In the second part of this thesis, in Chapter \ref{cha4}, we will show in a more direct manner that the simplex solves the CWIs by using new scalar representations of the simplex.

\subsection{4-point Ward identities: an example}
In the previous section we showed that the general 4-point function depends on an arbitrary function $\hat{f}$ of momentum-space cross ratios. This means that different 4-point functions associated to different $\hat{f}$ exist. 
Certain classes of solutions for the 4-point CWIs are known, including Witten diagrams and free fields \cite{Baumann:2019oyu, Bzowski:2022rlz, Arkani-Hamed:2018kmz, Coriano:2019sth, Raju:2012zs, Serino:2020pyu}. The corresponding simplex representations of these solutions including the form of the function $\hat{f}$ have been analysed in \cite{Bzowski:2022rlz, Bzowski:2020kfw}.
Amongst the conformal integrals appearing in the second part of this thesis, we study contact and exchange Witten diagrams \cite{Witten:1998qj, DHoker:1998ecp}. Therefore, we conclude this chapter with an exercise: proving that these 4-point Witten diagrams solve the CWIs.
Both diagrams consist of integrals of multiple Bessel functions.\\First let us derive the 4-point CWIs in terms of the scalar variables $p_1,p_2,p_3,p_4,s,t$, with
\[
s^2=(\bs{p}_1+\bs{p}_2)^2,\qquad t^2=(\bs{p}_2+\bs{p}_3)^2.
\]
Using the chain rule,
\begin{align}\label{chain4}
&\frac{\p}{\p p_{1\mu}}=\frac{p_1^{\mu}}{p_1}\frac{\p}{\p p_1}+\frac{p_1^{\mu}+p_2^{\mu}+p_3^{\mu}}{p_4}\frac{\p}{\p p_4}+\frac{p_1^{\mu}+p_2^\mu}{s}\frac{\p}{\p s},\notag\\
&\frac{\p}{\p p_{2\mu}}=\frac{p_2^{\mu}}{p_2}\frac{\p}{\p p_2}+\frac{p_1^{\mu}+p_2^{\mu}+p_3^{\mu}}{p_4}\frac{\p}{\p p_4}+\frac{p_1^{\mu}+p_2^\mu}{s}\frac{\p}{\p s}+\frac{p_2^{\mu}+p_3^\mu}{t}\frac{\p}{\p t},\notag\\
&\frac{\p}{\p p_{3\mu}}=\frac{p_3^{\mu}}{p_3}\frac{\p}{\p p_3}+\frac{p_1^{\mu}+p_2^{\mu}+p_3^{\mu}}{p_4}\frac{\p}{\p p_4}+\frac{p_2^{\mu}+p_3^\mu}{t}\frac{\p}{\p t},
\end{align}
we obtain the DWI 
\begin{align}\label{4ptDWI}
0=\left[-\Delta_t+3d+\sum_{i=1}^4p_i\p_i+s\p_s+t\p_t\right]\lla \O_1(\bs{p}_1)\O_2(\bs{p}_2)\O_3(\bs{p}_3)\O_4(\bs{p}_4)\rra
\end{align}
and the SCWIs \cite{Arkani-Hamed:2018kmz}
\[\label{4ptSCWI}
0=\mathcal{D}_{ij}\lla \O_1(\bs{p}_1)\O_2(\bs{p}_2)\O_3(\bs{p}_3)\O_4(\bs{p}_4)\rra,\quad i,j=1,...,4,
\]
where
\begin{align}
&\mathcal{D}_{12}=K_{12}+(L_1-L_2-L_3+L_4)\frac{1}{t}\p_t+(-p_3^2+p_4^2)\frac{1}{st}\p_s\p_t,\\
&\mathcal{D}_{23}=K_{23}+(L_1+L_2-L_3-L_4)\frac{1}{s}\p_s+(p_1^2-p_4^2)\frac{1}{st}\p_s\p_t,\\
&\mathcal{D}_{34}=K_{34}+(-L_1+L_2+L_3-L_4)\frac{1}{t}\p_t+(-p_1^2+p_2^2)\frac{1}{st}\p_s\p_t,
\end{align}
are the independent special conformal operators and we defined $L_i=p_i\p_i-\Delta_i$.
\\Now, let us assume we seek a solution to these equations that does not depend on $s$ and $t$. Then the SCWIs simplify to
\[
0=\K_{ij}\lla \O_1(\bs{p}_1)\O_2(\bs{p}_2)\O_3(\bs{p}_3)\O_4(\bs{p}_4)\rra,\quad i,j=1,...,4.
\]
By comparing these equations with the 3-point SCWIs in \eqref{cwisum}, it becomes evident that they can also be solved by separation of variables, giving an integral of four Bessel-K functions. Since the DWI must also be satisfied, we find
\[\label{4K}
\lla\O_1(\bs{p}_1)\O_2(\bs{p}_2)\O_3(\bs{p}_3)\O_4(\bs{p}_4)\rra=c\int_0^{\infty} \dd  x \: x^{d-1}\prod_{j=1}^4p_j^{\b_j}K_{\b_j}(p_jx).
\]
This indeed coincides with the 4-point contact Witten diagram, $i_{[d;\,\Delta_1, \Delta_2; \,\Delta_3, \Delta_4; ]}$, when
\[
c=\left(\prod_{j=1}^4 2^{\b_j-1}\Gamma\left(\b_j\right)\right)^{-1}.
\]
Let us now move to the $4$-point $s$-channel exchange diagram. This is also an integral of multiple Bessel functions and it reads 
\begin{align}\label{iexch0}
 i_{[d;\,\Delta_1, \Delta_2; \,\Delta_3, \Delta_4; \,\Delta_x]}
 &= \int_0^\infty \dd z \, z^{-d-1}\mathcal{K}_{[\Delta_1]}(z, p_1) \mathcal{K}_{[\Delta_2]}(z, p_2) 
\\ &\qquad \times
 \int_0^\infty \dd \z \, \z^{-d-1} \mathcal{G}_{[\Delta_x]}(z, s; \z) \mathcal{K}_{[\Delta_3]}(\z, p_3) \mathcal{K}_{[\Delta_4]}(\z, p_4),\nn
\end{align}
where $\Delta_x$ is the dimension of the exchanged operator and $\mathcal{G}_{[\Delta_x]}$ denotes the bulk-to-bulk propagator 
\begin{align} \label{Gprop0}
	\mathcal{G}_{[\Delta_x]}(z, s; \z) = \left\{ \begin{array}{ll}
		(z \z)^{\frac{d}{2}} I_{\beta_x}(s z) K_{\beta_x}(s \z) & \text{ for } z < \z, \\
		(z \z)^{\frac{d}{2}} K_{\beta_x}(s z) I_{\beta_x}(s \z) & \text{ for } z > \z,
	\end{array} \right.	
\end{align}
with $I_{\beta}$ and $K_{\beta}$ representing modified Bessel functions and $\beta_x=\Delta_x-d/2$. We will not derive this result here, which is a result known in holographic CFT. Our goal here is, instead, to show that \eqref{iexch0} solves the 4-point CWIs \eqref{4ptDWI}, \eqref{4ptSCWI}. First, we note that $s$-channel exchange diagrams don't depend on $t$. Hence the SCWIs operators reduces to
\begin{align}\label{CWIred}
&\mathcal{D}_{12}=K_{12},\nn\\
&\mathcal{D}_{23}=K_{23}+(L_1+L_2-L_3-L_4)\frac{1}{s}\p_s ,\nn\\
&\mathcal{D}_{34}=K_{34}.
\end{align}
Taking into account that $\K_{12}=K_1-K_2$, with $K_i$ the Bessel operator in \eqref{Keq}, by direct computation we find that the following SCWIs are satisfied
\begin{align}
0=\K_{12} i_{[d;\,\Delta_1, \Delta_2; \,\Delta_3, \Delta_4; \,\Delta_x]}=\K_{34} i_{[d;\,\Delta_1, \Delta_2; \,\Delta_3, \Delta_4; \,\Delta_x]}.
\end{align}
To prove the second SCWI involving $\mathcal{D}_{23}=0$, we first introduce the Casimir operator. This is also useful to derive the action of the operator $\mathcal{W}_{12}^{--}$ on both contact and exchange diagrams. The quadratic Casimir operator in momentum-space reads
\begin{align}
\begin{aligned}
\mathcal{C}_{12}=&(\bs{p}_1\cdot\bs{p}_2\delta^{\mu\nu}-2p_1^\mu p_2^\nu)\p_{12}^\mu\p_{12}^\nu+2[(\Delta_1-d)p_2^\mu-(\Delta_2-d)p_1^\mu]\p_{12,\mu}\\&+(\Delta_1+\Delta_2-2d)(\Delta_1+\Delta_2-d).
\end{aligned}
\end{align}
In Mandelstam variables, omitting terms involving derivatives with respect to $t$, this is
\[
\mathcal{C}_{12}=\frac{1}{2}(s^2+p_1^2-p_2^2)K_1+\frac{1}{2}(s^2+p_2^2-p_1^2)K_2-(L_1+L_2+\frac{3d}{2})^2+\frac{d^2}{4}+O(\p_t).
\]
Then the action of the Casimir operator on the exchange diagram amounts to the action of the following reduced operator \cite{Bzowski:2022rlz}
\[\label{C12red}
\tilde{\mathcal{C}}_{12}=\frac{s^2}{2}(K_1+K_2)-(L_1+L_2+\frac{3d}{2})^2+\frac{d^2}{4},
\]
since the exchange diagram satisfies $K_{12} i_{[d;\,\Delta_1, \Delta_2; \,\Delta_3, \Delta_4; \,\Delta_x]}=\K_{34} i_{[d;\,\Delta_1, \Delta_2; \,\Delta_3, \Delta_4; \,\Delta_x]}=0$. This reduced operator has the property that it sends an exchange diagram to a contact diagram as follows 
\[\label{C12redact}
(\tilde{\mathcal{C}}_{12}+m_x^2)i_{[d;\,\Delta_1, \Delta_2; \,\Delta_3, \Delta_4; \,\Delta_x]}=i_{[d;\,\Delta_1, \Delta_2; \,\Delta_3, \Delta_4]},
\]
where $m_x^2=\Delta_x(\Delta_x-d)$. We are now ready to prove that the $s$-channel exchange diagram satisfies the remaining SCWI $\mathcal{D}_{23}i_{[d;\,\Delta_1, \Delta_2; \,\Delta_3, \Delta_4; \,\Delta_x]}=0$. Taking into account $\mathcal{D}_{23}$ in \eqref{CWIred} and using the DWI (dropping the derivative with respect to $t$) to eliminate the derivative with respect to $s$, we have
\[
s^2\mathcal{D}_{23}=s^2 K_{23} + (L_1+L_2-L_3-L_4)(-3d - L_1-L_2-L_3-L_4),
\]
and by rearranging, we find
\[
s^2\mathcal{D}_{23}= s^2 K_{23} - (L_1 + L_2+3d/2)^2 + (L_3+L_4+3d/2)^2 = \tilde{C}_{12} - \tilde{C}_{34},
\]
where we used \eqref{C12red}. Thus, when acting on the exchange diagram
\[
s^2\mathcal{D}_{23}i_{[d;\,\Delta_1, \Delta_2; \,\Delta_3, \Delta_4; \,\Delta_x]}=(\tilde{C}_{12} - \tilde{C}_{34})i_{[d;\,\Delta_1, \Delta_2; \,\Delta_3, \Delta_4; \,\Delta_x]}=0,
\]
where in the last equality we used the action of the reduced Casimir operator \eqref{C12redact} and took into account that $\tilde{C}_{12}$ and $\tilde{C}_{34}$ when acting on the exchange diagram give the same contact diagram, hence the action of their difference on the latter vanishes. 

In section \ref{wshiftop} we introduced the weight-shifting operators $\mathcal{W}_{12}^{\pm\pm}$ acting on any $n$-point functions. We computed the action of $\mathcal{W}_{12}^{--}$ on the 3-point function \eqref{W12on3pt}, showing that it generates a shifted 3-point function. While the 3-point function is unique, 4-point functions are not. For instance, here we considered two types of 4-point functions, the contact and $s$-channel Witten diagrams. As a consequence, the action of the weight-shifting operator $\mathcal{W}_{12}^{\pm\pm}$ on a 4-point function does not a priori generate the same function with shifted parameters. In fact, in \cite{Bzowski:2022rlz} it has been shown that the operator $\mathcal{W}_{12}^{\pm\pm}$ acts on an exchange Witten diagram to generate a linear combination of a shifted exchange and a shifted contact diagrams, or equivalently a shifted exchange diagram but with derivative vertices. Hence, it does not generate the same function with shifted parameters. A natural question then arises: is there a weight-shifting operator that when acting on 4-point Witten diagrams preserves the form of the function and only shifts the parameters? We show the answer in Chapter \ref{cha5}.
\section{Discussion}
In this chapter we gave an overview of conformal field theory in momentum space, focusing on the scalar sector. We derived the $n$-point CWIs and discussed their solutions. We devoted considerable space to the 3-point function, presenting its equivalent representations and singularities. We constructed the shift operators $\mathcal{L}_i$ and $\mathcal{R}_i$ \eqref{LRdef00}, which connect 3-point functions in spacetime dimensions $d$ differing by two. Moreover, we presented the shift operators $\mathcal{W}_{ij}^{\pm\pm}$ that connect $n$-point functions with shifted scaling dimensions but same $d$. We then derived the general solutions of 4- and $n$-point CWIs that were found recently in terms of the simplex integral \eqref{simplex0}.

While various studies of $n$-point functions yielded special classes of solutions to the 4-point CWIs, the simplex integral provides the \emph{general} solution. Several questions arise. Is there a scalar representation of the simplex integral that simplifies the study of $n$-point functions? What is the generalisation of the shift operators $\mathcal{L}_i$ at $n$ points? Is there a representation of the simplex integral that helps us to find this class of operators? Moreover, proving that the simplex integral satisfies the CWIs was cumbersome. Is there a representation that simplifies this computation? We address these questions in the next chapter, where we find new scalar parametrisations of the integral by using insights from the physics of electrical circuits.

We concluded this chapter by showing that 4-point contact and exchange Witten diagrams solve the CWIs and quoted the action of the shift operators $\mathcal{W}_{ij}^{\pm\pm}$ on such solutions. Unlike the 3-point case, these operators act on 4-point functions connecting solutions of CWIs with shifted parameters but do not leave the form of the functions unchanged. Thus, further questions about shift operators arise. Is there a shift operator that when acting on Witten diagrams (and more generally on a certain integral) preserves the form of the integral while shifting the parameters? Does the inverse operator of $\mathcal{L}_i$ exist? We discuss these questions and find a class of such operators in Chapter \ref{cha5}, using the formalism of a class of multivariable hypergeometric functions known as GKZ systems.

\end{chapter}
\clearpage{} 

  \part{Integral representations and shift operators}

  \clearpage{}\begin{chapter}{\label{cha4}Shift operators from the simplex representation in momentum-space CFT}
\allowdisplaybreaks
\section{Introduction}

Understanding 
the general form of correlation functions in momentum-space conformal field theory is an important goal.  
Working in momentum space is natural for many applications, particularly 
 inflationary cosmology (see, {\it e.g.,} \cite{Antoniadis:2011ib,Maldacena:2011nz, Creminelli:2011mw, Bzowski:2012ih, Mata:2012bx,Kehagias:2012pd, McFadden:2013ria,Ghosh:2014kba, Anninos:2014lwa,Arkani-Hamed:2015bza, Arkani-Hamed:2018kmz, Baumann:2019oyu,Baumann:2020dch,Sleight:2019hfp}), 
 and reveals  features inherited from scattering amplitudes that would otherwise be hidden, for example double-copy structure and colour/kinematics duality \cite{Raju:2012zr, Farrow:2018yni,Lipstein:2019mpu, Armstrong:2020woi,Albayrak:2020fyp}.  
Momentum-space methods are moreover well suited for renormalisation \cite{Bzowski:2015pba,Bzowski:2017poo,Bzowski:2018fql,Bzowski:2022rlz}, and are of growing interest
 for the conformal 
 bootstrap 
 \cite{Gillioz:2018mto, Gillioz:2019iye,Gillioz:2020wgw}. 
 
In position space, the structure of general scalar $n$-point  functions has been understood for over fifty years \cite{Polyakov:1970xd}. 
A correspondingly general solution in momentum space was  proposed only
 recently in \cite{Bzowski:2019kwd, Bzowski:2020kfw}.   
This takes the form of a generalised 
Feynman integral with the topology of an $(n-1)$-simplex,
\begin{align} \label{simplex}
&\< \O_1(\bs{p}_1) \ldots \O_n(\bs{p}_n) \> =
 \prod_{1 \leq i < j \leq n} \int \frac{\dd^d \bs{q}_{ij}}{(2 \pi)^d} \frac{f(\hat{\bs{q}})}{q_{ij}^{2 \alpha_{ij} + d}}  \prod_{k=1}^n (2\pi)^d \delta \Big( \bs{p}_k + \sum_{l=1}^n \bs{q}_{lk} \Big),
\end{align}
where  the integration is taken over the internal momenta $\bs{q}_{ij}$ running between vertices of the simplex.  Here $\bs{q}_{ij}=-\bs{q}_{ji}$ runs from vertex $i$ to $j$, while the external momenta  $\bs{p}_i$ enter only via momentum conservation as imposed by the delta function inserted at each vertex.    Each propagator corresponds to an edge of the simplex, as illustrated in figure \ref{simplexfig}, and is raised to a power specified by the parameter $\alpha_{ij}$.  Together, these satisfy the constraints
\begin{equation} \label{alphaijdef}
\Delta_i = - \sum_{j=1}^n \alpha_{ij}, 
\end{equation}
where $\Delta_i$ is the scaling dimension of the  operator $\mathcal{O}_i$.  To simplify the writing of such sums we define $\alpha_{ii} = 0$ and $\alpha_{ji} = \alpha_{ij}$. 
Euclidean signature will be assumed throughout.

The distinguishing  
feature of the simplex representation 
\eqref{simplex} is the presence of an {\it arbitrary function} $f(\hat{\bs{q}})$  of the independent 
momentum-space cross ratios  
\[ \label{conf_ratio_q}
\hat{q}_{[ijkl]} = \frac{q_{ij}^2 q_{kl}^2}{q_{ik}^2 q_{jl}^2},
\]
denoted collectively  by the vector $\hat{\bs{q}}$. 
As the simplex representation can be derived  from the general position-space solution \cite{Bzowski:2019kwd, Bzowski:2020kfw}, the number of independent cross ratios is the same in both cases, {\it i.e.,} $n(n-3)/2$ for $n\le d+2$ and $nd - (d+2)(d+1)/2$  for $n>d+2$. 
For $n\ge 4$, the solution of the constraints \eqref{alphaijdef} for the $\alpha_{ij}$ is not unique, but making a different choice simply multiplies $f(\hat{\bs{q}})$ by a product of powers of the cross ratios \eqref{conf_ratio_q}.  Since $f(\hat{\bs{q}})$ is arbitrary,  the solution of \eqref{alphaijdef}  chosen is therefore immaterial.

\begin{figure}[t]
\centering
\resizebox{0.55\textwidth}{!}{
\def\r{6pt}
\tikzset{d/.style={draw,circle,fill=YellowGreen,minimum size=\r,inner sep=0pt, anchor=center}}

\begin{tikzpicture}[decoration={markings, 
    mark= at position 0.5 with {\arrow{latex}},
    }]

\pgfmathtruncatemacro{\Ncorners}{5}
\node[draw=white, regular polygon,regular polygon sides=\Ncorners,minimum size=8cm] 
(poly\Ncorners) {};
\node[regular polygon,regular polygon sides=\Ncorners,minimum size=8.7cm] 
(outerpoly\Ncorners) {};
\foreach\x in {1,...,\Ncorners}{
    \pgfmathtruncatemacro{\y}{90-(\x-1)*360/\Ncorners}
    \node[d] (poly\Ncorners-\x) at (\y:3.2){};
    \node (outerpoly\Ncorners-\x) at (\y:3.6){$\bs{p}_\x$};
}

\foreach \x in {1,...,5} {
  \foreach \y in {\x,...,5} {
    \ifthenelse{\x=\y}{}{\draw [postaction={decorate}](poly\Ncorners-\x)--node[auto,sloped]{$\hspace{-4mm}\textbf{q}_{\x\y}$}(poly\Ncorners-\y);}
}}

\end{tikzpicture}
}
\caption{Structure of the simplex integral, illustrated for the  $5$-point function.   \label{simplexfig}}
\end{figure}
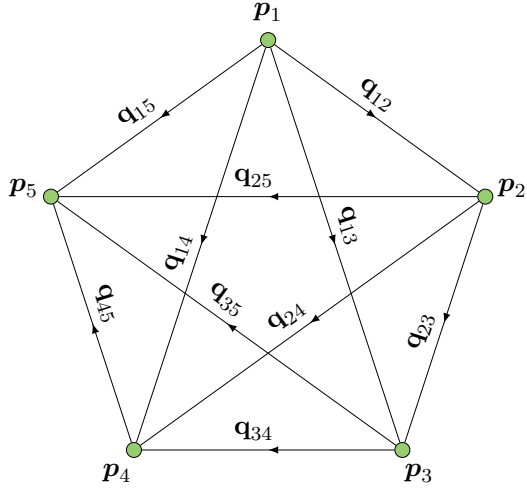

In this chapter, we explore scalar parametric representations of the simplex integral \eqref{simplex} obtained by integrating out the internal momenta.  This offers several advantages:

\begin{itemize}
\item The original integral  \eqref{simplex} 
features $n(n-1)/2$ $d$-dimensional loop integrations and we have $(n-1)$ delta functions to help us, with one remaining behind to enforce overall momentum conservation.  This leaves the equivalent of $(n-1)(n-2)d/2$ scalar integrals to perform. 
In contrast, the  parametrisations we derive feature fewer integrals: only $n(n-1)/2$ scalar parametric integrals,
one for each edge of the simplex.

\item By inverting the graph  polynomials that arise, 
we construct novel weight-shifting operators connecting  solutions of the conformal Ward identities in spacetime dimension $d$ to new solutions in dimension $d+2$.  Remarkably, these operators have a determinantal structure 
based on the Cayley-Menger matrix familiar from 
distance geometry.  
In contrast, the well-known  weight-shifting operators 
 introduced in \cite{Karateev:2017jgd} preserve the spacetime dimension.  
Operators mapping  $d\rightarrow d+2$ are we believe 
known only for  $3$-point functions, where their existence can be seen from the  triple-$K$ representation in momentum space \cite{Bzowski:2015yxv}\footnote{These $d\rightarrow d+2$ operators also enable the construction of  $d$-dimensional {\it tensorial} correlators \cite{Bzowski:2013sza, Bzowski:2017poo,Bzowski:2018fql}.},
and for $4$-point  conformal blocks in position space  
(the operator $\mathcal{E}_+$ in \cite{Dolan:2011dv}).
The new $d\rightarrow d+2$ operators we obtain 
can be viewed as a natural generalisation of the $3$-point operators of \cite{Bzowski:2015yxv} 
to  arbitrary $n$-point correlators.
\end{itemize}

The plan of this chapter is as follows.  In section \ref{sec1}, we show that all  graph polynomials for the simplex integral \eqref{simplex} can be constructed from the corresponding Gram matrix.
The standard parametric representations for Feynman integrals then follow. 
Alternatively, by regarding the  Schwinger parameters as resistances in an electrical network, we can  compute  the {\it effective} resistances between all vertices of the simplex.  This latter set of variables dramatically simplifies the structure of the Schwinger exponential.  In section \ref{sec:wsops}, we use these effective resistances to construct new $d\rightarrow d+2$ shift operators for the general $n$-point function.  The cases  $n=3,4$  are discussed in detail, and we verify 
the action of all operators independently 
through computation of their intertwining relations with the conformal Ward identities.  
The actions of
the   $d$-preserving weight-shifting operators of \cite{Karateev:2017jgd} are also demonstrated from this scalar parametric perspective. 
In section \ref{sec:CWI}, we  prove that the new parametric representations 
indeed solve the conformal Ward identities.  In contrast to the vectorial representation \eqref{simplex} (for which the Ward identities  are analysed in \cite{Bzowski:2019kwd, Bzowski:2020kfw}), 
for  the new scalar parametric representations the Ward identities can be verified directly without use of recursive arguments in the number of points $n$. 
As we show in section \ref{sec:posnspace},
the validity of the conformal  Ward identities, as well as the action of the $d$-preserving weight-shifting operators, can also be seen from  the position-space counterpart of the simplex. Section \ref{sec:disc} concludes with a summary of results and open directions.

\section{Parametric representations of the simplex}
\label{sec1}

This section investigates scalar parametric representations for the simplex integral \eqref{simplex}.
In the following, we identify the necessary graph polynomials (section \ref{sec:graphpolys}), standard parametric representations (section \ref{sec:stdreps}), and introduce new variables analogous to the effective resistances between nodes of the simplex (section \ref{sec:effres}).  To re-formulate the simplex integral in these variables, we  solve the inverse problem to express the original Schwinger parameters in terms of the effective resistances (section \ref{sec:reparam}). The re-parametrised integral, which will be the basis of our new shift operators,  then follows (section \ref{sec:CayMeng}).

\subsection{Graph polynomials} \label{sec:graphpolys}

Exponentiating all propagators via Schwinger parametrisation, the internal momenta can be integrated out reducing the simplex integral   to various scalar parametrisations.  The structure of the resulting Symanzik polynomials is clearest however when expressed in terms of the {\it inverse} of the usual variables.
For this reason, we use the inverse Schwinger parametrisation
\[\label{Schwingerintegral}
\frac{1}{q_{ij}^{2\alpha_{ij}+d}} = \frac{1}{\Gamma(\alpha_{ij}+d/2)}\int_0^\infty \dd v_{ij} \,v_{ij}^{-d/2-\alpha_{ij}-1}e^{-q_{ij}^2/v_{ij}}.
\]
The resulting polynomials $\mathcal{U}$ and $\mathcal{F}$ are  then related to the standard Symanzik polynomials $U$ and $F$ by
\[\label{Kirchhofdef}
\mathcal{U}(v_{ij}) = \Big(\prod_{i<j}^n v_{ij}\Big) U\Big(\frac{1}{v_{ij}}\Big),
\qquad
\mathcal{F}(v_{ij}) = \Big(\prod_{i<j}^n v_{ij}\Big) F\Big( \frac{1}{v_{ij}}\Big).
\]
For the simplex, the  structure of  $\mathcal{U}$ and $\mathcal{F}$ can be expressed in terms of two matrices.  The first is the $(n-1)\times(n-1)$ Gram matrix $G_{ij}  = \bs{p}_i\cdot\bs{p}_j$.  For our purposes, the most convenient parametrisation is 
\[\label{Gramdef}
G_{ij} = \begin{cases} \sum_{k=1}^{n} V_{ik}\qquad &i=j \\
-V_{ij} \qquad & i\neq j
\end{cases}\
\qquad i,j = 1,\ldots, n-1,
\]
where 
\[\label{Vdef}
V_{ij} =\begin{cases} -\bs{p}_i\cdot \bs{p}_j \qquad & i\neq j\\
0\qquad & i=j
\end{cases}\qquad i,j = 1,\ldots, n.
\]
Here the $V_{ij}$  provide a full set of $n(n-1)/2$ symmetric and independent Lorentz invariants.
To write 
 the diagonal entries in the Gram matrix,  we used momentum conservation to express $p_i^2=-\sum_{k\neq i}^n \bs{p}_i\cdot\bs{p}_k$.
The second matrix  is simply the image of the Gram matrix under the mapping
$
V_{ij}\rightarrow v_{ij},
$ 
namely
\[\label{gdef}
g_{ij} = \begin{cases} \sum^{n}_{k=1} v_{ik} \qquad & i=j, \\
-v_{ij}\qquad & i\neq j, 
\end{cases}\qquad i,j=1,\ldots, n-1.
\]
Since the $v_{ij}$ correspond to the edges of the simplex we define,  as we did for the $V_{ij}$,
\[
v_{ij} = v_{ji}, \qquad v_{ii}=0.
\]

As shown in appendix \ref{incidence_app},
the graph polynomials are now  
\begin{empheq}[box=\nicebox]{align} 
\label{Kirchhoff1}
\mathcal{U}  = |g|, \qquad \mathcal{F}= \mathrm{tr}(\mathrm{adj}(g)\cdot G), \qquad \frac{\mathcal{F}}{\mathcal{U}} = \mathrm{tr}(g^{-1}\cdot G),
\end{empheq}
where $|g|=\mathrm{det}\,g$, $\mathrm{adj}\,g=|g| \,g^{-1}$ is the adjugate matrix and $g^{-1}$ the inverse matrix.  
The derivation proceeds by expressing the delta functions of \eqref{simplex} in Fourier form and integrating out all internal momenta.  Only  after this step has been performed are the Fourier integrals for the delta functions then  evaluated.
As the Gram determinant $|G|$ is proportional to the squared volume of the simplex spanned by the independent momenta, the polynomial $\mathcal{U}$ describes the image of this squared volume under the mapping $V_{ij}\rightarrow v_{ij}$.  Alternatively, by the matrix tree theorem (see {\it e.g.,} \cite{Weinzierl:2022eaz}), $\mathcal{U}$ is the Kirchhoff polynomial encoding the sum of  spanning trees on the simplex.

A second useful expression for $\mathcal{F}$ can be derived 
from  Jacobi's identity, 
\[\label{Jacobi}
\partial_{v_{ij}}|g| = \mathrm{tr}\big(\mathrm{adj}\,(g)\cdot \partial_{v_{ij}} g\big),
\] 
in combination with the relation
\[\label{bigGrel}
G_{ij} = \sum_{k<l}^n V_{kl} \frac{\partial G_{ij}}{\partial V_{kl}}= \sum_{k<l}^n V_{kl} \frac{\partial g_{ij}}{\partial v_{kl}}.
\]
This last relation  follows from the linearity of the $G_{ij}$ in the $V_{kl}$, as we saw in \eqref{Gramdef}, and the mapping of  $G_{ij}\rightarrow g_{ij}$ under $V_{kl}\rightarrow v_{kl}$.  The sums run over all $k$ and $l$ such that $k<l$, corresponding to all edges of the simplex. 
Substituting \eqref{bigGrel} into \eqref{Kirchhoff1} then using \eqref{Jacobi}, \[
\mathcal{F} = \sum_{i<j}^{n} \frac{\partial |g|}{\partial v_{ij}} \,V_{ij}  ,\qquad
\frac{\mathcal{F}}{\mathcal{U}} = \sum_{i<j}^{n}\frac{\partial\ln |g|}{\partial v_{ij}}\,V_{ij},
\]
or in terms of the raw momenta,
\begin{empheq}[box=\nicebox]{align} 
\label{Kirchhoff2}
\mathcal{F} = -\sum_{i<j}^{n}\frac{\partial |g|}{\partial v_{ij}}\,  \bs{p}_i\cdot\bs{p}_j,\qquad
\frac{\mathcal{F}}{\mathcal{U}} = -\sum_{i<j}^{n}\frac{\partial\ln |g|}{\partial v_{ij}}\,  \bs{p}_i\cdot\bs{p}_j.
\end{empheq}

\subsection[\texorpdfstring{Parametric representations of the
$n$-point correlator}{Parametric representations of the n-point correlator}]{\texorpdfstring{Parametric representations of the
\boldmath{$n$}-point correlator}{Parametric representations of the n-point correlator}}\label{sec:stdreps}

To express correlators compactly,  we extract the overall delta function of momentum conservation as
\[\label{redcorr}
\< \O_1(\bs{p}_1) \ldots \O_n(\bs{p}_n) \> = \lla \O_1(\bs{p}_1) \ldots \O_n(\bs{p}_n) \rra (2\pi)^d\delta(\sum_{i=1}^n \bs{p}_n).
\]
We also define an arbitrary function $f(\hat{\bs{v}})$ whose arguments, denoted collectively by the vector $\hat{\bs{v}}$, are the independent  inverse Schwinger parameter cross ratios
\[\label{vcrossratio}
v_{[ijkl]} =  \frac{v_{ij} v_{kl}}{v_{ik}v_{jl}}.
\]  
The simplex integral \eqref{simplex} can now be written in 
a variety of standard forms using the polynomials $\mathcal{U}$ and $\mathcal{F}$ defined in \eqref{Kirchhoff1} or \eqref{Kirchhoff2}.  Among the most useful  are: 
\begin{itemize}
\item[1.] {\it Schwinger parametrisation:}
\[\label{Schwrep}
\lla \O_1(\bs{p}_1) \ldots \O_n(\bs{p}_n) \rra = \Big(\prod_{i<j}^n\int_0^\infty \dd v_{ij}\, v_{ij}^{-\alpha_{ij}-1}\Big) f(\hat{\bs{v}})\,\mathcal{U}^{-d/2} e^{-\mathcal{F}/\mathcal{U}}
\]
Here, the $v_{ij}^{-d/2}$ factors in \eqref{Schwingerintegral} cancel with those associated with $\mathcal{U}^{-d/2}$ via \eqref{Kirchhofdef}.
\item[2.] {\it Lee-Pomeransky  parametrisation} \cite{Lee:2013hzt}:
\[\label{LPrep}
\lla \O_1(\bs{p}_1) \ldots \O_n(\bs{p}_n) \rra =\Big(\prod_{i<j}^n\int_0^\infty \dd v_{ij}\, v_{ij}^{-\alpha_{ij}-1} \Big)f(\hat{\bs{v}})\,(\mathcal{U}+\mathcal{F})^{-d/2}
\]
\item[3.] {\it Feynman  parametrisation:}
\[\label{Feynrep}
\lla \O_1(\bs{p}_1) \ldots \O_n(\bs{p}_n) \rra =\Big(\prod_{i<j}^n\int_0^\infty \dd v_{ij}\,v_{ij}^{-\alpha_{ij}-1} \Big)\delta\Big(1-\sum_{i<j}^n \kappa_{ij}v_{ij}\Big) f(\hat{\bs{v}})\,\mathcal{U}^{\omega-d/2} \mathcal{F}^{-\omega}
\]
where $\omega=(n-1)d/2+\sum_{i<j}^n\alpha_{ij} = (-\Delta_t+(n-1)d)/2$ and the constants $\kappa_{ij}\ge 0$ can be chosen arbitrarily provided they are not all zero.\footnote{
The Feynman parametrisation follows from the Schwinger parametrisation by setting $v_{ij}=y_{ij}/\sigma$ subject to the constraint $\sum_{i<j}^n \kappa_{ij}y_{ij}=1$. The  $\mathcal{U}$ and $\mathcal{F}$ are homogeneous polynomials of weights $(n-1)$ and $(n-2)$ respectively, meaning that $\mathcal{F}(v_{ij})/\mathcal{U}(v_{ij}) =\sigma\mathcal{F}(y_{ij})/\mathcal{U}(y_{ij})$ while the Jacobian can be evaluated as per appendix B of \cite{Bzowski:2020kfw}.  We then perform the scale integral over $\sigma$ and relabel the $y_{ij}\rightarrow v_{ij}$. }
If we choose all $\kappa_{ij}=1$ then the integration region is a simplex in the space spanned by the $v_{ij}$.  Alternatively, we can set a single $\kappa_{ij}$ to unity and the rest to zero which trivialises one of the integrations at the cost of obscuring permutation invariance.

\end{itemize}

These representations are all equivalent up to numerical factors; for clarity, we  have re-absorbed these into the arbitrary functions.  
For analysing the action of weight-shifting operators and verifying the conformal Ward identities, we will focus exclusively on the Schwinger parametrisation \eqref{Schwrep}.  Nevertheless, the Lee-Pomeransky representation \eqref{LPrep} is well suited for studying the Landau singularities, as discussed in appendix \ref{Landau_app}, and the Feynman parametrisation \eqref{Feynrep} has the  virtue that 
 one integral can be performed using the delta function.

\paragraph{Example:} As a quick illustration, the $4$-point function in Schwinger parametrisation is
\begin{align}
&\lla \O_1(\bs{p}_1)  \O_2(\bs{p}_2) \O_3(\bs{p}_3)\O_4(\bs{p}_4) \rra \nn\\& \qquad = \Big(\prod_{i<j}^4\int_0^\infty \dd v_{ij}\, v_{ij}^{-\alpha_{ij}-1}\Big) f\Big(\frac{v_{12}v_{34}}{v_{13}v_{24}},\,\frac{v_{14}v_{23}}{v_{13}v_{24}}\Big)\,|g|^{-d/2} e^{-\mathrm{tr}\,( g^{-1}\cdot G)}
\end{align}
where $G_{ij} = \bs{p}_i\cdot \bs{p}_j$ is the  $3\times 3$ Gram matrix and $g$ is its image
\begin{align}
g =\left( \begin{matrix}v_{12}+v_{13}+v_{14} & -v_{12} & -v_{13}\\ -v_{12} & v_{12}+v_{23}+v_{24} & -v_{23}\\
-v_{13} & -v_{23} & v_{13}+v_{23}+v_{34}\end{matrix}\right).
\end{align} 
The determinant is
\begin{align}\label{4ptU}
|g| &=  v_{12} v_{13} v_{14} + v_{12} v_{14} v_{23} + v_{13} v_{14} v_{23} + v_{12} v_{13} v_{24} \nn\\&\quad + v_{13} v_{14} v_{24} + 
 v_{12} v_{23} v_{24} + v_{13} v_{23} v_{24} + v_{14} v_{23} v_{24}\nn\\&\quad + v_{12} v_{13} v_{34} + v_{12} v_{14} v_{34} +
  v_{12} v_{23} v_{34} + v_{13} v_{23} v_{34} \nn\\&\quad + v_{14} v_{23} v_{34} + v_{12} v_{24} v_{34} + 
 v_{13} v_{24} v_{34} + v_{14} v_{24} v_{34}
\end{align}
and the equivalence of \eqref{Kirchhoff1} and \eqref{Kirchhoff2} can be verified directly.

\subsection{The effective resistances}\label{sec:effres}

Thus far, we have expressed the Kirchhoff polynomial
$\mathcal{U}$ as the determinant of $g$, the image under $V_{ij}\rightarrow v_{ij}$ of the 
 Gram matrix, where $\bs{p}_n$ is eliminated using momentum conservation.
However, since all vertices of the simplex are equivalent, $\mathcal{U}$ ought also to be expressible in terms of the $n\times n$ matrix $\tilde{g}$ corresponding to the image 
of the {\it extended} Gram matrix $\tilde{G}_{ij} = \bs{p}_i\cdot\bs{p}_j$ for $i,j=1,\ldots,n$.  
This is simply the Laplacian matrix for the simplex: 
\[\label{tildeg}
\tilde{g}_{ij} = \begin{cases} \sum^{n}_{k=1} v_{ik}, \qquad & i=j, \\
-v_{ij},\qquad & i\neq j, 
\end{cases}\qquad i,j=1,\ldots, n.
\]
As every row and column sum of the Laplacian matrix is zero its determinant 
 vanishes identically,  but its cofactors ({\it i.e.,} signed first minors) 
are in fact all equal to $\mathcal{U}$.
To see this, consider 
the diagonal minor 
 $|\tilde{g}^{(n,n)}|$  formed by deleting row $n$ and column $n$ then taking the determinant.   Comparing with \eqref{gdef}, we then see that  $|\tilde{g}^{(n,n)}|=|g|=\mathcal{U}$.  As any diagonal minor is equal to its cofactor, $\mathcal{U}$ is likewise the $(n,n)$ cofactor.   However, by elementary row and column operations one can show that all cofactors of the Laplacian matrix are equal.\footnote{For example, add one to every element of $\tilde{g}_{ij}$ then add all rows to the first row, and all columns to the first column.  The top left entry is now $n^2$ while all remaining entries of the first row and column are $n$. Taking the determinant, we first extract an overall factor of $n$ from the top row, then subtract the new top row (whose leftmost entry is now $n$ with all other entries one) from all the other rows. The resulting matrix has zeros in all entries of the first column apart from the top one which is $n$, and all entries other than those in the first row and column are $\tilde{g}_{ij}$ (since we added one then subtracted one).  The determinant of $\tilde{g}_{ij}$ plus the all ones matrix is therefore $n^2$ times the $(1,1)$ cofactor of $\tilde{g}_{ij}$.  Repeating the exercise for any other choice of row and column yields the same result with the corresponding cofactor, hence all cofactors are equal.  Note this also shows that  $\mathcal{U}$ is $n^{-2}$ times the determinant of the Laplacian plus the all-ones matrix.}  
Thus, every cofactor (and every diagonal minor) is equal to  $\mathcal{U}$.  Note this also confirms that our choice of eliminating $\bs{p}_n$ in section \ref{sec:graphpolys} was immaterial.

Let us now turn to an electrical analogy involving a simplicial network of resistors.   Here, the Laplacian matrix naturally encodes the external current $\mathcal{I}_i$ 
flowing into node $i$, since 
\[
\mathcal{I}_i = \sum_{j\neq i} v_{ij}(\mathcal{V}_i-\mathcal{V}_j) = \sum_{j\neq i} \tilde{g}_{ij}\mathcal{V}_j,
\] 
where  $v_{ij}$ is the conductivity of the edge connecting nodes $i$ and $j$ and $\mathcal{V}_j$ is the voltage of node $j$. 
Given this identification of the $v_{ij}$ with the conductivities, a natural question to ask is what are the corresponding {\it effective resistances} 
between the nodes?   From Kirchhoff, the effective resistance $s_{ij}$ between nodes $i$ and $j$ is given by the ratio of minors \cite{kirchhoff1847,kirchhoff1958}
\[
s_{ij} = \frac{|\tilde{g}^{(ij,ij)}|}{|\tilde{g}^{(j,j)}|}, 
\]
where $|\tilde{g}^{(I,J)}|$ indicates the minor formed by deleting the set of rows $I$ and columns $J$ then taking the determinant.
Thus,  $|\tilde{g}^{(ij,ij)}|$ is the second minor formed by  deleting rows $i$ and $j$ as well as columns $i$ and $j$, while $|\tilde{g}^{(j,j)}|$ is  the first minor corresponding to deleting row and column $j$.
From \eqref{tildeg}, the element $v_{ij}$ appears only in the row and columns $(i,i)$, $(i,j)$, $(j,i)$ and $(j,j)$  of $\tilde{g}$.  Forming the  first minor $|\tilde{g}^{(j,j)}|$ by deleting row and column $j$, $v_{ij}$ then appears only once in the $(i,i)$ position. 
The derivative $\partial |\tilde{g}^{(j,j)}|/\partial v_{ij}$ is thus equal to the second minor $|\tilde{g}^{(ij,ij)}|$ formed by additionally deleting row and column $i$ in $|\tilde{g}^{(j,j)}|$.
Since $|\tilde{g}^{(j,j)}|=|g|$ as above, we have
\begin{empheq}[box=\nicebox]{align} 
\label{sdef}
s_{ij} = \frac{\partial \ln |g|}{\partial v_{ij}}, \qquad \frac{\mathcal{F}}{\mathcal{U} }= -\sum_{i<j}^n s_{ij}\,\bs{p}_i\cdot\bs{p}_j,
\end{empheq}
where the second result  follows immediately from \eqref{Kirchhoff2}.
The Schwinger exponent in \eqref{Schwrep} thus  encodes the effective resistances $s_{ij}$ between all vertices.  Moreover, both  $\mathcal{U}$ and $\mathcal{F}$ have been related to minors of the Laplacian: $\mathcal{U}$ is any diagonal first minor (or cofactor), while the coefficients of the $\mathcal{F}$ polynomial correspond to the second minors: from \eqref{Kirchhoff2}, the coefficient of $V_{ij}=-\bs{p}_i\cdot\bs{p}_j$ (for $i<j$) is  $
\partial |g|/\partial v_{ij} = |\tilde{g}^{(ij,ij)}|
$. 

Earlier, we noted that $\mathcal{U}=|g|$ is proportional to the squared volume of the $(n-1)$-simplex formed by the 
 independent momenta under the map $V_{ij}\rightarrow v_{ij}$.
By the same token, each coefficient $|\tilde{g}^{(ij,ij)}|$ of the $\mathcal{F}$ polynomial thus corresponds to the image of  $|\tilde{G}^{(ij,ij)}|$, the second minor of the extended Gram matrix.
However, this minor is simply the determinant of the reduced Gram matrix formed from all the momenta apart from $\bs{p}_i$ and $\bs{p}_j$.
Thus, the coefficient of $V_{ij}$ in the $\mathcal{F}$ polynomial
is 
proportional to the squared volume of the $(n-2)$-simplex, formed from all the momenta except for $\bs{p}_i$ and $\bs{p}_j$, under the map $V_{ij}\rightarrow v_{ij}$. 
Similarly, the effective resistance $s_{ij}$ is proportional to the ratio of the squared volume of this $(n-2)$-simplex to the squared volume of the full $(n-1)$-simplex.

\subsection{Re-parametrising the simplex}
\label{sec:reparam}

The original Schwinger parametrisation \eqref{Schwrep} is complicated by the non-linear dependence of  the exponent on the $v_{ij}$.   As we  saw in \eqref{sdef},  however, the coefficients of the $V_{ij}=-\bs{p}_i\cdot\bs{p}_j$ 
in $\mathcal{F}/\mathcal{U}$ are simply the effective resistances $s_{ij}$ between nodes.
The next step  is thus to invert the relation \eqref{sdef} to find the $v_{ij}$ in terms of the $s_{ij}$, {\it i.e.,} to express the conductivities in terms of the effective resistances. 
The simplex integral can then be fully re-parametrised in terms of the $s_{ij}$, with the linearity of the Schwinger exponent giving a Fourier-style duality between the $V_{ij}$ and the $s_{ij}$.  This duality means that all momentum derivatives acting on the simplex, and all momenta, can be trivially exchanged for operators constructed from the $s_{ij}$ and derivatives $\partial/\partial s_{ij}$.  The latter can then be integrated by parts.  This strategy will repeatedly prove useful to us later.

We start by applying Jacobi's relation to further evaluate \eqref{sdef},
\begin{align}
s_{ij}  =\frac{1}{|g|}\frac{\partial|g|}{\partial v_{ij}}=\mathrm{tr}\Big(g^{-1}\cdot\frac{\partial g}{\partial v_{ij}}\Big)=\begin{cases}(g^{-1})_{ii}+(g^{-1})_{jj}-2(g^{-1})_{ij}, \qquad &i<j< n \\ (g^{-1})_{ii},\qquad &i< j=n\end{cases}\label{srel1}
\end{align}
where the matrices $\partial g/\partial v_{ij}$ are easily evaluated from \eqref{gdef}. 
Defining $s_{ii}=0$  for convenience (as we similarly defined $v_{ii}=0$) and re-arranging, we find 
\[\label{ginvs}
(g^{-1})_{ij} = \frac{1}{2}(s_{in}+s_{jn}-s_{ij}), \qquad i,j=1,\ldots, n-1
\]
where the diagonal entries reduce to $(g^{-1})_{ii}=s_{in}$.
Inverting this matrix will now give us back the matrix $g$, as defined in \eqref{gdef}, but re-expressed in terms of the $s_{ij}$.  The  desired expressions for the $v_{ij}$ in terms of the $s_{ij}$ can then be read off from the appropriate entries.

In fact,  it is sufficient simply to know the determinant $|g^{-1}|$.  For $i<j<n$, the $(i,j)$ minor formed by deleting row $i$ and column $j$ of $g^{-1}$ 
is $|(g^{-1})^{(i,j)}|=-(-1)^{i+j}\partial |g^{-1}|/\partial s_{ij}$, since from \eqref{ginvs} $s_{ij}$ appears (with coefficient minus one-half) only in the positions $(i,j)$ and $(j,i)$ of the symmetric matrix $g^{-1}$.  
The off-diagonal entries of the adjugate matrix are thus
\[
\mathrm{adj}(g^{-1})_{ij} = (-1)^{i+j}|(g^{-1})^{(i,j)}| = - \frac{\partial |g^{-1}|}{\partial s_{ij}}, \qquad i<j<n
\] 
so 
\[\label{vfromginv}
v_{ij} = -g_{ij} = -\frac{1}{|g^{-1}|}\,\mathrm{adj}(g^{-1})_{ij} = \frac{\partial \ln |g^{-1}|}{\partial s_{ij}}, \qquad i<j<n.
\]
Similarly,  $s_{in}$ appears in every entry of the $i^{\mathrm{th}}$ row of $g^{-1}$, and in every entry of the $i^{\mathrm{th}}$ column.
The coefficients for the off-diagonal entries are all one-half, while that for the diagonal entry is one.
The derivative $\partial |g^{-1}|/\partial s_{in}$ then corresponds to summing one-half times the signed minors  both along the $i^{\mathrm{th}}$ row and down the $i^{\mathrm{th}}$ column such that the diagonal entry is counted twice.  As $g^{-1}$ is symmetric, however, these two sums are equal so we can simply sum along the $i^{\mathrm{th}}$ row only with coefficient one.
This gives
\[\label{vfromginvn}
\frac{\partial \ln |g^{-1}|}{\partial s_{in}} = \sum_{j=1}^{n-1}\frac{ (-1)^{i+j}}{|g^{-1}|}|(g^{-1})^{(i,j)}|=\sum_{j=1}^{n-1}\frac{1}{|g^{-1}|}\mathrm{adj} (g^{-1})_{ij} = \sum_{j=1}^{n-1} g_{ij} = v_{in},
\]
where in the final step we used \eqref{gdef} to identify the sum of the first $n-1$ entries along the $i^{\mathrm{th}}$ row of the Laplacian as $v_{in}$.   The relation \eqref{vfromginv} thus holds for all $i<j\le n$.

To simplify this formula further, we observe that  $|g^{-1}|$ can be re-expressed in terms of the determinant of the $(n+1)\times (n+1)$ Cayley-Menger matrix,
\[\label{mdef}
m = \left(\begin{matrix}
 0 & s_{12} & s_{13}  & \ldots & s_{1n} & 1\\
 s_{12} & 0 & s_{23}  & \ldots & s_{2n} & 1\\
 s_{13} & s_{23} & 0  & \ldots & s_{3n} & 1\\ 
  \vdots & \vdots & \vdots & &\vdots&\vdots\\
  s_{1n} & s_{2n} & s_{3n}  & \ldots & 0 & 1\\
 1 & 1 & 1&\ldots & 1& 0\\
\end{matrix}\right).
\]
When evaluating the determinant, if we subtract the $n^{\mathrm{th}}$ column from the first $(n-1)$  columns, and then the $n^{\mathrm{th}}$ row from the first $(n-1)$  rows, we find  
\[
|m| = \left|\begin{matrix}
 -2s_{1n}\quad  & s_{12}-s_{1n}-s_{2n}\quad  & s_{13}-s_{1n}-s_{3n} & \ldots & s_{1n} \quad & 0 & \\[1ex]
\,\,  s_{12} -s_{1n}-s_{2n}& -2s_{2n} & s_{23}-s_{2n}-s_{3n}  & \ldots & s_{2n} & 0\\[1ex]
 s_{13} -s_{1n}-s_{3n} & s_{23}-s_{2n}-s_{3n} & -2s_{3n}  & \ldots & s_{3n} & 0\\[1ex] 
  \vdots & \vdots & \vdots & &\vdots&\vdots\\[1ex]
  s_{1n} & s_{2n} & s_{3n}  & \ldots & 0 & 1\\[1ex]
 0 & 0 & 0&\ldots & 1& 0\\
\end{matrix}\right|.
\]
Comparing with \eqref{ginvs}, the upper-left $(n-1)\times (n-1)$ submatrix is $-2g^{-1}$.
Laplace expanding along the $(n+1)^\mathrm{th}$ row and then the $(n+1)^\mathrm{th}$ column thus gives
\[\label{mginvreln}
|m| = -(-2)^{n-1}|g^{-1}|.
\]
Equations \eqref{vfromginv} and \eqref{vfromginvn} can now be  cleanly re-expressed in terms of the Cayley-Menger determinant:
\begin{empheq}[box=\nicebox]{align} 
\label{vdef}
v_{ij} = \frac{\partial \ln |m|}{\partial s_{ij}}, \qquad i<j\le n.
\end{empheq}
This is our desired result expressing all the $v_{ij}$ in terms of the $s_{ij}$, inverting \eqref{sdef}.
A few additional relations also follow.  
 Jacobi's relation allows us to write 
\[
v_{ij} =\frac{1}{|m|} \frac{\partial  |m|}{\partial s_{ij}} =\frac{1}{|m|}\mathrm{tr}\Big(\mathrm{adj}(m)\cdot\frac{\partial m}{\partial s_{ij}}\Big)=\mathrm{tr}\Big(m^{-1}\cdot\frac{\partial m}{\partial s_{ij}}\Big) =  2(m^{-1})_{ij}, \label{vrel1}
\]
since  $\partial m_{kl}/\partial s_{ij} = 2\delta_{i(k}\delta_{l)j}$ from \eqref{mdef}.
As the off-diagonal entries of the Laplacian matrix are $\tilde{g}_{ij}=-v_{ij}$, this means that
\[\label{minvisLa1}
\tilde{g}_{ij} = -2 (m^{-1})_{ij}, \qquad  i,j\le n.
\]
In fact, as indicated, this equation also holds for the diagonal elements with $i=j\le n$, since if we multiply the $(n+1)^{\mathrm{th}}$ row of $m$ by column $i$ of $m^{-1}$ we find
\[
0 = \sum_{j=1}^n m^{-1}_{ij}, \qquad i\le n
\]
and since all row and column sums of the Laplacian matrix vanish, 
\[\label{minvisLa2}
\tilde{g}_{ii} = -\sum_{j\neq i}^n \tilde{g}_{ij} = \sum_{j\neq i}^n 2(m^{-1})_{ij} = -2(m^{-1})_{ii}.
\]
Thus, the $n\times n$ upper-left  submatrix of the inverse Cayley-Menger matrix is minus one-half the Laplacian matrix,
using either \eqref{sdef} or \eqref{vdef} to convert between the $s_{ij}$ and $v_{ij}$.\footnote{
To the best of our knowledge, this result, 
along with a geometrical interpretation of the  remaining $(n+1)^{\mathrm{th}}$ row and column of the Cayley-Menger inverse, was first obtained by Fiedler, see \cite{fiedler_2011, Devriendt_2022}.}

The appearance of the Cayley-Menger matrix in our analysis is not a total surprise: in Euclidean distance geometry, the Cayley-Menger determinant is proportional to the squared volume of the simplex whose squared side lengths are given by the $s_{ij}$.  Here, the map $v_{ij}\rightarrow V_{ij}$ sends $(g^{-1})_{ij}$ to the inverse Gram matrix $G^{-1}_{ij} = (\tilde{\bs{p}}_i\cdot\tilde{\bs{p}}_j)$, which is itself the Gram matrix formed from the independent {\it dual} momentum vectors $\tilde{\bs{p}}_i=G^{-1}_{ij}\bs{p}_j$ satisfying $\tilde{\bs{p}}_i\cdot \bs{p}_j=\delta_{ij}$.  
The determinant $|g^{-1}|$ is thus proportional to the squared volume of the dual $(n-1)$-simplex spanned by the independent $\tilde{\bs{p}}_i$, and
by \eqref{mginvreln}, the $s_{ij}$ are then the squared side lengths of this dual simplex. 
This provides an alternative (dual) geometrical interpretation for the $s_{ij}$, besides the volume ratio discussed at the end of section \ref{sec:effres}.

\paragraph{Example:}  All the relations above are easily checked for  small values of $n$, and the $s_{ij}$ are always rational functions of the $v_{ij}$ and vice versa.
For the 4-point function, we find, {\it e.g.,} 
\begin{align}
s_{12}&= \frac{\partial \ln |g|}{\partial v_{12}} = ( v_{13} v_{14} + v_{14} v_{23} + v_{13} v_{24} + v_{23} v_{24} + v_{13} v_{34} + v_{14} v_{34} + v_{23} v_{34} + 
 v_{24} v_{34})\, |g|^{-1} \nn\\[1ex]
v_{12}&= \frac{\partial \ln |m|}{\partial s_{12}} =(-2 s_{13} s_{23} + 2 s_{14} s_{23} + 2 s_{13} s_{24} - 2 s_{14} s_{24} - 4 s_{12} s_{34} 
\nn\\[-0.5ex]&\qquad\qquad\qquad 
+  2 s_{13} s_{34} 
 + 2 s_{14} s_{34} + 2 s_{23} s_{34} + 2 s_{24} s_{34} - 2 s_{34}^2)\,|m|^{-1},
\end{align}
where  $|g|$ was evaluated in \eqref{4ptU} and 
\begin{align}\label{m4}
|m| &= \left|\,\begin{matrix}
  0 & s_{12} & s_{13}  & s_{14} & 1\\
 s_{12} & 0 & s_{23}  & s_{24} & 1\\
 s_{13} & s_{23} & 0  & s_{34} & 1\\ 
  s_{14} & s_{24} & s_{34}  & 0 & 1\\
 1 & 1 & 1 & 1& 0
\end{matrix}\,\right|
\end{align}
evaluates to 
\begin{align}\label{detm4}
|m| &=-2 s_{12} s_{13} s_{23} + 2 s_{12} s_{14} s_{23} + 2 s_{13} s_{14} s_{23} - 2 s_{14}^2 s_{23}
 -  2 s_{14} s_{23}^2 + 2 s_{12} s_{13} s_{24} - 2 s_{13}^2 s_{24} 
 \nn\\
&\quad 
 - 2 s_{12} s_{14} s_{24} + 
 2 s_{13} s_{14} s_{24} + 2 s_{13} s_{23} s_{24} + 2 s_{14} s_{23} s_{24} - 2 s_{13} s_{24}^2 - 
 2 s_{12}^2 s_{34} 
\nn\\
&\quad 
  + 2 s_{12} s_{13} s_{34}+ 2 s_{12} s_{14} s_{34} - 2 s_{13} s_{14} s_{34} + 
 2 s_{12} s_{23} s_{34} + 2 s_{14} s_{23} s_{34} + 2 s_{12} s_{24} s_{34} 
\nn\\
&\quad 
+ 2 s_{13} s_{24} s_{34}  -  2 s_{23} s_{24} s_{34} - 2 s_{12} s_{34}^2.
 \end{align}

\subsection{\texorpdfstring{Cayley-Menger parametrisation of the $n$-point correlator}{Cayley-Menger parametrisation of the n-point correlator}}\label{sec:CayMeng}

Using the results above, we can re-express the various 
parametrisations of the 
simplex integral in terms of the effective resistances $s_{ij}$.  
If we write the external momenta in Cayley-Menger form,
\[
M = m\Big|_{s_{ij}\rightarrow \bs{p}_i\cdot \bs{p}_j}
\]
the Schwinger exponent can be written as
\[\label{CaySchwexp}
-\frac{\mathcal{F}}{\mathcal{U}} = \sum_{i<j}^n s_{ij}(\bs{p}_i\cdot\bs{p}_j) = \frac{1}{2}\mathrm{tr} (M\cdot m) + n,
\]
where the constant term $n$ just produces an overall scaling which can be re-absorbed into the arbitrary function of cross-ratios. 
Moreover, as shown in appendix \ref{Jac_app}, the determinant of the Jacobian is \[\label{Jac0}
\left|\frac{\partial s}{\partial v} \right| = \left|\frac{\partial^2 \ln |g|}{\partial v\partial v}\right|\propto |g|^{-n} \propto |m|^{n},
\]
where the constant of proportionality can again be absorbed into the arbitrary function.

The Schwinger form \eqref{Schwrep} of  the simplex integral  now  becomes
\[\label{Cay1}
\lla \O_1(\bs{p}_1) \ldots \O_n(\bs{p}_n) \rra  = \Big(\prod^n_{i<j} \int_0^\infty\mathrm{d} s_{ij} \, \Big(\frac{\partial \ln |m|}{\partial s_{ij}}\Big)^{-\alpha_{ij}-1} \Big)f(\hat{\bs{v}}) \,|m|^{d/2-n}\, e^{\frac{1}{2}\mathrm{tr} (M\cdot m)}
\]
where the cross-ratios $\hat{\bs{v}}$ are rational functions of the $s_{ij}$ as defined via \eqref{vcrossratio} and \eqref{vdef}.
An alternative expression can be given in terms of the Cayley-Menger minors, since
from Jacobi's relation
\[\label{dmdstominnors}
\frac{\partial |m|}{\partial s_{ij}} = 2(-1)^{i+j}\,|m^{(i,j)}|,
\]
where $|m^{(i,j)}|$ is the minor formed by taking the determinant after deleting row $i$ and column $j$.
After absorbing numerical factors into the arbitrary function, this gives
\[\label{Cay2}
\lla \O_1(\bs{p}_1) \ldots \O_n(\bs{p}_n) \rra  = \Big(\prod^n_{i<j} \int_0^\infty\mathrm{d} s_{ij} \, |m^{(i,j)}|^{-\alpha_{ij}-1} \Big)\,f(\hat{\bs{v}})\,|m|^{\alpha}\, e^{\frac{1}{2}\mathrm{tr} (M\cdot m)}
\]
where 
\[\label{alphadef}
\alpha=\frac{1}{2}\Big(d+n(n-3)-\sum_{i=1}^n\Delta_i\Big)
,\qquad
\hat{\bs{v}}_{[ijkl]} = 
\frac{v_{ij}v_{kl}}{v_{ik}v_{jl}} = \frac{|m^{(i,j)}| |m^{(k,l)}|}{|m^{(i,k)}||m^{(j,l)}|}.
\]
Analogous expressions can be obtained for the Lee-Pomeransky and Feynman representations \eqref{LPrep} and \eqref{Feynrep}, but the Schwinger parametrisations \eqref{Cay1} and \eqref{Cay2} are particularly convenient. 
As noted, 
the diagonal Schwinger exponent 
means differential operators in the momenta can easily be traded for equivalent differential operators in the $s_{ij}$ acting on the exponential, whose action can be further evaluated through integration by parts.

\section{Weight-shifting operators}
\label{sec:wsops}

New weight-shifting operators now follow 
from the Cayley-Menger parametrisation \eqref{Cay2}.  Acting on the Schwinger exponent \eqref{CaySchwexp} with an appropriate polynomial differential operator in the momenta pulls down a corresponding polynomial in the $s_{ij}$.
Choosing these polynomials to be the Cayley-Menger determinant and its minors, we obtain shift operators either increasing $\alpha$ or decreasing one of the $\alpha_{ij}$ by integer units. 
We discuss these new operators in section \ref{sec:dup2ops}, showing their effect is to increase the spacetime dimension by two while performing assorted shifts of the operator dimensions. 
 Further weight-shifting operators can then be constructed by conjugating these operators with shadow transforms as shown in section \ref{shadowsec}.
Explicit examples  are given for the $3$- and $4$-point functions in section \ref{3n4ptShifts}. 
 Then, in section \ref{sec:Wops}, we  turn to analyse the weight-shifting operators proposed in \cite{Karateev:2017jgd}.   These preserve the spacetime dimension but  their action can nevertheless be understood using our parametric representations.

\subsection[\texorpdfstring{New operators sending $d\rightarrow d+2$}{New operators sending d to d+2}]{\texorpdfstring{New operators sending \boldmath{$ d\rightarrow d+2$}}{New operators sending d to d+2}}
\label{sec:dup2ops}

Let us begin with the $V_{ij}$ defined in \eqref{Vdef} as our independent momentum variables. 
Acting on the Schwinger exponent \eqref{CaySchwexp}, for any $i<j$ 
\[
-\frac{\partial}{\partial V_{ij} }e^{\frac{1}{2}\mathrm{tr}(M\cdot m)} = s_{ij}\,e^{\frac{1}{2}\mathrm{tr}(M\cdot m)}, \qquad  
 -V_{ij} \,e^{\frac{1}{2}\mathrm{tr}(M\cdot m)} = \frac{\partial}{\partial s_{ij}}e^{\frac{1}{2}\mathrm{tr}(M\cdot m)}
\]
allowing differential operators in the momenta to be traded for  equivalent operators in the integration variables 
$s_{ij}$.
The shift operators 
\begin{empheq}[box=\nicebox]{align} 
\label{Sdef}
\mathcal{S}_{ij}^{++} = |m^{(i,j)}|\Big|_{s_{ij}\rightarrow -\partial/\partial V_{ij}}, \qquad \mathcal{S} = |m|\Big|_{s_{ij}\rightarrow -\partial/\partial V_{ij}},
\end{empheq}
then serve to pull down factors of $|m^{(i,j)}|$ and $|m|$ respectively, thus their action is to send 
\[
\mathcal{S}_{ij}^{++}: \quad \alpha_{ij}\rightarrow \alpha_{ij} -1, \qquad
\mathcal{S}:\quad \alpha\rightarrow \alpha+1.
\]
From \eqref{alphaijdef} and \eqref{alphadef}, this is equivalent to shifting
\begin{align}\label{Sijshifts}
\mathcal{S}_{ij}^{++}: \quad &d\rightarrow d+2,\quad \Delta_i\rightarrow \Delta_i+1,\quad
\Delta_j\rightarrow \Delta_j+1,\\\mathcal{S}:\quad & d\rightarrow d+2,
\end{align}
and so the superscript on $\mathcal{S}_{ij}^{++}$ is chosen to indicate its action of raising $\Delta_i$ and $\Delta_j$ by one. 

While the Cayley-Menger structure of $\mathcal{S}_{ij}^{++}$ and $\mathcal{S}$ is manifest in the variables $V_{ij}$, where convenient these operators can easily be rewritten in terms of other scalar invariants ({\it e.g.,} Mandelstam variables) via the chain rule.  We will discuss this for 3- and 4-point functions shortly in section \ref{3n4ptShifts}.  

Alternatively, we can  express  $\mathcal{S}_{ij}^{++}$ and $\mathcal{S}$   in terms of {\it vectorial} derivatives with respect to independent momentum  $\bs{p}_i$ for $i=1,\ldots n-1$.  For $\mathcal{S}$, we find 
\begin{empheq}[box=\nicebox]{align} \label{Sdef2}
\mathcal{S} = -\, \frac{(n-1)!}{|G|}\, p_{1}^{[\mu_1}\ldots p_{n-1}^{\mu_{n-1]}}\frac{\partial}{\partial p_1^{\mu_1}}\ldots \frac{\partial}{\partial p_{n-1}^{\mu_{n-1}}}
\end{empheq}
where 
$|G|=|\bs{p}_i\cdot\bs{p}_j|$ is the Gram determinant and the $\mu_i$ are Lorentz indices.  (We leave all Lorentz indices upstairs to avoid confusion with the momentum labels, given we are working on a flat Euclidean metric.) 
The equivalence of \eqref{Sdef2} to \eqref{Sdef} can be established either by direct calculation for specific $n$, or else by considering its action on the Schwinger exponential of the representation \eqref{Schwrep}.  
This representation is the appropriate one since, from \eqref{Kirchhoff1}, it involves only dot products of the {\it independent} momenta. 
Evaluating,  we find
\begin{align}\label{Scalc}
&\mathcal{S}\,\Big( e^{-\sum_{i,j}^{n-1} (g^{-1})_{ij}\bs{p}_i\cdot \bs{p}_j }\Big) 
 \nn\\&\quad
=-\frac{(n-1)!}{|G|}\, p_1^{[\mu_1}\ldots p_{n-1}^{\mu_{n-1}]}
\nn\\&\qquad\times
\Big(-2\sum_{j_1}^{n-1}(g^{-1})_{1{j_1}}p_{j_1}^{\mu_1}\Big)\ldots
\Big(-2\sum_{j_{n-1}}^{n-1}(g^{-1})_{n-1,{j_{n-1}}}p_{j_{n-1}}^{\mu_{n-1}}\Big)e^{-\sum_{i,j}^{n-1} (g^{-1})_{ij}\bs{p}_i\cdot \bs{p}_j }\nn\\
&\quad =
\frac{-(-2)^{n-1}(n-1)!}{|G|}\,\sum_{j_1,k_1}^{n-1}\ldots \sum_{j_{n-1},k_{n-1}}^{n-1}(g^{-1})_{1j_1}\ldots (g^{-1})_{n-1,j_{n-1}}(\bs{p}_1\cdot\bs{p}_{k_1})\ldots (\bs{p}_{n-1}\cdot\bs{p}_{k_{n-1}})
\nn\\[0ex]&\quad \qquad \qquad\qquad \qquad  \qquad\qquad \qquad \times\delta^{[k_1}_{j_1}\ldots \delta^{k_{n-1}]}_{j_{n-1}}e^{-\sum_{i,j}^{n-1} (g^{-1})_{ij}\bs{p}_i\cdot \bs{p}_j }\nn\\[1ex]
&\quad=-(-2)^{n-1}|g|^{-1}e^{-\sum_{i,j}^{n-1} (g^{-1})_{ij}\bs{p}_i\cdot \bs{p}_j }
\nn\\[1ex]
&\quad=|m|\,e^{-\sum_{i,j}^{n-1} (g^{-1})_{ij}\bs{p}_i\cdot \bs{p}_j },
\end{align}
where in the last step  we used the Levi-Civita identity $(n-1)!\,\delta^{[k_1}_{j_1}\ldots \delta^{k_{n-1}]}_{j_{n-1}}=\ep^{j_1\ldots j_{n-1}}\ep_{k_1\ldots k_{n-1}}$ to generate a product of determinants  $|g^{-1}| |G|$, with the $|G|$ then cancelling.
Referring back to \eqref{Schwrep}, since $\mathcal{U}^{-d/2}=|g|^{-d/2}$ we see the action of $\mathcal{S}$ is thus indeed to raise $d\rightarrow d+2$. 

Through similar manipulations, we find
\begin{empheq}[box=\nicebox]{align} \label{Sindef}
\mathcal{S}_{in}^{++} = (-1)^{i+n}\, \frac{(n-1)!}{|G|}\, p_{1}^{[\mu_1}\ldots p_{n-1}^{\mu_{n-1]}} p_{n}^{\mu_{i}}
\prod_{k\neq i}^{n-1}
\frac{\partial}{\partial p_k^{\mu_k}}.
\end{empheq}
Relative to \eqref{Sdef2},  the derivative  $\partial/\partial p_i^{\mu_i}$  has been replaced by the dependent momentum $p_n^{\mu_{i}} =- \sum_{j_i=1}^{n-1}p_{j_i}^{\mu_i}$ positioned to the left of all derivatives.  This leads to 
\begin{align}\label{Sijcalc}
&\mathcal{S}_{in}^{++}\,\Big( e^{-\sum_{i,j}^{n-1} (g^{-1})_{ij}\bs{p}_i\cdot \bs{p}_j }\Big) 
 \nn\\&
=\frac {(-1)^{i+n}(n-1)!}{|G|}\, p_1^{[\mu_1}\ldots p_{n-1}^{\mu_{n-1}]} \nn\\&{\times}
\Big({-}2\sum_{j_1}^{n-1}(g^{-1})_{1{j_1}}p_{j_1}^{\mu_1}\Big)\ldots\Big({-}2\sum_{j_{i}}^{n-1} p_{j_{i}}^{\mu_i}\Big)\ldots
\Big({-}2\sum_{j_{n-1}}^{n-1}(g^{-1})_{n-1,{j_{n-1}}}p_{j_{n-1}}^{\mu_{n-1}}\Big)e^{-\sum_{i,j}^{n-1} (g^{-1})_{ij}\bs{p}_i\cdot \bs{p}_j }\nn\\ &
=(-1)^{i}\,2^{n-2}\sum_{j_i=1}^{n-1}\frac{\partial |g^{-1}|}{\partial (g^{-1})_{i j_{i}}}\,e^{-\sum_{i,j}^{n-1} (g^{-1})_{ij}\bs{p}_i\cdot \bs{p}_j },
\end{align}
since, relative to our previous calculation, the matrix element $(g^{-1})_{ij_i}$ is missing in the product on the middle line.  Instead of obtaining the full determinant $|g^{-1}|$, we then get the derivative of this with respect to the missing element.
As in \eqref{vfromginvn}, we can now
rewrite
\[
\sum_{j_i=1}^{n-1}\frac{\partial |g^{-1}|}{\partial (g^{-1})_{i j_{i}}} = \sum_{j_i=1}^{n-1}(\mathrm{adj}\, g^{-1})_{ij_i} = \sum_{j_i=1}^{n-1}g_{i j_i} |g^{-1}|=v_{in} |g|^{-1}= (-1)^{i}\,2^{2-n}|m^{(i,n)}|
\]
using \eqref{dmdstominnors} in the last step.
The action of $\mathcal{S}_{in}^{++}$ in \eqref{Sindef} on the exponential is thus to pull down a factor of $v_{ij} |g|^{-1}$.  From the representation \eqref{Schwrep}, this has precisely the required action of sending $\alpha_{ij}\rightarrow \alpha_{ij}-1$ and $d\rightarrow d+2$.

Finally, since the choice of dependent momentum is  immaterial,  \eqref{Sindef} generalises to 
\begin{empheq}[box=\nicebox]{align} \label{Sijdef2}
\mathcal{S}_{ij}^{++} =(-1)^{i+j}\, \frac{(n-1)!}{|G|}\, p_{1}^{[\mu_1}\ldots \hat{p}_{j}^{\hat{\mu}_j} \ldots p_{n}^{\mu_{n]}} p_{j}^{\mu_{i}}
\prod_{k\neq i,j}^{n}
\frac{\partial}{\partial p_k^{\mu_k}}
\end{empheq}
where the hats $\hat{p}_{j}^{\hat{\mu}_j}$ indicates that this factor and index are omitted in the antisymmetrised product, and we take $p_j^{\mu_j} = -\sum_{k_j\neq j}^n p_{k_j}^{\mu_j}$ as the dependent momentum. 
In principle these last few derivations allow use of the $s_{ij}$ variables to be avoided entirely, although in practice the form of the operators 
\eqref{Sdef2} and \eqref{Sijdef2} would be hard to anticipate.

\subsection{Further shift operators from shadow conjugation}
\label{shadowsec}

Additional $d\rightarrow d+2$ shift operators can now be constructed -- at no expense -- by conjugating  $\mathcal{S}_{ij}^{++}$ and $\mathcal{S}$ by a pair of shadow transforms.  This idea was discussed recently  for  $d$-preserving weight-shifting operators  in 
\cite{Bzowski:2022rlz}.  

In momentum space, the shadow transform $\Delta_i\rightarrow d - \Delta_i$ (leaving $d$ invariant) simply corresponds to multiplying by $p_i^{d-2\Delta_i}$.  
First, notice that attempting to conjugate $\mathcal{S}_{ij}^{++}$ by shadow transforms on either of  $\Delta_i$ or $\Delta_j$ has no effect:
for example, the action of the operator $p_i^{2\Delta_i-d} \mathcal{S}_{ij}^{++} p_i^{d-2\Delta_i}$ corresponds to the successive parameter shifts
\begin{align}
(\Delta_i,\Delta_j,d) &\xrightarrow{p_i^{d-2\Delta_i}} (d-\Delta_i,\Delta_j,d)\nn\\& \xrightarrow{\mathcal{S}_{ij}^{++}} (d-\Delta_i+1,\Delta_j+1,d+2)\nn\\&
\xrightarrow{p_i^{(d+2)-2(d-\Delta_1+1)}\,=\,p_i^{2\Delta_i-d}} 
((d+2)-(d-\Delta_i+1), \Delta_j+1,d+2) \nn\\
&\qquad\qquad \qquad\qquad \qquad \quad=(\Delta_i+1,\Delta_j+1,d+2)
\end{align}
which is equivalent to the action of $\mathcal{S}_{ij}^{++}$ alone.  Further computations confirm that the shadow transform on $\Delta_i$ or $\Delta_j$ commutes with $\mathcal{S}_{ij}^{++}$.

However, we {\it do} obtain new operators if we shadow conjugate  $\mathcal{S}_{ij}^{++}$ on any index $k\neq i,j$.  For example, the action of 
\[ \label{simplexRop}
p_k^{2\Delta_k+2-d}\mathcal{S}_{ij}^{++} p_k^{d-2\Delta_k}
\] 
corresponds to the successive parameter shifts
\begin{align}\label{simplexRopshifts}
(\Delta_i,\Delta_j,\Delta_k,d) &\xrightarrow{p_k^{d-2\Delta_k}} (\Delta_i,\Delta_j,d-\Delta_k,d)\nn\\& \xrightarrow{\mathcal{S}_{ij}^{++}} (\Delta_i+1,\Delta_j+1,d-\Delta_k,d+2)\nn\\&
\xrightarrow{p_k^{(d+2)-2(d-\Delta_k)}\,=\,p_k^{2\Delta_k+2-d}} 
(\Delta_i+1, \Delta_j+1,\Delta_k+2,d+2).
\end{align}
Thus, in addition  to the shifts produced by $\mathcal{S}_{ij}^{++}$ alone, we have also shifted $\Delta_k$ up by two.
Shadow conjugating on further variables has the same effect, for example, 
\[
p_k^{2\Delta_k+2-d}p_l^{2\Delta_l+2-d}\mathcal{S}_{ij}^{++} p_k^{d-2\Delta_k}p_l^{d-2\Delta_l}
\] 
for  any $(k,l)\neq (i,j)$  sends $(\Delta_i,\Delta_j,\Delta_k,\Delta_l,d)\rightarrow (\Delta_i+1,\Delta_j+1,\Delta_k+2,\Delta_l+2,d+2)$. 

We can also apply similar considerations to $\mathcal{S}$.  The action of
\[
p_i^{2\Delta_i+2-d} \mathcal{S} p_i^{d-2\Delta_i}
\]
corresponds to the shifts
\begin{align}
(\Delta_i,d) \xrightarrow{p_i^{d-2\Delta_i}} (d-\Delta_i,d)
\xrightarrow{\mathcal{S}} (d-\Delta_i, d+2)
\xrightarrow{p_i^{(d+2)-2(d-\Delta_i)}=p_i^{2\Delta_i+2-d}} (\Delta_i+2, d+2).
\end{align}
Shadow conjugating on further momenta $p_k$ leads similarly to shifting $\Delta_k\rightarrow\Delta_k+2$. 

With all these operators obtained by shadow conjugation, notice  we can always obtain an equivalent  differential operator with purely polynomial coefficients ({\it i.e.,} an operator in the Weyl algebra) by commuting the inner $p_k^{d-2\Delta_k}$ shadow factors through the differential operator $\mathcal{S}$ or $\mathcal{S}_{ij}^{++}$, whereupon all non-integer powers cancel with those from the outer shadow transform.

\subsection{Examples at three and four points} 
\label{3n4ptShifts}

To illustrate the general discussion in the two preceding subsections, let us now compute the explicit form of  these $d\rightarrow d+2$ shift operators for $3$- and $4$-point functions. 

\subsubsection{3-point shift operators}

For the 3-point function, it is convenient to use the three squared momentum magnitudes as variables.  Defining
\[
P_i = p_i^2, \qquad D_i = \frac{\partial}{\partial P_i}=\frac{1}{2p_i}\frac{\partial}{\partial p_i}, \qquad i=1,2,3
\]
via momentum conservation we have
\[\label{3ptchain}
P_i = -\sum_{j\neq i}^3 \bs{p}_i\cdot\bs{p}_j = \sum_{j\neq i}^3 V_{ij},\qquad \frac{\partial}{\partial V_{ij}} =\sum_{k=1}^3 \frac{\partial P_k}{\partial V_{ij}}\frac{\partial}{\partial P_k} =  D_i+D_j
\]
From \eqref{Sdef}, writing $D_iD_j=D_{ij}$ for short, we then find
\begin{align}
&\qquad \qquad \mathcal{S} = -4(D_{12}+D_{23}+D_{13}), \nn\\ 
&\mathcal{S}_{12}^{++} = -2D_3, \quad \mathcal{S}_{23}^{++} = -2D_1, \quad \mathcal{S}_{13}^{++} = 2D_2.
\end{align}
The various signs on the second line reflect our choice to use the Cayley minors in \eqref{Cay2} and \eqref{Sdef}: had we used instead the cofactors or $\partial|m|/\partial s_{ij}$ as per \eqref{dmdstominnors} then all signs would be the same. 
Generally, any overall coefficient in $\mathcal{S}$ or the  $\mathcal{S}_{ij}^{++}$ can be eliminated by rescaling the corresponding prefactor in the definition of the simplex integral. 

As noted in the introduction,   these 3-point operators (and their shadow conjugates) are already known from the triple-$K$ representation of the 3-point function.  
In \cite{Bzowski:2013sza, Bzowski:2015yxv}, the Bessel shift operators
\[\label{LRdef}
\mathcal{L}_i = -\frac{1}{p_i}\frac{\partial}{\partial p_i}, \qquad
\mathcal{R}_i = 2\beta_i - p_i\frac{\partial}{\partial p_i} = p_i^{2\beta_i+2}\,\mathcal{L}_i \,p_i^{-2\beta_i},\qquad \beta_i=\Delta_i-\frac{d}{2}
\]
where shown to act on the 3-point function by sending
\begin{align}\label{LRaction}
\mathcal{L}_i: \quad\beta_i\rightarrow \beta_i -1,\quad d\rightarrow d+2, \qquad
\mathcal{R}_i:\quad \beta_i\rightarrow\beta_i+1,\quad d\rightarrow d+2,
\end{align}
or equivalently, 
\begin{align}
&\mathcal{L}_1:\quad  (d,\Delta_1,\Delta_2,\Delta_3)\rightarrow (d+2,\Delta_1,\Delta_2+1,\Delta_3+1), \\
&\mathcal{R}_1:\quad (d,\Delta_1,\Delta_2,\Delta_3)\rightarrow (d+2,\Delta_1+2,\Delta_2+1,\Delta_3+1), 
\end{align}
and similarly under permutations.  This is consistent with our analysis here, since 
\[
(\mathcal{L}_1,\mathcal{L}_2,\mathcal{L}_3)=
(\mathcal{S}_{23}^{++},-\mathcal{S}_{13}^{++},\mathcal{S}_{12}^{++})
\]
and $\mathcal{S}_{ij}^{++}$ augments $\Delta_i$ and $\Delta_j$ by one and $d$ by two.  The $\mathcal{R}_i$ operators are then their shadow conjugates as defined in \eqref{simplexRop}, producing the expected shifts \eqref{simplexRopshifts}.
Finally,
\[
\mathcal{S} = -\mathcal{L}_1\mathcal{L}_2-\mathcal{L}_2\mathcal{L}_3-\mathcal{L}_3\mathcal{L}_1
\]
does not appear explicitly in \cite{Bzowski:2015yxv}, but can be derived as follows.  Writing the 3-point function as the triple-$K$ integral $I_{d/2-1,\{\beta_1,\beta_2,\beta_3\}}$, from \eqref{LRaction} we have
\begin{align}
-\mathcal{S}I_{d/2-1,\{\beta_1,\beta_2,\beta_3\}}& =
I_{d/2+1,\{\beta_1-1,\beta_2-1,\beta_3\}}+
I_{d/2+1,\{\beta_1,\beta_2-1,\beta_3-1\}}+
I_{d/2+1,\{\beta_1-1,\beta_2,\beta_3-1\}}\nn\\[1ex]&=
(\mathcal{R}_1+\mathcal{R}_2+\mathcal{R}_3)I_{d/2,\{\beta_1-1,\beta_2-1,\beta_3-1\}}\nn\\&
=\Big(\frac{d}{2}+\beta_t+4\Big)I_{d/2,\{\beta_1-1,\beta_2-1,\beta_3-1\}}
\end{align}
where the final line follows by eliminating the sum of $\mathcal{R}_i$ operators using the dilatation Ward identity.  The effect of $\mathcal{S}$ is thus to increase $d\rightarrow d+2$ and all $\beta_i\rightarrow \beta_i-1$.  All dimensions $\Delta_i=\beta_i+d/2$ are then preserved, consistent with \eqref{Sijshifts}.

\subsubsection{4-point shift operators}

The 3-point calculations above provide a first consistency check, but to obtain genuinely new shift operators  we now turn to the 4-point function.

To write our results, we introduce the  Mandelstam variables,
\[\label{4ptMands}
P_I = \{p_1^2,p_2^2,p_3^2,p_4^2,s^2,t^2\}, \qquad I=1,\ldots, 6
\]
where $s^2=(\bs{p}_1+\bs{p}_2)^2$ and $t^2=(\bs{p}_2+\bs{p}_3)^2$, and define the derivative operators 
\[
D_I = \frac{\partial}{\partial P_I}, \qquad D_{IJ} = D_I D_J, \qquad D_{IJK} = D_I D_J D_K.
\]
Defining $\mathcal{S}_{ij}^{++} = 4(-1)^{i+j}S_{ij}^{++}$ and $\mathcal{S} = -8 S$ to suppress trivial numerical factors, from \eqref{Sdef} and \eqref{m4}, using the chain rule analogous to \eqref{3ptchain},  we obtain the operators
\begin{align}\label{4ptSij}
&S_{12}^{++} = D_{34}+D_{45}+D_{35}+D_{56}, \nn\\
&S_{13}^{++} = D_{24} - D_{56},\nn\\
&S_{14}^{++} = D_{23}+D_{26}+D_{36}+D_{56},\nn\\
&S_{23}^{++} = D_{14}+D_{16}+D_{46}+D_{56},\nn\\
&S_{24}^{++} = D_{13} - D_{56},\nn\\
&S_{34}^{++} = D_{12}+D_{15}+D_{25}+D_{56},
\end{align}
and 
\begin{align}
S& =D_{456}+D_{356}+D_{346}+D_{256}+D_{246}+D_{245}+D_{235}+D_{234}\nn\\ & \quad +D_{156}+D_{145}+D_{136}+D_{135}+D_{134}+D_{126}+D_{124}+D_{123}.
\end{align}
As per \eqref{Sijshifts}, the $S_{ij}^{++}$ increase $\Delta_i$ and $\Delta_j$ by one and $d$ by two, while $S$ increases $d$ by two.

Following section \ref{shadowsec}, we can obtain further shift operators by shadow conjugation. 
As noted earlier, shadow conjugating each $S_{ij}^{++}$ on either of the $(i,j)$ indices has no effect: from \eqref{4ptSij},  $S_{ij}^{++}$ contains neither $D_i$ or $D_j$ hence these shadow factors commute through the operator.
Instead, we must shadow conjugate each $S_{ij}^{++}$ with respect to indices other than $(i,j)$. 
At four points, once a pair of insertions $(i,j)$ is specified, the remaining set also form a pair $(k,l)\neq (i,j)$. Shadow conjugating each $S_{ij}^{++}$ on the opposite pair $(k,l)$ then  defines
\[
\bar{S}_{ij}^{++}=p_k^{2(\b_k+1)}p_l^{2(\b_l+1)}S_{ij}^{++}p_k^{-2\b_k}p_l^{-2\b_l}, \qquad (k,l)\neq (i,j)
\]
where $\beta_i=\Delta_i-d/2$.  Expressed in terms of the variables \eqref{4ptMands}, we find
\begin{align}
&\bar{S}_{12}^{++}=\b_3\b_4-\b_4P_3D_3-\b_3P_4D_4-(\b_3P_4+\b_4P_3)D_5+P_3P_4S_{12}^{++}, \nn\\
&\bar{S}_{13}^{++}=\b_2\b_4-\b_4P_2D_2-\b_2P_4D_4+P_2P_4S_{13}^{++}, \nn\\
&\bar{S}_{14}^{++}=\b_2\b_3-\b_3P_2D_2-\b_2P_3D_3-(\b_2P_3+\b_3P_2)D_6+P_2P_3S_{14}^{++},\nn\\
&\bar{S}_{23}^{++}=\b_1\b_4-\b_4P_1D_1-\b_1P_4D_4-(\b_1P_4+\b_4P_1)D_6+P_1P_4S_{23}^{++},\nn\\
&\bar{S}_{24}^{++}=\b_1\b_3-\b_3P_1D_1-\b_1P_3D_3+P_1P_3S_{24}^{++},\nn\\
&\bar{S}_{34}^{++}=\b_1\b_2-\b_2P_1D_1-\b_1P_2D_2-(\b_1P_2+\b_2P_1)D_5+P_1P_2S_{34}^{++}.
\end{align}
The action of each operator $\bar{S}_{ij}^{++}$ is to shift $d\rightarrow d+2,~\Delta_{i,j}\rightarrow \Delta_{i,j}+1$ and $\Delta_{k,l}\rightarrow \Delta_{k,l}+2$.  This leaves $\beta_i$ and $\beta_j$ invariant while raising $\beta_k$ and $\beta_l$ by one. 
Heuristically, these $\bar{S}_{ij}^{++}$ are then the 4-point generalisation of the 3-point $\mathcal{R}_i$ operators in \eqref{LRdef}.  Likewise, the $S_{ij}^{++}$ in \eqref{4ptSij} 
leave $\beta_i$ and $\beta_j$ invariant but lower $\beta_{k}$ and $\beta_l$ by one, and represent the 4-point generalisation of the 3-point $\mathcal{L}_i$ operators.

Besides shadow conjugating $S_{ij}^{++}$ with respect to the pair $(k,l)$, one can of course also conjugate with respect to only a single index $k$ to find operators sending $d\rightarrow d+2$, $\Delta_{i,j}\rightarrow \Delta_{i,j}+1$ and $\Delta_k\rightarrow\Delta_k+2$ only.  One can also apply the shadow conjugation procedure to the $d\rightarrow d+2$ operator $S$.  All these operators can be evaluated similarly to the $\bar{S}_{ij}^{++}$ above and we will not write them explicitly.
One case of particular interest, however, corresponds to acting with $\bar{S}_{ij}^{++}$ followed by $S_{ij}^{++}$, which produces an overall shift of $d\rightarrow d+4$ while increasing all operator dimensions by two.  The same shift is produced when acting with these operators in the opposite order (remembering to shift $\beta_{k,l}\rightarrow \beta_{k,l}-1$ in $\bar{S}_{ij}^{++}$  to account for the prior action of $S_{ij}^{++}$).  By subtracting, we then obtain a shift operator of only {\it second} order in derivatives, rather than fourth.  
For example,
\[
\bar{S}_{24}^{++}\Big|_{\beta_1-1,\beta_3-1} S_{24}^{++} - S_{24}^{++} \bar{S}_{24}^{++}\Big|_{\beta_1,\beta_3} = (\beta_1+\beta_3)D_{56}
\]
and so  $D_{56}$ shifts $d\rightarrow d+4$ while sending all $\Delta_i\rightarrow\Delta_i+2$ and preserving the $\beta_i$.

Finally, let us emphasise that the action of all these shift operators  is  general and not in any way tied to the simplex representation: {\it any} solution of the 4-point conformal Ward identities is mapped to an appropriately shifted solution.\footnote{Up to a technical caveat (common to all shift operators) that where divergences occur, one must work in a suitable dimensional regularisation scheme.  In some cases the shift operator then only yields the leading divergences of the shifted correlator, see the discussion in \cite{Bzowski:2022rlz}.} 
We have confirmed this explicitly by computing all the relevant intertwining relations between the shift operators in this section and the conformal Ward identities, whose form in Mandelstam variables can be found in {\it e.g.,} \cite{Bzowski:2022rlz, Arkani-Hamed:2018kmz}.  Thus, for example,
\[\label{intertwiner1}
\mathcal{K}(\Delta_1+1,\Delta_2+1,\Delta_3,\Delta_4,d+2)S_{12}^{++} = S_{12}^{++}\mathcal{K}(\Delta_1,\Delta_2,\Delta_3,\Delta_4,d)
\]
where $\mathcal{K}(\{\Delta_i\},d)$ represents schematically any of the special conformal or dilatation Ward identities with the operator and spacetime dimensions as indicated.  Applying this relation to any CFT correlator with dimensions $(\{\Delta_i\},d)$, the right-hand side vanishes and the left-hand side then indicates that the action of $S_{12}^{++}$ produces a solution of the shifted Ward identities.
Intertwining relations such as these\footnote{More generally,  the right-hand side of \eqref{intertwiner1} could feature {\it any} operator in the left ideal of the conformal Ward identities, since all that matters is that it vanishes when acting on a solution with dimensions $(\{\Delta_i\},d)$.}  allow the shift action of operators to be established independently of any integral representation for the correlator.

\subsection[{\texorpdfstring{Operators preserving $d$}{Operators preserving d}}]{\texorpdfstring{Operators preserving \boldmath{$d$}}{Operators preserving d}}
\label{sec:Wops}

A different class of weight-shifting operators that preserve the spacetime dimension $d$ while shifting the $\Delta_i$ was identified in \cite{Karateev:2017jgd}.
In momentum space, these operators have been applied to de Sitter correlators in \cite{Arkani-Hamed:2018kmz, Baumann:2019oyu}.
With the aid of shadow conjugation, we can write them in the compact form  \cite{Bzowski:2022rlz}
\begin{align}
\label{Wops}
\mathcal{W}^{--}_{ij} &=\frac{1}{2}\, \Big(\frac{\partial}{\partial p_i^\mu}-\frac{\partial}{\partial p_j^\mu}\Big)\Big(\frac{\partial}{\partial p_{i\mu}}-\frac{\partial}{\partial p_{j \mu}}\Big)
\nn\\
\mathcal{W}^{+-}_{ij}&= p_i^{2(\beta_i+1)}\mathcal{W}^{--}_{ij} p_i^{-2\beta_i}\nn\\
\mathcal{W}^{-+}_{ij} &= p_j^{2(\beta_j+1)}\mathcal{W}^{--}_{ij} p_j^{-2\beta_j} \nn\\
\mathcal{W}_{ij}^{++}  &=
p_i^{2(\beta_i+1)}p_j^{2(\beta_j+1)} \mathcal{W}_{ij}^{--}p_i^{-2\beta_i}p_j^{-2\beta_j},
\end{align}
where $\beta_i=\Delta_i-d/2$ and $1\le i<j\le n-1$ so $\bs{p}_n$ is taken as the dependent momentum.  
Their action is to shift
\[\label{Waction}
\mathcal{W}^{\sigma_i\sigma_j}_{ij}:\quad \Delta_i\rightarrow\Delta_i+\sigma_i,\quad \Delta_j\rightarrow\Delta_j+\sigma_j,\quad d\rightarrow d, \quad \{\sigma_i,\sigma_j\}\in \pm1.
\]

In this section, our goal is to understand the action of the simplest of these operators, $\mathcal{W}_{ij}^{--}$, from the simplex perspective.  The action of the others then follows via shadow conjugation, or else can be shown explicitly: for example, we analyse  $\mathcal{W}_{ij}^{-+}$ in section \ref{sec:Wmp}.

We begin by writing the Schwinger exponential \eqref{Kirchhoff1} in the form
\[\label{SchwExpRes}
-\mathrm{tr}(g^{-1}\cdot G)  
=\sum_{k<l}^n s_{kl}\,\bs{p}_k\cdot\bs{p}_l
=-\sum_{k<l}^{n-1} (s_{kn}+s_{ln}-s_{kl})(\bs{p}_k\cdot\bs{p}_l) -\sum_{k}^{n-1}s_{kn}\,p_k^2.
\]
As only the independent momenta feature in this last expression, the action of $\mathcal{W}_{ij}^{--}$ on the Schwinger exponential can  be rewritten as a differential operator in the $s_{kl}$.  We will do this in several steps.  First, notice that 
\begin{align}
\frac{\partial}{\partial p_i^\mu} e^{-\mathrm{tr}(g^{-1}\cdot G) } &=-\Big( 2s_{in}p_i^\mu
+\sum_{k\neq i}^{n-1}(s_{in}+s_{kn}-s_{ik})p_k^\mu \Big) e^{-\mathrm{tr}(g^{-1}\cdot G) } \nn\\&
=-\sum_{k}^{n-1}(s_{in}+s_{kn}-s_{ik})\,p_k^\mu\, e^{-\mathrm{tr}(g^{-1}\cdot G) },
\end{align}
where in the second line  $s_{ik}$ vanishes for $i=k$.   This gives
\begin{align}
\Big(\frac{\partial}{\partial p_i^\mu} -\frac{\partial}{\partial p_j^\mu} \Big)e^{-\mathrm{tr}(g^{-1}\cdot G) } &=
\sum_{k}^{n-1}(s_{ik}-s_{jk}-s_{in}+s_{jn})\,p_k^\mu \,e^{-\mathrm{tr}(g^{-1}\cdot G) },
\end{align}
and hence
\begin{align}
&\mathcal{W}_{ij}^{--}e^{-\mathrm{tr}(g^{-1}\cdot G) }\\  &\quad =
\Big(-d s_{ij} + \frac{1}{2}\sum_{k,l}^{n-1}(s_{ik}-s_{jk}-s_{in}+s_{jn})(s_{il}-s_{jl}-s_{in}+s_{jn})\,\bs{p}_k\cdot\bs{p}_l\Big)\,e^{-\mathrm{tr}(g^{-1}\cdot G) }.\nn
\end{align}
To rewrite these momentum dot products as derivatives with respect to the $s_{kl}$, we now rearrange this sum as follows.  Using momentum conservation $p_k^2 =-\sum_{l\neq k}^n\bs{p}_k\cdot \bs{p}_l$, for any generic coefficient $A_k$ such that $A_n=0$, we have
\begin{align}\label{Atrick}
\sum_{k,l}^{n-1}A_k A_l \,\bs{p}_k\cdot\bs{p}_l &= \sum_{\substack{k,l\\k\neq l}}^{n-1}A_k A_l \,\bs{p}_k\cdot\bs{p}_l
+\sum_{k}^{n-1}A_k^2\, p_k^2 \nn\\
& = \sum_{\substack{k,l\\k\neq l}}^{n-1}A_k (A_l-A_k) \,\bs{p}_k\cdot\bs{p}_l -\sum_{k}^{n-1}A_k^2 \,\bs{p}_k\cdot\bs{p}_n
\nn\\
& = -\frac{1}{2}\sum_{\substack{k,l\\k\neq l}}^{n-1} (A_l-A_k)^2 \,\bs{p}_k\cdot\bs{p}_l -\sum_{k}^{n-1}A_k^2 \,\bs{p}_k\cdot\bs{p}_n
\nn\\
& = -\sum_{k<l}^{n} (A_l-A_k)^2 \,\bs{p}_k\cdot\bs{p}_l 
\end{align}
where in the final line the sum runs up to $n$.
Setting $A_k = s_{ik}-s_{jk}-s_{in}+s_{jn}$, we find
\begin{align}\label{Wmmid}
\mathcal{W}_{ij}^{--}e^{-\mathrm{tr}(g^{-1}\cdot G) } &=
\Big(-d s_{ij} - \frac{1}{2}\sum_{k<l}^{n} (s_{ik}-s_{jk}-s_{il}+s_{jl})^2\,\bs{p}_k\cdot\bs{p}_l
\Big)\,e^{-\mathrm{tr}(g^{-1}\cdot G) }\nn\\
&
=
\Big(-d s_{ij} - \frac{1}{2}\sum_{k<l}^{n} (s_{ik}-s_{jk}-s_{il}+s_{jl})^2\,\partial_{s_{kl}}
\Big)\,e^{-\mathrm{tr}(g^{-1}\cdot G) }\nn\\&
=(-d s_{ij} + 2\partial_{v_{ij}})\,e^{-\mathrm{tr}(g^{-1}\cdot G) }\nn\\[1ex]
& = 2\, |g|^{d/2} \,\partial_{v_{ij}} \,\Big(|g|^{-d/2}\,e^{-\mathrm{tr}(g^{-1}\cdot G) }\Big).
\end{align}
In the second line here, we exchanged $\bs{p}_k\cdot\bs{p}_l$ for $\partial_{s_{kl}}$ using the first expression in \eqref{SchwExpRes}.
The change of variables from $\partial_{s_{kl}}$ to $\partial_{v_{ij}}$ in the third line then comes from the Jacobian evaluated in appendix \ref{sec:Jacobimatrixels}, and in the final line we used \eqref{sdef}.

The action of $\mathcal{W}_{ij}^{--}$ on the full simplex integral \eqref{Schwrep} now follows.  First, the outer factor of $|g|^{d/2}$ in \eqref{Wmmid} cancels with the factor $\mathcal{U}^{-d/2}=|g|^{-d/2}$ in \eqref{Schwrep}.  Integrating by parts with respect to $v_{ij}$, assuming the boundary terms vanish,\footnote{For the upper limit this is automatic for momentum configurations with non-vanishing Gram determinant thanks to the decaying exponential.  The lower limit vanishes provided $\alpha_{ij}<0$.} 
the derivative then acts on the prefactors  as
\[
-2\partial_{v_{ij}}\Big(\prod_{k<l}^n v_{kl}^{-\alpha_{kl}-1} f(\hat{\bs{v}}) \Big)= v_{ij}^{-1}\prod_{k<l}^n v_{kl}^{-\alpha_{kl}-1} \tilde{f}(\hat{\bs{v}}).  
\]
Here, the terms coming from $\partial_{v_{ij}}$ hitting the cross-ratios \eqref{vcrossratio} inside the arbitrary function $f(\hat{\bs{v}})$, as well as those from hitting $v_{ij}^{-\alpha_{ij}-1}$, have been repackaged  in the form $v_{ij}^{-1}\tilde{f}(\hat{\bs{v}})$ for some new function of cross-ratios  $\tilde{f}(\hat{\bs{v}})$.  Thus, overall, we find 
\begin{align}
&\mathcal{W}_{ij}^{--} \Big(\prod_{k<l}^n \int_0^\infty \mathrm{d}v_{kl}\, v_{kl}^{-\alpha_{kl}-1}\Big) \,f(\hat{\bs{v}}) |g|^{-d/2} e^{-\mathrm{tr}(g^{-1}\cdot G)} \nn\\
&\qquad \quad =\Big(
\prod_{k<l}^n \int_0^\infty \mathrm{d}v_{kl}\,  v_{kl}^{-\alpha_{kl}-1} \Big)\,v_{ij}^{-1}\tilde{f}(\hat{\bs{v}})|g|^{-d/2} e^{-\mathrm{tr}(g^{-1}\cdot G)}.
\end{align}
The action of $\mathcal{W}_{ij}^{--}$ on the simplex is therefore to send $\alpha_{ij}\rightarrow \alpha_{ij}+1$, up to changes of the arbitrary function.  The latter is of no account as far as mapping one solution of the conformal Ward identities to another is concerned.\footnote{An exception is if  $\mathcal{W}_{ij}^{--}$ maps us from a finite correlator to a singular one, corresponding to a solution of the conditions $d+\sum_{i=1}^n \sigma_i(\Delta_i-d/2)=-2k$ for some non-negative integer $k$ and a choice of signs $\{\sigma_i\}\in \pm 1$, see \cite{Bzowski:2019kwd}.  In such cases, the arbitrary function $\tilde{f}(\hat{\bs{v}})$ vanishes.  In dimensional regularisation, this zero then cancels the pole coming from the divergent correlator such that the result is finite, see \cite{Bzowski:2022rlz}.}    
From \eqref{alphaijdef}, we now confirm that sending $\alpha_{ij}\rightarrow \alpha_{ij}+1$ while keeping the remaining $\alpha_{kl}$ fixed is equivalent to sending
$\Delta_i\rightarrow \Delta_i-1$ and $\Delta_j\rightarrow \Delta_j-1$ while preserving $d$, in perfect agreement with \eqref{Waction}.

\section{Verifying the conformal Ward identities}
\label{sec:CWI}

In this section, we 
prove that the  parametric representation of the simplex integral \eqref{Schwrep} satisfies the conformal Ward identities for any arbitrary function of cross-ratios.
The corresponding result for the vectorial simplex integral \eqref{simplex} was established in  \cite{Bzowski:2019kwd, Bzowski:2020kfw}.
Working purely in momentum space,
our  approach is 
to show that the action of the Ward identities on the simplex integral reduces to a total derivative.  With a degree of hindsight, the structure of this total derivative, obtained in  \eqref{CWItotalderiv}, 
can also be understood from somewhat simpler position-space arguments.  We will return to these in section \ref{sec:posnspacecwis}.

As the dilatation Ward identity can be verified by power counting, we focus on the special conformal Ward identities
\[\label{scwi}
0 =  \sum_{j=1}^{n-1}\Big(p_j^\mu \frac{\partial}{\partial p_j^\nu}\frac{\partial}{\partial p_j^\nu}-2p_j^\nu \frac{\partial}{\partial p_j^\nu}\frac{\partial}{\partial p_j^\mu}+2(\Delta_j-d)\frac{\partial}{\partial p_j^\mu}\Big)\lla \O_1(\bs{p}_1) \ldots \O_n(\bs{p}_n) \rra,
\]
treating $\bs{p}_n$ as the dependent momentum.
As a first step, we rewrite the action of each individual term in \eqref{scwi} on the Schwinger exponential as an equivalent differential operator in $v_{ij}$.   
From \eqref{Kirchhoff1}, we have
\begin{align}
&\sum_{j}^{n-1}2(\Delta_j-d)\frac{\partial}{\partial p_j^\mu} \,e^{-\mathrm{tr}\,(g^{-1}\cdot G)} = \sum_{j}^{n-1} p_j^\mu\Big( -4\sum_{k}^{n-1} (\Delta_k-d)g^{-1}_{jk} \Big)\,e^{-\mathrm{tr}\,(g^{-1}\cdot G)} ,\\
&\sum_{j}^{n-1}p_j^\mu \frac{\partial}{\partial p_j^\nu}\frac{\partial}{\partial p_j^\nu}\,e^{-\mathrm{tr}\,(g^{-1}\cdot G)} =\sum_{j}^{n-1} p_j^\mu\Big(-2d g^{-1}_{jj}+ 4\sum_{k,l}^{n-1}\,g^{-1}_{jk}g^{-1}_{jl} \,\bs{p}_k\cdot\bs{p}_l\Big)\,e^{-\mathrm{tr}\,(g^{-1}\cdot G)}. 
\end{align}
Using \eqref{ginvs} for the inverse metric and the manipulation \eqref{Atrick}, this last expression can be rewritten analogously to \eqref{Wmmid}:
\begin{align}
\sum_{j}^{n-1}p_j^\mu \frac{\partial}{\partial p_j^\nu}\frac{\partial}{\partial p_j^\nu}\,e^{-\mathrm{tr}\,(g^{-1}\cdot G)} &=\sum_{j}^{n-1} p_j^\mu\Big(-2d s_{jn}-\sum_{k<l}^{n}\,(s_{jk}-s_{kn}-s_{jl}+s_{ln})^2  \,\bs{p}_k\cdot\bs{p}_l\Big)\,e^{-\mathrm{tr}\,(g^{-1}\cdot G)}\nn\\
 &=\sum_{j}^{n-1} p_j^\mu\Big(-2d s_{jn}-\sum_{k<l}^{n}\,(s_{jk}-s_{kn}-s_{jl}+s_{ln})^2 \,\partial_{s_{kl}}\Big)\,e^{-\mathrm{tr}\,(g^{-1}\cdot G)}\nn\\
  & =\sum_{j}^{n-1} p_j^\mu\Big(-2d s_{jn}+4\partial_{v_{jn}}\Big)\,e^{-\mathrm{tr}\,(g^{-1}\cdot G)}. 
\end{align}
Next, we must deal with
\begin{align}
&\sum_{j}^{n-1}\Big(-2p_j^\nu \frac{\partial}{\partial p_j^\nu}\frac{\partial}{\partial p_j^\mu}\Big)\,e^{-\mathrm{tr}\,(g^{-1}\cdot G)} \nn\\&\qquad\qquad =4\sum_{j}^{n-1} p_j^\mu\,\Big( g_{jj}^{-1}-\sum_{k,l}^{n-1}(g^{-1}_{jk}+g^{-1}_{jl})g^{-1}_{kl}\,\bs{p}_k\cdot\bs{p}_l\Big)\,e^{-\mathrm{tr}\,(g^{-1}\cdot G)}\nn\\&
\qquad\qquad 
=4\sum_{j}^{n-1} p_j^\mu\,\Big( g_{jj}^{-1}-2\sum_{k<l}^{n-1}(g^{-1}_{jk}+g^{-1}_{jl})g^{-1}_{kl}\,\bs{p}_k\cdot\bs{p}_l
-2\sum_{k}^{n-1} g^{-1}_{jk}g^{-1}_{kk} p_k^2
\Big)\,e^{-\mathrm{tr}\,(g^{-1}\cdot G)}. \label{gkkpiece1}
\end{align}
Using \eqref{ginvs} and momentum conservation, the $p_k^2$ terms in this final sum can be rewritten 
\begin{align}
-2\sum_{k}^{n-1} g^{-1}_{jk}g^{-1}_{kk} p_k^2 &=
\sum_{k}^{n}s_{kn}(s_{jn}+s_{kn}-s_{jk}) \sum_{l\neq k}^n \bs{p}_k\cdot\bs{p}_l \nn\\
& = \sum_{k<l}^n \Big(s_{kn}(s_{jn}+s_{kn}-s_{jk})+s_{ln}(s_{jn}+s_{ln}-s_{jl})\Big)\bs{p}_k\cdot\bs{p}_l. \label{gkkpiece2}
\end{align}
In the first line here, notice we extended the sum over $k$ to run up to $n$, which is possible since the additional term with $k=n$ vanishes as $s_{nn}=0$.  To get the second line, we then re-expressed the terms for which $k>l$ by swapping $k\leftrightarrow l$. 
For convenience, it is useful to define
\[\label{hatgdef}
\hat{g}^{-1}_{ij} = \frac{1}{2}(s_{in}+s_{jn}-s_{ij}) = \begin{cases}
g^{-1}_{ij}\qquad & i,j\le n-1,\\
0\qquad & i=n \,\,\mathrm{and/or}\,\, j=n,
\end{cases}
\]
effectively extending the $(n-1)\times(n-1)$ matrix $g^{-1}_{ij}$ to an $n\times n$ matrix $\hat{g}^{-1}_{ij}$ by adding a final row and column of zeros. 
This allows us to compactly rewrite \eqref{gkkpiece1} and \eqref{gkkpiece2} as
\begin{align}
&\sum_{j}^{n-1}\Big(-2p_j^\nu \frac{\partial}{\partial p_j^\nu}\frac{\partial}{\partial p_j^\mu}\Big)\,e^{-\mathrm{tr}\,(g^{-1}\cdot G)} \nn\\&\quad
=\sum_{j}^{n-1} p_j^\mu\,\Big( 4\hat{g}_{jj}^{-1}+8\sum_{k<l}^{n}\Big(-(\hat{g}^{-1}_{jk}+\hat{g}^{-1}_{jl})\hat{g}^{-1}_{kl}+\hat{g}_{kk}^{-1}\hat{g}^{-1}_{jk}+\hat{g}^{-1}_{ll}\hat{g}^{-1}_{jl}\Big)\,\partial_{s_{kl}}
\Big)\,e^{-\mathrm{tr}\,(g^{-1}\cdot G)}.
\end{align}
Here, the sum over $l$  for  the $\bs{p}_k\cdot\bs{p}_l$ terms in \eqref{gkkpiece1} has similarly been extended to run up to $n$, noting the additional $l=n$ term vanishes.  We then replaced $\bs{p}_k\cdot\bs{p}_l$  by a derivative with respect to $s_{kl}$ using \eqref{SchwExpRes}.  
The result now simplifies  further upon exchanging 
\[
\partial_{s_{kl}} = \sum_{a<b}^n \frac{\partial v_{ab}}{\partial s_{kl}}\, \partial_{v_{ab}}=  -\sum_{a<b}^n \tilde{g}_{a(k}\tilde{g}_{l)b}\,
\partial_{v_{ab}},
\]
where $\tilde{g}_{ab}$ is the Laplacian matrix \eqref{tildeg} and the Jacobian is evaluated in appendix \ref{sec:Jacobimatrixels}.
 First, we write
\begin{align}
&8\sum_{k<l}^{n}\Big(-(\hat{g}^{-1}_{jk}+\hat{g}^{-1}_{jl})\hat{g}^{-1}_{kl}+\hat{g}_{kk}^{-1}\hat{g}^{-1}_{jk}+\hat{g}^{-1}_{ll}\hat{g}^{-1}_{jl}\Big)\,\partial_{s_{kl}}\nn\\
&\quad =
4\sum_{a<b}^n\sum_{k,l}^{n}\tilde{g}_{ak}\tilde{g}_{lb}\,\Big((\hat{g}^{-1}_{jk}+\hat{g}^{-1}_{jl})\hat{g}^{-1}_{kl}-\hat{g}_{kk}^{-1}\hat{g}^{-1}_{jk}-\hat{g}^{-1}_{ll}\hat{g}^{-1}_{jl}\Big)\,\partial_{v_{ab}} \label{inter1}
\end{align}
where the sum over $k<l$ of $(k,l)$-symmetric terms has been rewritten as half the sum over all $k$ and $l$, noting the terms with $k=l$ explicitly cancel.  
The final two terms now vanish since all row and column sums of the Laplacian matrix $\tilde{g}$ are zero: 
\begin{align}
 \sum_{k,l}^{n}\tilde{g}_{ak}\,\tilde{g}_{lb}\,\hat{g}^{-1}_{kk}\,\hat{g}^{-1}_{jk}&=
  \sum_{k}^{n}\tilde{g}_{ak}\,\hat{g}^{-1}_{kk}\,\hat{g}^{-1}_{jk}
 \sum_{l}^n \,\tilde{g}_{lb} =0,\\
  \sum_{k,l}^{n}\tilde{g}_{ak}\,\tilde{g}_{lb}\,\hat{g}^{-1}_{ll}\,\hat{g}^{-1}_{jl}&=
  \sum_{k}^{n}\tilde{g}_{lb}\,\hat{g}^{-1}_{ll}\,\hat{g}^{-1}_{jl}
 \sum_{k}^n \,\tilde{g}_{ak} =0. 
\end{align}
For the first two terms in \eqref{inter1}, we use the identity
\[
\sum_{k}^n \tilde{g}_{ik}\,\hat{g}^{-1}_{kj}=\delta_{ij}-\delta_{in}, \qquad i,j\le n.
\]
To derive this, note the sum over $k$ restricts to $k\le n-1$ from  \eqref{hatgdef}, then for $i,j\le n-1$ we have $\tilde{g}_{ik}\hat{g}^{-1}_{kj}=g_{ik}g^{-1}_{kj}$.  For $i=n$, $j\le n-1$ we use $\tilde{g}_{nk}\hat{g}^{-1}_{kj}=-\sum_{l}^{n-1}g_{lk}g^{-1}_{kj}$ and for $j=n$ and any $i$ the sum vanishes from \eqref{hatgdef}.  
With the aid of this identity, we then find
\begin{align}
&\sum_{j}^{n-1}\Big(-2p_j^\nu \frac{\partial}{\partial p_j^\nu}\frac{\partial}{\partial p_j^\mu}\Big)\,e^{-\mathrm{tr}\,(g^{-1}\cdot G)} 
\nn\\&\qquad
=4\sum_{j}^{n-1} p_j^\mu\,\Big( s_{jn} -\partial_{v_{jn}}-\sum_{a<b}^{n}(\hat{g}^{-1}_{ja}+\hat{g}^{-1}_{jb})\theta_{v_{ab}}
\Big)\,e^{-\mathrm{tr}\,(g^{-1}\cdot G)} 
\end{align}
where we used $\tilde{g}_{ab}=-v_{ab}$ for $a<b$ to obtain the Euler operator $\theta_{v_{ab}}=v_{ab}\partial_{v_{ab}}$.

Assembling the pieces above, the action of the conformal Ward identity is now
\begin{align}
&\sum_{j=1}^{n-1}\Big(p_j^\mu \frac{\partial}{\partial p_j^\nu}\frac{\partial}{\partial p_j^\nu}-2p_j^\nu \frac{\partial}{\partial p_j^\nu}\frac{\partial}{\partial p_j^\mu}+2(\Delta_j-d)\frac{\partial}{\partial p_j^\mu}\Big)\,e^{-\mathrm{tr}\,(g^{-1}\cdot G)} 
\nn\\&\quad
=4\sum_{j}^{n-1}p_j^\mu\Big(\Big(1-\frac{d}{2}\Big) s_{jn} - \sum_{k}^{n-1}(\Delta_k-d)g^{-1}_{jk}
-\sum_{a<b}^{n}(\hat{g}^{-1}_{ja}+\hat{g}^{-1}_{jb})\theta_{v_{ab}}
\Big)\,e^{-\mathrm{tr}\,(g^{-1}\cdot G)} 
\nn\\&\quad
=4\sum_{j}^{n-1}p_j^\mu\Big(\Big(1-\frac{d}{2}\Big)  s_{jn}  + d\sum_{a}^{n}\hat{g}^{-1}_{ja}
+\sum_{a\neq b}^{n}\hat{g}^{-1}_{ja}(\alpha_{ab}-\theta_{v_{ab}})
\Big)\,e^{-\mathrm{tr}\,(g^{-1}\cdot G)} 
\label{CWIres1}
\end{align}
using \eqref{alphaijdef} in the last line.
Finally, we need two further identities:
\[\label{twoids}
\sum_{a\neq b}^n \theta_{v_{ab}} \hat{g}^{-1}_{ja}  = -s_{jn}, \qquad
 \sum_{a\neq b}^n \hat{g}^{-1}_{ja}v_{ab}s_{ab} = -s_{jn}+2\sum_a^{n}\hat{g}^{-1}_{ja}
\]
To establish the first of these, we write
\begin{align}
\sum_{a\neq b}^n \theta_{v_{ab}} \hat{g}^{-1}_{ja}  
= 
-\sum_{a\neq b}^{n-1} g_{ab}\frac{\partial g_{ja}^{-1}}{\partial v_{ab}}+\sum_{a}^{n-1}v_{an}\frac{\partial g^{-1}_{ja}}{\partial v_{an}}
= 
-\sum_{a\neq b}^{n-1} g_{ab}\frac{\partial g_{ja}^{-1}}{\partial v_{ab}}+\sum_{a,b}^{n-1}g_{ab}\frac{\partial g^{-1}_{ja}}{\partial v_{an}}
\label{thetaid1}
\end{align}
then use the chain rule, which for $i,j,k,l\le n-1$ gives 
\begin{align}
\frac{\partial g^{-1}_{ij}}{\partial v_{kl}} &= -\sum_{a,b}^{n-1}g^{-1}_{i(a}g^{-1}_{b)j}\,\frac{\partial g_{ab}}{\partial v_{kl}}
= (g^{-1}_{ik}-g^{-1}_{il})(g^{-1}_{jl}-g^{-1}_{jk}), \\
\frac{\partial g^{-1}_{ij}}{\partial v_{kn}} &= -\sum_{a,b}^{n-1}g^{-1}_{i(a}g^{-1}_{b)j}\,\frac{\partial g_{ab}}{\partial v_{kn}} = - g^{-1}_{ik}g^{-1}_{kj}.
\end{align}
Inserting these into \eqref{thetaid1}, the sum over $a\neq b$ can be extended to run over all $a,b$ since the term with $a=b$ vanishes.  The only non-cancelling term is then $-g^{-1}_{jj}=-s_{jn}$ as required. 

For the second identity in \eqref{twoids}, we use  \eqref{ginvs} to rewrite
\begin{align}
\sum_{a\neq b}^n \hat{g}^{-1}_{ja}v_{ab}s_{ab} &= -\sum_{a\neq b}^{n-1} g^{-1}_{ja}g_{ab}s_{ab} +\sum_a^{n-1}g^{-1}_{ja}v_{an}s_{an} \nn\\&
=-\sum_{a\neq b}^{n-1} g^{-1}_{ja}g_{ab}(g^{-1}_{aa}+g^{-1}_{bb}-2g^{-1}_{ab})+ \sum_a^{n-1}g^{-1}_{ja}\Big(\sum_b^{n-1}g_{ab}\Big)g^{-1}_{aa}.
\end{align}
The sum over $a\neq b$ can then be extended to run over all $a,b$ as the term with $a=b$ cancels, after which the  first and the last terms cancel and the result follows.

With the aid of the identities \eqref{twoids}, we find that \eqref{CWIres1} becomes
\begin{align}
&\sum_{j=1}^{n-1}\Big(p_j^\mu \frac{\partial}{\partial p_j^\nu}\frac{\partial}{\partial p_j^\nu}-2p_j^\nu \frac{\partial}{\partial p_j^\nu}\frac{\partial}{\partial p_j^\mu}+2(\Delta_j-d)\frac{\partial}{\partial p_j^\mu}\Big)\,e^{-\mathrm{tr}\,(g^{-1}\cdot G)} 
\nn\\&\quad
=-4\sum_{j}^{n-1}p_j^\mu\,
|g|^{d/2}\Omega^{-1}\sum_{a\neq b}^{n}\partial_{v_{ab}}\Big(v_{ab}\,\hat{g}^{-1}_{ja}|g|^{-d/2}\Omega
\,e^{-\mathrm{tr}\,(g^{-1}\cdot G)} \Big)
\end{align}
where 
$
\Omega=\prod_{k<l}^n v_{kl}^{-\alpha_{kl}-1}.
$
Recalling that the simplex representation \eqref{Schwrep} is 
\begin{align}
\lla \O_1(\bs{p}_1) \ldots \O_n(\bs{p}_n) \rra=\Big(\prod_{k<l}^n\int_0^\infty \dd v_{kl}\,v_{kl}^{-\alpha_{kl}-1}\Big) f(\hat{\bs{v}}) |g|^{-d/2} \,e^{-\mathrm{tr}\,(g^{-1}\cdot G)}, 
\end{align}
we note that
\[\label{sumofEuleronf}
\sum_{\substack{b\\b\neq a}}^n \theta_{v_{ab}}f(\hat{\bs{v}})=0
\]
since whenever the index $a$ appears in a cross ratio $\hat{v}_{[acde]} = v_{ac}v_{de}/v_{ad}v_{ce}$ it enters with equal weight in the numerator and the denominator producing a cancellation.
Acting with the Ward identity thus yields a total derivative:
\begin{align}
&\sum_{j=1}^{n-1}\Big(p_j^\mu \frac{\partial}{\partial p_j^\nu}\frac{\partial}{\partial p_j^\nu}-2p_j^\nu \frac{\partial}{\partial p_j^\nu}\frac{\partial}{\partial p_j^\mu}+2(\Delta_j-d)\frac{\partial}{\partial p_j^\mu}\Big)\,\lla \O_1(\bs{p}_1) \ldots \O_n(\bs{p}_n) \rra\nn\\&\quad
=-4\sum_{j}^{n-1}p_j^\mu\Big(\prod_{k<l}^n\int_0^\infty \dd v_{kl}\Big)  \sum_{a\neq b}^{n}\partial_{v_{ab}}\Big(v_{ab}\,\hat{g}^{-1}_{ja} \,f(\hat{\bs{v}})|g|^{-d/2}\Omega
\,e^{-\mathrm{tr}\,(g^{-1}\cdot G)} \Big).
\label{CWItotalderiv}
\end{align}
The boundary terms vanish under reasonable assumptions: for generic  momentum configurations with non-vanishing Gram determinant, the upper limit is suppressed by the decay of the Schwinger exponential; the lower limit is zero 
provided $v_{ab}^{-\alpha_{ab}}f(\hat{\bs{v}})$ vanishes as $v_{ab}\rightarrow 0$, which is satisfied whenever the simplex representation itself converges.
The simplex integral thus solves the special conformal Ward identity.

\section{Insight from position space}
\label{sec:posnspace}

Thus far, our analysis has been entirely in momentum space.  However, as noted above, the form of the total derivative produced by the action of the special conformal Ward identity in  \eqref{CWItotalderiv} 
can also  be understood through independent position-space arguments.  
We present these in section \ref{sec:posnspacecwis}.  Then, in section \ref{sec:Wmp}, we  show how similar position-space arguments can  be applied to verify the action of $d$-preserving shift operators such as $\mathcal{W}_{12}^{-+}$.

\subsection{The conformal Ward identities}
\label{sec:posnspacecwis}

To Fourier transform the simplex representation \eqref{Schwrep}  to position space, we  compute
\begin{align}
&\<\O(\x_1)\ldots \O(\x_n)\> = \prod_{k}^{n-1}\int\frac{\dd^d\bs{p}_k}{(2\pi)^d}\,e^{i\bs{p}_k\cdot \bs{x}_{kn}} \lla \O(\bs{p}_1)\ldots \O(\bs{p}_n)\rra\nn\\&\quad =
\Big(\prod_{i<j}^n\int\dd v_{ij}\, v_{ij}^{-\alpha_{ij}-1}\Big)f(\hat{\bs{v}}) |g|^{-d/2} \Big(\prod_{k}^{n-1}\int\frac{\dd^d\bs{p}_k}{(2\pi)^d}\Big)\exp\Big(\sum_{k}^{n-1}i\bs{p}_k\cdot \bs{x}_{kn} - \sum_{k,l}^{n-1}g^{-1}_{kl} \,\bs{p}_k\cdot \bs{p}_l\Big)\nn\\
&\quad =
\Big(\prod_{i<j}^n\int\dd v_{ij}\, v_{ij}^{-\alpha_{ij}-1}\Big)\tilde{f}(\hat{\bs{v}}) \exp\Big(-\frac{1}{4}\sum_{k,l}^{n-1}g_{kl}\,\bs{x}_{kn}\cdot\bs{x}_{ln}\Big) \label{possp1}
\end{align}
where $\bs{x}_{ij}=\bs{x}_i-\bs{x}_j$, and for the Gaussian integral over momenta we completed the square: 
\begin{align}
&\sum_{k}^{n-1}i\bs{p}_k\cdot \bs{x}_{kn} - \sum_{k,l}^{n-1}g^{-1}_{kl} \,\bs{p}_k\cdot \bs{p}_l \nn\\\qquad  &=-\sum_{k,l}^{n-1}g^{-1}_{kl}(\bs{p}_k-\frac{i}{2}\sum_{a}^{n-1}g_{ka}\,\bs{x}_{an})\cdot(\bs{p}_l-\frac{i}{2}\sum_{b}^{n-1}g_{lb}\,\bs{x}_{bn}) -\frac{1}{4}\sum_{k,l}^{n-1}g_{kl}\,\bs{x}_{kn}\cdot\bs{x}_{ln}.
\end{align}
The numerical factor from the integration can then be re-absorbed into the arbitrary function by setting $(4\pi)^{(1-n)d/2} f(\hat{\bs{v}})=\tilde{f}(\hat{\bs{v}})$.
The exponent in \eqref{possp1} now simplifies to\footnote{Recall the analogous relation  in a resistor network of simplex topology, namely, that the power dissipated is $\sum_{i<j}^n v_{ij}(V_i-V_j)^2 = \sum_{i,j}^n \tilde{g}_{ij}V_i V_j$, where $v_{ij}$ is the conductivity and $V_i$ the voltage at node $i$.}
\begin{align}
-\frac{1}{4}\sum_{k,l}^{n-1}g_{kl}\,\bs{x}_{kn}\cdot\bs{x}_{ln}
=-\frac{1}{4}\sum_{k,l}^{n}\tilde{g}_{kl}\,\bs{x}_{k}\cdot\bs{x}_{l}=-\frac{1}{4}\sum_{i<j}^{n} v_{ij}x_{ij}^2,
\end{align}
and hence the simplex representation in position space is
\begin{align}\label{posnsimplex}
&\<\O(\x_1)\ldots \O(\x_n)\> =
\Big(\prod_{i<j}^n\int\dd v_{ij}\, v_{ij}^{-\alpha_{ij}-1} e^{-\frac{1}{4}v_{ij} x_{ij}^2}\Big)\tilde{f}(\hat{\bs{v}}). 
\end{align}

If the arbitrary function $\tilde{f}(\hat{\bs{v}}) $ is a product of powers, this expression reduces to the conformal correlator $\prod_{i<j}^n x_{ij}^{2\tilde{\alpha}_{ij}}$ where the $\tilde{\alpha}_{ij}$ satisfy $\sum_{j\neq i} \tilde{\alpha}_{ij}=-\Delta_i$.
More generally, wherever $\tilde{f}(\hat{\bs{v}})$ admits a Mellin-Barnes representation, 
we recover $\prod_{i<j}^n x_{ij}^{2\alpha_{ij}}$ times a function of position-space cross ratios as shown in  \cite{Bzowski:2020kfw}.  However, the most straightforward way to check that  \eqref{posnsimplex} solves the conformal Ward identities is to note that, when acting on a function $F=F(\{ x_{kl}^2\})$ of the squared coordinate separations, 
\begin{align}
&\sum_{i}^{n} \Big(2 x_i^\mu x_i^\nu \frac{\partial}{\partial x_i^\nu}-x_i^2 \frac{\partial}{\partial x_i^\mu}+2\Delta_i x_i^\mu\Big) F =
\sum_{i}^n 2x_i^\mu \Big(\Delta_i+\sum_{\substack{j\\j\neq i}}^n x_{ij}^2 \frac{\partial}{\partial (x_{ij}^2)} \Big) F.
\end{align}
It then follows that
\begin{align}
&\sum_{i}^{n} \Big(2 x_i^\mu x_i^\nu \frac{\partial}{\partial x_i^\nu}-x_i^2 \frac{\partial}{\partial x_i^\mu}+2\Delta_i x_i^\mu\Big) \Big(\prod_{k<l}^n\int\dd v_{kl}\, v_{kl}^{-\alpha_{kl}-1} e^{-\frac{1}{4}v_{kl} x_{kl}^2}\Big)\tilde{f}(\hat{\bs{v}}) \nn\\ &\qquad
=\sum_{i}^n 2x_i^\mu \,\Big(\prod_{k<l}^n\int\dd v_{kl}\, v_{kl}^{-\alpha_{kl}-1}\Big) \tilde{f}(\hat{\bs{v}}) 
\Big(\Delta_i+\sum_{\substack{j\\j\neq i}}^n v_{ij} \frac{\partial}{\partial v_{ij}} \Big)  e^{-\frac{1}{4}\sum_{k<l}^{n}v_{kl} x_{kl}^2}\nn\\
&\qquad
=\sum_{i}^n 2x_i^\mu \,\Big(\prod_{k<l}^n\int\dd v_{kl}\, v_{kl}^{-\alpha_{kl}-1}\Big) \tilde{f}(\hat{\bs{v}}) 
\Big(\Delta_i+\sum_{j}^n \alpha_{ij}\Big)  e^{-\frac{1}{4}\sum_{k<l}^nv_{kl} x_{kl}^2} = 0
\end{align}
where in the last line we integrated by parts\footnote{As previously, the boundary terms vanish
provided $v_{kl}^{-\alpha_{kl}}\tilde{f}(\hat{\bs{v}})$ as $v_{kl}\rightarrow 0$. }  then used \eqref{alphaijdef}.
The middle line here accounts for the form of the total derivative we found earlier in 
\eqref{CWItotalderiv}.
Multiplying by $-i$ and Fourier transforming, the first line yields the momentum-space conformal Ward identity acting on the momentum-space simplex representation ({\it i.e.,} the left-hand side of \eqref{CWItotalderiv}), while the middle line yields
\begin{align}
& \sum_i^{n-1}2\frac{\partial}{\partial p_i^\mu}\Big(\Big(\prod_{k<l}^n \int_0^\infty \dd v_{kl}\,v_{kl}^{-\alpha_{kl}-1}\Big)f(\hat{\bs{v}})\Big(\Delta_i+\sum_{\substack{j\\j\neq i}}^n v_{ij}\frac{\partial}{\partial v_{ij}}\Big) |g|^{-d/2}  e^{-\sum_{a,b}^{n-1}g^{-1}_{ab}\,\bs{p}_a\cdot\bs{p}_b}\Big)
\nn\\&=
\sum_i^{n-1}  \Big(\prod_{k<l}^n \int_0^\infty \dd v_{kl}\Big)\,\sum_{\substack{j\\j\neq i}}^n \frac{\partial}{\partial v_{ij}} \Big(v_{ij} \,\Omega \,f(\hat{\bs{v}})|g|^{-d/2} \Big(\sum_a^{n-1} -4g^{-1}_{ia}p_a^\mu\Big)e^{-\sum_{a,b}^{n-1}g^{-1}_{ab}\,\bs{p}_a\cdot\bs{p}_b}\Big)
\nn\\&=
-4\sum_a^{n-1} p_a^\mu \Big(\prod_{k<l}^n \int_0^\infty \dd v_{kl}\Big)\,\sum_{i\neq j}^n \frac{\partial}{\partial v_{ij}}\Big(v_{ij}\,\hat{g}^{-1}_{ia}  \Omega\, f(\hat{\bs{v}})|g|^{-d/2}e^{-\mathrm{tr}\,(g^{-1}\cdot G)}\Big)
\end{align}
where in the second line we evaluated the momentum derivative of the exponential and  pushed the factors of $\Omega=\prod_{k<l} v_{kl}^{-\alpha_{kl}-1}$, $f(\hat{\bs{v}})$ and $v_{ij}$  inside the $v_{ij}$-derivative which cancels the $\Delta_i$ term via \eqref{alphaijdef}.
In the final line, we extended the sum over $i$ to run up to $n$ by replacing $g^{-1}_{ia}$ with $\hat{g}^{-1}_{ia}$ and combined it with the sum over $j$.  Up to a relabelling of indices, this final line is now the total derivative appearing on the right-hand side of \eqref{CWItotalderiv}.

The manipulations above illustrate a  general theme: given
the simplicity of the position-space simplex representation \eqref{posnsimplex}, it is often  profitable to work with the position-space equivalents of differential operators in order to evaluate their action in terms of the $v_{ij}$ variables.  Both sides can then be Fourier transformed back to momentum space in order to deduce the action of the corresponding momentum-space operator on the momentum-space simplex in terms of the $v_{ij}$ variables.   In many cases this is more straightforward than working in momentum space throughout.

\subsection[\texorpdfstring{Action of $\mathcal{W}_{12}^{-+}$}{Action of W(-+)}]{\texorpdfstring{Action of \boldmath{$\mathcal{W}_{12}^{-+}$}}{Action of W(-+)}}
\label{sec:Wmp}

As a further illustration of this approach, let us evaluate the action of the shift operator $\mathcal{W}_{12}^{-+}$ defined in \eqref{Wops}.
After expanding out the derivative, this operator can easily be Fourier transformed to position space where it reads
\[
\mathcal{W}_{12}^{-+}=\frac{1}{2}x_{12}^2 \frac{\partial}{\partial x_2^\mu}\frac{\partial}{\partial x_{2\mu}}+2(\beta_2+1)\Big(\beta_2+\frac{d}{2}-x_{12}^\mu \frac{\partial}{\partial x_2^\mu}\Big).
\]
Acting on a function $F=F(\{ x_{kl}^2\})$ of the squared coordinate separations, we find via the chain rule
\begin{align}
\mathcal{W}_{12}^{-+} F &= 
\sum_{\substack{i,j\\i,j\neq 2}}^n x_{12}^2 (x_{2i}^2+x_{2j}^2-x_{ij}^2)\frac{\partial^2F}{\partial (x_{2i}^2)\partial (x_{2j}^2)}\nn\\&\quad
+\sum_{i\neq 2}^n\Big(2(\beta_2+1)(x_{12}^2-x_{1i}^2+x_{2i}^2)+dx_{12}^2\Big)\frac{\partial F}{\partial (x_{2i}^2)}
+2(\beta_2+1)\Big(\beta_2+\frac{d}{2}\Big)F
\end{align}
Acting on the Schwinger exponent appearing in the position-space simplex representation \eqref{posnsimplex}, this can be translated into $v_{ij}$-derivatives  as
\begin{align}
&\mathcal{W}_{12}^{-+} \Big(\prod_{k<l}^n\int\dd v_{kl}\, v_{kl}^{-\alpha_{kl}-1} e^{-\frac{1}{4}v_{kl} x_{kl}^2}\Big)\tilde{f}(\hat{\bs{v}}) \nn\\
&\quad = 2\,\Big(\prod_{k<l}^n\int\dd v_{kl}\, v_{kl}^{-\alpha_{kl}-1} \Big)\tilde{f}(\hat{\bs{v}}) \Big[\,(\beta_2+1)\Big(\beta_2+\frac{d}{2}\Big)+
\Big(2\beta_2+1+\frac{d}{2}+\theta_{v_{12}}\Big)\theta_{v_{12}}
\nn\\&\qquad\qquad\qquad
+\sum_{i=3}^n  v_{2i}\Big((\beta_2+1+\theta_{v_{12}})(\partial_{v_{12}}+\partial_{v_{2i}}-\partial_{v_{1i}})+\Big(\frac{d}{2}+\theta_{v_{2i}}\Big)\partial_{v_{12}}\Big)\nn\\&\qquad\qquad\qquad
+\sum_{3\le i<j}^n v_{2i}v_{2j}(\partial_{v_{2i}}+\partial_{v_{2j}}-\partial_{v_{ij}})\partial_{v_{12}}\Big] e^{-\frac{1}{4}\sum_{k<l}^nv_{kl} x_{kl}^2}
\end{align}
where $\partial_{v_{ij}}=\partial/\partial v_{ij}$ and $\theta_{v_{ij}} = v_{ij}\partial_{v_{ij}}$. 
Integrating by parts, we find
\begin{align}
&\mathcal{W}_{12}^{-+} \Big(\prod_{k<l}^n\int\dd v_{kl}\, v_{kl}^{-\alpha_{kl}-1} e^{-\frac{1}{4}v_{kl} x_{kl}^2}\Big)\tilde{f}(\hat{\bs{v}}) \nn\\&
\quad = 2\,\Big(\prod_{k<l}^n\int\dd v_{kl}\,e^{-\frac{1}{4}v_{kl} x_{kl}^2}\Big)
\Big[(\theta_{v_{12}}-\beta_2)\Big(\theta_{v_{12}}-\beta_2-\frac{d}{2}-1+n\Big)
\nn\\&\qquad\qquad
+\sum_{i=3}^n v_{2i}\Big((\theta_{v_{12}}-\beta_2)(\partial_{v_{12}}+\partial_{v_{2i}}-\partial_{v_{1i}})+\Big(n-\frac{d}{2}+\theta_{v_{2i}}\Big)\partial_{v_{12}}\Big)\nn\\&\qquad
\qquad+
\sum_{3\le i<j}^n v_{2i}v_{2j}(\partial_{v_{2i}}+\partial_{v_{2j}}-\partial_{v_{ij}})\partial_{v_{12}}\Big] \Omega \tilde{f}(\hat{\bs{v}}).
\label{step0}
\end{align}
We now rewrite the first part of the last line as 
\begin{align}
&\Big[\sum_{3\le i<j}^n v_{2i}v_{2j}(\partial_{v_{2i}}+\partial_{v_{2j}})\partial_{v_{12}}\Big]\Omega \tilde{f}(\hat{\bs{v}})
=\Big[\sum_{i=3}^n \Big(\sum_{\substack{j=3\\j\neq i}}^n v_{2j}\Big)\theta_{v_{2i}}\partial_{v_{12}}
\Big]\Omega \tilde{f}(\hat{\bs{v}})\nn\\&\qquad 
=\Big[
\Big(\sum_{j=3}^n v_{2j}\Big)\partial_{v_{12}}\Big(-\theta_{v_{12}}+\sum_{i\neq 2}^n \theta_{v_{2i}}\Big)-\sum_{i=3}^n v_{2i}\theta_{v_{2i}}\partial_{v_{12}}\Big]\Omega \tilde{f}(\hat{\bs{v}})\nn\\&\qquad 
=\Big[-
\sum_{i=3}^n v_{2i}\Big((\theta_{v_{12}} +1)-\beta_2-\frac{d}{2}+(n-1)+
\theta_{v_{2i}}\Big)\partial_{v_{12}}\Big]\Omega \tilde{f}(\hat{\bs{v}})
\label{step1}
\end{align}
where in the final step we rewrote $\partial_{v_{12}}\theta_{v_{12}}=(\theta_{v_{12}}+1)\partial_{v_{12}}$ and
used $\sum_{i\neq 2}^n\theta_{v_{2i}}\tilde{f}(\hat{\bs{v}})=0$, as follows from \eqref{sumofEuleronf}, along with \eqref{alphaijdef} with $\Delta_2=\beta_2+d/2$ to replace
\[\label{step2}
\Big(\sum_{i\neq 2}^n\theta_{v_{2i}}\Big)\Omega \tilde{f}(\hat{\bs{v}})=(\beta_2+d/2-(n-1))\Omega  \tilde{f}(\hat{\bs{v}}).
\]
Substituting \eqref{step1} into \eqref{step0} and making further use of \eqref{step2}, we find the result
\begin{align}
&\mathcal{W}_{12}^{-+} \Big(\prod_{k<l}^n\int\dd v_{kl}\, v_{kl}^{-\alpha_{kl}-1} e^{-\frac{1}{4}v_{kl} x_{kl}^2}\Big)\tilde{f}(\hat{\bs{v}}) \nn\\&
\quad = -2\,\Big(\prod_{k<l}^n\int\dd v_{kl}\,e^{-\frac{1}{4}v_{kl} x_{kl}^2}\Big)
\Big[
(\theta_{v_{12}}-\beta_2)\sum_{i=3}^n v_{2i}\partial_{v_{1i}}
+
\sum_{3\le i<j}^n v_{2i}v_{2j}\partial_{v_{ij}}\partial_{v_{12}}\Big] \Omega \tilde{f}(\hat{\bs{v}}).
\end{align}
Equivalently, acting on the position-space simplex with $\mathcal{W}_{12}^{-+}$ corresponds to acting on the arbitrary function $\tilde{f}(\hat{\bs{v}})$ with the operator
\[\label{tildeWdef}
\tilde{\mathcal{W}}_{12}^{-+} =-2 \Omega^{-1}\Big[
(\theta_{v_{12}}-\beta_2)\sum_{i=3}^n v_{2i}\partial_{v_{1i}}
+
\sum_{3\le i<j}^n v_{2i}v_{2j}\partial_{v_{ij}}\partial_{v_{12}}\Big] \Omega. 
\]
The same remains true when we Fourier transform back to momentum space, giving
\begin{align}
&\mathcal{W}_{12}^{-+}\Big(\prod_{k<l}^n\int_0^\infty\dd v_{kl}\, v_{kl}^{-\alpha_{kl}-1}\Big)|g|^{-d/2}e^{-\mathrm{tr}(g^{-1}\cdot G)}f(\hat{\bs{v}})\nn\\&\qquad 
=\Big(\prod_{k<l}^n\int_0^\infty\dd v_{kl}\, v_{kl}^{-\alpha_{kl}-1}\Big)|g|^{-d/2}e^{-\mathrm{tr}(g^{-1}\cdot G)}(\tilde{\mathcal{W}}_{12}^{-+}f(\hat{\bs{v}})).
\end{align}

Finally, it remains to check that the action of $\tilde{\mathcal{W}}_{12}^{-+}$ on the arbitrary function produces the required shift in dimensions $\Delta_1\rightarrow \Delta_1-1$ and $\Delta_2\rightarrow \Delta_2+1$.  
Since 
\[
\partial_{v_{ij}}\Omega=-(\alpha_{ij}+1)\frac{\Omega}{v_{ij}},\qquad \partial_{v_{ij}}f(\hat{\bs{v}})=\frac{ h(\hat{\bs{v}})
}{v_{ij}},
\] 
where $h(\hat{\bs{v}})$ is also function of the cross ratios, we see that 
\begin{align}\label{tildeWaction}
&\tilde{\mathcal{W}}_{12}^{-+} f(\hat{\bs{v}}) = 
\sum_{i=3}^n \frac{v_{2i}}{v_{1i}} h_i(\hat{\bs{v}})
+
\sum_{3\le i<j}^n \frac{v_{2i}v_{2j}}{v_{ij}v_{12}}h_{ij}(\hat{\bs{v}})
\end{align}
where $h_i(\hat{\bs{v}})$ and $h_{ij}(\hat{\bs{v}})$ are specific functions of the cross ratios.   Each term in the first sum then corresponds to a simplex integral with the shifts 
\[\label{shiftpattern1}
\alpha_{2i}\rightarrow \alpha_{2i}-1,\qquad
\alpha_{1i}\rightarrow \alpha_{1i}+1,
\]
while each term in the second sum  corresponds to a simplex integral with the shifts
\[\label{shiftpattern2}
\alpha_{2i}\rightarrow \alpha_{2i}-1, \qquad \alpha_{2j}\rightarrow \alpha_{2j}-1,\qquad
\alpha_{ij}\rightarrow \alpha_{ij}+1,\qquad
\alpha_{12}\rightarrow \alpha_{12}+1.
\]
From \eqref{alphaijdef}, both \eqref{shiftpattern1} and \eqref{shiftpattern2}  correspond to shifting $\Delta_1\rightarrow \Delta_1-1$ and $\Delta_2\rightarrow \Delta_2+1$ leaving all other operator dimensions fixed. 
The action of $\mathcal{W}_{12}^{-+}$ on the simplex thus produces an appropriately shifted simplex integral, whose function of cross ratios is obtained through the action of the operator \eqref{tildeWdef}.

\section{Discussion}
\label{sec:disc}

Our analysis has furnished
useful
parametric representations for the general momentum-space conformal $n$-point function.  
Starting from the generalised simplex Feynman integral of \cite{Bzowski:2019kwd, Bzowski:2020kfw}, 
we showed how all graph polynomials can be obtained from the corresponding Laplacian matrix, or the Gram matrix to which it reduces once momentum conservation has been enforced. 
With the graph polynomials to hand, all the usual scalar parametrisations of Feynman integrals can  be adapted to represent the simplex solution.   Only $n(n-1)/2$ integrals over Schwinger parameters remain to be performed --  one for each leg of the simplex -- in contrast to the $(n-1)(n-2)d/2$ scalar integrals we started with.

Building on  the analogy between Feynman graph polynomials and those  of electrical circuits, we then formulated a second class of parametric representations.  For these, the integration variables represent the {\it effective} resistances between vertices of the simplex, 
rather than   
the conductivities 
({\it i.e.,} the inverse Schwinger parameters) used previously.  
This change of variables immediately diagonalises the Schwinger exponential, expressing the $n$-point function as a standard Laplace transform of a product of polynomials raised to generalised powers.  These polynomials correspond to the determinant and  first minors of the Cayley-Menger matrix for the simplex, which 
plays an analogous role to the Gram matrix for this second class of parametrisations.  From the form of these polynomials, new weight-shifting operators can  immediately be constructed to raise the power of these polynomials, with further shift operators  following by shadow conjugation.   Besides shifting the scaling dimensions of external operators, these new weight-shifting operators  raise the spacetime dimension by two.
They therefore generalise the $3$-point shift operators of \cite{Bzowski:2013sza, Bzowski:2015yxv} to $n$-points, and 
 constitute a distinct class of operators to those 
 identified in  \cite{Karateev:2017jgd}.

Our results suggest several interesting directions for further pursuit:
\begin{itemize}
\item Given  we now have  weight-shifting operators that both preserve and raise the spacetime dimension, is it also possible to construct operators that {\it lower} the spacetime dimension?  
One approach we have explored, explained in appendix \ref{app:Bernstein},  is to find   so-called {\it Bernstein-Sato} operators which act to lower the powers to which the  various polynomials of interest are raised.  In this case, the relevant polynomials are
the Cayley-Menger determinant and its minors appearing in the parametrisation \eqref{Cay2}.  
We found, for example,  that  replacing $v_{ij}\rightarrow\partial_{s_{ij}}$ in the Kirchhoff polynomial $\mathcal{U}=|g|$ yields an operator 
\[
\mathcal{B}_{|m|} = (|g|)\Big|_{v_{ij}\rightarrow \partial_{s_{ij}}}
\]
which lowers by one the power to which the Cayley-Menger determinant is raised: 
\[\label{CMBernstein}
\mathcal{B}_{|m|}\, |m|^a = b_{|m|}(a) |m|^{a-1}, \qquad b_{|m|}(a) =- \prod_{k=1}^{n-1}(1-k-2a).
\] 
For the simplex representation \eqref{Cay2}, $a$ is the parameter $\alpha$ given in \eqref{alphadef} and so lowering $\alpha$ by one corresponds to sending $d\rightarrow d-2$ if all the operator dimensions are kept fixed.   In principle, one would then integrate by parts to obtain an operator acting solely on the Schwinger exponential, which, due to its diagonal structure, could be 
translated into a differential operator in the external momenta.
In practice, however, this approach 
 is complicated by the presence of all the 
remaining powers of Cayley-Menger minors present in \eqref{Cay2}.

\item In sections \ref{sec:CWI} and \ref{sec:posnspacecwis}, we saw how the action of the special conformal Ward identity on the simplex reduces  to a total derivative.   This followed directly from the scalar parametric representation, without any recourse to the recursive arguments developed in \cite{Bzowski:2019kwd, Bzowski:2020kfw}.
Nevertheless, these arguments, and the recursion relation between  $n$- and $(n+1)$-point simplices on which they are based, are  of  considerable
interest in their own right and could be reformulated in the scalar-parametric language used here.
The deletion/contraction relations of graph polynomials (see, {\it e.g.,} \cite{Weinzierl:2022eaz}) and Kron reduction, corresponding to taking the Schur complement of a subset of vertices in the simplex Laplacian (see {\it e.g.,} \cite{dorfler2012kron}), may also yield relevant identities.

\item Starting from the general simplex solution, the arbitrary function of momentum-space cross ratios can be restricted by imposing additional conditions of interest: for example, dual conformal invariance \cite{Bzowski:2015pba, Coriano:2019sth, Loebbert:2020hxk, Rigatos:2022eos}, or the Casimir equation for conformal blocks.  For such investigations, the connection with position-space developed in section \ref{sec:posnspace}  provides a very simple link between the action of a given differential operator in the external momenta or coordinates, and its corresponding action on the arbitrary function of the simplex representation.

\item For holographic $n$-point functions, bulk scalar Witten diagrams have the interesting property that their form is invariant under the action of a shadow transform on any of the external legs.  In momentum space, shadow transforming the operator $\O_{i}$ corresponds to multiplying the correlator by $p_i^{-2\beta_i}$, where $\beta_i=\Delta_i-d/2$, which  has the effect of replacing $\beta_i\rightarrow - \beta_i$ in the   bulk-boundary propagator 
$z^{d/2} p_i^{\beta_i} K_{\beta_i}(p_iz)$.  
It would be interesting to understand the restriction this condition places on the function of cross-ratios appearing in the  simplex representation.

\item Finally, the  parametric  representations we have developed may provide a useful starting point for  the construction of general spinning $n$-point correlators via the action of spin-raising operators \cite{Karateev:2017jgd, Arkani-Hamed:2018kmz, Baumann:2019oyu}, and for bootstrapping cosmological correlators in de Sitter spacetime.   

\end{itemize}

\end{chapter}
\clearpage{} 
  \clearpage{}\begin{chapter}{\label{cha5}GKZ integrals
and creation operators for Feynman and Witten diagrams}
\allowdisplaybreaks

\section{Introduction}

It has long been suspected that Feynman integrals represent 
a multi-variable generalisation of 
hypergeometric functions \cite{Regge, Kashiwara:1977nf}.  
Recently \cite{Vanhove:2018mto, de_la_Cruz_2019, Klausen:2019hrg,Klausen:2021yrt,  Klausen:2023gui,Klemm_2020,Feng_2020,Chestnov:2022alh,Ananthanarayan:2022ntm,Zhang:2023fil}, this connection has been sharpened by writing Feynman integrals as Gel'fand-Kapranov-Zelevinksy (GKZ) or $\mathcal{A}$-hypergeometric functions  \cite{GKZ_1, GKZ_2,GKZ_3, GKZ_book}.   
As shown in \cite{de_la_Cruz_2019, Klausen:2019hrg}, this can be achieved simply
 by expressing Feynman integrals in Lee-Pomeransky form \cite{Lee:2013hzt}, where only a single denominator polynomial appears, followed by uplifting to a higher-dimensional space of generalised momenta. 
$\mathcal{A}$-hypergeometric functions are well-studied 
in the mathematics literature  \cite{Stienstra:2005nr, BeukersNotes, cattani2006three, takayama2020hypergeometric, saito2013grobner, reichelt2021algebraic} and satisfy a set of linear partial differential equations whose form can be read off in systematic fashion from a certain matrix -- the $\mathcal{A}$-matrix -- which encodes both the structure of the integral 
as well as all kinematic and spectral singularities.

A task of great practical interest is then 
to construct hypergeometric 
{\it shift operators} connecting integrals of different parameter values. 
These operators enable a known `seed' integral to be converted, by simple differentiation, into an entire series of new integrals.
For Feynman integrals, the parameters are typically  the powers of various propagators and the spacetime dimension.  Here we will also study Witten diagrams in anti-de Sitter spacetime
for which the relevant parameters, besides the spacetime dimension, are 
the scaling dimensions of operators in the holographically dual conformal field theory. 

While various techniques for constructing 
shift operators for Feynman integrals \cite{Tarasov:1996br, Lee:2010wea,  Bytev:2009kb, Bytev:2013gva, Bitoun:2017nre} and Witten diagrams \cite{Dolan:2000ut,Bzowski:2015yxv,Karateev:2017jgd, Costa:2018mcg, Baumann:2019oyu,Rigatos:2022eos, Bzowski:2022rlz} are known, the GKZ formalism offers a more powerful and unified approach.  Besides the elementary shift operators,  known as `annihilation' operators in the mathematics literature, their {\it inverses} -- a highly non-trivial class of operators known as `creation' operators -- can be systematically constructed \cite{Saito_param_shift, Saito_restrictions, saito_sturmfels_takayama_1999}.
Together, these creation and annihilation operators 
form a full set of shift  operators connecting $\mathcal{A}$-hypergeometric functions of different parameter values,  
just as the ordinary creation and annihilation (or ladder) operators 
connect different eigenstates of the quantum harmonic oscillator.

A key aim of this chapter is to 
show that  creation operators can be constructed 
directly 
from knowledge of the spectral singularities of an $\mathcal{A}$-hypergeometric function, namely, 
the special set of parameter values for which the corresponding GKZ integral representation diverges.
These singularities can be computed directly from the $\mathcal{A}$-matrix of the integral.  Remarkably, they correspond geometrically to an infinite series of hyperplanes parallel to the 
co-dimension one facets of the Newton polytope  associated with the integral's denominator \cite{nilsson2010mellin,berkesch2013eulermellin}.  (See figure \ref{3Kpolytope}.)
Standard convex hulling algorithms exist for computing such facets 
allowing a simple identification of all  singularities.  

\begin{figure}
\centering
\begin{tikzpicture}[scale=2]
   \draw[thick,black] (1,0,0) -- (0,1,0) -- (0,0,1); 
  \draw[thick,black]  (-1,0,0) -- (0,-1,0) ;
   \draw[thick,black] (-1,0,0) -- (0,1,0) ;
    \draw[thick,black] (-1,0,0) -- (0,0,1)--(1,0,0) ;
    \draw[thick,black] (0,0,1) -- (0,-1,0)--(1,0,0) ;
   \draw[style=dashed, color=black] (0,-1,0)-- (0,0,-1)--(-1,0,0);
   \draw[style=dashed, color=black] (1,0,0)-- (0,0,-1)--(0,1,0); 
 
   \draw[thick,black,fill=YellowGreen, fill opacity=0.2] (-1,0,0)--(0,-1,0) -- (1,0,0) -- (0,1,0) --cycle;

   \end{tikzpicture}
   \caption{The Newton polytope for a 3-point contact Witten diagram in momentum space 
   is an octahedron as shown.  At $n$-points, we obtain an $n$-dimensional cross-polytope.  The spectral singularities consist of an infinite series of hyperplanes parallel to the facets of the Newton polytope, while the integral is convergent for parameter values lying inside the polytope.   Identification of the singularities enables a systematic construction of all creation-type shift operators. \label{3Kpolytope}}
\label{fig:octa}
\end{figure}
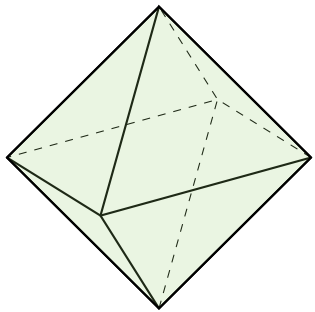

To construct creation operators, we start with a pair of integrals 
 connected by an annihilation operator.  As we will review, this annihilator consists of   a single derivative with respect to one of the GKZ generalised momenta. 
Specifically, we are interested in cases with parameters such that the starting integral is {\it divergent} while the resulting integral is  {\it finite}.  
(To regulate divergences, we assume a dimensional scheme where  parameters are infinitesimally shifted away from their singular values.)
The divergences are thus projected out by the action of the annihilator.  
As the inverse of the annihilator, the creation operator must then produce the reverse shift, from the finite integral to the divergent one. 
Clearly, however, this cannot be achieved directly: the result of acting with a finite differential operator on a finite integral must necessarily be finite.  
Instead, the outcome must be a finite product of 
 the divergent integral multiplied by a vanishing function of the parameters. 
This  function,  whose zeros serve to cancel out the divergence, is known as the {\it $b$-function} and holds the key to the construction of creation operators.  

From a knowledge of the singular parameter values,  
we can predict the necessary zeros of the $b$-function and hence its minimal form as a polynomial.  
Then, acting on an integral with both the annihilator {\it and} the (as yet unknown) creation operator, we must recover the original integral multiplied by the $b$-function.
In the GKZ formalism, however,  any polynomial in the parameters can  be  traded for an equivalent polynomial in Euler operators acting on the generalised  momenta. 
Applying this procedure to the $b$-function, the resulting differential operator must thus be factorisable into a product of the annihilation and the creation operator.  As the annihilator is just a single derivative, this factorisation is easily performed (with the aid of a further set of PDEs known as the toric equations) revealing the identity of the creation operator. 
As a final step, one then projects back from the higher-dimensional GKZ space of generalised momenta to that of the physical variables (the external momenta and masses), with the aid of an auxiliary set of Euler equations.

We hope this simple physical approach, based on the  spectral singularities of the GKZ integral, 
will facilitate
the application 
of creation operators to a range of  physical
systems. 
As an initial demonstration of the possibilities, we have used the formalism to construct new shift operators for a range of simple Feynman integrals, as well as Witten diagrams encoding momentum-space correlators in holographic conformal field theories.  These latter objects are intimately related to cosmological correlators in de Sitter spacetime, and the new shift operators we construct can also be applied 
in this context.
In particular, we have found new shift operators connecting both exchange and contact 4-point Witten diagrams, with arbitrary external scaling dimensions, to corresponding diagrams with shifted scaling dimensions but the same spacetime dimension.
Until now, such operators were only available in the case where diagrams with {\it non-derivative} vertices are mapped to those with {\it derivative} vertices, and for a restricted set of scaling dimensions at that \cite{Baumann:2019oyu,Bzowski:2022rlz}.  In contrast, the new shift operators we find can  be applied for any scaling dimensions, and moreover map non-derivative  to non-derivative vertices.  This  enlarges the available arsenal of shift operators for Witten diagrams (and by extension, cosmological correlators), and as such is a useful and nontrivial result.  
We believe these examples provide a first proof of principle that the creation operator method, and the GKZ formalism more generally, holds 
promise for a variety of physical applications.

An outline of this chapter is as follows.  Section \ref{sec:GKZformalism} introduces $\mathcal{A}$-hypergeometric functions and the GKZ formalism.  We summarise the PDEs these functions obey, their construction, and their invariance under affine reparametrisations.  In section \ref{sec:sings}, we relate the spectral singularities of GKZ integrals to the Newton polytope of the denominator.
In section \ref{sec:shiftops}, we introduce creation operators and detail their construction based on the spectral singularities of the integral.  In section \ref{sec:Witten}, we construct creation operators for 3- and 4-point contact Witten diagrams in momentum space, as well as a further set of shift operators that preserve the spacetime dimension.  Using these results, we then derive novel shift operators for exchange diagrams.
Section \ref{sec:Feynman}  constructs creation operators for a variety of simple Feynman integrals introducing the use of Gr{\"o}bner bases and convex hulling algorithms to automate the computation.  We conclude in section \ref{sec:Discussion} with a summary of results and open directions.  In the appendices we discuss the conversion of Feynman  to GKZ integrals, creation operators for position-space contact Witten diagrams, and an extension of the minimal construction algorithm outlined above.

\section{\texorpdfstring{$\mathcal{A}$-hypergeometric functions}{A-hypergeometric functions}}
\label{sec:GKZformalism}

The application of the GKZ formalism to Feynman integrals has been explored in a number of recent works \cite{Vanhove:2018mto, de_la_Cruz_2019,Klausen:2019hrg, Klausen:2021yrt,  Klausen:2023gui, Klemm_2020, Feng_2020,Chestnov:2022alh,Ananthanarayan:2022ntm,Zhang:2023fil}.  In addition, many excellent expositions are available in the mathematics literature  \cite{Stienstra:2005nr, BeukersNotes,  cattani2006three, takayama2020hypergeometric,saito2013grobner, reichelt2021algebraic}.
Here, we focus on providing a simple and self-contained summary of the key material needed to understand the construction of creation operators.

\subsection{GKZ integrals}

An $\mathcal{A}$-hypergeometric function (or equivalently,  GKZ integral), is a multi-variable hypergeometric function depending on a set of real parameters $\bs{\gamma}=(\gamma_0,\gamma_1,\ldots,\gamma_N)$ and independent variables $\x = (x_1,\ldots, x_n)$, where $n\ge N+1$.  The integral takes the form
\[\label{GKZint}
\mathcal{I}_{\bs{\gamma}} =\Big(\prod_{i=1}^N \int_0^\infty\dd z_i\, z_i^{\gamma_i-1} \Big)\mathcal{D}^{-\gamma_0},
\]
where the `denominator' $\mathcal{D}$ can be expressed as a polynomial in the integration variables  $z_i$.  Every term in this polynomial is moreover multiplied by a {\it nonzero} coefficient $x_j$:
\[\label{GKZden}
\mathcal{D} = \sum_{j=1}^n x_j \prod_{i=1}^N z_i^{a_{ij}}
\]
The parameters  $a_{ij}\in \mathbb{Z}^+$ specifying the powers can be assembled into an $N\times n$ matrix $A$,
\[
(A)_{ij} = a_{ij}.
\]
Thus, the $j$th term in the denominator $\mathcal{D}$ corresponds to the column $j$ of the matrix $A$, whose entries are then the powers of the variables $z_i$ appearing in that particular term.
(We will return  to the relation between this matrix $A$ and the larger $\mathcal{A}$-matrix shortly.)

For Feynman integrals, it is useful to consider the Lee-Pomeransky representation \cite{Lee:2013hzt} in which the denominator $\mathcal{G}=\mathcal{U}+\mathcal{F}$ is formed from the sum of the first and second Symanzik polynomials $\mathcal{U}$ and $\mathcal{F}$.  To uplift this to the GKZ integral \eqref{GKZint},
we simply promote the coefficient of every term in $\mathcal{G}$ to a generalised independent variable $x_j$  \cite{de_la_Cruz_2019, Klausen:2019hrg}, as summarised in appendix \ref{LPrepapp}.   The original Lee-Pomeransky integral can then be restored by returning the $x_j$ to their physical values, namely, unity for any of the terms in $\mathcal{U}$, and the appropriate function of the masses and external momenta for every term in $\mathcal{F}$.

\paragraph{Example:}  
As discussed in appendix \ref{LPrepapp}, 
the massless triangle Feynman integral
\[\label{masslesstri}
I= 
\int\frac{\mathrm{d}^d\bs{q}}{(2\pi)^d}\frac{1}{q^{2\gamma_3}|\bs{q}-\bs{p}_1|^{2\gamma_2}|\bs{q}+\bs{p}_2|^{2\gamma_1}}
\]
has the Lee-Pomeransky representation
\[
I = c_\gamma\,\Big( \prod_{i=1}^3\int_0^\infty \dd z_i\, z_i^{\gamma_i-1}\Big)(p_1^2 z_2 z_3+p_2^2 z_1 z_3+p_3^2 z_1 z_2+z_1+z_2+z_3)^{-d/2}
\]
 where the coefficient
\[\label{Cdef}
c_{\bs{\g}}=(4\pi)^{-d/2}\frac{\Gamma(d/2)}{\Gamma(d-\gamma_t)\prod_{i=1}^3\Gamma(\gamma_i)},\qquad \gamma_t=\sum_{i=1}^3\gamma_i.
\] 
The corresponding  GKZ integral is 
\[\label{triGKZ}
\mathcal{I}_{\bs{\gamma}} = \Big(\prod_{i=1}^3\int_0^\infty\dd z_i z_i^{\gamma_i-1}\Big) \mathcal{D}^{-\gamma_0}
\]
where the denominator
\[\label{triD}
\mathcal{D} =  x_1 z_2 z_3 + x_2 z_1 z_3+x_3 z_1 z_2+x_4 z_1 + x_5 z_2 + x_6 z_3
\]
corresponds to the matrix
\[
A=  \left(\begin{matrix}
0&1&1&1&0&0\\
1&0&1&0&1&0\\
1&1&0&0&0&1
\end{matrix}
\right).
\]
To recover the original Lee-Pomeransky integral, we project to the physical subspace
\[\label{triphys}
\x = (p_1^2,p_2^2,p_3^2,1,1,1),\qquad \bs{\gamma} = (d/2,\gamma_1,\gamma_2,\gamma_3),
\]
after which $I = c_{\bs{\g}}\, \mathcal{I}_{\bs{\g}}$. 

\subsection{The Euler and toric equations}

The primary advantage of uplifting from the original masses and momenta to the generalised GKZ space parametrised by the  variables $\x$ is that the integral now obeys a systematic set of linear partial differential equations.  These can be grouped into two categories, known as the Euler equations and the toric equations.

\subsubsection{Euler equations}

The Euler equations arise  from integrating by parts with respect to the  variables $z_i$, under the assumption that all boundary terms vanish.
For $z_1$, for example, we have
\begin{align}
0 &=\int_0^\infty\dd z_1\, \frac{\partial}{\partial z_1}\Big(z_1^{\gamma_1} \Big(\prod_{i=2}^N\int_0^\infty\dd z_i\,z_i^{\gamma_i-1}\Big)\mathcal{D}^{-\gamma_0}\Big)\nn\\
&= \gamma_1 \mathcal{I}_{\bs{\gamma}}+\Big(\prod_{i=1}^N\int_0^\infty \dd z_i\,z_i^{\gamma_i-1}\Big) z_1\frac{\partial}{\partial z_1} \mathcal{D}^{-\gamma_0}.
\end{align}
In the second term here, we can trade derivatives with respect to the integration variable $z_1$ for derivatives with respect to the external variables $x_j$:
\begin{align}
 z_1\frac{\partial}{\partial z_1} \mathcal{D}^{-\gamma_0} =
-\gamma_0 \mathcal{D}^{-\gamma_0-1}\Big(\sum_{j=1}^n a_{1j}x_j \prod_{i=1}^N z_i^{a_{ij}}\Big)=  
 \Big( \sum_{j=1}^n a_{1j}\theta_j\Big) \mathcal{D}^{-\gamma_0} 
\end{align}
where, here and throughout the chapter, we define the Euler operators
\[
\theta_j = x_j\frac{\partial}{\partial x_j}, \qquad j=1,\ldots, n.
\]
Pulling these Euler operators outside the integrals, we obtain the equation
\[
0 = \Big(\gamma_1 + \sum_{j=1}^n a_{1j}\theta_j\Big)\mathcal{I}_{\bs{\gamma}}.
\]
Repeating this exercise for  the remaining $z_i$ then leads to the set of {\it Euler equations}
\[\label{EulereqnsN}
0 = \Big(\gamma_i + \sum_{j=1}^n a_{ij}\theta_j\Big)\mathcal{I}_{\bs{\gamma}}, \qquad i=1,\ldots, N.
\]
We are not quite done, however, since  in addition we have the general identity 
\[
\Big(\sum_{j=1}^n \theta_j \Big) \mathcal{D}^{-\gamma_0} = -\gamma_0 \mathcal{D}^{-\gamma_0}
\]
which, when applied to the GKZ integral, yields 
\[\label{DWI}
0 = \Big(\gamma_0+\sum_{j=1}^n\theta_j\Big)\mathcal{I}_{\bs{\gamma}}.
\]
This equation is effectively a dilatation Ward identity (or DWI, as we will use for short)
encoding the scaling behaviour of the GKZ integral under a dilatation $\bs{x}\rightarrow \lambda \bs{x}$ of the external variables. 

Evidently this dilatation Ward identity can be placed on the same footing as the Euler equations \eqref{EulereqnsN} by enlarging the matrix $A$ to include a top row consisting of all $1$s.  This construction defines the $\mathcal{A}$-matrix mentioned in the introduction,
\[\label{AtoA}
\mathcal{A} = \left(\begin{matrix} \bs{1}\\ A\end{matrix}\right)\!,
\]
where $\bs{1}$ is the $n$-dimensional row vector with all-$1$ entries, or  equivalently, 
\[
(\mathcal{A})_{0j}=1,\qquad (\mathcal{A})_{ij}=a_{ij}, \qquad i=1,\ldots, N,\qquad  j=1,\ldots, n,
\]
where we henceforth adopt the convention that the top row of $\mathcal{A}$ always carries index $0$.
The $\mathcal{A}$-matrix is thus $(N+1)\times n$ dimensional, and 
 the Euler equations and DWI together correspond to the $(N+1)$ equations
\[\label{allEulers}
0 = \Big(\gamma_i + \sum_{j=1}^n \mathcal{A}_{ij}\theta_j\Big)\mathcal{I}_{\bs{\gamma}}, \qquad i=0,\ldots, N.
\]
This is in effect a single matrix equation,
\[
0 = \Big(\bs{\gamma}+\mathcal{A}\cdot\bs{\theta}\Big)\mathcal{I}_{\bs{\gamma}},
\]
regarding $\bs{\theta}=(\theta_1,\ldots,\theta_n)^T$ and $\bs{\gamma}=(\gamma_0,\gamma_1,\ldots,\gamma_N)^T$ as  $n$- and $(N+1)$-component column vectors respectively.

\paragraph{Example:}  Returning to the  massless triangle integral above, the $\mathcal{A}$-matrix is
\[\label{Atri}
\mathcal{A}=  \left(\begin{matrix}
1&1&1&1&1&1\\
0&1&1&1&0&0\\
1&0&1&0&1&0\\
1&1&0&0&0&1
\end{matrix}
\right)
\]
and the GKZ integral satisfies  the  Euler equations
\begin{align}\label{triEulers}
0=(\gamma_1+\theta_2+\theta_3+\theta_4)\mathcal{I}_{\bs{\gamma}},\quad 0=(\gamma_2+\theta_1+\theta_3+\theta_5)\mathcal{I}_{\bs{\gamma}},\quad 0=(\gamma_3+\theta_1+\theta_2+\theta_6)\mathcal{I}_{\bs{\gamma}}
\end{align}
and DWI
\[\label{triDWI}
0 = (\gamma_0+\sum_{j=1}^6 \theta_j)\mathcal{I}_{\bs{\gamma}}.
\]
Notice the form of these equations can be directly read off from the  rows of the $\mathcal{A}$-matrix.

\subsubsection{Toric equations}  The toric equations arise from  vectors in the {\it kernel} of the $\mathcal{A}$-matrix, and are closely related to the corresponding toric ideal \cite{saito2013grobner}.\footnote{The kernel is the space of vectors $\bs{u}$ such that  $\mathcal{A}\cdot\bs{u}=\bs{0}$, obtained {\it e.g.,} via $\tt{NullSpace}[\mathcal{A}]$  in Mathematica.  The full toric ideal, though not needed here, can be constructed using {\it Singular}  \cite{singular}: see  section \ref{toricidealdisc}.}  
Their origin can 
 easily be grasped using the example of the massless triangle integral above.  
Defining
\[
\partial_j = \frac{\p}{\p x_j},\qquad j=1,\ldots, n
\]
in all that follows, the  denominator \eqref{triD} obeys the  relations
\[
\p_1\p_4\mathcal{D}^{-\gamma_0}=
\p_2\p_5\mathcal{D}^{-\gamma_0}=
\p_3\p_6\mathcal{D}^{-\gamma_0}=-\gamma_0 (-\gamma_0-1)z_1 z_2 z_3\mathcal{D}^{-\gamma_0-2},
\]
giving rise to the two independent (toric) equations
\[\label{tritorics}
0=(\p_1\p_4-\p_3\p_6)\mathcal{I}_{\bs{\gamma}},\qquad
0 = (\p_2\p_5-\p_3\p_6)\mathcal{I}_{\bs{\gamma}}.
\]
For comparison, the kernel of the $\mathcal{A}$-matrix \eqref{Atri} is spanned by two independent vectors, $\bs{u}_{(1)}$ and $\bs{u}_{(2)}$, which we can choose to be
\[\label{trikern}
\bs{u}_{(1)} = (1,0,-1,1,0,-1)^T,\qquad \bs{u}_{(2)} = (1,-1,0,1,-1,0)^T.
\]
Notice that since the top row of the $\mathcal{A}$-matrix is all  $1$s, the sum of the components of any kernel vector is always zero.  
There is now a one-to-one match between kernel vectors and toric equations \eqref{tritorics} as follows. First, for each kernel vector $\bs{u}$, we form a vector $\bs{u}^+$ composed only of the {\it positive} components of $\bs{u}$, and a vector $\bs{u}^-$ composed of only the {\it negative} components.  The components of $\bs{u}^\pm$, for each $j=1,\ldots, n$, are thus 
\[
u^\pm_j=\mathrm{max}(\pm u_j,0).
\]
By inspection, the toric equation corresponding to the kernel vector  $\bs{u} = \bs{u}^+ -\bs{u}^-$ is now
\[\label{torics}
0 = \Big(\prod_{j=1}^n \p_j^{u_j^+} -\prod_{j=1}^n \p_j^{u_j^-}\Big)\,\mathcal{I}_{\bs{\gamma}}.
\]
For example, for $\bs{u}_{(1)}$ in \eqref{trikern}, $\bs{u}_{(1)}^+=(1,0,0,1,0,0)^T$ while $\bs{u}_{(1)}^-=(0,0,1,0,0,1)^T$ hence \eqref{torics} reduces to the first equation in \eqref{tritorics}.

Some investigation shows this construction is a general one. 
First, the action of each differential operator is
\[
\prod_{j=1}^n \p_j^{u_j^\pm}\mathcal{D}^{-\gamma_0} = (-\gamma_0)(-\gamma_0-1)\ldots(-\gamma_0-\mathfrak{u}^\pm+1) \mathcal{D}^{-\gamma_0-\mathfrak{u}^\pm}\Big(\prod_{i=1}^N z_i^{\sum_{j=1}^n a_{ij}u_j^\pm}
\Big)
\]
where $\mathfrak{u}^\pm=\sum_{j=1}^n u_j^\pm$.    Moreover, since the sum of components in any kernel vector vanishes (as the top row of the $\mathcal{A}$-matrix is all $1$s), we have that $\mathfrak{u}^+=\mathfrak{u}^-=\mathfrak{u}$.  Thus, 
\begin{align}\label{toriceval}
& \Big(\prod_{j=1}^n \p_j^{u_j^+} -\prod_{j=1}^n \p_j^{u_j^-}\Big)\,\mathcal{I}_{\bs{\gamma}}\\[-1ex]
 &\qquad =(-\gamma_0)(-\gamma_0-1)\ldots(-\gamma_0-\mathfrak{u}+1) \mathcal{D}^{-\gamma_0-\mathfrak{u}}\Big(\prod_{i=1}^N z_i^{\sum_{j=1}^n a_{ij}u_j^+}-\prod_{i=1}^N z_i^{\sum_{j=1}^n a_{ij}u_j^-}
\Big).\nn
\end{align}
However, for any kernel vector we have $\mathcal{A}\cdot\bs{u} = \mathcal{A}\cdot(\bs{u}^+-\bs{u}^-)=\bs{0}$ and hence
\[
\sum_{j=1}^n a_{ij}u_j^+=\sum_{j=1}^n a_{ij}u_j^-, \qquad i=1,\ldots, N.
\]
The two terms appearing within the final factor of \eqref{toriceval} are thus exactly equal producing a cancellation.
In general, as the $\mathcal{A}$-matrix is $(N+1)\times n$, there are $(n-N-1)$ independent vectors in the kernel, and hence this same number of independent toric equations.

To summarise,  given a GKZ integral defined by an $\mathcal{A}$-matrix and parameters $\bs{\gamma}$,  we have two sets of linear partial differential equations: 
the Euler equations (and DWI) \eqref{allEulers}, and the toric equations \eqref{torics}.  
We can also go in reverse: the Euler equations and DWI fix $\bs{\gamma}$ and the $\mathcal{A}$-matrix, and hence the toric equations and the GKZ integral.  
Note the Euler equations all commute among themselves, as do the toric equations, but an Euler and a toric equation do not in general commute.

\subsection{Projection to physical variables} 

The systematic structure of the Euler and toric equations above is a consequence of uplifting from the Lee-Pomeransky  to the GKZ denominator \eqref{GKZden}.
To recover a set of PDEs satisfied by the original Lee-Pomeransky integral we need to reverse this process.
This requires projecting the Euler and toric equations back to the {\it physical hypersurface} where the $\x$ variables take their true physical values.
Derivatives in directions not tangential to this hypersurface (which therefore cannot be expressed purely in terms of physical variables) can be exchanged for purely tangential derivatives through use of the Euler equations and DWI.  
Together these provide $N+1$ equations, and so for {\it all} unphysical ({\it i.e.,} non-tangential) derivatives to be removable requires the original Lee-Pomeransky polynomial to contain at least $n-N-1$ independent physical variables ({\it i.e.,} masses and external momenta).  This will generally be the case for the examples we consider, but does not hold universally -- particularly for higher-loop Feynman integrals -- as we discuss  in section \ref{sec:Discussion}.

\paragraph{Example:}  For the massless triangle integral, the physical hypersurface is the $3$-dimensional subspace spanned by the momenta in \eqref{triphys}, namely $x_1=p_1^2$, $x_2=p_2^2$ and $x_3=p_3^2$, with $x_4=x_5=x_6=1$.
On this hypersurface, the Euler equations \eqref{triEulers} reduce to
\[\label{linderivs}
0=(\gamma_1+\theta_2+\theta_3+\partial_4)\mathcal{I}_{\bs{\gamma}},\quad 0=(\gamma_2+\theta_1+\theta_3+\partial_5)\mathcal{I}_{\bs{\gamma}},\quad 0=(\gamma_3+\theta_1+\theta_2+\partial_6)\mathcal{I}_{\bs{\gamma}},
\]
where, as always, $\partial_j=\partial/\partial x_j$.
These equations allow us to eliminate the unphysical derivatives $\partial_4$, $\partial_5$ and $\partial_6$ from all remaining equations in which they appear linearly.\footnote{More generally, we can rewrite $\p_4^m = x_4^{-m}\t_4(\t_4-1)\ldots(\t_4-m+1)$, {\it etc.}, then use the full Euler equations to eliminate $\t_4$, $\t_5$ and $\t_6$ before setting $x_4=x_5=x_6=1$.
 Alternatively, we can supplement \eqref{linderivs} with derivatives of the Euler equations (and DWI) evaluated on the physical hypersurface.}
For example, evaluating the first toric equation in \eqref{tritorics} on the physical hypersurface,
\begin{align}\label{physeq1}
0 &= (\partial_1\partial_4-\partial_3\partial_6)\mathcal{I}_{\bs{\gamma}} \nn\\
&= \Big(\partial_1(-\gamma_1-\theta_2-\theta_3) - \partial_3(-\gamma_3-\theta_1-\theta_2)\Big)\mathcal{I}_{\bs{\gamma}} \nn\\&
=\frac{1}{4}\Big[-\Big(2\gamma_1+p_2\frac{\partial}{\partial p_2}+p_3\frac{\partial}{\partial p_3}\Big)\frac{1}{p_1}\frac{\partial}{\partial p_1}+\Big(2\gamma_3+p_1\frac{\partial}{\partial p_1}+p_2\frac{\partial}{\partial p_2}\Big)\frac{1}{p_3}\frac{\partial}{\partial p_3}\Big]\mathcal{I}_{\bs{\gamma}} ,
\end{align}
while for the second toric equation,
\begin{align}
0 &= (\partial_2\partial_5-\partial_3\partial_6)\mathcal{I}_{\bs{\gamma}} \nn\\
&= \Big(\partial_2(-\gamma_2-\theta_1-\theta_3) - \partial_3(-\gamma_3-\theta_1-\theta_2)\Big)\mathcal{I}_{\bs{\gamma}} \nn\\&
=\frac{1}{4}\Big[-\Big(2\gamma_2+p_1\frac{\partial}{\partial p_1}+p_3\frac{\partial}{\partial p_3}\Big)\frac{1}{p_2}\frac{\partial}{\partial p_2}+\Big(2\gamma_3+p_1\frac{\partial}{\partial p_1}+p_2\frac{\partial}{\partial p_2}\Big)\frac{1}{p_3}\frac{\partial}{\partial p_3}\Big]\mathcal{I}_{\bs{\gamma}} .
\end{align}
Finally,  on the physical hypersurface, the DWI \eqref{triDWI} reduces to
\begin{align}\label{physeq3}
0 &= \Big(\frac{d}{2}+\theta_1+\theta_2+\theta_3+\partial_4+\partial_5+\partial_6\Big)\mathcal{I}_{\bs{\gamma}}\nn\\
&=
\Big(\frac{d}{2}-\gamma_1-\gamma_2-\gamma_3-\theta_1-\theta_2-\theta_3\Big)\mathcal{I}_{\bs{\gamma}}\nn\\
&=\frac{1}{2}\Big(d-2\gamma_1-2\gamma_2-2\gamma_3-p_1\frac{\partial}{\partial p_1}-p_2\frac{\partial}{\partial p_2}-p_3\frac{\partial}{\partial p_3}\Big)\mathcal{I}_{\bs{\gamma}}
\end{align}
Equations \eqref{physeq1}-\eqref{physeq3} involve only physical variables, namely, the momentum magnitudes.

\subsection{Affine reparametrisations}

As we have seen, the set of Euler equations associated with a given GKZ integral can be read off from the rows of the  $\mathcal{A}$-matrix: in the $i$th Euler equation \eqref{EulereqnsN}, the coefficient of the operator $\theta_j$ is $a_{ij}=(\mathcal{A})_{ij}$ where $1\le i\le N$ and $1\le j\le n$.  (Recall we are labelling the top all-$1$s row of the $\mathcal{A}$-matrix as $i=0$.)
Viewed in reverse, the set of Euler equations determines both the $\mathcal{A}$-matrix and the set of parameters $\bs{\gamma}$, and hence the GKZ integral.

What happens if we now form a {\it new} set of Euler equations by taking  linear combinations of the old ones?  In the process, we could simultaneously add to each Euler equation some multiple of the DWI.
Together, these operations correspond to left-multiplying  the $\mathcal{A}$-matrix by an $(N+1)\times(N+1)$ matrix 
\[\label{affineM}
\mathcal{M}=
\left(\begin{matrix} 1 & \bs{0}\\
\bs{b} & M\end{matrix}\right)\!,
\]
where $\bs{0}$ is an $N$-dimensional row vector of zeros, $\bs{b}$ is an $N$-dimensional column vector and $M$ an $N\times N$ matrix.  This yields
\[\label{affineA}
\mathcal{A}'=\mathcal{M}\mathcal{A} =
\left(\begin{matrix} 1 & \bs{0}\\
\bs{b} & M\end{matrix}\right)\left(\begin{matrix}\bs{1}\\A\end{matrix}\right)=\left(\begin{matrix}\bs{1}\\A'\end{matrix}\right)\!,
\]
where the components of $A$ undergo the  affine transformation
\[\label{aprime}
(A')_{ij}=a'_{ij}=b_i+ \sum_{k=1}^N m_{ik}a_{kj}.
\]
The new set of Euler equations now corresponds to the rows of $\mathcal{A}'$: the $i$th new Euler equation  is  the sum of $m_{ik}$ times the $k$th old Euler equation 
 plus $b_i$ times the DWI (for which the coefficient of every $\theta_j$ is one).    
In order to have $a'_{ij}\in\mathbb{Z}^+$, so as to form a new denominator polynomial $\mathcal{D}'$ via \eqref{GKZden}, we will restrict the entries of $\mathcal{M}$ to 
 $m_{ij}\in\mathbb{Z}^+$ and $b_i\in\mathbb{Z}^+$.
Note the transformation  \eqref{affineA} leaves the DWI unchanged.

The new set of Euler equations now takes the form
\[
0 = \Big(\bs{\gamma}'+\mathcal{A}'\cdot\bs{\theta}\Big)\mathcal{I}_{\bs{\gamma}'},
\]
where
\begin{align}\label{gammaprime}
\bs{\gamma}' = \left(\begin{matrix}\gamma_0\\\gamma'_1\\\vdots\\\gamma'_N\end{matrix}\right) =\left(\begin{matrix} 1 & \bs{0}\\
\bs{b} & M\end{matrix}\right) \left(\begin{matrix}\gamma_0\\\gamma_1\\\vdots\\\gamma_N\end{matrix}\right)= \mathcal{M}\bs{\gamma}
\end{align}
so that  $\gamma'_i = \gamma_0 b_i+\sum_{k=1}^N m_{ik}\gamma_k$ for $1\le i\le N$ while the DWI \eqref{DWI} remains unchanged.
Provided that $\mathrm{det}(\mathcal{M})$ is nonzero,  the toric equations are also unchanged since the kernel of $\mathcal{A}$ is preserved under multiplication by an invertible matrix.

What is now the relation of this new GKZ integral, defined by $\mathcal{A}'$, to the original?
The new integral is
\[
\mathcal{I}_{\bs{\gamma}'} =\Big(\prod_{i=1}^N \int_0^\infty\dd z'_i\, (z'_i)^{\gamma'_i-1} \Big)(\mathcal{D}')^{-\gamma_0},
\]
where 
\[\label{GKZdenprime}
\mathcal{D}' = \sum_{j=1}^n x_j \prod_{i=1}^N (z'_i)^{a'_{ij}}.
\]
Using \eqref{aprime}, and making the identification
\[
z_k = \prod_{i=1}^N (z'_i)^{m_{ik}},
\]
we find
\begin{align}
\mathcal{D}' &= \sum_{j=1}^n x_j \prod_{i=1}^N (z'_i)^{b_i+\sum_{k=1}^N m_{ik}a_{kj}} 
= \Big(\prod_{l=1}^N (z'_l)^{b_l}\Big)\Big(\sum_{j=1}^n x_j
\prod_{i=1}^N\prod_{k=1}^N (z'_i)^{m_{ik}a_{kj}}\Big) 
\nn\\&
=\Big(\prod_{l=1}^N (z'_l)^{b_l}\Big)\Big(\sum_{j=1}^n x_j
\prod_{k=1}^Nz_k^{a_{kj}}\Big)=\Big(\prod_{l=1}^N (z'_l)^{b_l}\Big)\mathcal{D}.
\end{align}
Moving the factor of $\prod_{l=1}^N (z'_l)^{b_l}$ from the denominator to the numerator and using \eqref{gammaprime} then gives
\begin{align}
\mathcal{I}_{\bs{\gamma}'} &=\Big(\prod_{i=1}^N \int_0^\infty\dd z'_i\, (z'_i)^{\gamma'_i-\gamma_0 b_i-1} \Big) \mathcal{D}^{-\gamma_0} = \Big(\prod_{i=1}^N \int_0^\infty\dd z'_i\, (z'_i)^{\sum_{k=1}^Nm_{ik}\gamma_k-1} \Big) \mathcal{D}^{-\gamma_0} \nn\\
&= \Big(\prod_{i=1}^N \int_0^\infty\frac{\dd z'_i}{z'_i}\,\prod_{k=1}^N (z'_i)^{m_{ik}\gamma_k} \Big) \mathcal{D}^{-\gamma_0}=
\Big(\prod_{i=1}^N \int_0^\infty\frac{\dd z'_i}{z'_i}\, z_i^{\gamma_i} \Big)\mathcal{D}^{-\gamma_0}.
\end{align}
Finally, since
\[
\frac{\dd z_i}{z_i}=\sum_{j=1}^N m_{ji}\frac{\dd z_j'}{z'_j}, \qquad \qquad \prod_i \int_0^\infty\frac{\dd z_i}{z_i} =  |\mathrm{det}\,M|\prod_i \int_0^\infty\frac{\dd z'_i}{z'_i}, 
\]
we find
\[\label{IgIgp}
\mathcal{I}_{\bs{\gamma}'} = |\mathrm{det}\,M|^{-1}\mathcal{I}_{\bs{\gamma}}. 
\]
Thus, choosing a new basis for the Euler equations by taking linear combinations of the old Euler equations and the DWI only rescales the GKZ integral by a constant factor.  As the GKZ system of equations is linear, this overall scaling is in any case not fixed and the solution is effectively unchanged.

\paragraph{Example:} \label{tripleKex} The affine reparametrisation above can be used to show the equivalence  of the massless triangle integral \eqref{masslesstri} with the {\it triple-K integral} (see also \cite{Bzowski:2013sza,Bzowski:2020kfw})
\[\label{tripleKdef}
I_{\alpha,\{\beta_1,\beta_2,\beta_3\}} = \int_0^\infty \dd z\, z^{\alpha}\prod_{i=1}^3 p_i^{\beta_i} K_{\beta_i}(p_i z).
\]
For  $\alpha=d/2-1$ and $\beta_i=\Delta_i-d/2$, this integral represents the momentum-space 3-point function of scalars $\O_{\Delta_i}$ in any $d$-dimensional CFT.
The triple-$K$ integral can be put into GKZ form by first Schwinger parametrising the modified Bessel functions as 
\[
p_i^{\beta_i} K_{\beta_i}(p_i z)=\frac{1}{2} \int_0^\infty \dd z_i' \, (z_i')^{\beta_i-1}\exp\Big[-\frac{z}{2}\Big(z_i'+\frac{p_i^2}{z_i'}\Big)\Big]
\]
then performing the $z$ integral.  This gives
\[
I_{\alpha,\{\beta_1,\beta_2,\beta_3\}} = 2^{\alpha-2}\Gamma(\alpha+1)\Big(\prod_{i=1}^3\int_0^\infty \dd z_i'\,(z'_i)^{\beta_i-1}\Big)\Big[\sum_{j=1}^3\Big( z_j'+\frac{p_j^2}{z_j'}\Big)\Big]
^{-\alpha-1}
\]
which uplifts to the GKZ integral
\[\label{GKZ3rep00}
I_{\alpha,\{\beta_1,\beta_2,\beta_3\}} = 
2^{\alpha-2}\Gamma(\alpha+1)\Big(\prod_{i=1}^3\int_0^\infty \dd z_i'\,(z'_i)^{\gamma_i'-1}\Big)(\mathcal{D}')^{-\gamma_0'}
\]
where
\[\label{3Kden}
\mathcal{D}' =  \frac{x_1}{z_1'}+ \frac{x_2}{z_2'}+ \frac{x_3}{z_3'}+x_4 z_1'+x_5 z_2'+x_6 z_3'. 
\]
The physical hypersurface ({\it i.e.,} the original triple-$K$ integral) corresponds to
\[\label{3Kphys}
\gamma_i'=\beta_i,\qquad \gamma_0'=\alpha+1, \qquad
\x = (p_1^2,p_2^2,p_3^2,1,1,1).
\]
Here, we are using primes to distinguish the parameters of the triple-$K$ integral from those of the massless triangle integral earlier.
Also, while the denominator \eqref{3Kden} is not a polynomial, this simple generalisation will nevertheless   turn out to be the most convenient representation for us later.\footnote{
Should a  purely polynomial denominator be required, one can simply pull out an overall factor of $(z_1' z_2' z_3')^{-1}$ from the right-hand side of \eqref{3Kden} then transfer this to the numerator by shifting the $\gamma_i'$.}
The $\mathcal{A}$-matrix corresponding to the triple-$K$ integral is then 
\[\label{3KA}
\mathcal{A}_{\mathrm{3K}} 
=\left(\begin{matrix}
1&1&1&1&1&1\\
-1&0&0&1&0&0\\
0&-1&0&0&1&0\\
0&0&-1&0&0&1
\end{matrix}
\right).
\]
Comparing with the massless triangle $\mathcal{A}$-matrix \eqref{Atri}, we find that 
\[
\mathcal{M}\mathcal{A}_{\mathrm{triangle}} =\mathcal{A}_{\mathrm{3K}},
\]
where
\[\label{Mtrito3K}
\mathcal{M} = \left(\begin{matrix} 1 & 0 & 0 & 0\\1 & 0&-1&-1\\
1&-1&0&-1\\1&-1&-1&0 \end{matrix}\right).
\]
The parameters of the triangle integral are  connected to those of the triple-$K$ integral by
\[\label{affinetransfofg}
\mathcal{M}\bs{\g}_{\mathrm{triangle}} =  \mathcal{M}\left(\begin{matrix}d/2\\ \gamma_1\\ \gamma_2\\ \gamma_3\end{matrix}\right) = \left(\begin{matrix} d/2\\d/2-\gamma_2-\gamma_3\\d/2- \gamma_1-\gamma_3\\
d/2-\gamma_1-\gamma_2\end{matrix}\right)=\left(\begin{matrix} \alpha+1\\ \beta_1\\ \beta_2\\ \beta_3\end{matrix}\right) = \bs{\gamma}_{3K}.
\]
Putting everything together, 
 from \eqref{IgIgp} with $\mathrm{det}\,\mathcal{M}=2$ and \eqref{Cdef}, we have
\[
I_{d/2-1,\{d/2-\gamma_2-\gamma_3,\,d/2-\gamma_1-\gamma_3,\,d/2-\gamma_1-\gamma_2\}} =C'\int\frac{\mathrm{d}^d\bs{q}}{(2\pi)^d}\frac{1}{q^{2\gamma_3}|\bs{q}-\bs{p}_1|^{2\gamma_2}|\bs{q}+\bs{p}_2|^{2\gamma_1}}
\]
where
\[
C' = \pi^{d/2}2^{3d/2-4}\Gamma(d-\gamma_t)\prod_{i=1}^3\Gamma(\gamma_i).
\]
As we saw above, the matrix multiplication here is  just a slick way of executing the change of variables 
\[
z_1 = \frac{1}{z_2'z_3'}, \qquad z_2 =\frac{1}{ z_1'z_3'},\qquad z_3 = \frac{1}{z_1'z_2'},
\]
on the triangle GKZ representation, followed by moving a factor
of $(z_1'z_2'z_3')^{-\gamma_0}$ from the denominator to the numerator.

\section{Spectral singularities and the Newton polytope}
\label{sec:sings}

We now turn to examine the singularities of GKZ integrals arising for special values  of the parameters $\bs{\g}$.  As we will see, these can be viewed geometrically in terms of the {\it Newton polytope} of the GKZ denominator $\mathcal{D}$.

\subsection{The Newton polytope}

A defining feature of the GKZ representation is that only a single denominator \eqref{GKZden} is present:
\[
\mathcal{D} = \sum_{j=1}^n x_{j} \prod_{i=1}^N z_i^{a_{ij}}.
\] 
The exponents of the $j$th term in this denominator 
define a vector $\bs{a}_j$ living in an $N$-dimensional space, whose components  are
\[
(\bs{a}_j)_i = a_{ij}, \qquad i=1,\ldots N.
\]
Thus, $\bs{a}_j$ is  the $j$th column of the $\mathcal{A}$-matrix after stripping off the top row of all $1$s.
Constructing the convex hull of these exponent vectors then defines the $N$-dimensional  Newton polytope of $\mathcal{D}$:
\[
\mathrm{Newt}(\mathcal{D}) =  \sum_{j=1}^n\alpha_j \bs{a}_j, \quad\mathrm{with}\quad \sum_{j=1}^n \alpha_j = 1, \,\quad \alpha_j\ge 0\,\,\, \forall\,\, \, j.
\]
For the denominator \eqref{3Kden} of the triple-$K$ integral, for example,  we obtain the regular octahedron shown on the left of  figure \ref{fig:tripolytope}.  For the denominator of the massless triangle integral \eqref{triD}, we also obtain an octahedron, but now with vertices as shown on the right of the figure.  The  vertices of each polytope are related by the affine transformation  \eqref{aprime},
\[
\bs{a}_j^{(3K)} = \bs{b}+M \bs{a}_j^{\mathrm{(triangle)}}, \qquad j=1,\ldots, 6
\]
where, from \eqref{affineM} and  \eqref{Mtrito3K}, 
\[
\bs{b} = \left(\begin{matrix} 1\\1\\1\end{matrix}\right),
\qquad M = \left(\begin{matrix} 0 & -1 &-1\\- 1 & 0 &- 1\\ -1 & -1&0\end{matrix}\right)\!.
\]
As we saw above,  for any two $\mathcal{A}$-matrices (and hence any two Newton polytopes) related by an affine transformation, the corresponding GKZ integrals are proportional to each other and satisfy the same system of  equations ({\it i.e.,} DWI, Euler and toric equations). 
Thus, Newton polytopes such as these related by affine transformations  are effectively equivalent.

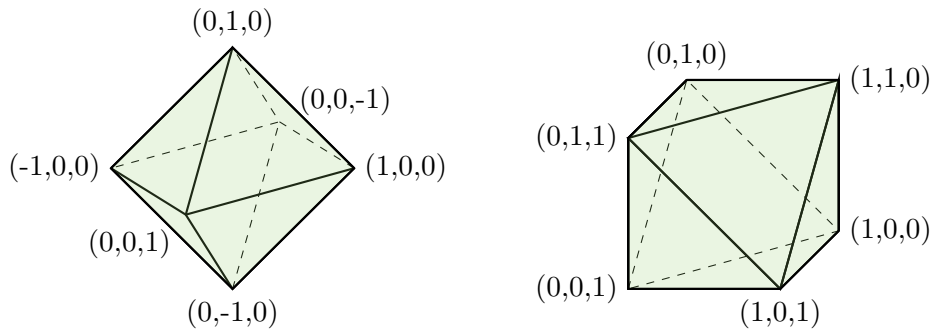
\begin{figure}[t]
\centering
\begin{tikzpicture}[scale=1.6]
   \draw[thick,black] (1,0,0) -- (0,1,0) -- (0,0,1); 
  \draw[thick,black]  (-1,0,0) -- (0,-1,0) ;
   \draw[thick,black] (-1,0,0) -- (0,1,0) ;
    \draw[thick,black] (-1,0,0) -- (0,0,1)--(1,0,0) ;
    \draw[thick,black] (0,0,1) -- (0,-1,0)--(1,0,0) ;
   \draw[style=dashed, color=black] (0,-1,0)-- (0,0,-1)--(-1,0,0);
   \draw[style=dashed, color=black] (1,0,0)-- (0,0,-1)--(0,1,0); 
     \draw node[right] at (1,0,0) {(1,0,0)};   
   \draw node[above] at (0,1,0) {(0,1,0)};
   \draw node[left] at (0,-0.2,1.1) {(0,0,1)};
   \draw node[left] at (-1,0,0) {(-1,0,0)};
    \draw node[below] at (0,-1,0) {(0,-1,0)};
   \draw node[right] at (0,0.1,-1.2) {(0,0,-1)};

\draw[thick,black,fill=YellowGreen, fill opacity=0.2] (-1,0,0)--(0,-1,0) -- (1,0,0) -- (0,1,0) --cycle;

   \end{tikzpicture}
\qquad
\begin{tikzpicture}[scale=2]

   \draw[thick,black] (1,0,0) -- (1,0,1) -- (0,1,1) -- (0,1,0); 
   \draw[thick,black] (0,0,1) -- (0,1,1) -- (0,1,0) ;
   \draw[thick,black] (1,1,0) -- (0,1,1) -- (1,0,1)--(1,1,0) ;
   \draw[thick,black] (1,0,1) -- (1,1,0) -- (1,0,0) ;
   \draw[thick,black] (1,0,1) -- (0,0,1) -- (0,1,1) ;
   \draw[thick,black] (1,1,0) -- (1,0,0) ;
   \draw[thick,black] (0,1,0)--(1,1,0) ;
 \draw[style=dashed, color=black] (0,0,1) -- (0,1,0) -- (1,0,0)--(0,0,1);
    \draw node[right] at (1,0,0) {(1,0,0)};   
   \draw node[above] at (0,1,0) {(0,1,0)};
   \draw node[left] at (0,0,1) {(0,0,1)};
   \draw node[right] at (1,1,0) {(1,1,0)};
    \draw node[below] at (1,0,1) {(1,0,1)};
   \draw node[left] at (0,1,1) {(0,1,1)};

\draw[thick,black,fill=YellowGreen, fill opacity=0.2] (0,0,1)--(1,0,1) -- (1,0,0) -- (1,1,0) --(0,1,0)--(0,1,1)--cycle;

   \end{tikzpicture}
   \caption{The Newton polytopes corresponding to the denominators of the triple-$K$ integral  \eqref{3Kden} (left) and the massless triangle integral \eqref{triD} (right).}
\label{fig:tripolytope}
\end{figure}

\subsection{Spectral singularities}

The physical significance of the Newton polytope becomes apparent when we consider the {\it spectral singularities} of the GKZ integral.
These are the divergences that arise for special values of the parameters $\bs{\gamma}$, with general kinematics, and are distinct from the {\it kinematic} (or Landau) singularities (discussed, {\it e.g.,} in \cite{Klausen:2021yrt}) which arise for general $\bs{\gamma}$ but special kinematics. 
Remarkably, it can be shown \cite{nilsson2010mellin,berkesch2013eulermellin} that the spectral singularities 
are closely related to the 
facets ({\it i.e.,} co-dimension one faces) 
  of the Newton polytope. 
As this polytope lives in an $N$-dimensional space, let us first define the $N$-dimensional parameter vector 
\[
\hat{\bs{\gamma}}=(\gamma_1,\ldots,\gamma_N)^T,
\]
where the hat serves to distinguish from the $(N+1)$-dimensional parameter vector $\bs{\g} = (\g_0,\hat{\bs{\g}})^T$.
In addition, we define the {\it rescaled} Newton polytope to be the convex hull of the  vertex vectors $\g_0\bs{a}_j$.  This corresponds to a linear rescaling\footnote{The significance of this rescaling can be anticipated by noting that the Newton polytope of the GKZ denominator $\mathcal{D}^{\g_0}$, in the special cases where $\g_0\in\mathbb{N}$ so that $\mathcal{D}^{\g_0}$ is itself a polynomial when expanded out, is simply the Newton polytope of $\mathcal{D}$ linearly rescaled by $\g_0$. } of the original Newton polytope by a factor of $\g_0$.
 The GKZ integral is then finite for all parameter values $\hat{\bs{\g}}$ lying {\it within} this rescaled Newton polytope.  On the hyperplanes corresponding to the facets of the rescaled Newton polytope, as well as on an infinite set of further hyperplanes both parallel and exterior to these facets, the integral is singular.

An exact formula for all  singular hyperplanes will be derived below in \eqref{hypsings}.  The location of these singularities will then be the main ingredient in our subsequent construction of creation operators.
Two key steps are needed to establish the result \eqref{hypsings}. 
 The first is to show that the GKZ integral converges for all $\hat{\bs{\gamma}}$ values lying in the interior of the rescaled Newton polytope.  Rather than 
recounting  the 
formal proof of \cite{nilsson2010mellin,berkesch2013eulermellin},  we will instead pursue a more informal approach based on a tropical analysis of the GKZ integral \cite{Arkani-Hamed:2022cqe,Matsubara-Heo:2023ylc}.   Many closely related constructions appear in sector decomposition, see {\it e.g.,} \cite{Kaneko:2009qx, Schultka:2018nrs}.
 The second step in the analysis is to construct a series of  meromorphic continuations across each of the singular hyperplanes.  This can be achieved by a scaling argument due to  \cite{nilsson2010mellin,berkesch2013eulermellin}.
 Here, we present a further variation of this argument involving a special linear combination of the Euler equations and DWI.

\paragraph{Example:}  \label{Ex1} As an initial check of the picture  above, we recall that the spectral singularities of the triple-$K$ integral \eqref{tripleKdef} are already known from conformal field theory \cite{Bzowski:2015pba}.\footnote{The argument in \cite{Bzowski:2015pba} involves  expanding the integrand of the triple-$K$ integral about its lower limit and looking for the appearance of $z^{-1}$ poles.}
The condition for  the triple-$K$ integral $I_{\alpha,\{\beta_1,\beta_2,\beta_3\}}$ to be singular is
\[
\alpha+1\pm\beta_1\pm\beta_2\pm\beta_3 = -2m,\qquad m\in \mathbb{Z}^+
\]
where any independent choice of the three $\pm$ signs can be made, and any value $m=0,1,2,\ldots$ is permitted.
(Throughout this chapter, we will take $\mathbb{Z}^+$ to be the set of all  non-negative integers {\it including} zero.)
Re-expressing this condition in terms of the $\bs{\gamma}$ parameters  \eqref{3Kphys} appearing in the GKZ integral, and dropping the primes, this is
\[\label{3Kgammasings}
\gamma_0 \pm \gamma_1\pm\gamma_2\pm\gamma_3=-2m.
\]
We see immediately that the $m=0$ singularities indeed correspond to 
the equations of the hyperplanes containing the eight facets of the regular octahedron on the left of figure \ref{fig:tripolytope}, where the vertices in the figure correspond to $(\gamma_1,\gamma_2,\gamma_3) = \gamma_0(\pm 1,0,0),$ $\gamma_0(0,\pm 1,0)$ and $\gamma_0(0,0,\pm 1)$.
 The remaining singularities for  $m>0$ then correspond to an infinite series of regularly spaced hyperplanes, both parallel, and exterior, to the facets of the octahedron.

\subsubsection{Tropical analysis: an example}

To appreciate the role of the Newton polytope, let us start with a simple example introduced in \cite{nilsson2010mellin}.  This is the  GKZ integral 
\[\label{houseint}
\mathcal{I}_{\bs{\gamma}} = \int_0^\infty\dd z_1\int_0^\infty\dd z_2 \,z_1^{\g_1-1}z_2^{\g_2-1} (x_1+x_2 z_2+x_3 z_1^2+x_4 z_1 z_2^2)^{-\gamma_0},
\]
whose $\mathcal{A}$-matrix is
\[\label{houseA}
\mathcal{A} = \left(\begin{matrix} 1 & 1 & 1& 1\\ 0 & 0 & 2 & 1\\ 0 & 1 & 0 & 2\end{matrix}\right)\!.
\]
The singularities of the integral derive from regions where the $z_i$ (for $i=1,2$) either vanish or tend to infinity.  Setting $z_i=e^{\tau_i}$, these regions are mapped to $|\tau_i|\rightarrow \infty$ and 
\[
\mathcal{I}_{\bs{\gamma}} = \int_{-\infty}^\infty\dd \tau_1\int_{-\infty}^\infty\dd \tau_2 \,e^{\gamma_1\tau_1+\gamma_2\tau_2}(x_1+x_2e^{\tau_2}+x_3 e^{2\tau_1}+x_4 e^{\tau_1+2\tau_2})^{-\gamma_0}.
\]
For large $|\tau_i|$, we can approximate this integral by  its {\it tropicalisation} as discussed in  \cite{Arkani-Hamed:2022cqe},
\[\label{trophouse}
\mathcal{I}_{\bs{\gamma}}^{\mathrm{trop.}} =x_j^{-\gamma_0} \int_{-\infty}^\infty\dd \tau_1\int_{-\infty}^\infty\dd \tau_2 \exp\Big[\gamma_1\tau_1+\gamma_2\tau_2- \gamma_0\, \mathrm{max}(0,\,\tau_2,\,2\tau_1,\,\tau_1+2\tau_2)\Big],
\]
which corresponds to retaining only the leading exponential in the GKZ denominator.  Which term this is will depend on which sector of the  $(\tau_1,\tau_2)$ plane we are in.  If the dominant term is, say, the $j$th one, then the overall prefactor is $x_j^{-\gamma_0}$ as shown. 
If all $x_k>0$ for $k=1,\ldots, 4$,  the tropicalisation of the denominator in fact provides a lower bound and so, for real $\gamma_0>0$ and real $\gamma_1$ and $\gamma_2$, we have $\mathcal{I}_{\bs{\gamma}}< \mathcal{I}_{\bs{\gamma}}^{\mathrm{trop.}}$.   The convergence of $\mathcal{I}_{\bs{\gamma}}^{\mathrm{trop.}}$ then establishes that of $ \mathcal{I}_{\bs{\gamma}}$.   (For rigorous bounds allowing complex $\gamma_i$, see \cite{nilsson2010mellin,berkesch2013eulermellin}.)

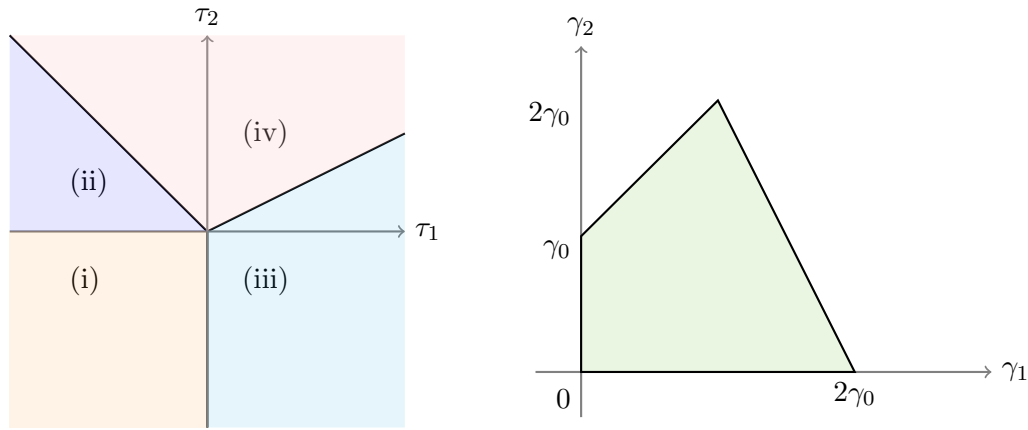
\begin{figure}[t]
\centering
\begin{tikzpicture}[scale=0.65]

\draw[thick](-4,4)--(0,0);
\draw[thick](0,-4)--(0,0);
\draw[thick](0,0)--(4,2)
;\draw[->,thick,gray] (-4,0)--(4,0);
\draw[->,thick,gray] (0,-4)--(0,4);
    \draw node[above] at (0,4) {$\tau_2$};
     \draw node[right] at (4,0) {$\tau_1$};
      \draw node[right] at (-3,-1) {(i)};
            \draw node[right] at (-3,1) {(ii)};
                  \draw node[right] at (0.5,2) {(iv)};
                        \draw node[right] at (0.5,-1) {(iii)};
                        
\fill[blue,opacity=0.1](-4,0)--(0,0)--(-4,4);
\fill[orange,opacity=0.1](-4,0)--(0,0)--(0,-4)--(-4,-4);
\fill[Cerulean,opacity=0.1](0,0)--(0,-4)--(4,-4)--(4,2);
\fill[pink,opacity=0.2](-4,4)--(0,0)--(4,2)--(4,4);

\end{tikzpicture}
\qquad
\begin{tikzpicture}[scale=1.8]
\draw[->,thick,gray] (2,0) -- (3,0); 
\draw[thick,gray] (-1/3,0) -- (0,0); 
\draw[->,thick,gray] (0,1) -- (0,2.4); 
\draw[thick,gray] (0,0) -- (0,-1/3); 
\draw[white] (0,-1/3) -- (0,-1/3-0.08);   

\draw[thick,black,fill=YellowGreen, fill opacity=0.2] (0,0)--(2,0) -- (1,2) -- (0,1) --cycle;

     \draw node[left] at (0,-0.2) {0};   
   \draw node[left] at (0,1-0.1) {$\gamma_0$};
   \draw node[left] at (0,2-0.1) {$2\gamma_0$};
   \draw node[below] at (2,0) {$2\gamma_0$};
    \draw node[right] at (3,0) {$\gamma_1$};
     \draw node[above] at (0,2.4) {$\gamma_2$};
   
  \end{tikzpicture}
\caption{{\it Left:}  The integration sectors for the tropicalised GKZ integral \eqref{trophouse}, where each sector corresponds to the dominance of a different term in the denominator. 
The sector boundaries are simultaneously the normals to the facets of the Newton polytope shown on the right.  {\it Right:} Combining the conditions on $\gamma_1$ and $\gamma_2$ for the convergence of each sector, we obtain the interior of the Newton polytope (rescaled by $\gamma_0$) as shaded.
 \label{fig:sectors}}
\end{figure}

The various integration sectors, as illustrated in figure \ref{fig:sectors}, are then as follows:
\begin{itemize}
\item[(i)] $\tau_1<0$ and $\tau_2<0$ so $j=1$ and $\mathrm{max}(0,\,\tau_2,\,2\tau_1,\,\tau_1+2\tau_2)=0$.
\item[(ii)] $\tau_1+\tau_2<0$ and $\tau_2>0$ so $j=2$ and $\mathrm{max}(0,\,\tau_2,\,2\tau_1,\,\tau_1+2\tau_2)=\tau_2$.
\item[(iii)] $\tau_1>0$ and $\tau_1-2\tau_2>0$ so $j=3$ and $\mathrm{max}(0,\,\tau_2,\,2\tau_1,\,\tau_1+2\tau_2)=2\tau_1$.
\item[(iv)] $\tau_1+\tau_2>0$ and $\tau_1-2\tau_2<0$ so
$j=4$ and $\mathrm{max}(0,\,\tau_2,\,2\tau_1,\,\tau_1+2\tau_2)=\tau_1+2\tau_2$.
\end{itemize}
Each sector forms a cone within which we can reparametrise  $\bs{\tau}=(\tau_1,\tau_2)$ as 
\[
\bs{\tau} = \lambda_1\bs{n}_1+ \lambda_2\bs{n}_2,\qquad \lambda_1,\lambda_2\ge 0
\]
where $\bs{n}_1$ and $\bs{n}_2$ are the outward-pointing vectors forming the boundary of that particular sector.   By inspection, these are simultaneously the normal vectors to the facets of the Newton polytope  shown in the right-hand panel of figure \ref{fig:sectors}, where the two normals chosen are those for the two facets  containing the leading vertex $j$. 
For the sector $j=3$, for example, we have $\bs{n}_1 = (0,-1)$ and $\bs{n}_2 = (2,1)$ and so $\tau_1=2\lambda_2$ and $\tau_2=-\lambda_1+\lambda_2$.  This third sector of the tropicalised integral is then 
\[
\mathcal{I}_{\bs{\gamma}}^{\mathrm{trop.}}\Big|_{j=3} = 2x_3^{-\gamma_0}\int_0^\infty\dd\lambda_1\int_0^\infty\dd\lambda_2 \exp\Big[ - \gamma_2\lambda_1+(2\gamma_1+\gamma_2-4\gamma_0)\lambda_2\Big]. 
\]
The linearity of the tropicalised exponent means that the integrals over $\lambda_1$ and $\lambda_2$ factorise, and for convergence as $\lambda_i\rightarrow \infty$, both exponents must separately be negative:
\[\label{houserules}
\gamma_2>0,\qquad -2\gamma_1-\gamma_2+4\gamma_0>0.
\]
This corresponds to the interior region bounded by the two lines intersecting the vertex $(\gamma_1,\gamma_2)=(2\gamma_0,0)$ in the right-hand panel of figure \ref{fig:sectors}.  This vertex is precisely that corresponding to the dominant $j=3$ term  (namely, $x_3z_1^2$) in the GKZ denominator, after rescaling by $\gamma_0$.  On the boundary of the convergence region, 
the integral has either a single or a double pole according to how many of the inequalities in \eqref{houserules} are saturated.

Repeating this exercise for the remaining  sectors,
we obtain the additional constraints
\[\label{houserules2}
\gamma_1>0, \qquad \gamma_1-\gamma_2+\gamma_0>0.
\]
Combining all these conditions, the full integral $\mathcal{I}_{\bs{\gamma}}^{\mathrm{trop.}}$ then converges for $(\gamma_1,\gamma_2)$ within the polytope shown in the figure.  This is indeed the Newton polytope for the GKZ denominator after rescaling all vertex vectors by $\gamma_0$.

\subsubsection{Tropical analysis: general case}

The  analysis above clearly  generalises. 
Setting again $z_i = e^{\tau_i}$, the general GKZ integral \eqref{GKZint} has the tropical approximation
\[\label{GKZtropical}
\mathcal{I}_{\bs{\gamma}}^{\mathrm{trop.}} =
\int_{\mathbb{R}^N}\dd\bs{\tau}\,
\exp\Big[\sum_{i=1}^N \gamma_i\tau_i - \gamma_0\,\mathrm{max}_k\,\Big(\ln x_k+\sum_{i=1}^N a_{ik}\tau_i\Big)\Big].
\]
In particular, this is a good approximation precisely for the large $|\tau_i|$ regions where any singularities of the GKZ integral must arise, and so convergence of the tropical approximation implies convergence of the full GKZ integral.\footnote{
For real $\gamma_i$, $\gamma_0>0$ and $x_j>0$, the tropical approximation provides an upper bound on the GKZ integral as noted in the previous example.  Cases where the $\gamma_i$ can be complex and the $x_j$ are not constrained to be positive can be handled by establishing a rigorous bound on the GKZ denominator, see \cite{nilsson2010mellin,berkesch2013eulermellin}.}

The different integration sectors of the tropical integral \eqref{GKZtropical} correspond to when  different terms dominate and are selected as the maximum  in the exponent.  For sufficiently large $|\tau_i|$, this depends only on the {\it direction} in the $\bs{\tau}=(\tau_1,\ldots,\tau_N)$ plane and we can neglect any contribution from the $\ln x_k$ terms.
Let us consider then the sector  where, say, the $j$th term  forms the maximum.  This sector can be parametrised as
\[\label{gentau}
\bs{\tau} = \sum_{J\in \Phi_j} \lambda^{(J)}\bs{n}^{(J)},\qquad \lambda^{(J)}\ge 0,
\]
where $\Phi_j$ denotes the set of facets containing the vertex $j$, the $\lambda^{(J)}$ are the new integration variables, and
\[\label{nJdef}
\bs{n}^{(J)}=(n_1^{(J)},\ldots, n_N^{(J)})^T
\] 
is the outward-pointing normal to the facet $J$.  We will assume that $\Phi_j$ contains precisely $N$ facets so that \eqref{gentau} holds.\footnote{If there are fewer than this, we can factor out a finite integral over a transverse subspace following appendix A of \cite{Arkani-Hamed:2022cqe} then apply the argument above for the remaining integral over a lower-dimensional cone.} 
The contribution of this sector is then
\[
\mathcal{I}_{\bs{\gamma}}^{\mathrm{trop.}}\Big|_j = x_j^{-\gamma_0}
\prod_{J\in\Phi_j}\int_0^\infty\dd\lambda^{(J)}
\exp\Big[\lambda^{(J)}\sum_{i=1}^N n_i^{(J)}(\gamma_i- \gamma_0 a_{ij})\Big].
\]
As in the previous example, convergence then requires each of these exponents to be negative giving
\[
\sum_{i=1}^N n_i^{(J)}(\gamma_i- \gamma_0 a_{ij})<0 \qquad \forall \, J\in \Phi_j.
\]
Viewed geometrically, these conditions state that the  parameter vector $\hat{\bs{\gamma}}$ lies to the {\it inside} of the $(N-1)$-dimensional hyperplane containing  facet $J$ of
the rescaled Newton polytope,
\[\label{inside}
\bs{n}^{(J)}\cdot (\hat{\bs{\g}}-\g_0\bs{a}_j) <0,
\]
and that this holds for all facets $J$ containing the $j$th vertex vector $\g_0\bs{a}_j$.
Convergence of the {\it full} tropicalised GKZ integral requires convergence in every integration sector, and hence for every vertex $j$ of the rescaled Newton polytope. The condition \eqref{inside} must thus hold for {\it all} facets $J$, meaning  $\hat{\bs{\gamma}}$ must lie completely inside the rescaled Newton polytope.

\subsection{Meromorphic continuation}

Having shown  the convergence of GKZ integrals for $\hat{\bs{\gamma}}$ lying within the rescaled Newton polytope, the existence of further infinite sets of singular hyperplanes parallel to each facet  can be established by meromorphic continuation \cite{nilsson2010mellin,berkesch2013eulermellin}.
Once again, the idea is most easily seen in the context of an example, after which we resume our  general analysis.

\subsubsection{Example}

Returning the GKZ integral \eqref{houseint}, let us construct a continuation across, say,  the upper-right facet of the Newton polytope shown on the right of figure \ref{fig:sectors}.  The relevant outward normal is $\bs{n}=(2,1)$.  Following \cite{nilsson2010mellin}, we perform a  rescaling $z_i\rightarrow \lambda^{-n_i}z_i$, namely $z_1\rightarrow \lambda^{-2}z_1$ and $z_2\rightarrow\lambda^{-1}z_2$, where $\lambda$ is some fixed parameter. The integral \eqref{houseint} becomes
\[
\mathcal{I}_{\bs{\gamma}} = \lambda^{-2\gamma_1-\gamma_2+4\gamma_0}\int_0^\infty\dd z_1\int_0^\infty\dd z_2 \,z_1^{\g_1-1}z_2^{\g_2-1} (x_1\lambda^4+x_2 z_2\lambda^3+x_3 z_1^2+x_4 z_1 z_2^2)^{-\gamma_0}
\]
but its {\it value} remains unchanged.  We can therefore differentiate to find
\[\label{lambdadiff}
0 = \frac{\dd}{\dd\lambda}\mathcal{I}_{\bs{\gamma}} \Big|_{\lambda=1}=(-2\gamma_1-\gamma_2+4\gamma_0)\mathcal{I}_{\bs{\gamma}} - 4\gamma_0 x_1\mathcal{I}_{\bs{\gamma}'}-3\gamma_0 x_2\mathcal{I}_{\bs{\gamma}''} 
\]
where
\begin{align}
\gamma_0'&=\gamma_0+1,\qquad \gamma_1'=\gamma_1,\qquad \gamma_2'=\gamma_2\nn\\
\gamma_0''&=\gamma_0+1,\qquad \gamma_1''=\gamma_1,\qquad \gamma_2''=\gamma_2+1.
\end{align}
Alternatively, \eqref{lambdadiff} can be obtained by taking a linear combination of the  Euler equations and DWI for \eqref{houseint}, namely
\begin{align}
0 &= \Big(-2(\gamma_1+2\theta_3+\theta_4)-(\gamma_2+\theta_2+2\theta_4)+4\big(\gamma_0+\sum_{j=1}^4\theta_j\big)\Big)\mathcal{I}_{\bs{\gamma}}\nn\\
&=\Big(4\gamma_0-2\gamma_1-\gamma_2+4\theta_1+3\theta_3\Big)\mathcal{I}_{\bs{\gamma}},
\end{align}
where evaluating the action of the $\theta_i=x_i\partial_{x_i}$ yields \eqref{lambdadiff}.

As  both $\mathcal{I}_{\bs{\gamma}'}$ and $\mathcal{I}_{\bs{\gamma}''}$ in \eqref{lambdadiff} take the same form as the original integral $\mathcal{I}_{\bs{\gamma}}$, except with shifted parameters, the convergence regions are given by \eqref{houserules} and \eqref{houserules2} replacing $\bs{\gamma}$ with $\bs{\gamma}'$ or $\bs{\gamma}''$.  In terms of the unshifted parameters, $\mathcal{I}_{\bs{\gamma}'}$ thus converges for 
\begin{align}\label{gamshift1}
\gamma_1>0, \qquad \gamma_2>0,\qquad \gamma_0+\gamma_1-\gamma_2+1>0,\qquad 4\gamma_0-2\gamma_1-\gamma_2+4>0,
\end{align}
while
$\mathcal{I}_{\bs{\gamma}''}$  converges for 
\begin{align}\label{gamshift2}
\gamma_1>0, \qquad \gamma_2+1>0,\qquad \gamma_0+\gamma_1-\gamma_2>0,\qquad 4\gamma_0-2\gamma_1-\gamma_2+3>0.
\end{align}
In each case,  the size of the Newton polytope is rescaled from $\gamma_0\rightarrow\gamma_0+1$, while for $\mathcal{I}_{\bs{\gamma}''}$ we also translate by the vector $(0,-1)$ as shown in figure \ref{shiftedhouses}.  Note that neither of these operations change the normals to the facets.
Re-arranging \eqref{lambdadiff}, we now have
\[\label{rearrlambdadiff}
\mathcal{I}_{\bs{\gamma}} = (4\gamma_0-2\gamma_1-\gamma_2)^{-1}\big(4\gamma_0 x_1\mathcal{I}_{\bs{\gamma}'}+3\gamma_0 x_2\mathcal{I}_{\bs{\gamma}''} \big),
\]
where the sum of shifted integrals on the right-hand side converges only for the {\it intersection} of the two shifted polytopes \eqref{gamshift1} and \eqref{gamshift2}, namely
\begin{align}\label{gamshift3}
\gamma_1>0, \qquad \gamma_2>0,\qquad \gamma_0+\gamma_1-\gamma_2>0,\qquad 4\gamma_0-2\gamma_1-\gamma_2+3>0.
\end{align}
Comparing with the original polytope formed by \eqref{houserules} and \eqref{houserules2}, only the final inequality has changed.  Now, the region of convergence  extends across the facet with normal $(2,1)$ as shown in figure \ref{shiftedhouses}.  Equation \eqref{rearrlambdadiff} thus gives a meromorphic continuation of $\mathcal{I}_{\bs{\gamma}}$ around the pole at $4\gamma_0-2\gamma_1-\gamma_2=0$ (corresponding to the facet of the original Newton polytope normal to $(2,1)$) to the larger region \eqref{gamshift3}.  

This process can then be repeated for the boundary of the new region \eqref{gamshift3} by applying the same procedure (namely, rescaling $z_i\rightarrow \lambda^{-n_i}z_i$, differentiating with respect to $\lambda$ then setting $\lambda=1$) to the integrals on the right-hand side of \eqref{rearrlambdadiff}.  Alternatively, we can extend \eqref{rearrlambdadiff}  iteratively by using shifted analogues of \eqref{rearrlambdadiff} 
to replace $\mathcal{I}_{\bs{\gamma}'}$ and $\mathcal{I}_{\bs{\gamma}''}$ on the right-hand side of  \eqref{rearrlambdadiff} itself.
Repeating such calculations for all the facet normals of the original Newton polytope, we obtain an infinite set of singular hypersurfaces parallel to the facets of the Newton polytope.  The integral \eqref{houseint} is thus singular on the hyperplanes 
\begin{align}\label{housesing}
\gamma_1=-m_1,\qquad \gamma_2= -m_2,\qquad \gamma_0+\gamma_1-\gamma_2 = -m_3,\qquad
4\gamma_0-2\gamma_1-\gamma_2=-3m_4
\end{align}
for any (independent) choice of non-negative integers $m_i\in \mathbb{Z}^+$, as illustrated in the right-hand panel of figure \ref{shiftedhouses}.

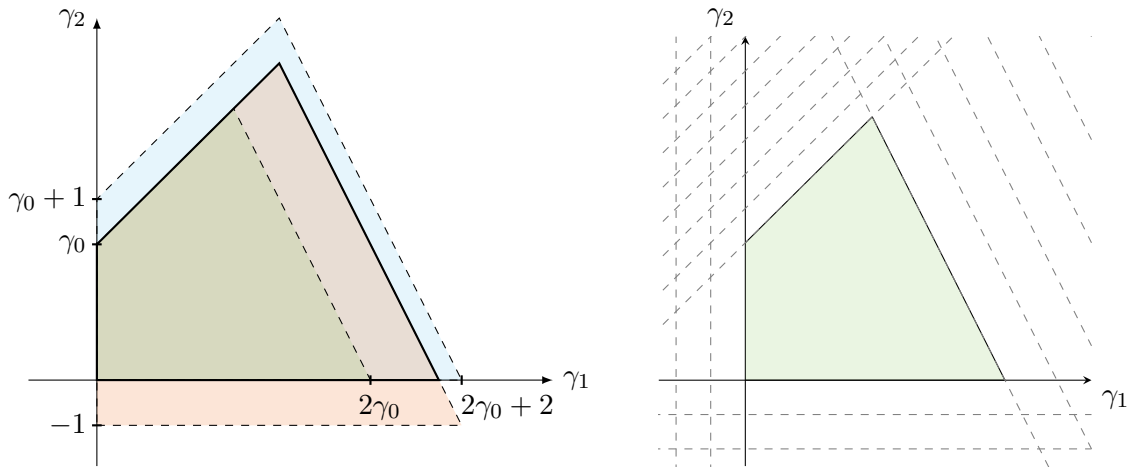
\begin{figure}[t]
\centering
\hspace{-0.5cm}\begin{tikzpicture}[scale=0.6]
\draw[-latex] (-1.5,0) -- (10,0) node[right]{$\gamma_1$};
\draw[-latex] (0,-1.9) -- (0,8) node[left]{$\gamma_2$};

\draw[dashed,black,fill=Cerulean, fill opacity=0.1] (0,0)--(8,0) -- (4,8) -- (0,4) --cycle;
\draw[dashed,black,fill=Orange, fill opacity=0.2] (0,-1)--(8,-1) -- (4,7) -- (0,3) --cycle;
\draw[dashed,black,fill=YellowGreen, fill opacity=0.2] (0,0)--(6,0) -- (3,6) -- (0,3) --cycle;

\draw[thick,black] (0,0)--(7.5,0) -- (4,7) -- (0,3)--cycle ;
\draw node[below] at (9,-0.1) {$2\gamma_0+2$};
\draw node[below] at (6.2,-0.1) {$2\gamma_0$};
\draw node[left] at (0,3) {$\gamma_0$};
\draw node[left] at (0,4) {$\gamma_0+1$};
\draw node[left] at (0,-1) {$-1$};

\foreach \x in {6,8} {    \draw [thick]($(\x,0) + (0,-\TickSize)$) -- ($(\x,0) + (0,\TickSize)$);
 }

 \foreach \y in {-1,3,4} {    \draw [thick]($(0,\y) + (-\TickSize,0)$) -- ($(0,\y) + (\TickSize,0)$);
}

\end{tikzpicture}
\hspace{5mm}
\begin{tikzpicture}[scale=1][>=latex]
\begin{axis}[
  axis x line=middle,
  axis y line=middle,
  xtick,
  ytick,
  xlabel={$\gamma_1$},
  ylabel={$\gamma_2$},
  xlabel style={below right},
  ylabel style={above left},
  xmin=-2.5,
  xmax=10,
  ymin=-2.5,
  ymax=10,
  axis equal image]

\addplot [mark=none,domain=0:11/3,name path=A] {x+3+1};
\addplot [mark=none,domain=11/3:7.5] {-2*x+12+3};
 \addplot [mark=none,domain=0:7.5,name path=B] {0};

\addplot [YellowGreen, fill opacity=0.2] fill between[of=A and B,soft clip={domain=0.01:8}];

\foreach \k in {1,...,8} {
 
\addplot [mark=none,dashed,gray,domain=-5:10] {x+3+\k};
\addplot [mark=none,dashed,gray,domain=-5:10] {-2*x+12+3*\k};
\addplot [mark=none,dashed,gray,domain=-5:10] {-\k};
\addplot [mark=none,dashed,gray,domain=-5:10] (-\k,x);

}
\end{axis}

\end{tikzpicture}

\caption{ {\it Left:} Convergence regions of $\mathcal{I}_{\bs{\gamma}}$ (green), $\mathcal{I}_{\bs{\gamma}'}$ (blue) and $\mathcal{I}_{\bs{\gamma}''}$ (orange). The meromorphic continuation \eqref{rearrlambdadiff} holds for the intersection of the convergence regions for $\mathcal{I}_{\bs{\gamma}'}$ and $\mathcal{I}_{\bs{\gamma}''}$.
This extends the domain of convergence of  $\mathcal{I}_{\bs{\gamma}}$ across the upper-right facet with normal $(2,1)$ to form the region bounded by the solid line. {\it Right:} Dashed lines indicate the complete set of singular hyperplanes \eqref{housesing} of the GKZ integral \eqref{houseint}.
\label{shiftedhouses}}
\end{figure}

\subsubsection{General analysis}\label{sec:genmero}

The analysis in this last example readily extends to general GKZ integrals.
We begin by defining a few useful quantities.  First, we have
 the $(N+1)$-dimensional vectors 
\[
\bs{\g} = \left(\begin{matrix}\g_0\\[0.5ex]\hat{\bs{\g}}\end{matrix}\right),  \qquad \bs{\mathcal{A}}_j = \left(\begin{matrix}1,\\[0.3ex]\bs{a}_j\end{matrix}\right), \qquad \bs{N}^{(J)} = \left(\begin{matrix}n_0^{(J)}\\[0.5ex]\bs{n}^{(J)}\end{matrix}\right),
\]
where $\bs{\g}$ is the usual GKZ parameter vector, $\bs{\mathcal{A}}_j$ is the $j$th column of the full $\mathcal{A}$-matrix (including the top row of $1$s), and, as above, $\bs{n}^{(J)}$ is the $N$-dimensional outwards-pointing normal to facet $J$ of the Newton polytope.  The additional component $n_0^{(J)}$ is fixed by requiring that
\[\label{NdotA}
0=\bs{N}^{(J)}\cdot\bs{\mathcal{A}}_j, \qquad  j\in\vphi_J
\]
where $\vphi_J$ denotes the set of vertices lying within the facet $J$, giving
\[\label{muJdef}
n_0^{(J)} =- \bs{n}^{(J)}\cdot\bs{a}_j,  \qquad  j\in\vphi_J.
\] 
The condition that $\hat{\bs{\g}}$ lies in the hyperplane containing facet $J$ of the rescaled Newton polytope,
\[
0=\bs{n}^{(J)}\cdot(\hat{\bs{\g}}-\g_0\bs{a}_j)
\] 
can now be compactly re-expressed as 
\[\label{gdotN}
0 = \bs{\g}\cdot \bs{N}^{(J)}
\]
and the domain of convergence \eqref{inside} corresponds to $\bs{\g}\cdot \bs{N}^{(J)}<0$ for all facets $J$.  (From an $(N+1)$-dimensional perspective, the Newton polytope therefore corresponds to a cone.)   
In addition, we define the distance function
\[\label{ddef}
d_i^{(J)} = -\bs{\mathcal{A}}_i\cdot \bs{N}^{(J)}
=\bs{n}^{(J)}\cdot(\bs{a}_j-\bs{a}_i),\qquad j\in\vphi_J.
\]
If  $\bs{n}^{(J)}$ is a unit vector, $d_i^{(J)}$ is the normal distance from vertex $i$  to facet $J$ of the Newton polytope.
Rather than choosing $\bs{n}^{(J)}$ to be a unit vector, however, it will be more convenient in practice to choose  $\bs{n}^{(J)}$ (and hence $\bs{N}^{(J)}$) to have integer components.

We now proceed to construct a meromorphic continuation of the GKZ integral across a chosen facet $J$ of the rescaled Newton polytope.  
To this end, we form a linear combination of 
 $n_0^{(J)}$ times the DWI plus the sum of $n_k^{(J)}$ times the $k$th Euler equation, namely
\begin{align}\label{euldwisum}
0&= \Big[n_0^{(J)}\Big(\gamma_0+\sum_{l=1}^n\theta_l\Big)+\sum_{k=1}^N n_k^{(J)}\Big(\gamma_k+\sum_{l=1}^n a_{kl}\theta_l\Big)\Big]\mathcal{I}_{\bs{\gamma}}\nn\\
&=\Big(\bs{\g}\cdot\bs{N}^{(J)}-
\sum_{l=1}^nd_l^{(J)}\, \theta_l \Big)\mathcal{I}_{\bs{\gamma}}.
\end{align}
The sum over $l$ on the second line here can be restricted to values $l\notin\vphi_J$, corresponding to vertices $l$ not in the facet $J$, since $d_l^{(J)}$ vanishes for all $l\in\vphi_J$.
Moreover, by direct differentiation of the GKZ integral as we will discuss further in section \ref{sec:annihilate}, one can show that 
\[
\theta_l \mathcal{I}_{\bs{\gamma}} 
= -\gamma_0 x_l \mathcal{I}_{\bs{\gamma}+\bs{\mathcal{A}}_l}.
\]
Here, the parameter vector of the right-hand integral has been shifted from $\bs{\g}\rightarrow\bs{\g}+\bs{\mathcal{A}}_l$.
Rearranging, this immediately gives the desired meromorphic continuation:\footnote{Alternatively, this equation can be derived by rescaling all $z_i\rightarrow \lambda^{-n_i^{(J)}}z_i$ in $\mathcal{I}_{\bs{\gamma}}$ and extracting a prefactor of $\lambda^{-\bs{\g}\cdot\bs{N}^{(J)}}$.  We then differentiate with respect to $\lambda$ and set $\lambda=1$ analogously to in \eqref{lambdadiff}.}
\begin{align}\label{genmero}
\mathcal{I}_{\bs{\gamma}} = -\frac{\g_0}{\bs{\g}\cdot\bs{N}^{(J)}}
\Big(\sum_{l\notin\vphi_J} \,x_l \, d_l^{(J)}\,\mathcal{I}_{\bs{\gamma}+\bs{\mathcal{A}}_l}\Big).
\end{align}
The denominator $\bs{\g}\cdot\bs{N}^{(J)}$ generates a pole at the hyperplane containing the facet $J$, while the sum of shifted integrals has a larger domain of convergence extending across the facet $J$ of the original rescaled Newton polytope for $\mathcal{I}_{\bs{\gamma}}$.

To see this,
for each shifted integral labelled by an $l\notin\vphi_J$ in the sum \eqref{genmero}, the domain of convergence \eqref{inside} is
\[
(\bs{\g}+\bs{\mathcal{A}}_l)\cdot\bs{N}^{(K)} = \bs{\g}\cdot\bs{N}^{(K)}-d_l^{(K)}<0 \qquad \forall\,\, K.
\]
This is equivalent to 
\[
\bs{n}^{(K)}\cdot((\hat{\bs{\gamma}}+\bs{a}_{l})-(\gamma_0+1)\bs{a}_{k})<0, \qquad k\in\vphi_K,
\]
{\it i.e.,} for every facet $K$, the shifted parameter vector $\hat{\bs{\gamma}}'=\hat{\bs{\gamma}}+\bs{a}_{l}$ must lie inside the Newton polytope rescaled by $\gamma_0'=\gamma_0+1$.
The common overlap of these domains for every $l\notin\vphi_J$ then corresponds to
\[\label{fulldom}
\bs{\g}\cdot\bs{N}^{(K)}<\delta^{(K)} \qquad \forall \,\, K ,
\]
where 
\[
\delta^{(K)} =\mathrm{min}_{l\notin\vphi_J}\,\big[ d_{l}^{(K)}\big] \ge 0.
\]
For any facet $K\neq J$, the set of vertices $l\notin\vphi_J$ includes vertices $l\in\vphi_K$ lying {\it in} the facet $K$.
For such vertices, $d_l^{(K)}$ and hence $\delta^{(K)}$ is then zero.  
Just as in our earlier example, the domain of convergence for the sum of shifted integrals in \eqref{genmero} is therefore unchanged for all facets $K\neq J$, 
\[\label{deltaKneqJ}
\delta^{(K)}=0 \quad \forall\,\,K\neq J.
\]  
The only facet across which the domain of convergence is extended is the facet $K=J$, for which we obtain an extension 
\begin{align}\label{Jshift}
\delta^{(J)} 
&= \mathrm{min}_{l\notin\vphi_J}\,\big[ d_{l}^{(J)}\big].
\end{align}
Geometrically, $\delta^{(J)}>0$ is the normal distance to the facet $J$ of the  (non-rescaled) Newton polytope starting from the {\it nearest} vertex $l$ not belonging to $J$, multiplied by $|\bs{n}^{(J)}|$.  If we choose $\bs{n}^{(J)}$ to have integer components, then as the components of the $\mathcal{A}$-matrix are also integer, $\delta^{(J)}$ will be a positive integer.

Equation \eqref{fulldom},  together with \eqref{deltaKneqJ} and \eqref{Jshift}, thus give us the domain of convergence of the meromorphic continuation \eqref{genmero}.  
Repeating the argument to construct further meromorphic continuations, one  finds that the GKZ integral $\mathcal{I}_{\bs{\gamma}}$ has an infinite series of singular hyperplanes lying parallel to each facet $J$ of the original Newton polytope.  These hyperplanes are given by 
\[\label{hypsings}
\bs{\g}\cdot\bs{N}^{(J)}=m_J\,\delta^{(J)}, \qquad m_J\in\mathbb{Z}^+,
\]
where $m_J$ is any non-negative integer $m=0,1,2,\ldots$.

\paragraph{Example:} Let us check \eqref{Jshift}
against our previous example.  Taking $J$ to be the facet with outward normal $\bs{n}^{(J)}=(2,1)$, we have $\vphi_J = \{3,4\}$ and so using the $\mathcal{A}$-matrix \eqref{houseA}, 
\[
\delta^{(J)} = \mathrm{min}_{l\in \{1,2\}} \sum_{i=1}^2 n_i^{(J)}(a_{i3}-a_{il}) =  \mathrm{min}(4,3) = 3.
\]
The sole shifted boundary
\[
\bs{\g}\cdot\bs{N}^{(J)} =
\gamma_0 n_0^{(J)}+\hat{\bs{\gamma}}\cdot\bs{n}^{(J)}<\delta^{(J)},
\]
where $n_0^{(J)} = -\sum_{i=1}^2 n_i^{(J)}a_{i3}=-4$, 
then evaluates to
\[
-4\gamma_0+2\gamma_1+\gamma_2-3<0
\]
in agreement with \eqref{gamshift3}, and the singular hyperplanes in \eqref{hypsings} match those in \eqref{housesing}.

\subsubsection{Implementation}

In higher-dimensional examples, a convenient way to determine the singular hyperplanes \eqref{hypsings} is to apply a convex hulling algorithm (see, {\it e.g.,} \cite{Nhull}) to identify which sets of vertex vectors $\bs{a}_{j}$ form the facets of the Newton polytope.  We will discuss this explicitly in section \ref{sec:hulling}.
The condition \eqref{gdotN} that $\hat{\bs{\g}}$ lies in the hyperplane containing facet $J$ of the rescaled Newton polytope is then equivalent to 
\[\label{fullgammaAdet}
0=\bs{\g}\cdot\bs{N}^{(J)}=\mathrm{\det}\,(\bs{\g}\,|\,\bs{\mathcal{A}}_{j_1}\,|\ldots |\,\bs{\mathcal{A}}_{j_N}),
\]
where $j_1,\ldots, j_N\in\vphi_J$ are the $N$ vertices belonging to  facet $J$, and the $\bs{\mathcal{A}}_{j}$ are the corresponding $\mathcal{A}$-matrix columns.  To see this, note that from \eqref{NdotA}  we have $\bs{\mathcal{A}}_{j}\cdot \bs{N}^{(J)}=0$ for all the $N$ vectors $j\in\vphi_J$.  As the total dimension of the vector space is $N+1$, the condition $\bs{\g}\cdot\bs{N}^{(J)}=0$ implies that $\bs{\g}$ lies in the span of the $\bs{\mathcal{A}}_{j}$ with $j\in\vphi_J$, and hence the determinant above vanishes.
The components  $n_i^{(J)}$ 
of $\bs{N}^{(J)}$, for $i=0,\ldots,N$, can thus be identified by expanding out the determinant and extracting the coefficient of $\g_i$. 
This tells us that $n_i^{(J)}$ is given by the $(i,1)$th cofactor of the matrix, for example
\[
n_0^{(J)} = \mathrm{det}\,(\bs{a}_{j_1}\,|\ldots|\,\bs{a}_{j_N}).
\]
 One must however also check that $\bs{n}^{(J)}$ corresponds to the outwards-pointing normal
by verifying that $d_k^{(J)}=-\bs{\mathcal{A}}_k\cdot\bs{N}^{(J)}>0$ for some $k\notin\vphi_J$, and swapping two  columns of \eqref{fullgammaAdet} if not. 
The spacing $\delta^{(J)}$ of the singular hyperplanes can then be computed using  \eqref{Jshift} and \eqref{ddef}.

\section{Shift operators}
\label{sec:shiftops}

Let us now examine the shift operators associated with $\mathcal{A}$-hypergeometric functions. 
Two natural classes present themselves: the `annihilation' operators which correspond to the simple derivative $\partial_j = \partial/\partial x_j$, and the `creation' operators which are purely polynomial differential operators ({\it i.e.,} operators in the Weyl algebra) that invert this operation.  

\subsection{Annihilation operators}
\label{sec:annihilate}

From the GKZ integral \eqref{GKZint} and denominator \eqref{GKZden}, we see by direct differentiation that 
\[\label{annihilateeq}
\partial_j \mathcal{I}_{\bs{\gamma}} = -\gamma_0 \mathcal{I}_{\bs{\gamma}'},\qquad j=1,\ldots, n
\]
where
\[\label{gshift1}
\gamma_0' = \gamma_0+1,\qquad \gamma_i' =\gamma_i+a_{ij},\qquad i=1,\ldots, N.
\]
In other words, differentiating with respect to $x_j$ increases the power of the denominator by one, and adds to the numerator all powers of $z_i$ multiplying $x_j$ in the denominator.
From the $\mathcal{A}$-matrix perspective, the shift of the parameter vector $\bs{\gamma}$ is given by the $j$th column of the full $\mathcal{A}$-matrix, 
\[
\bs{\gamma}' = \bs{\gamma} + \bs{\mathcal{A}}_{j},
\]
combining the two formulae in \eqref{gshift1}.

One can naturally think of the toric equations \eqref{torics} as representing the difference of two products of annihilation operators, such that the total shift generated by each product is the same leading to a cancellation.  Namely, each factor 
\[
\prod_{j=1}^n \partial_j^{u_j^\pm}
\]
produces an overall parameter shift
\[
\bs{\gamma}\rightarrow \bs{\gamma} + \sum_{j=1}^n \bs{\mathcal{A}}_{j}u_{j}^\pm,
\]
but since
\[
\sum_{j=1}^n \bs{\mathcal{A}}_{j}u_j^+= \sum_{j=1}^n\bs{\mathcal{A}}_{j}u_j^-
\]
the final shifted integral is the same in both cases and the difference vanishes.

Notice also that knowledge of the full set of $n$ annihilation operators, plus the parameter shifts they produce, is equivalent to knowledge of all columns of the $\mathcal{A}$-matrix and hence the full GKZ integral itself.\footnote{
Prior to the work of GKZ, this approach was pioneered by Miller {\it et al}  \cite{MillerBook, Miller} 
for various  Lauricella and Horn-type  
hypergeometric functions for which the annihilators can be identified from the series definition.}

\paragraph{Example:} The annihilation operators for the GKZ uplift \eqref{GKZ3rep00} of the triple-$K$ integral \eqref{tripleKdef} are $\partial_j$ for $j=1,\ldots, 6$.
The triple-$K$ integral  itself corresponds to evaluating the GKZ integral on the physical hypersurface $\x = (p_1^2,p_2^2,p_3^2,1,1,1)$ according to \eqref{3Kphys}.  The first three annihilators thus become 
\[
\partial_j= \frac{\p}{\p x_j}=\frac{\partial}{\partial p_j^2} =\frac{1}{p_j}\frac{\partial}{\partial p_j},\qquad j = 1,2,3.
\]
while for the remaining three we need to use the Euler equations
following from the $\mathcal{A}$-matrix \eqref{3KA}.  These are
\[
0=\beta_1 -\theta_1+\theta_4,\qquad0= \beta_2-\theta_2+\theta_5,\qquad0=\beta_3-\theta_3+\theta_6,
\]
and projecting to the physical hypersurface by setting $x_4=x_5=x_6=1$ gives
\[
\partial_4 = \theta_1-\beta_1 = \frac{p_1}{2}\frac{\partial}{\partial p_1} - \beta_1, \quad
\partial_5 = \theta_2-\beta_2 = \frac{p_2}{2}\frac{\partial}{\partial p_2} - \beta_2, \quad
\partial_6 = \theta_3-\beta_3 = \frac{p_3}{2}\frac{\partial}{\partial p_3} - \beta_3.
\]
Up to trivial numerical factors, these are the shift operators 
\[\label{LRdef0}
\mathcal{L}_j = -\frac{1}{p_j}\frac{\partial}{\partial p_j},\qquad
\mathcal{R}_j = 2\beta_j-p_j\frac{\partial}{\partial p_j}, \qquad j=1,2,3,
\]
introduced in \cite{Bzowski:2013sza, Bzowski:2015yxv}.
The action of these operators on the triple-$K$ integral \eqref{tripleKdef} can be obtained from their action  on the individual Bessel functions in the integrand giving
\[
\mathcal{L}_1 I_{\alpha,\{\beta_1,\beta_2,\beta_3\}} = -(\alpha+1)I_{\alpha+1,\{\beta_1-1,\beta_2,\beta_3\}},\qquad 
\mathcal{R}_1 I_{\alpha,\{\beta_1,\beta_2,\beta_3\}} =-(\alpha+1) I_{\alpha+1,\{\beta_1+1,\beta_2,\beta_3\}},
\]
with the others following by permutation.
This is consistent with the expected action for the annihilation operators: from the columns of the $\mathcal{A}$-matrix \eqref{3KA}, this is 
\[
\mathcal{L}_j:\quad\gamma_0'\rightarrow\gamma_0'+1,\qquad \gamma_j' \rightarrow \gamma_j' -1,\qquad
\mathcal{R}_j:\quad \gamma_0'\rightarrow\gamma_0'+1,\qquad\gamma_j' \rightarrow\gamma_j'+1,
\]
which from \eqref{3Kphys} is 
\[
\mathcal{L}_j:\quad\alpha\rightarrow\alpha+1,\qquad
\beta_j\rightarrow \beta_j-1,\qquad
\mathcal{R}_j:\quad\alpha\rightarrow\alpha+1,\qquad
\beta_j\rightarrow \beta_j+1.
\]

\subsection{Creation operators}
\label{sec:balgorithm}

Over the next three subsections, we present a construction of creation operators motivated by consideration of the spectral singularities.  These ideas are then illustrated using  the Gauss hypergeometric function.   
Originally, creation operators were first proposed by Saito in 
\cite{Saito_param_shift, Saito_restrictions}; for further discussion, see  \cite{saito_sturmfels_takayama_1999, smeets2000}.

By definition,  when acting on a GKZ integral, the creation operator $\mathcal{C}_j$ produces the {\it inverse} parameter shift to the annihilation operator $\partial_j=\partial/\partial x_j$.
If we act with one operator followed by the other, therefore, we must  arrive back at the original integral up to some function  of the parameters:
\[\label{bfndef}
\mathcal{C}_j\partial_j \mathcal{I}_{\bs{\gamma}}= b_j(\bs{\gamma}) \mathcal{I}_{\bs{\gamma}}.
\]
As we will see shortly, this `$b$-function' $b_j(\bs{\gamma})$ is a polynomial whose 
zeros correspond to a specific subset of the singular hyperplanes of  $\mathcal{I}_{\bs{\gamma}}$ given in \eqref{hypsings}.
First, however,  let us sketch how knowing $b_j(\bs{\gamma})$ enables a direct construction of the creation operator $\mathcal{C}_j$.

The first step is to replace all the parameters $\bs{\gamma}$ appearing in the $b$-function with linear combinations of Euler operators using the DWI and Euler equations \eqref{allEulers}.  
This defines a new polynomial $B_j(\theta)$ in the Euler operators, 
\[
 B_j(\theta)=b_j(\bs{\g})\Big|_{\bs{\gamma}\rightarrow -\sum_{k=1}^n \bs{\mathcal{A}}_{k}\theta_k}
\]
such that 
\[\label{CdB}
\mathcal{C}_j\partial_j \mathcal{I}_{\bs{\gamma}}= B_j(\theta) \mathcal{I}_{\bs{\gamma}}.
\]
As all Euler operators commute with one another, there are no ordering ambiguities here.

Next, we expand out $B_j(\theta)$  and re-arrange so that, in every term, all factors of  $x_k$ are to  the left of all derivatives $\partial_k$.  Up to a constant coefficient, each term of $B_j(\theta)$ is thus of the form
\[\label{Bterm}
\prod_{k=1}^n x_k^{\mathfrak{b}_k}\partial_k^{\mathfrak{b}_k}
\]
for some  set of powers $\mathfrak{b}_k$.  In certain cases, the  product  $\prod_{k}\partial_k^{\mathfrak{b}_k}$ will already contain an explicit factor of $\partial_j$.  
Otherwise, we can use the toric equations \eqref{torics} to replace the product  $\prod_{k}\partial_k^{\mathfrak{b}_k}$  (which acts on the GKZ integral $\mathcal{I}_{\bs{\gamma}}$ as per \eqref{CdB}) with an equivalent product that {\it does} contain an explicit factor of $\partial_j$.   Such a replacement will always be possible provided the $b$-function is correctly chosen.  After completing this operation for every term,  the right-hand side of \eqref{CdB}  now matches the form of the left-hand side allowing the operator $\mathcal{C}_j$ to be read off.
Thus, with the aid of the toric equations,   $B_j(\theta)$ acting on $\mathcal{I}_{\bs{\gamma}}$ can be explicitly factorised into the form $\mathcal{C}_j\partial_j$.

As a final step, the creation operator  $\mathcal{C}_j$, which is a differential operator 
with polynomial coefficients 
defined in the $n$-dimensional GKZ space, must be projected back to the physical hypersurface.  For this,  we restore all $x_k$ to their physical values (noting the $x_k$ are  positioned to the left of all derivatives), and use the Euler equations evaluated on the physical hypersurface to replace derivatives in directions lying off the physical hypersurface with derivatives tangential to this hypersurface.  This replacement also restores a dependence on the parameters $\bs{\gamma}$.
Many examples of this projection procedure will appear in subsequent sections.

\subsection{Action of the creation operator}

Returning to \eqref{bfndef} and using the action of the annihilator $\partial_j$ as given in \eqref{annihilateeq}, the action of the creation operator is
\[\label{creationeq}
\mathcal{C}_j\mathcal{I}_{\bs{\gamma}'} = -\gamma_0^{-1} b_j(\bs{\gamma}) \mathcal{I}_{\bs{\gamma}}.
\]
As the shift here is acting in the direction $\bs{\gamma}'\rightarrow\bs{\gamma}=\bs{\g}'-\bs{\mathcal{A}}_j$, rearranging \eqref{gshift1} we have
\[\label{gshift2}
\gamma_0 = \gamma_0'-1,\qquad \gamma_i = \gamma_i' - a_{ij},\qquad i=1,\ldots N.
\]
We will retain this allocation of prime and unprimed variables in the following  for compatibility with the algorithm  in the previous section based on \eqref{bfndef}.

Before discussing the $b$-function itself, 
a crucial point to notice is that the parameter shift \eqref{gshift2} can potentially take us from a {\it finite} to a {\it divergent} GKZ integral.  In contrast, the reverse shift \eqref{gshift1} associated with the annihilation operator $\partial_j$, when acting on a finite integral, will always produce another finite integral.

To see this, let us start with an integral 
$\mathcal{I}_{\bs{\gamma}'}$ for which 
the vector $\hat{\bs{\gamma}}'=(\gamma_1',\ldots,\gamma_N')$ lies  inside the rescaled Newton polytope with vertices $ \gamma_0' \bs{a}_{j}$.  In the notation of section \ref{sec:genmero}, this means that for every facet $K$ we  have 
\[\label{fincond}
\bs{\gamma}'\cdot\bs{N}^{(K)} < 0
\]
and the GKZ representation for $\mathcal{I}_{\bs{\gamma}'}$ converges without meromorphic continuation.  
For the {\it shifted} integral $\mathcal{I}_{\bs{\gamma}}$ in \eqref{creationeq},  we then have
\[\label{shiftedcond}
\bs{\gamma}\cdot\bs{N}^{(K)} =\bs{\gamma}'\cdot\bs{N}^{(K)} +d_j^{(K)}
\]
where
\[
d_j^{(K)}=\bs{n}^{(K)}\cdot (\bs{a}_{k} -\bs{a}_{j}),\qquad k\in\vphi_K
\] 
 is proportional to the 
normal distance from vertex $j$ to facet $K$ of the (non-rescaled) Newton polytope.
Now, for any facet $K$ {\it containing} the vertex $j$, 
$d_j^{(K)}$  vanishes and hence  $\bs{\g}\cdot\bs{N}^{(K)}<0$.
For the remaining facets {\it not} containing the vertex $j$, 
however, $d_j^{(K)}>0$ since $\bs{n}^{(K)}$ is the outward normal and  $j$ lies to the inside of the facet.  
Consequently, we  cannot be sure that $\bs{\g}\cdot\bs{N}^{(K)}<0$ 
for all facets $K$, and hence that $\mathcal{I}_{\bs{\gamma}}$ is finite.
Rather, if there are any facets for which $\bs{\g}\cdot\bs{N}^{(K)}\ge 0$, the shifted integral $\mathcal{I}_{\bs{\gamma}}$ will diverge whenever  the singularity condition \eqref{hypsings}, 
\[\label{Ksings}
\bs{\g}\cdot\bs{N}^{(K)}=m_K\delta^{(K)}, \qquad m_K\in\mathbb{Z}^+ 
\]
is satisfied for some non-negative integer $m_K$.
Combined with \eqref{shiftedcond}, this condition allows us to identify the initial parameter values $\bs{\g}'$ for which the shifted integral $\mathcal{I}_{\bs{\g}}$ diverges.

For the annihilation operator $\partial_j$, the direction of the parameter shifts  are reversed and so if the starting integral is finite,  the shifted integral is also necessarily finite.

\subsection{Finding the b-function}
\label{sec:findingb}

An apparent puzzle now arises for cases where the shifted integral $\mathcal{I}_{\bs{\gamma}}$ in \eqref{creationeq} is divergent, since the action of a differential operator $\mathcal{C}_j$ with polynomial coefficients  on any finite integral $\mathcal{I}_{\bs{\gamma}'}$ must clearly be finite.  The resolution is that, for such cases, the $b$-function in \eqref{creationeq} must have a {\it zero} cancelling the divergence in $\mathcal{I}_{\bs{\gamma}}$  such that the right-hand side is finite.\footnote{In a `dimensional' regularisation scheme where all parameters  are shifted infinitesimally $\bs{\gamma}\rightarrow \bs{\gamma}+\ep\,\bar{\bs{\gamma}}$, this requires $b_j(\bs{\gamma})\sim\ep^k$ while $\mathcal{I}_{\bs{\gamma}}\sim \ep^{-k}$ for some $k\in \mathbb{Z}^+$ such that $b_j(\bs{\g})\mathcal{I}_{\bs{\g}}$ is finite as $\ep\rightarrow 0$.}

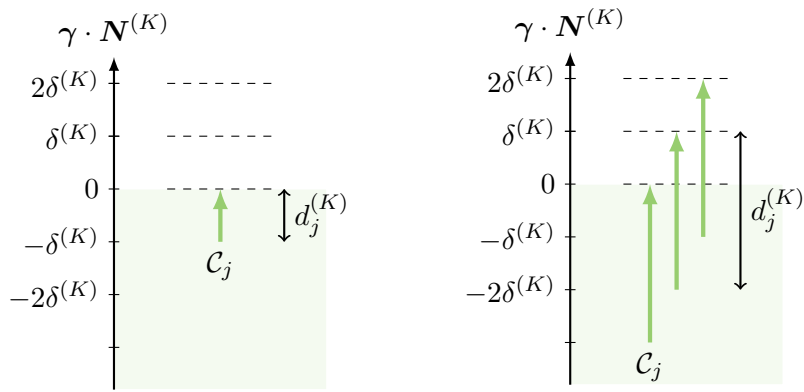
\begin{figure}[t]
\centering
\hspace{-1cm}
\begin{tikzpicture}[scale=0.7]
\draw[white,fill=YellowGreen, fill opacity=0.1]   (-2,0) -- (-2,-3.8)--(2,-3.8)--(2,0)--cycle ;
\draw[-latex,thick] (-2,-3.8) -- (-2,2.5) node[above]{$\bs{\g}\cdot \bs{N}^{(K)}$};
\draw (-2.1,2) -- (-1.9,2) node[left]{$2\delta^{(K)}\,\,$};
\draw (-2.1,1) -- (-1.9,1) node[left]{$\delta^{(K)}\,\,$};
\draw (-2.1,0) -- (-1.9,0) node[left]{$0\,\,$};
\draw (-2.1,-1) -- (-1.9,-1) node[left]{$-\delta^{(K)}\,\,$};
\draw (-2.1,-2) -- (-1.9,-2) node[left]{$-2\delta^{(K)}\,\,$};
\draw (-2.1,-3) -- (-1.9,-3) ;
\draw[dashed] (-1,0) -- (1,0);
\draw[dashed] (-1,1) -- (1,1);
\draw[dashed] (-1,2) -- (1,2);
\draw (-2.1,1) -- (-1.9,1); 
\draw (-2.1,2) -- (-1.9,2);  
\draw[-latex,ultra thick,YellowGreen] (0,-1) -- (0,0);
\draw[<->,thick] (1.2,-1)--(1.2,0);
\draw  node[right] at (1.2,-0.5){$d_j^{(K)}$};
\draw[black] node[below] at (0,-1) {$\mathcal{C}_j$} ;
\end{tikzpicture}
\hspace{1cm}
\begin{tikzpicture}[scale=0.7]
\draw[white,fill=YellowGreen, fill opacity=0.1]   (-2,0) -- (-2,-3.8)--(2,-3.8)--(2,0)--cycle ;
\draw[-latex,thick] (-2,-3.8) -- (-2,2.5) node[above]{$\bs{\g}\cdot \bs{N}^{(K)}$};
\draw (-2.1,2) -- (-1.9,2) node[left]{$2\delta^{(K)}\,\,$};
\draw (-2.1,1) -- (-1.9,1) node[left]{$\delta^{(K)}\,\,$};
\draw (-2.1,0) -- (-1.9,0) node[left]{$0\,\,$};
\draw (-2.1,-1) -- (-1.9,-1) node[left]{$-\delta^{(K)}\,\,$};
\draw (-2.1,-2) -- (-1.9,-2) node[left]{$-2\delta^{(K)}\,\,$};
\draw (-2.1,-3) -- (-1.9,-3) ;
\draw[dashed] (-1,0) -- (1,0);
\draw[dashed] (-1,1) -- (1,1);
\draw[dashed] (-1,2) -- (1,2);
\draw (-2.1,1) -- (-1.9,1); 
\draw (-2.1,2) -- (-1.9,2); 
\draw[-latex,ultra thick,YellowGreen] (-0.5,-3) -- (-0.5,0);
\draw[black] node[below] at (-0.5,-3) {$\mathcal{C}_j$} ;
\draw[-latex,ultra thick,YellowGreen] (0,-2) -- (0,1);
\draw[<->,thick] (1.2,-2)--(1.2,1);
\draw  node[right] at (1.2,-0.5){$d_j^{(K)}$};
\draw[-latex,ultra thick,YellowGreen] (0.5,-1) -- (0.5,2);
\end{tikzpicture}
\caption{Mapping of  finite  to divergent integrals under the action of the creation operator $\mathcal{C}_j$ as per \eqref{shiftedcond}, and construction of the corresponding $b$-functions.
{\it Left:} If $d_j^{(K)}=\delta^{(K)}$, facet $K$ contributes only the factor $\bs{\g}\cdot \bs{N}^{(K)}$ to the $b$-function.  The zero of this factor cancels the pole of the only singular integral (dashed line) that can be reached starting from a finite integral.  {\it Right:} If $d_j^{(K)}=3\delta^{(K)}$, the facet contributes three factors, $\prod_{m_K=0}^2(\bs{\g}\cdot \bs{N}^{(K)}-m_K\delta^{(K)})$ whose zeros cancel the poles of the three singular integrals reachable from a finite starting integral.  The shaded region indicates the rescaled Newton polytope.
\label{Cshiftfig}}
\end{figure} 

The $b$-function for the creation operator $\mathcal{C}_j$ must thus have zeros corresponding to  every singular hyperplane that can be reached by a single application of $\mathcal{C}_j$ to any finite starting integral, as illustrated in figure \ref{Cshiftfig}. 
The minimal $b$-function, containing just these factors alone, is 
\[\label{stdbfn}
b_j(\bs{\gamma}) = \prod_{K\notin \Phi_j}\prod_{m_K=0}^{F_j^{(K)}-1}(\bs{\g}\cdot\bs{N}^{(K)}-m_K\delta^{(K)})
\]
where the first product runs over all facets $K$ not containing the vertex $j$ and the upper limit in the second product is set by 
\[\label{Fdef}
F_j^{(K)} =\frac{\vphantom{\big|} d_j^{(K)}}{\vphantom{\big|} \delta^{(K)}}.
\]
This counts by how many  steps (in units of $\delta^{(K)}$, the spacing between singular hyperplanes) the creation operator $\mathcal{C}_j$ 
raises $\bs{\g}'\cdot\bs{N}^{(K)}$ according to \eqref{shiftedcond}.  
 Effectively, if we {\it define} an initial $m_K'$  by the relation $\bs{\g}'\cdot\bs{N}^{(K)}=m_K'\delta^{(K)}$, the creation operator $\mathcal{C}_j$ acts to raise this to $m_K = m_K'+F_j^{(K)}$.
Thus, if $F_j^{(K)}=1$ for some particular facet $K$, only the singularity in \eqref{Ksings} with $m_K=0$ can be reached by the action of $\mathcal{C}_j$ on a finite starting integral (namely, that  with $m_K'=-1$).  The product over $m_K$ in \eqref{stdbfn} is thus capped at $F_j^{(K)}-1=0$.  Alternatively, if $F_j^{(K)}=2$ for some facet, both the $m_K=0$ and $m_K=1$ singularities can be reached by acting with $\mathcal{C}_j$ on the finite starting integrals  with $m_K'=-2$ and $m_K'=-1$ respectively.  The product over $m_K$ in \eqref{stdbfn} then runs up to $F_j^{(K)}-1=1$, and so on.

For all the Feynman and Witten diagrams we analyse in the remainder of the chapter,  $F_j^{(K)}$ is an integer for all $K$ and 
 the minimal $b$-function \eqref{stdbfn} (containing only the zeros necessary to cancel out the singularities of $\mathcal{I}_{\bs{\gamma}}$) is sufficient to find all creation operators.  These operators are moreover of the lowest possible order in derivatives, since the $b$-function has the fewest factors. 
Nevertheless, in certain exceptional cases, the factorisation step of the algorithm in section \ref{sec:balgorithm} can fail when using
the minimal $b$-function.  Such cases, which arise when the associated toric ideal is non-normal \cite{Saito_param_shift, Saito_restrictions, saito_sturmfels_takayama_1999, smeets2000}, can be handled by supplementing  \eqref{stdbfn} with additional factors.  An example, which also features a non-integer $F_j^{(K)}$, is discussed in appendix \ref{sec:nonnormal}.

Despite its formal appearance, the formula \eqref{stdbfn} is straightforward to evaluate in practice as will become clear in the examples to follow. All that is required is to identify the singular hyperplanes \eqref{hypsings} for a given GKZ integral, along with the shift produced by the creation operator $\mathcal{C}_j$, 
and then to form the $b$-function from the product of all singular hyperplanes that can be reached by one application of $\mathcal{C}_j$ on any finite starting integral.  
We emphasise too that, while consideration of singular cases has been used to deduce the form of the $b$-function, the creation operators we subsequently obtain can be used to map finite integrals to finite integrals.

\subsection{Example}
As a simple illustration before turning our efforts to Witten diagrams and Feynman integrals in the following sections, we
compute creation operators for  the GKZ integral \cite{nilsson2010mellin, de_la_Cruz_2019}
\begin{align}\label{2f1example}
\mathcal{I}_{\bs{\gamma}} &= \int_{\mathbb{R}_+^2}\dd z_1 \dd z_2 \frac{z_1^{\gamma_1-1} z_2^{\gamma_2-1}}{(x_1+x_2 z_1+x_3 z_2+x_4 z_1 z_2)^{\gamma_0}}.
\end{align}
On the hypersurface $(x_1,x_2,x_3,x_4) = (1,1,1,y)$, this can be directly evaluated in terms of the Gauss hypergeometric function 
\begin{align}\label{2f1eval}
\mathcal{I}_{\bs{\gamma}}(y)  &= \frac{\Gamma(\gamma_1) \Gamma(\gamma_2)
\Gamma(\gamma_0-\gamma_1) \Gamma(\gamma_0-\gamma_2)}{\Gamma(\gamma_0)^2}\; {}_2F_1(\gamma_1,\gamma_2,\gamma_0;1-y).
\end{align}
Since all shift operators for the Gauss hypergeometric function are known this will allow an easy check of our calculations.

Evaluating the $\mathcal{A}-$matrix, 
\begin{align}
\mathcal{A}= \left(\begin{matrix}
    1&1&1&1\\
    0&1&0&1\\
    0&0&1&1
   \end{matrix}\right)\!,
 \end{align}
we can read off the DWI and Euler equations
\[\label{2f1Eulers}
0 = \left(\gamma_0 +\t_1+\t_2+\t_3+\t_4\right)\mathcal{I}_{\bs{\gamma}},\quad 0 = \left(\gamma_1+\t_2+\t_4\right)\mathcal{I}_{\bs{\gamma}},\quad 0=\left(\gamma_2+\t_3+\t_4\right)\mathcal{I}_{\bs{\gamma}},
\]
where $\theta_i=x_i\partial_i$ and, from the kernel of the $\mathcal{A}$-matrix, we find a single toric equation,
\[\label{2f1toric}
0=\left(\p_2\p_3-\p_1\p_4\right)\mathcal{I}_{\bs{\gamma}}.
\]
From \eqref{hypsings}, the singular hyperplanes are
\[\label{2f1sings}
\gamma_1=-m_1,\qquad \gamma_2=-m_2,\qquad \gamma_0-\gamma_1 = -m_3,\qquad \gamma_0-\gamma_2 = -m_4, \qquad m_i\in \mathbb{Z}^+
\]
as displayed in figure \ref{fig:2f1sings}.  As expected, these singularities coincide with the poles of the gamma functions in the numerator of the projected integral 
\eqref{2f1eval}.

\begin{figure}[t]
\centering
 \begin{tikzpicture}[scale=1.1][>=latex]
\begin{axis}[
  axis x line=middle,
  axis y line=middle,
  xtick={3,5},
  ytick={3,4},
  xticklabels={$\gamma_0$, $\gamma_0+1$},
  yticklabels={$\gamma_0$, $\gamma_0+1$},
   xlabel={$\gamma_1$},
  ylabel={$\gamma_2$},
  xlabel style={below right},
  ylabel style={above left},
  xmin=-3.5,
  xmax=6.5,
  ymin=-3.5,
  ymax=6.5,
  axis equal image]

\foreach \k in {0,...,8} {
 
\addplot [mark=none,dashed,gray,domain=-5:10] {3+\k};
\addplot [mark=none,dashed,gray,domain=-5:10] (\k+3,x);
\addplot [mark=none,dashed,gray,domain=-5:10] {-\k};
\addplot [mark=none,dashed,gray,domain=-5:10] (-\k,x);

}

\addplot [mark=none,domain=0:3,name path=A] {3};
\addplot [mark=none,domain=0:3] (0,x);
\addplot [mark=none,domain=0:3] (3,x);
\addplot [mark=none,domain=0:3,name path=B] {0};

\addplot [YellowGreen, fill opacity=0.2] fill between[of=A and B,soft clip={domain=0.01:8}];

\addplot[holdot] coordinates{(3,0)(4,0)};
\addplot[holdot] coordinates{(0,3)(0,4)};

\end{axis}

\end{tikzpicture}
   \caption{The singular hyperplanes of \eqref{2f1example}.  \label{fig:2f1sings}}
\end{figure}
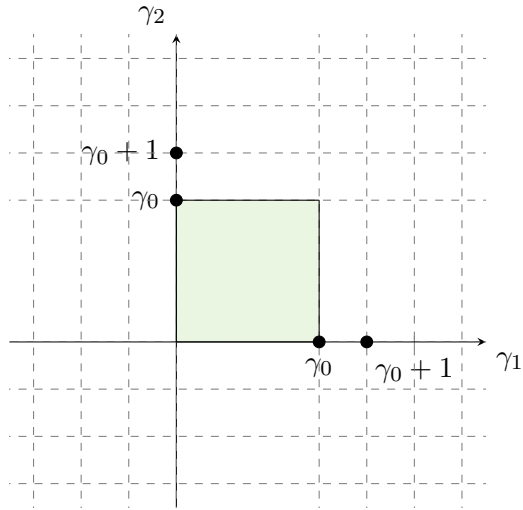

The annihilation operators $\partial_j$ send $\bs{\gamma}\rightarrow\bs{\gamma}'$ while the creation operators $\mathcal{C}_j$ send $\bs{\gamma}'\rightarrow\bs{\gamma}$,
where for each $j$ these parameters are related by
\begin{align}\label{2f1Cshifts}
j&=1: & \gamma_0'&=\gamma_0+1,&\gamma_1'&=\gamma_1, & \gamma_2'&=\gamma_2 \nn\\
j&=2: & \gamma_0'&=\gamma_0+1, &\gamma_1'&= \gamma_1+1, & \gamma_2'&=\gamma_2\nn\\
j&=3: & \gamma_0'&=\gamma_0+1, & \gamma_1'&=\gamma_1,&\gamma_2'&=\gamma_2+1,\nn\\
j&=4: & \gamma_0'&=\gamma_0+1, & \gamma_1'&=\gamma_1+1, & \gamma_2'&=\gamma_2+1.
\end{align}
The corresponding $b$-functions are
\begin{align}\label{2f1bs}
b_1&=(\gamma_0-\gamma_1)(\gamma_0-\gamma_2),\nn\\
b_2&=\gamma_1(\gamma_0-\gamma_2),\nn\\
b_3&=\gamma_2(\gamma_0-\gamma_1),\nn\\
b_4&=\gamma_1\gamma_2.
\end{align}
In each case, the factors that appear are the zeros needed to cancel the poles arising when the creation operator moves us from a finite to a singular integral.  For $b_1$, for example, the shift produced by $\mathcal{C}_1$ can take a finite integral with $\gamma_0'-\gamma_i'=1$ to a singular integral with $\gamma_0-\gamma_i=0$ for both $i=1$ and $i=2$, as we see from \eqref{2f1sings}.   The zeros of $b_1$ then cancel these singularities so that the action \eqref{creationeq}  of $\mathcal{C}_1$ on a finite integral is always finite.
For $b_2$, the shifts  produced by $\mathcal{C}_2$ 
can take a finite integral with $\gamma_2'=1$ to a singular integral with $\gamma_2=0$, and a finite integral with $\gamma_0'-\gamma_1'=1$ to a singular integral with $\gamma_0-\gamma_1=0$, with these singularities again being cancelled by the zeros of $b_2$.  Note that the action of $\mathcal{C}_2$ leaves $\gamma_1'$ and $\gamma_0'-\gamma_2'$ unchanged hence no further singularities arise, and hence no further factors in $b_2$.
One can further check that the $b$-functions \eqref{2f1bs} are consistent with the general formula \eqref{stdbfn}.

From \eqref{CdB} plus the DWI and Euler equations \eqref{2f1Eulers}, we now have, for example,
\begin{align}
\mathcal{C}_1 \partial_1 \mathcal{I}_{\bs{\gamma}} &= (\gamma_0-\gamma_1)(\gamma_0-\gamma_2)\mathcal{I}_{\bs{\gamma}} =(\theta_1+\theta_3)(\theta_1+\theta_2)\mathcal{I}_{\bs{\gamma}} \nn\\
&=(x_1\p_1+x_1^2\p_1^2+x_1x_2\p_1\p_2+x_1x_3\p_1\p_3+x_2x_3\p_2\p_3)
\mathcal{I}_{\bs{\gamma}}.
\end{align}
By inspection, every term in the final line contains an explicit factor of $\p_1$ except for the last, but this can be replaced by $x_2x_3\p_1\p_4$ using the toric equation \eqref{2f1toric}.
This gives us the desired factorisation
\begin{align}
\mathcal{C}_1 &=x_1+ x_1^2\p_1+x_1x_2\p_2+x_1x_3\p_3
+x_2x_3\p_4 \nn\\ &= x_1(1+\theta_1+\theta_2+\theta_3)+x_2x_3\p_4.
\end{align}
In an identical fashion, we obtain
\begin{align}
\mathcal{C}_2 &=x_2(1+\theta_1+\theta_2+\theta_4)+x_1x_4\p_3,\nn\\
\mathcal{C}_3 &=x_3(1+\theta_1+\theta_3+\theta_4)+x_1x_4\p_2,\nn\\
\mathcal{C}_4 &=x_4(1+\theta_2+\theta_3+\theta_4)+x_2x_3\p_1.
\end{align}

Finally, in order to understand their action on \eqref{2f1eval}, these creation operators can be projected to the `physical' hypersurface $(x_1,x_2,x_3,x_4) = (1,1,1,y)$.  For this we use the DWI and Euler equations \eqref{2f1Eulers} evaluated on this hypersurface,
which can be re-arranged so as to eliminate all derivatives apart from $\p_y$:
\begin{align}
\p_1\mathcal{I}_{\bs{\gamma}'}(y) &= (-\gamma_0'+\gamma_1'+\gamma_2'+\theta_y)\mathcal{I}_{\bs{\gamma}'}(y),\nn\\
\p_2\mathcal{I}_{\bs{\gamma}'}(y) &= -(\gamma_1'+\theta_y)\mathcal{I}_{\bs{\gamma}'}(y),\nn\\
\p_3\mathcal{I}_{\bs{\gamma}'}(y) &= -(\gamma_2'+\theta_y)\mathcal{I}_{\bs{\gamma}'}(y).
\end{align}
Notice here that as the creation operators act on the integral with parameters $\bs{\gamma}'$ by our definition \eqref{creationeq}, we need to use these parameters here.
With the aid of these equations, the creation operators project to
\begin{align}
\mathcal{C}_1^{\mathrm{ph}} &= 1-\gamma_0' + (1-y)\p_y,\nn\\
\mathcal{C}_2^{\mathrm{ph}} &= 1-\gamma_0' + (1-y)(\gamma_2'+\theta_y),\nn\\
\mathcal{C}_3^{\mathrm{ph}} &=1-\gamma_0' + (1-y)(\gamma_3'+\theta_y),\nn\\
\mathcal{C}_4^{\mathrm{ph}} &=1-\gamma_0' +(1-y)(\gamma_1'+\gamma_2'-1+\theta_y),
\end{align}
where the `ph' superscript indicates the operators expressed in physical variables.  
From \eqref{creationeq}, we then have, for example,
\begin{align}
\mathcal{C}_1^{\mathrm{ph}}\mathcal{I}_{\gamma_0',\gamma_1',\gamma_2'}(y)&=
-\gamma_0^{-1}(\gamma_0-\gamma_1)(\gamma_0-\gamma_2)
\mathcal{I}_{\gamma_0,\gamma_1,\gamma_2}(y)\nn\\&=
-(\gamma_0'-1)^{-1}(\gamma_0'-\gamma_1'-1)(\gamma_0'-\gamma_2'-1)\mathcal{I}_{\gamma_0'-1,\gamma_1',\gamma_2'},
\end{align}
since here the creation operator shifts $\gamma_0'\rightarrow\gamma_0=\gamma_0'-1$ while  $\gamma_i'=\gamma_i$ for $i=1,2$.  Noting the presence of the gamma functions in \eqref{2f1eval}, this corresponds to
\[
\big(1-\gamma_0'+(1-y)\p_y)\,{}_2F_1(\gamma_1',\gamma_2',\gamma_0';1-y) = (1-\gamma_0')\,{}_2F_1(\gamma_1',\gamma_2',\gamma_0'-1;1-y)
\]
which indeed follows from standard relations for ${}_2F_1$ (see {\it e.g.,} equation 15.5.4 of  \cite{NIST:DLMF}).

Taking into account the shifts \eqref{2f1Cshifts}, for the remaining operators we find
\begin{align}
\mathcal{C}_2^{\mathrm{ph}}\mathcal{I}_{\gamma_0',\gamma_1',\gamma_2'}(y)&=-\gamma_0^{-1}\gamma_1(\gamma_0-\gamma_2)\mathcal{I}_{\gamma_0,\gamma_1,\gamma_2}\nn\\&
=-(\gamma_0'-1)^{-1}(\gamma_1'-1)(\gamma_0'-\gamma_2'-1)\mathcal{I}_{\gamma_0'-1,\gamma_1'-1,\gamma_2'}(y)
,\nn\\
\mathcal{C}_3^{\mathrm{ph}}\mathcal{I}_{\gamma_0',\gamma_1',\gamma_2'}(y)&=-\gamma_0^{-1}\gamma_2(\gamma_0-\gamma_1)\mathcal{I}_{\gamma_0,\gamma_1,\gamma_2}\nn\\&
=-(\gamma_0'-1)^{-1}(\gamma_2'-1)(\gamma_0'-\gamma_1'-1)\mathcal{I}_{\gamma_0'-1,\gamma_1',\gamma_2'-1}(y)
,\nn\\
\mathcal{C}_4^{\mathrm{ph}}\mathcal{I}_{\gamma_0',\gamma_1',\gamma_2'}(y)&=-\gamma_0^{-1}\gamma_1\gamma_2\mathcal{I}_{\gamma_0,\gamma_1,\gamma_2}\nn\\&=
-(\gamma_0'-1)^{-1}(\gamma_1'-1)(\gamma_2'-1)\mathcal{I}_{\gamma_0'-1,\gamma_1'-1,\gamma_2'-1}(y).
\end{align}
These can again be verified using standard  shift identities and contiguity relations for the Gauss hypergeometric function.

\section{Creation operators for Witten diagrams}
\label{sec:Witten}

The correlators of holographic conformal field theories 
at strong coupling can be  computed via Witten diagrams in anti-de Sitter spacetime.  
As the evaluation of these diagrams is nontrivial, particularly in momentum space, it is  important to identify classes of shift operators connecting known `seed' solutions to a broader family of correlators.

In this section, we construct novel creation operators for Witten diagrams in momentum space. (Results for the position-space contact diagram, or holographic $D$-function, are given in appendix \ref{sec:Dfn}.)  Starting with the contact diagram, we derive explicit creation operators at  $3$- and $4$-points, though the method extends to higher points.  We also show,  again at $3$- and $4$-points, how to construct operators that shift the scaling dimensions while preserving the spacetime dimension.

A case of particular interest, given the close connection to cosmological correlators,  is the 4-point exchange  diagram.
Here, a class of weight-shifting operators is known connecting exchange diagrams with different external scaling dimensions  \cite{Karateev:2017jgd, Baumann:2019oyu}, but subject to two  restrictions \cite{Bzowski:2022rlz}: first, these operators map an exchange diagram with {\it non-derivative} vertices to one with {\it derivative} vertices; and second, they work only for a special set of initial scaling dimensions.  While these results are sufficient for cosmologies where the inflaton is a derivatively-coupled massless scalar, finding further generalisations  is highly desirable.

A key problem, therefore, is 
to find a  shift operator connecting exchange diagrams with {\it non-derivative} vertices to new exchange diagrams, with shifted operator dimensions, but still with {\it non-derivative} vertices. 
This operator should moreover be applicable for diagrams of arbitrary initial scaling dimensions.
In section \ref{sec:exch}, we derive such an operator.

\subsection{Definitions}

In momentum space, the $n$-point contact Witten diagram is
\begin{align}\label{icont}
& i_{[d;\,\Delta_1, \,\ldots,\, \Delta_n]}
 = \int_0^\infty \dd z \, z^{-d-1} \prod_{i=1}^n \mathcal{K}_{[\Delta_i]}(z, p_i) 
\end{align}
where $d$ is the boundary spacetime dimension of the CFT, $\Delta_i$ is the scaling dimension of the scalar operator $\O_i$, 
and the bulk-to-boundary propagator
\begin{align} \label{Kprop}
	\mathcal{K}_{[\Delta_i]}(z, p_i) = \frac{ z^{\frac{d}{2}} p_i^{\beta_i}K_{\beta_i}(p_i z)}{2^{\beta_i - 1} \Gamma (\beta_i)}, \qquad \beta_i=\Delta_i-\frac{d}{2}.
\end{align}
Since the modified Bessel-$K$ function is invariant under changing the sign of its index, note we have the shadow relation
\begin{align}\label{contshadow}
i_{[d;\,\Delta_1,\,\ldots, \,\Delta_n]}\Big|_{\Delta_i\rightarrow d-\Delta_i} =\frac{4^{\beta_i}\Gamma(\beta_i)}{\Gamma(-\beta_i)}p_i^{-2\beta_i} i_{[d;\,\Delta_1,\,\ldots,\,\Delta_n]}.
\end{align}

In addition to the contact diagram, we will discuss the $4$-point $s$-channel exchange diagram shown in figure \ref{Wittendias},
\begin{align}\label{iexch}
 i_{[d;\,\Delta_1, \Delta_2; \,\Delta_3, \Delta_4; \,\Delta_x]}
 &= \int_0^\infty \dd z \, z^{-d-1}\mathcal{K}_{[\Delta_1]}(z, p_1) \mathcal{K}_{[\Delta_2]}(z, p_2) 
\\ &\qquad \times
 \int_0^\infty \dd \z \, \z^{-d-1} \mathcal{G}_{[\Delta_x]}(z, s; \z) \mathcal{K}_{[\Delta_3]}(\z, p_3) \mathcal{K}_{[\Delta_4]}(\z, p_4),\nn
\end{align}
where $\Delta_x$ is the dimension of the exchanged operator and  $s^2 = (\bs{p}_1+\bs{p}_2)^2$.  The bulk-to-bulk propagator in this expression is
\begin{align} \label{Gprop}
	\mathcal{G}_{[\Delta_x]}(z, s; \z) = \left\{ \begin{array}{ll}
		(z \z)^{\frac{d}{2}} I_{\beta_x}(s z) K_{\beta_x}(s \z) & \text{ for } z < \z, \\
		(z \z)^{\frac{d}{2}} K_{\beta_x}(s z) I_{\beta_x}(s \z) & \text{ for } z > \z,
	\end{array} \right.	
\end{align}
with $I_{\beta}$ and $K_{\beta}$ representing modified Bessel functions and $\beta_x=\Delta_x-d/2$.  Where necessary, these integrals can be regulated by infinitesimally shifting the operator dimensions and spacetime dimension $d$ so as to ensure convergence \cite{Bzowski:2022rlz}.

\begin{figure}[t]
\begin{tikzpicture}[scale=0.8]
\draw [darkgray] (0,0) circle [radius=3];
\draw [fill=black] (-2.121,-2.121) circle [radius=0.1];
\draw [fill=black] (-2.121, 2.121) circle [radius=0.1];
\draw [fill=black] ( 2.121,-2.121) circle [radius=0.1];
\draw [fill=black] ( 2.121, 2.121) circle [radius=0.1];
\draw [fill=black] ( 0, 0) circle [radius=0.1];
\draw (-2.121,-2.121) -- ( 2.121, 2.121);
\draw ( 2.121,-2.121) -- (-2.121, 2.121);	
\node [left] at (-2.121, 2.2) {$\O_1(\bs{p}_1)$}; 
\node [left] at (-2.121,-2.2) {$\O_2(\bs{p}_2)$}; 	
\node [right] at ( 2.121,-2.2) {$\O_3(\bs{p}_3)$}; 
\node [right] at ( 2.121, 2.2) {$\O_4(\bs{p}_4)$}; 	
\node [above] at (-0.9, 1.06) {$\mathcal{K}_{[\Delta_1]}$};
\node [above] at (-1.3,-1.06) {$\mathcal{K}_{[\Delta_2]}$};
\node [above] at ( 1.2,-1.06) {$\mathcal{K}_{[\Delta_3]}$};
\node [above] at ( 0.8, 1.06) {$\mathcal{K}_{[\Delta_4]}$};
\end{tikzpicture}
\qquad
\begin{tikzpicture}[scale=0.8]
\draw [darkgray] (0,0) circle [radius=3];
\draw [fill=black] (-2.121,-2.121) circle [radius=0.1];
\draw [fill=black] (-2.121, 2.121) circle [radius=0.1];
\draw [fill=black] ( 2.121,-2.121) circle [radius=0.1];
\draw [fill=black] ( 2.121, 2.121) circle [radius=0.1];
\draw [fill=black] (-1, 0) circle [radius=0.1];
\draw [fill=black] ( 1, 0) circle [radius=0.1];
\draw (-2.121,-2.121) -- (-1,0) -- (-2.121, 2.121);
\draw ( 2.121, 2.121) -- ( 1,0) -- ( 2.121,-2.121);
\draw (-1,0) -- (1,0);
\node [left] at (-2.121, 2.2) {$\O_1(\bs{p}_1)$}; 
\node [left] at (-2.121,-2.2) {$\O_2(\bs{p}_2)$}; 	
\node [right] at ( 2.121,-2.2) {$\O_3(\bs{p}_3)$}; 
\node [right] at ( 2.121, 2.2) {$\O_4(\bs{p}_4)$}; 	
\node [right] at (-1.5, 1.2) {$\mathcal{K}_{[\Delta_1]}$};
\node [right] at (-1.5, -1.2) {$\mathcal{K}_{[\Delta_2]}$};
\node [left] at ( 1.5, -1.2) {$\mathcal{K}_{[\Delta_3]}$};
\node [left] at ( 1.5, 1.2) {$\mathcal{K}_{[\Delta_4]}$};
\node [above] at (0,0) {$\mathcal{G}_{[\Delta_x]}$};
\end{tikzpicture}
\centering
\caption{Witten diagrams representing the contact and exchange 4-point diagram $\ino_{[\Delta_1 \Delta_2 \Delta_3 \Delta_4]}$ and $\ino_{[\Delta_1 \Delta_2, \Delta_3 \Delta_4 x \Delta_x]}$  given by the integrals \eqref{icont} and \eqref{iexch}.\label{Wittendias}}
\end{figure}
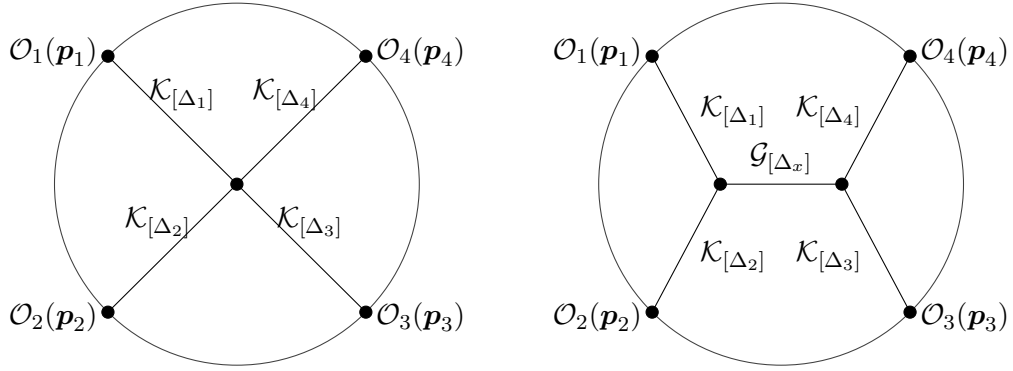

\subsection{GKZ representation of the contact diagram}

The GKZ representation for the $n$-point momentum-space contact diagram can be evaluated analogously to that for the triple-$K$ integral (see page \pageref{tripleKex}).
This yields the GKZ integral
\begin{align}\label{contactGKZ}
 \mathcal{I}_{\bs{\gamma}} =\Big(\prod_{i=1}^n\int_0^\infty \dd z_i \,z_i^{\gamma_i-1}\Big)\Big[\sum_{j=1}^n \Big(\frac{x_j}{ z_j} + \bar{x}_j z_j\Big)\Big]^{-\gamma_0}
\end{align}
with the contact diagram  being
\[\label{icontIg}
 i_{[d;\,\Delta_1, \,\ldots, \,\Delta_n]} =  2^{\gamma_0}\Gamma(\gamma_0)\Big(\prod_{i=1}^n \frac{1}{2^{\gamma_i}\Gamma(\gamma_i)}\Big)  \mathcal{I}_{\bs{\gamma}}
 \]
with parameters
\[\label{contparams}
\gamma_0 = \big(\frac{n}{2}-1\big)d,\qquad 
\gamma_i = \Delta_i - \frac{d}{2} = \beta_i,
\]
and physical hypersurface
\[\label{contphys}
x_i = p_i^2, \qquad \bar{x}_i = 1, \qquad i=1,\ldots, n.
\]
Our notation $\x = (x_i,\bar{x}_i)$ for the GKZ variables here and in \eqref{contactGKZ} is simply a convenience designed to simplify the form of the Euler and toric equations as we will see below; $\bar{x}_i$  should be regarded as an independent dynamical variable equivalent to $x_{i+n}$ in the notation of the previous section.

The $(n+1)\times 2n$ dimensional $\mathcal{A}$-matrix for the integral \eqref{contactGKZ}  is now
\begin{align}
\mathcal{A} &= \left(\begin{matrix}
&\bs{1} & \bs{1}& \\
 &-\mathbb{I}_n &\mathbb{I}_n&
\end{matrix}\right)
\end{align}
where $\bs{1}$ is the $n$-dimensional row vector of $1$s and $\mathbb{I}_n$ is the $n\times n$ identity matrix.  (Again, 
we are departing from the notation of the previous section where $n$ referred to the number of columns in the $\mathcal{A}$-matrix, reserving $n$ now for the number of points.)
Writing
\[
\partial_i = \frac{\partial}{\partial x_i}, \qquad \bar{\partial}_i = \frac{\partial}{\partial \bar{x}_i},\qquad \theta_i = x_i\partial_i,\qquad \bar{\theta}_i = \bar{x}_i\bar{\partial}_i,
\]
the Euler equations are
\[\label{eulercont}
0 = (\gamma_i -  \theta_i +\bar{\theta}_i)\mathcal{I}_{\bs{\gamma}},\qquad i=1,\ldots,n,
\]
while the DWI is 
\[\label{DWIcont}
0 = \Big(\gamma_0 + \sum_{i=1}^n (\theta_i+\bar{\theta}_i)\Big)\mathcal{I}_{\bs{\gamma}}.
\]
In addition, we have the toric equations
\[\label{toriccont}
0 = (\p_i\bar{\p}_i - \p_j\bar{\p}_j)\mathcal{I}_{\bs{\gamma}},\qquad i\neq j=1,\ldots,n.
\]
These can easily be verified by noting that $\partial_i\bar{\partial}_i$ sends $\gamma_0\rightarrow\gamma_0+2$ but makes no change to the power of $z_i$ appearing in the numerator of \eqref{contactGKZ}, hence the two terms in \eqref{toriccont} cancel.

It is well known that the contact diagram satisfies the equation, 
\[
0=(K_i-K_j)i_{[d,\,\Delta_1,\ldots,\Delta_n]}\qquad \forall \,\,i\neq j
\]
where $K_i$ is the Bessel operator 
\[
K_i = \partial_{p_i}^2 + \frac{(1-2\g_i)}{p_i}\p_{p_i}= \p_i(\t_i-\g_i).
\]
To see this from a GKZ perspective, we use the Euler and toric equations to show that 
\[
(K_i-K_j)\mathcal{I}_{\bs{\g}} = 
(\p_i\tb_i-\p_j\tb_j)\mathcal{I}_{\bs{\g}}=(\bar{x}_i\p_i\bar{\p}_i-\bar{x}_j\p_j\bar{\p}_j)\mathcal{I}_{\bs{\g}}=(\bar{x}_i-\bar{x}_j)\p_i\bar{\p}_i\mathcal{I}_{\bs{\g}}.
\]
Upon projecting to the physical hypersurface \eqref{contphys}, the right-hand side now vanishes.

Finally, we observe that the shadow relation \eqref{contshadow} uplifts to
\[\label{GKZshadow}
\mathcal{I}_{\bs{\g}}\Big|_{\gamma_i\rightarrow - \gamma_i} =
\Big(\frac{\bar{x}_i}{x_i}\Big)^{\g_i}\mathcal{I}_{\bs{\g}}
\]
for any $i=1,\ldots, n$ in GKZ variables. This  can be seen by evaluating the right-hand side with the substitution $z_i =  x_i/(\bar{x}_i z'_i)$ in \eqref{contactGKZ}.

\subsection{Creation and annihilation operators}

The action of the annihilation operators is
\[\label{contannih}
\partial_i \mathcal{I}_{\bs{\g}} =-\g_0 \mathcal{I}_{\bs{\g}}\Big|_{\g_0\rightarrow \gamma_0+1,\,\g_i\rightarrow\gamma_i-1},\qquad
\bar{\partial}_i \mathcal{I}_{\bs{\g}} =-\g_0 \mathcal{I}_{\bs{\g}}\Big|_{\g_0\rightarrow \gamma_0+1,\,\g_i\rightarrow\gamma_i+1}
\]
for any $i=1,\ldots, n$.  After projecting to the physical hypersurface \eqref{contphys}, up to numerical factors $\partial_i$ and $\bar{\partial}_i$ become the operators $\mathcal{L}_i$ and $\mathcal{R}_i$ respectively, as defined in \eqref{LRdef0}.
Note that due to the shadow relation \eqref{GKZshadow} (or  re-arranging the Euler equation \eqref{eulercont}), we have
\[
\bar{\partial}_i \mathcal{I}_{\bs{\g}}= \Big(\frac{\bar{x}_i}{x_i}\Big)^{-\g_i-1} \partial_i\, \Big(\frac{\bar{x}_i}{x_i}\Big)^{\g_i}\mathcal{I}_{\bs{\g}}. 
\]
In physical variables, this projects  to 
\[
\mathcal{R}_i = p_i^{2(\beta_i+1)}\mathcal{L}_i \,p_i^{-2\beta_i}.
\]

The action of the creation operators is the inverse of that in   \eqref{contannih}, namely
\[\label{actionofCi}
\mathcal{C}_i: \quad \gamma_0\rightarrow\gamma_0-1,\quad\gamma_i\rightarrow\gamma_i +1,\qquad
\bar{\mathcal{C}}_i: \quad\gamma_0\rightarrow\gamma_0-1,\quad\gamma_i\rightarrow\gamma_i-1,
\]
where all remaining $\gamma_j$ for $j\neq i$ stay the same. 
By virtue of the shadow relation \eqref{contshadow}, however, it suffices to construct only $\mathcal{C}_i$ since
\[\label{Cshadow}
\bar{\mathcal{C}}_i \,\mathcal{I}_{\bs{\g}}= \Big(\frac{\bar{x}_i}{x_i}\Big)^{1-\g_i} \mathcal{C}_i\Big|_{\gamma_i\rightarrow -\gamma_i}\, \Big(\frac{\bar{x}_i}{x_i}\Big)^{\g_i}\mathcal{I}_{\bs{\g}}.
\]

To construct $\mathcal{C}_i$, we first need to identify the singular hyperplanes of $\mathcal{I}_{\bs{\g}}$.   These can be found either by expanding the integrand of \eqref{icont} about the lower limit $z=0$ and looking for the appearance of $z^{-1}$ pole terms (see \cite{Bzowski:2015pba}, and the  example on page \pageref{Ex1}),  or by using the formula \eqref{hypsings} based on the Newton polytope.  Here, the Newton polytope takes the form of an $n$-dimensional cross-polytope with vertices at 
$\pm \bs{e}_j$ for every basis vector $(\bs{e}_j)_k=\delta_{jk}$ and $2^n$ facets with outward normals
$\bs{n}= (\sigma_1, \ldots, \sigma_n)^T$ for every possible independent choice of  $\sigma_j=\pm 1$.  From \eqref{muJdef} and \eqref{Jshift}, $n_0=-1$ and $\delta=2$ 
for every facet, hence the singular hyperplanes are
\[\label{contsing}
0 = -\gamma_0 +\sum_{j=1}^n \sigma_j\gamma_j  -2m, \qquad m\in \mathbb{Z}^+.
\]

Given the action of $\mathcal{C}_i$ in \eqref{actionofCi},
the only way this operator can shift us from a finite to a singular integral is if $\sigma_i=+1$ so that 
$m$ increases by one.  
The corresponding $b$-function is then
\[\label{littlebdef0}
b_i(\bs{\gamma}) = \prod_{\{\sigma_j=\pm 1\}}\frac{1}{2}(-\gamma_0+\gamma_i +\sum_{j\neq i}\sigma_j\gamma_j),
\]
where the product runs over every possible choice of signs for all $j\neq i$.  
Using the Euler equations,  this gives
\begin{align}\label{bigBdef}
B_i(\theta,\bar{\theta}) &= \prod_{\{\sigma_j=\pm 1\}}\Big(\theta_i + \sum_{j\neq i}( \delta_{\sigma_j,+1}\theta_j+\delta_{\sigma_j,-1}\bar{\theta}_j)\Big).
\end{align}
For convenience, to eliminate an overall numerical factor in this expression we inserted factors of one-half  in \eqref{littlebdef0}. 
Overall, this is simply a trivial  rescaling of both the creation operator and the $b$-function.

For the 3-point function, for example, these formulae evaluate to
\[\label{3ptcontb}
b_1(\bs{\gamma}) =\frac{1}{16}(-\gamma_0+\gamma_1+\gamma_2+\gamma_3)(-\gamma_0+\gamma_1-\gamma_2+\gamma_3)(-\gamma_0+\gamma_1+\gamma_2-\gamma_3)(-\gamma_0+\gamma_1-\gamma_2-\gamma_3).
\]
and
\begin{align}\label{3ptbigBex}
B_1(\theta,\bar{\theta}) = (\theta_1+\theta_2+\theta_3) (\theta_1+\bar{\theta}_2+\theta_3) (\theta_1+\theta_2+\bar{\theta}_3) (\theta_1+\bar{\theta}_2+\bar{\theta}_3).
\end{align}
Recalling now the creation operators obey
\begin{align}
\mathcal{C}_i\partial_i \mathcal{I}_{\bs{\gamma}} = b_i(\bs{\gamma}) \mathcal{I}_{\bs{\gamma}}  = B_i(\theta,\bar{\theta}) \mathcal{I}_{\bs{\gamma}}
\end{align}
the idea is to expand out as\footnote{A factorisation of this form always exists as can be seen recursively in the number of points $n$.  
Once all $\t_i$-dependence has been gathered into $Q_i\t_i$, let us write the remainder at $n$-points as $b_i^{(n)} = B_i^{(n)}|_{\t_i\rightarrow 0}$.   Multiplying out all the factors containing $\t_n$, and, separately, all the factors containing $\tb_n$, we obtain  $b_i^{(n)} = ((\ldots)\t_n+b_i^{(n-1)})((\ldots)\tb_n + b_i^{(n-1)}) = (\ldots)\t_n\tb_n + (\ldots)b_i^{(n-1)}$, where $b_i^{(n-1)}$ is independent of $\t_n$ and $\tb_n$.  Thus, if the decomposition $b_i^{(n-1)}=\sum_{j\neq i}^{n-1} Q^{(n-1)}_j\t_j\tb_j$ exists at $(n-1)$-points, then it also exists at $n$-points: $b_i^{(n)}=\sum_{j\neq i}^n Q^{(n)}_j\t_j\tb_j$.}  
\begin{align}
B_i(\theta,\bar{\theta}) &= Q_i(\theta,\bar{\theta})\theta_i + \sum_{j\neq i}Q_j(\theta,\bar{\theta}) \theta_j\bar{\theta}_j,
\end{align}
where without loss of generality we can choose all $Q_j(\theta,\bar{\theta})$  
for $j\neq i$ to be  independent of  both $\theta_i$ and $\bar{\theta}_i$.  (Note from \eqref{bigBdef} 
that $B_i(\theta,\bar{\theta})$ is automatically independent of $\bar{\theta}_i$.) 
We then use the toric equations \eqref{toriccont} to re-express
\[\label{useoftorics}
\theta_j\bar{\theta}_j \mathcal{I}_{\bs{\gamma}}= x_j\bar{x}_j\partial_j\bar{\partial}_j\mathcal{I}_{\bs{\gamma}}= x_j\bar{x}_j\partial_i\bar{\partial}_i\mathcal{I}_{\bs{\gamma}}
\]
so that 
\begin{align}
B_i(\theta,\bar{\theta})\mathcal{I}_{\bs{\gamma}} &= \Big[Q_i(\theta,\bar{\theta})x_i + \sum_{j\neq i}Q_j(\theta,\bar{\theta}) x_j\bar{x}_j \bar{\partial}_i\Big]\partial_i\mathcal{I}_{\bs{\gamma}},
\end{align}
This yields the creation operators
\begin{align}
\mathcal{C}_i &= Q_i(\theta,\bar{\theta})x_i + \sum_{j\neq i}Q_j(\theta,\bar{\theta}) x_j\bar{x}_j \bar{\partial}_i.
\end{align}
To project 
from the GKZ space to the physical space spanned by the momenta, we first re-write (suppressing  arguments  for clarity)
\begin{align}\label{re1}
Q_ix_i &= x_i Q_i\Big|_{\theta_i\rightarrow\theta_i+1}\\
Q_j x_j\bar{x}_j \bar{\partial}_i &= 
x_j\bar{x}_j\bar{\partial}_iQ_j\Big|_{\theta_j\rightarrow\theta_j+1,\,\bar{\theta}_j\rightarrow\bar{\theta}_j+1}
\label{re2}
\end{align} 
where for  \eqref{re2} we recall the $Q_j$ are independent of both $\theta_i$ and $\bar{\theta}_i$. 
We then project to the physical hypersurface \eqref{contphys} by
using the Euler equations \eqref{eulercont} to replace $\bar{\theta}_k \rightarrow  \theta_k-\gamma_k$ for all $k=1,\ldots, n$ (which is justified since after the re-arrangements \eqref{re1}-\eqref{re2} all $\bar{\theta}_k$ act directly on $\mathcal{I}_{\bs{\gamma}}$) and set all $\bar{x}_k\rightarrow 1$.
Note also that  $\bar{\partial}_i=\bar{\theta}_i$ on the physical hypersurface since $\bar{x}_i=1$, hence we can also replace $\bar{\partial}_i\rightarrow \theta_i-\gamma_i$. 
 The result is
 \begin{align}\label{Cres}
\mathcal{C}_i^{\mathrm{ph}} &= x_i Q_i\Big|_{\theta_i\rightarrow\theta_i+1,\,\bar{\theta}_k\rightarrow \theta_k-\gamma_k} + (\theta_i-\gamma_i)\sum_{j\neq i}x_j Q_j\Big|_{\theta_j\rightarrow\theta_j+1,\,\bar{\theta}_k\rightarrow\theta_k-\gamma_k+\delta_{kj}}
\end{align}
 where the replacement on $\bar{\theta}_k$ applies to all the $\bar{\theta}$ variables present.   As previously, the superscript `ph' denotes the operator expressed in physical variables. 
From the shadow relation \eqref{contshadow}, we also have
\begin{align}\label{Cbarshadow}
\bar{\mathcal{C}}_i{}^{\mathrm{ph}}&= x_i^{\g_i-1} \mathcal{C}_i^{\mathrm{ph}}\Big|_{\g_i\rightarrow-\g_i}\, x_i^{-\g_i}\nn\\&
=Q_i\Big|_{\t_i\rightarrow\t_i-\g_i+1,\,
\bar{\theta}_k\rightarrow \theta_k-\gamma_k} + \sum_{j\neq i}x_j\partial_i Q_j\Big|_{\theta_j\rightarrow\theta_j+1,\,\bar{\theta}_k\rightarrow\theta_k-\gamma_k+\delta_{kj}}.
\end{align}
Together, these expressions gives us the creation operators in terms of the physical variables 
\[\label{badnotation}
x_k=p_k^2, \qquad \theta_k = x_k\partial_k = \frac{1}{2}p_k\partial_{p_k},
\]  
From \eqref{creationeq}, their action is 
\begin{align}\label{creationactioncont}
\mathcal{C}_i\mathcal{I}_{\{\gamma_0,\,\gamma_i\}} &= -(\gamma_0-1)^{-1}b_i(\gamma_0-1,\g_i+1)\mathcal{I}_{\{\gamma_0-1,\,\gamma_i+1\}},\nn\\[0.5ex]
\bar{\mathcal{C}}_i\mathcal{I}_{\{\gamma_0,\,\gamma_i\}} &= -(\gamma_0-1)^{-1}\bar{b}_i(\gamma_0-1,\,\g_i-1)\mathcal{I}_{\{\gamma_0-1,\,\gamma_i-1\}}.
\end{align}
The shift in $b$-function arguments on the right-hand sides here reflects the fact that, in replacing $\tb_k\rightarrow\t_k-\g_k$ in the projection step above, we are taking the creation operator to act on the integral $\mathcal{I}_{\{\gamma_0,\,\gamma_i\}}$.  This is equivalent to eliminating $\bs{\g}$ from \eqref{creationeq} using  \eqref{gshift2} then relabelling $\bs{\g}'\rightarrow\bs{\g}$.

\subsubsection{3-point creation operator}
\label{sec:3KC}

Let us find the creation operator $\mathcal{C}_1$ for the 3-point function via the procedure outlined above.
Starting from the expression for 
 $B_1(\theta,\bar{\theta})$ in \eqref{3ptbigBex}, we  decompose
\begin{align}
B_1(\theta,\bar{\theta}) = Q_1 \theta_1+Q_2\theta_2\bar{\theta}_2+Q_3\theta_3\bar{\theta}_3
\end{align}
where 
\begin{align}
&Q_1=(\t_1+u_2+u_3)\bigl( (\t_1+u_2)(\t_1+u_3)+2(v_2+v_3)\bigr),\\
&Q_2=(u_2+u_3)u_3+v_2-v_3,\\
&Q_3=(u_2+u_3)u_2-v_2+v_3,
\end{align}
with 
\[
u_i = \theta_i+\bar{\theta}_i,\qquad v_i= \theta_i\bar{\theta}_i,\qquad i=2,3.
\]
Note that $Q_2$ and $Q_3$ are independent of $\theta_1$ and all coefficients are independent of $\bar{\theta}_1$.  We have also chosen $Q_2$ and $Q_3$ to preserve the $2\leftrightarrow 3$ symmetry though this is not essential.

Re-iterating the steps above, making use of \eqref{useoftorics}, we have
\begin{align}
B_1(\theta,\bar{\theta})\mathcal{I}_{\bs{\gamma}}
=\big(Q_1 x_1 + Q_2 x_2\bar{x}_2\bar{\partial}_1+Q_3 x_3\bar{x}_3\bar{\partial}_1\big)\p_1\mathcal{I}_{\bs{\gamma}}=\mathcal{C}_1\p_1 \mathcal{I}_{\bs{\gamma}}
\end{align}
yielding the creation operator $\mathcal{C}_1$ in GKZ space.  Moving the $Q_k$ to the right, this can equivalently be written
\begin{align}
\mathcal{C}_1&= x_1 Q_1\Big|_{\theta_1\rightarrow\theta_1+1} +  x_2\bar{x}_2\bar{\partial}_1 Q_2\Big|_{\theta_2\rightarrow\theta_2+1,\,\bar{\theta}_2\rightarrow\bar{\theta}_2+1}+  x_3\bar{x}_3\bar{\partial}_1 Q_3\Big|_{\theta_3\rightarrow\theta_3+1,\,\bar{\theta}_3\rightarrow\bar{\theta}_3+1} \end{align}
Since shifting $\theta_i\rightarrow \theta_i+1$ and $\bar{\theta}_i\rightarrow\bar{\theta}_i+1$ is equivalent to $u_i\rightarrow u_i+2$ and $v_i\rightarrow 1+u_i+v_i$, this is 
\begin{align} \label{3KC1gkz}
\mathcal{C}_1&
 = x_1(\t_1+1+u_2+u_3)\big((\t_1+1+u_2)(\t_1+1+u_3)+2(v_2+v_3)\big)\notag\\
&\quad +x_2\bar{x}_2\bar{\p}_1\bigl(
1+u_2+v_2-v_3+(u_2+u_3+2)u_3\bigr)\notag\\
&\quad +x_3\bar{x}_3\bar{\p}_1\bigl(1+u_3+v_3-v_2+(u_2+u_3+2)u_2\bigr).
\end{align}
Finally, to project to the physical hypersurface, we set
\[\label{3Kphysrepl}
\bar{x}_i\rightarrow 1,\qquad
\bar{\theta}_i\rightarrow \theta_i-\gamma_i,\qquad
\bar{\p}_i\rightarrow \theta_i-\gamma_i
\]
which sends $u_i\rightarrow 2\theta_i-\gamma_i$ and $v_i\rightarrow \theta_i(\theta_i-\gamma_i)$, yielding
\begin{align}
&\mathcal{C}_1 =x_1(\t_1+1+2\t_2+2\t_3-\g_2-\g_3)\times \\&\quad \times \bigl[(\t_1+1+2\t_2-\g_2)(\t_1+1+2\t_3-\g_3)
 +2\t_2(\t_2-\g_2)+2\t_3(\t_3-\g_3)
\bigr]+(\t_1-\g_1)\times\notag\\
&\quad\times\bigl[x_2\bigl(1+2\t_2-\g_2+\t_2(\t_2-\g_2)-\t_3(\t_3-\g_3)+(2\t_2-\g_2+2\t_3-\g_3+2)(2\t_3-\g_3)\bigr)\notag\\
&\quad\,\, +x_3\bigl(1+2\t_3-\g_3+\t_3(\t_3-\g_3)-\t_2(\t_2-\g_2)+(2\t_2-\g_2+2\t_3-\g_3+2)(2\t_2-\g_2)\bigr)\bigr]\nn
\end{align}
Of course, this result also follows from \eqref{Cres} directly. 
We can simplify somewhat further by using the DWI evaluated on the physical hypersurface,
\[
0 = \Big(\t_1+\t_2+\t_3+\frac{1}{2}(\g_0-\g_1-\g_2-\g_3)\Big)\mathcal{I}_{\bs{\g}}.
\]
This gives the alternative form
\begin{align}\label{3KC1}
&\mathcal{C}_1
=
- \frac{x_1}{2}(\t_1-1+\g_0-\g_1)\Big[2\t_1^2+2(\g_0-\g_1)\t_1+(\g_0-\g_1-1)^2+1-\g_2^2-\g_3^2\Big] \\&\qquad
+(\t_1-\g_1)
\times\nn\\&\qquad \times
\Big[x_2\bigl(1+2\t_2-\g_2+\t_2(\t_2-\g_2)-\t_3(\t_3-\g_3)-(2\t_1-\g_1+\g_0-2)(2\t_3-\g_3)\bigr)\notag\\
&\qquad +x_3\bigl(1+2\t_3-\g_3+\t_3(\t_3-\g_3)-\t_2(\t_2-\g_2)-(2\t_1-\g_1+\g_0-2)(2\t_2-\g_2)\bigr)\Big]. \nn
\end{align}
The action of this creation operator is
\[
\mathcal{C}_1\mathcal{I}_{\{\g_0,\g_1\}} = -(\g_0-1)^{-1}b_1(\g_0-1,\g_1+1)\mathcal{I}_{\{\g_0-1,\g_1+1\}}
\]
where 
\[\label{evalbfn3pt}
b_1(\g_0-1,\g_1+1)= b_1(\bs{\g})\Big|_{\g_0\rightarrow\g_0-1,\,\g_1\rightarrow\g_1+1}
=
\frac{1}{16} \Big[ \big((2 -\g_0+\g_1)^2 - \g_2^2-\g_3^2\big)^2 - 4 \g_2^2\g_3^2\Big]
\]
using $b_1(\bs{\g})$ as given in \eqref{3ptcontb}.

\subsubsection{4-point creation operator}

From \eqref{littlebdef0}, the 4-point $b$-function is 
\begin{align}\label{b1contdef}
b_1(\bs{\gamma}) &= 2^{-8}(-\gamma_0+\gamma_1+\gamma_2+\gamma_3+\g_4)(-\gamma_0+\gamma_1-\gamma_2+\gamma_3+\g_4)
(-\gamma_0+\gamma_1+\gamma_2-\gamma_3+\g_4)
\nn\\&\qquad \times(-\g_0+\g_1+\g_2+\g_3-\g_4)
(-\gamma_0+\gamma_1-\gamma_2-\gamma_3+\g_4)
(-\gamma_0+\gamma_1-\gamma_2+\gamma_3-\g_4)
\nn\\&\qquad \times(-\g_0+\g_1+\g_2-\g_3-\g_4)
(-\g_0+\g_1-\g_2-\g_3-\g_4)
\end{align}
which, after use of the Euler equations and DWI, corresponds to
\begin{align}\label{Bfacs}
B_1 &=\,
 (\theta_1+\theta_2+\theta_3+\t_4) (\theta_1+\bar{\theta}_2+\theta_3+\t_4) (\theta_1+\theta_2+\bar{\theta}_3+\t_4)
 (\theta_1+\theta_2+\t_3+\bar{\t}_4)\nn\\&\quad\times
  (\theta_1+\bar{\theta}_2+\bar{\theta}_3+\t_4)
   (\theta_1+\bar{\theta}_2+\theta_3+\bar{\t}_4)
    (\theta_1+\theta_2+\bar{\theta}_3+\bar{\t}_4)
       (\theta_1+\bar{\theta}_2+\bar{\theta}_3+\bar{\t}_4)
\end{align}
consistent with \eqref{bigBdef}.
We wish to decompose this as
\begin{align}\label{4ptBdec}
B_1(\theta,\bar{\theta}) = Q_1 \theta_1+Q_2\theta_2\bar{\theta}_2+Q_3\theta_3\bar{\theta}_3+Q_4\theta_4\bar{\theta}_4
\end{align}
where $Q_2$, $Q_3$ and $Q_4$ are independent of $\theta_1$.

Let us deal with the $Q_1$ term first.
Denoting the eight factors in \eqref{Bfacs} as $R_m$ for $m=1,\ldots, 8$, we have
\[
B_1\Big|_{\t_1=0} = \prod_{m=1}^8 (-\t_1+R_m) = \sum_{m=0}^8 \sigma_{(m)}(R) (-\t_1)^{8-m} = B_1 +  \sum_{m=0}^7 \sigma_{(m)}(R) (-\t_1)^{8-m} 
\]
where $\sigma_{(m)}(R)$ is the $m$th elementary symmetric polynomial in the $R_m$.  Rearranging then gives
\[
Q_1 = \t_1^{-1}\big(B_1-B_1\big|_{\t_1=0}\big) = \sum_{m=0}^7 \sigma_{(m)}(R) (-\t_1)^{7-m},
\]
and since $\t_1$ appears in each of the factors in \eqref{Bfacs},
\[
Q_1 x_1 = x_1 Q_1\Big|_{\theta_1\rightarrow\theta_1+1} = \sum_{m=0}^7 \sigma_{(m)}(1+R) (-1-\t_1)^{7-m}.
\]
When acting on the GKZ integral $\mathcal{I}_{\bs{\g}} $, 
we can now use the Euler equations \eqref{eulercont} and DWI  \eqref{DWIcont} to rewrite this expression in terms of elementary symmetric polynomials  of just the parameters $\bs{\g}$ alone, namely
\[
x_1 Q_1\Big|_{\theta_1\rightarrow\theta_1+1}\mathcal{I}_{\bs{\g}} = x_1\sum_{m=0}^7 \sigma_{(m)}(r) (-1-\t_1)^{7-m}\mathcal{I}_{\bs{\g}},
\]
where the eight variables
\[
r_{\{m\}} = 1-\g_0+\g_1 \pm \g_2 \pm \g_3 \pm \g_4
\]
are formed by making all possible independent choices of $\pm$ signs.

We now turn to the remaining $Q_k$ coefficients in \eqref{4ptBdec} for $k=2,3,4$.
Defining the auxiliary functions
\begin{align}
S(\t)&=(\t+\t_3+\t_4)(\t+\tb_3+\t_4)(\t+\t_3+\tb_4)(\t+\tb_3+\tb_4),\\
T(\t) &= \t^{-1}(S(\t)-S(0)) = (\t+u_3+u_4)\big((\t+u_3)(\t+u_4)+2v_3+2v_4\big),
\end{align}
where
\[
u_k = \t_k+\tb_k,\qquad v_k=\t_k\tb_k,\qquad k=3,4
\]
we can decompose
\begin{align}
Q_2 &= T(\t_2)T(\tb_2),\\
Q_3 &= \big((u_3+u_4)u_4+v_3-v_4)(S(\t_2)+S(\tb_2)-S(0)),\\
Q_4 &= \big((u_3+u_4)u_3+v_4-v_3)(S(\t_2)+S(\tb_2)-S(0)).
\end{align}
Noting that for $k=3,4$,
\[
(S(\t_2)+S(\tb_2)-S(0))\Big|_{\t_k\rightarrow\t_k+1,\,\tb_k\rightarrow\tb_k+1} = (S(\t_2+1)+S(\tb_2+1)-S(1)),
\]
and using \eqref{Cres}, the creation operator is then
\begin{align}\label{4KC1}
&\mathcal{C}_1^{\mathrm{ph}} = 
x_1\sum_{m=0}^7 \sigma_{(m)}(r) (-1-\t_1)^{7-m}
 +(\t_1-\g_1)\Big[ x_2\, \hat{T}(\t_2+1)\hat{T}(\t_2-\g_2+1) \\&\quad\,
+\Big(x_3\big((2+\hat{u}_3+\hat{u}_4)\hat{u}_4+\hat{u}_3+1+\hat{v}_3-\hat{v}_4\big)
+x_4\big((2+\hat{u}_3+\hat{u}_4)\hat{u}_3+\hat{u}_4+1+\hat{v}_4-\hat{v}_3\big)\Big)
\nn\\[0.5ex]&\quad\,
\times \Big(\hat{S}(\t_2+1)+\hat{S}(\t_2-\g_2+1)-\hat{S}(1)\Big)\Big] \nn
\end{align}
where all hatted quantities are defined by replacing $\tb_k\rightarrow \t_k-\g_k$ for  $k=3,4$ in the corresponding unhatted quantities.
Its action is 
\[
\mathcal{C}_1\mathcal{I}_{\{\g_0,\g_1\}} = -(\g_0-1)^{-1}b_1(\g_0-1,\g_1+1)
\mathcal{I}_{\{\g_0-1,\g_1+1\}}
\]
where, using $b_1(\bs{\g})$ as given in \eqref{b1contdef}, 
\begin{align}
b_1(\g_0-1,\g_1+1)&= b_1(\bs{\g})\Big|_{\g_0\rightarrow\g_0-1,\,\g_1\rightarrow\g_1+1}\nn\\
&=
\frac{1}{256 }\Big[\Big((2 -\g_0+\g_1)^2 - \mathcal{S}_{(1)}\Big)^4 - 8 \Big((2 -\g_0+\g_1\big)^2 - \mathcal{S}_{(1)}\Big)^2 \mathcal{S}_{(2)} \nn\\&\qquad + 16 \mathcal{S}_{(2)}^2 - 
   64 (2 -\g_0+\g_1)^2 \mathcal{S}_{(3)}\Big]
\end{align}
with the $\mathcal{S}_{(m)}$ being elementary symmetric polynomials in $\g_2^2$, $\g_3^2$ and $\g_4^2$.

\subsection{Examples}

Taking into account the additional gamma function factors in \eqref{icontIg}, the action of these creation operators on contact diagrams is
\begin{align}
\mathcal{C}_1 \,i_{[d;\,\Delta_1,\,\ldots,\,\Delta_n]} &= -4\g_1 b_1(\g_0-1,\g_1+1) \,i_{[\tilde{d};\,\tilde{\Delta}_1,\Delta_2\,\ldots,\,\Delta_n]} 
\end{align}
where
\[\label{Cdshift}
\tilde{d} = d - \frac{2}{n-2},\qquad \tilde{\Delta}_1 = \Delta_1+\frac{n-3}{n-2},
\]
Alternatively, in terms of the multiple-Bessel integral 
\[\label{nKdef0}
I_{\g_0\,\{\g_1,\,\ldots,\,\g_n\}} = 
\int_0^\infty \dd z\, z^{\g_0}\prod_{i=1}^n p_i^{2\g_i}K_{\g_i}(p_iz),
\]
from \eqref{icont} and \eqref{icontIg} we have
\[\label{nKIrel}
\mathcal{I}_{\bs{\g}}=\frac{2^{n-\g_0}}{\Gamma(\g_0)}\,I_{\g_0\,\{\g_1,\,\ldots,\,\g_n\}} 
\]
and hence
 \begin{align}\label{C13K}
\mathcal{C}_1 I_{\g_0\,\{\g_1,\,\g_2,\,\ldots,\,\g_n\}}  &= -2b_1(\g_0-1,\g_1+1) \,I_{\g_0-1\,\{\g_1+1,\,\g_2,\,\ldots,\,\g_n\}}.
\end{align}
Here we can either use \eqref{badnotation} to rewrite $\mathcal{C}_1$ in terms of the momenta $p_i$, or more easily, re-express \eqref{nKdef0} using $p_i=\sqrt{x_i}$ then convert back to $p_i$ after acting with $\mathcal{C}_1$.

A quick check of these results can be obtained by examining cases where  all the  Bessel indices $\g_i$ take half-integer values allowing direct evaluation of the contact diagrams.  (We restrict to cases where both the initial and the shifted integral are finite; for the analysis of renormalised cases see \cite{Bzowski:2022rlz}.)
For example,
at three points, the triple-$K$ integrals  \eqref{nKdef0}
\begin{align}
&I_{4\{\frac{1}{2}\frac{1}{2}\frac{1}{2}\}}=\frac{15 \pi^2}{16\sqrt{2}}\left(p_1+p_2+p_3\right)^{-7/2},\qquad
I_{3\{\frac{3}{2}\frac{1}{2}\frac{1}{2}\}}=\frac{\pi^2(5p_1+2p_2+2p_3)}{8\sqrt{2}(p_1+p_2+p_3)^{5/2}},
\end{align}
and one can verify that 
\[
\mathcal{C}_1 I_{4\{\frac{1}{2}\frac{1}{2}\frac{1}{2}\}}=
-\frac{45}{128}\,
I_{3\{\frac{3}{2}\frac{1}{2}\frac{1}{2}\}},
\]
consistent with \eqref{C13K} using \eqref{evalbfn3pt} for the 3-point $b$-function.  We have performed many similar checks at both $3$- and $4$-points.

More non-trivially, many  triple-$K$ integrals with integer indices can be evaluated \cite{Bzowski:2015yxv}
by acting with the annihilators $\mathcal{L}_i$ and $\mathcal{R}_i$ 
given in \eqref{LRdef0}
on the known `seed' integral $I_{1\,\{000\}}$
which can be evaluated in terms of 
the Bloch-Wigner dilogarithm.  These relations enable computation of all the necessary triple-$K$ integrals arising in 3-point functions of conserved currents and stress tensors in even spacetime dimensions \cite{Bzowski:2017poo, Bzowski:2018fql}.
Since the creation operators $\mathcal{C}_i$ and $\bar{\mathcal{C}}_i$ are the inverse of $\mathcal{L}_i$ and $\mathcal{R}_i$, this allows us to reverse the direction of all operations linking different triple-$K$ integrals within the reduction scheme.  Thus, for example, we find
\[
\mathcal{R}_1 I_{1\{000\}} = I_{2\{100\}},\qquad
-8 C_1  I_{2\{100\}} =   I_{1\{000\}}
\]
where the integrals 
\begin{align}
 I_{1\{000\}} &= \frac{1}{2p_3^2(z-\bar{z})}\Big[\mathrm{Li}_2\, z
-\mathrm{Li}_{2} \, \bar{z}+\frac{1}{2}\ln (z\bar{z}) \ln\Big(\frac{1-z}{1-\bar{z}}\Big)\Big],\\
I_{2\{100\}}&= 
\frac{1}{2 p_3^2 (z - \bar{z})^2}\Big[  
  4 p_3^2 z\bar{z} (-2 + z + \bar{z})  I_{1\{000\}}  - 2 z \bar{z} \ln(z \bar{z})\nn\\ &\qquad\qquad\qquad\quad - (z + \bar{z} - 
      2 z\bar{z})  \ln[(1 - z) (1 - \bar{z})] \Big]
\end{align}
and the variables 
\[\label{zexp}
z = \frac{1}{2p_3^2}\Big(p_1^2-p_2^2+p_3^2+\sqrt{-J^2}\Big), \qquad
\bar{z} = \frac{1}{2p_3^2}\Big(p_1^2-p_2^2+p_3^2-\sqrt{-J^2}\Big)
\]
or equivalently
\[
\frac{p_1^2}{p_3^2} = z\bar{z}, \qquad
\frac{p_2^2}{p_3^2}= (1-z)(1-\bar{z})
\]
with 
\begin{align}
J^2 & = (p_1 + p_2 + p_3) (- p_1 + p_2 + p_3) (p_1 - p_2 + p_3) (p_1 + p_2 - p_3)\nn\\
&= -p_1^4-p_2^4-p_3^4+2p_1^2p_2^2+2p_2^2p_3^2+2p_3^2p_1^2. 
\end{align}

\subsection{Shift operators preserving the spacetime dimension} 

The creation operators  constructed above decrease the spacetime dimension according to \eqref{Cdshift}.
 For many applications, we would prefer an operator capable of changing the operator dimensions of a contact diagram while preserving the spacetime dimension.  Thus, we seek an operator $W_{12}^{\sigma_1,\sigma_2}$ such that 
\[\label{desiredaction}
W_{12}^{\sigma_1,\sigma_2}i_{[d;\,\Delta_1, \Delta_2, \,\Delta_3,\,\ldots,\, \Delta_n]} \propto\, i_{[d;\,\Delta_1+\sigma_1,\, \Delta_2+\sigma_2, \,\Delta_3,\,\ldots,\, \Delta_n]} 
\] 
for any independent choice of  signs $\sigma_1=\pm 1$ and $\sigma_2=\pm 1$.
Operators of this type are known at three points \cite{Karateev:2017jgd,Baumann:2019oyu}, but their analogue at four points acts on contact diagrams to generate shifted contact diagrams with derivative vertices \cite{Bzowski:2022rlz}. Instead,  our discussion of creation operators above can be modified to enable operators of this type to be identified.\footnote{The shift operators that we identify will moreover be of minimal order, unlike the $d$-preserving combination of an annihilator $\p_i$ or $\bar{\p}_i$ followed by a creation operator  $\mathcal{C}_j$ or $\bar{\mathcal{C}}_j$. For example, the combination $\bar{\mathcal{C}}_1\p_2-\bar{\mathcal{C}}_2\p_1$ produces the same shift as $W_{12}^{--}$ but is of seventh order in derivatives for the 4-point function, since each product is eighth order and taking the difference lowers the order by one.  In contrast, the 4-point operator  $W_{12}^{--}$ we find will be of only fourth order.}
  At three points we will see these coincide with the operators of \cite{Karateev:2017jgd,Baumann:2019oyu}, but at four points and above they are novel.  Using these operators will then enable further new shift operators to be constructed for exchange diagrams.

Our starting point is the observation that, for the GKZ integral \eqref{contactGKZ} corresponding to the contact diagram, 
\[\label{Weqn}
W_{12}^{--}\bar{\p}_1 \mathcal{I}_{\bs{\g}}= b_W(\bs{\g}) \p_2  \mathcal{I}_{\bs{\g}}.
\]
Recalling the parameter identifications \eqref{contparams}, the action of the  operators here is
\begin{align}\label{Wshift0}
&W_{12}^{--}:& \gamma_0&\rightarrow\gamma_0, & \gamma_1&\rightarrow\g_1-1, &\g_2&\rightarrow\g_2-1,\nn\\
&\bar{\p}_1: & \gamma_0&\rightarrow\gamma_0+1, & \gamma_1&\rightarrow\g_1+1, & \g_2&\rightarrow\g_2,\nn\\
&\p_2: & \gamma_0&\rightarrow\gamma_0+1, & \gamma_1&\rightarrow\g_1, & \g_2&\rightarrow\g_2-1,
\end{align}
with all remaining $\g_k$ for $k=3,\ldots, n$ staying  the same.
As  the shifts produced by the operators on each side of \eqref{Weqn} are the same, both sides involve the same integral $\mathcal{I}_{\bs{\g}}$.   As previously, the $b$-function $b_W(\bs{\g})$ should be a product of linear factors that vanishes whenever $W_{12}^{--}$ maps us from a finite to a singular integral.  
Taking into account the action \eqref{contannih} of the annihilators in \eqref{Weqn}, we have
\begin{align}\label{Weqn2}
W_{12}^{--}\bar{\p}_1 \mathcal{I}_{\bs{\g}}
&=-\g_0 W_{12}^{--}\mathcal{I}_{\bs{\g}}\Big|_{\g_0\rightarrow\g_0+1,\,\g_1\rightarrow\g_1+1}\nn\\
&=-\g_0 b_W(\bs{\g})   \mathcal{I}_{\bs{\g}}\Big|_{\g_0\rightarrow\g_0+1,\g_2\rightarrow \g_2-1}
= b_W(\bs{\g}) \p_2  \mathcal{I}_{\bs{\g}}
\end{align}
and so the zeros of $b_W(\bs{\g})$ must cancel the singularities of 
$ \mathcal{I}_{\bs{\g}}|_{\g_0\rightarrow\g_0+1,\g_2\rightarrow \g_2-1}$.
From \eqref{contsing},  this means
\begin{align}\label{bWres}
b_W(\bs{\g}) &=\prod_{\sigma_k\in\pm1}\frac{1}{2} \Big(-(\g_0+1)-\g_1-(\g_2-1)+ \sigma_3\g_3+\ldots \sigma_n\g_n\Big)\nn\\
&=\prod_{\sigma_k\in\pm1}\frac{1}{2} \Big(-\g_0-\g_1-\g_2+ \sigma_3\g_3+\ldots \sigma_n\g_n\Big).
\end{align}
Only the singularities with $\sigma_1=\sigma_2=-1$ in \eqref{contsing} appear here since these are the only cases for which $ \mathcal{I}_{\bs{\g}}|_{\g_0\rightarrow\g_0+1,\g_2\rightarrow \g_2-1}$ is singular but the integral $ \mathcal{I}_{\bs{\g}}|_{\g_0\rightarrow\g_0+1,\g_1\rightarrow \g_1+1}$ on which $W_{12}^{--}$ acts is finite.  Every possible independent choice of $\sigma_k\in\pm 1$ for all $k=3,\ldots, n$ is permitted, however, and gives rise to a corresponding factor in \eqref{bWres}.
Once again, we have also chosen to include trivial factors of one-half in $b_W(\bs{\g})$ to simplify the subsequent form of $W_{12}^{--}$.
Replacing the parameters $\bs{\g}$ in $b_W(\bs{\g})$ using the Euler equations \eqref{eulercont} and DWI \eqref{DWIcont}, we find
\begin{align}\label{Weqn3}
W_{12}^{--}\bar{\p}_1 \mathcal{I}_{\bs{\g}}
=\p_2 \big(b_W(\bs{\g})\mathcal{I}_{\bs{\g}}\big) = \p_2 B_W(\t,\tb)\mathcal{I}_{\bs{\g}} 
\end{align}
where
\begin{align}
 B_W(\t,\tb)&=\prod_{\sigma_k\in\pm1}\frac{1}{2} \Big(\sum_{j=1}^n (\t_j+\tb_j)-(\t_1-\tb_1)-(\t_2-\tb_2)+ \sigma_3(\t_3-\tb_3)+\ldots \sigma_n(\t_n-\tb_n)\Big)\nn\\
&= \prod_{\sigma_k\in\pm1}\Big(\tb_1+\tb_2+(\delta_{\sigma_3,+1}\t_3+\delta_{\sigma_3,-1}\tb_3)+\ldots +(\delta_{\sigma_n,+1}\t_n+\delta_{\sigma_n,-1}\tb_n)\Big).
\end{align}
Since $B_W(\t,\tb)$ is in fact independent of $\t_2$ the ordering of $\p_2$ and $B_W(\t,\tb)$ on the right-hand side of \eqref{Weqn3} is in fact immaterial, but had this not been the case the ordering shown would be the correct one when using the unshifted Euler equations and DWI to replace the $\bs{\g}$ parameters.

To identify $W_{12}^{--}$, all that is then needed is to start with $\p_2B_W(\t,\tb)$ and, using the toric equations \eqref{toriccont}, pull out a right factor of $\bar{\p}_1$ according to \eqref{Weqn3}.
As usual, the resulting operator can then be projected down to the physical hypersurface using the Euler equations and DWI.  These procedures are illustrated for the 3- and 4-point function below. 
Finally, given $W_{12}^{--}$ in physical variables, all the remaining operators in \eqref{desiredaction} can be found by shadow conjugation using \eqref{GKZshadow}, namely
\begin{align}\label{Wshadows}
(W_{12}^{+-})_{\mathrm{ph}} &= x_1^{1+\g_1} (W_{12}^{--})_{\mathrm{ph}}\,  x_1^{-\g_1},\\
(W_{12}^{-+})_{\mathrm{ph}}\,  &= x_2^{1+\g_2} (W_{12}^{--})_{\mathrm{ph}}\,  x_2^{-\g_2},\\
(W_{12}^{++})_{\mathrm{ph}}\,  &= x_1^{1+\g_1} x_2^{1+\g_2} (W_{12}^{--})_{\mathrm{ph}}\,  x_1^{-\g_1}x_2^{-\g_2}.
\end{align}

\subsubsection{3-point function}
\label{sec:3KW}

To illustrate the above discussion, for the 3-point function we have
\[
b_W(\bs{\g}) = \frac{1}{4}(-\g_0-\g_1-\g_2+\g_3)(-\g_0-\g_1-\g_2-\g_3)
\]
and 
\[
B_W(\t,\tb) = (\tb_1+\tb_2+\t_3)(\tb_1+\tb_2+\tb_3).
\]
The operator $W_{12}^{--}$ can now be extracted from
\[
W_{12}^{--}\bar{\p}_1\mathcal{I}_{\bs{\g}}  = \p_2 B_W(\t,\tb)\mathcal{I}_{\bs{\g}}.
\]
For this, we write
\begin{align}
&\p_2(\tb_1+\tb_2+\t_3)(\tb_1+\tb_2+\tb_3)\mathcal{I}_{\bs{\g}}\nn\\
&\qquad=\p_2\big[(\tb_1+\tb_2+\tb_3+\t_3)(\tb_1+\tb_2)+\t_3\tb_3]\mathcal{I}_{\bs{\g}}\nn\\
&\qquad=\big[(\tb_1+\tb_2+\tb_3+\t_3)(\bar{x}_1\p_2\bar{\p}_1+\bar{x}_2\p_2\bar{\p}_2)+x_3\bar{x}_3 \p_2\p_3\bar{\p}_3]\mathcal{I}_{\bs{\g}}\nn\\
&\qquad= \big[(\tb_1+\tb_2+\tb_3+\t_3)(\bar{x}_1\p_2+\bar{x}_2\p_1)+x_3\bar{x}_3 \p_2\p_1]\bar{\p}_1 \mathcal{I}_{\bs{\g}}
\end{align}
where in the penultimate line we used the toric equations \eqref{toriccont}.  Thus
\begin{align}\label{W3ptgkz}
W_{12}^{--}&=(\tb_1+\tb_2+\tb_3+\t_3)(\bar{x}_1\p_2+\bar{x}_2\p_1)+x_3\bar{x}_3 \p_2\p_1 \nn\\
&=(\bar{x}_1\p_2+\bar{x}_2\p_1)(1+\tb_1+\tb_2+\tb_3+\t_3)+x_3\bar{x}_3 \p_2\p_1, 
\end{align}
and using the DWI \eqref{DWIcont} to project  to the physical hypersurface \eqref{contphys},  we obtain
\begin{align}\label{W3ptphys}
(W_{12}^{--})_{\mathrm{ph}}&=
(\p_2+\p_1)(1-\g_0-\t_1-\t_2)+x_3\p_2\p_1\nn\\&=
-(\g_0+\t_1+\t_2)(\p_1+\p_2)+x_3\p_1\p_2
\end{align}
where for the 3-point function $\g_0=d/2$ from \eqref{contparams}.
A short calculation shows that
\[\label{curlyWdef}
(W_{12}^{--})_{\mathrm{ph}} = -\frac{1}{4}\Big( \p_{p_1}^2+\partial_{p_2}^2+\frac{(d-1)}{p_1}\p_{p_1}+\frac{(d-1)}{p_2}\p_{p_2}+(p_1^2+p_2^2-p_3^2)\frac{1}{p_1p_2}\p_{p_1}\p_{p_2}\Big)
\]
which, up to a factor of $-2$, is the 3-point shift operator studied in \cite{Baumann:2019oyu, Bzowski:2022rlz}.

The action of $W_{12}^{--}$ is
\begin{align}\label{Weqn4}
W_{12}^{--}\mathcal{I}_{\bs{\g}} = b_W(\bs{\g})\Big|_{\g_0\rightarrow\g_0-1,\,\g_1\rightarrow\g_1-1}\mathcal{I}_{\bs{\g}}\Big|_{\g_1\rightarrow\g_1-1,\,\g_2\rightarrow\g_2-1}
\end{align}
where the shift on the $b$-function derives from the fact that, in the projection step going from \eqref{W3ptgkz} to \eqref{W3ptphys}, we have chosen that $W_{12}^{--}$ acts on the integral $\mathcal{I}_{\bs{\g}}$ requiring us to shift the $\bs{\g}$ parameters present in \eqref{Weqn2}.  Evaluating, this gives
\begin{align}
b_W(\bs{\g})\Big|_{\g_0\rightarrow\g_0-1,\,\g_1\rightarrow\g_1-1} &= \frac{1}{4}(2-\g_0-\g_1-\g_2+\g_3)(2-\g_0-\g_1-\g_2-\g_3)\nn\\&
=\frac{1}{4}\Big((\g_0+\g_1+\g_2-2)^2-\g_3^2\Big)
\end{align}
such that \eqref{Weqn4} is consistent with the action of $\mathcal{W}_{12}^{--}$ obtained  in  \cite{Bzowski:2022rlz}.
Acting on the holographic contact diagram, from \eqref{icontIg}
we have
\[
W_{12}^{--}i_{[d,\,\Delta_1,\,\Delta_2,\,\Delta_3]} = 
\frac{1}{4(\g_1-1)(\g_2-1)}b_W(\bs{\g})\Big|_{\g_0\rightarrow\g_0-1,\,\g_1\rightarrow\g_1-1}
i_{[d,\,\Delta_1-1,\,\Delta_2-1,\,\Delta_3]}. 
\]

\subsubsection{4-point function}

At 4-points, we find
\begin{align}
b_W(\bs{\g}) &= \frac{1}{16}(-\g_0-\g_1-\g_2+\g_3+\g_4)
(-\g_0-\g_1-\g_2-\g_3+\g_4)\nn\\&\qquad\times  
(-\g_0-\g_1-\g_2+\g_3-\g_4)
(-\g_0-\g_1-\g_2-\g_3-\g_4)
\end{align}
and hence
\begin{align}
B_W(\t,\tb) &= (\tb_1+\tb_2+\t_3+\t_4) (\tb_1+\tb_2+\tb_3+\t_4)
 (\tb_1+\tb_2+\t_3+\tb_4)
  (\tb_1+\tb_2+\tb_3+\tb_4).
\end{align}
Once again, to find $W_{12}^{--}$ we must factorise
\[
W_{12}^{--}\bar{\p}_1\mathcal{I}_{\bs{\g}}  = \p_2 B_W(\t,\tb)\mathcal{I}_{\bs{\g}}.
\]
As a first step, we expand
\[
B_W(\t,\tb) = Q_0(\tb_1+\tb_2)+Q_3\t_3\tb_3+Q_4\t_4\tb_4
\]
where the coefficients 
\begin{align}\label{WQco}
Q_0 &=(u_3 + u_4 + \tb_1+\tb_2) \Big(2 (v_3 + v_4) + (u_3 + \tb_1+\tb_2) (u_4 + \tb_1+\tb_2)\Big),\nn\\
Q_3 &= (u_3+u_4)u_4+v_3-v_4,\nn\\
Q_4 &= (u_3+u_4)u_3-v_3+v_4,
\end{align}
and 
\[
u_k=\t_k+\tb_k,\qquad v_k=\t_k\tb_k, \qquad k=3,4.
\]
Now, since all coefficients are independent of  $\t_2$, 
\begin{align}
\p_2B_W(\t,\tb)\mathcal{I}_{\bs{\g}} &=\Big[
Q_0(\bar{x}_1\p_2\bar{\p}_1
+ \bar{x}_2 \p_2\bar{\p_2})+Q_3 x_3\bar{x}_3\p_2\bar{\p}_3\p_3
+Q_4 x_4\bar{x}_4\p_2\bar{\p}_4\p_4\Big]\mathcal{I}_{\bs{\g}} 
\\&
=\Big[\bar{x}_1 Q_0\Big|_{\tb_1\rightarrow\tb_1+1}\p_2
+\bar{x}_2  Q_0\Big|_{\tb_2\rightarrow\tb_2+1}\p_1
\nn\\&\qquad
+
x_3 \bar{x}_3 Q_3\Big|_{\t_3\rightarrow\t_3+1,\,\tb_3\rightarrow\tb_3+1}\p_2\p_1
+x_4 \bar{x}_4 Q_4\Big|_{\t_4\rightarrow\t_4+1,\,\tb_4\rightarrow\tb_4+1}\p_2\p_1\Big]\bar{\p}_1\mathcal{I}_{\bs{\g}} \nn
\end{align}
where in the second line we used the toric equations \eqref{toriccont}.
We thus have
\begin{align}
W_{12}^{--}&=(\bar{x}_1 \p_2 +\bar{x}_2\p_1)Q_0\Big|_{\tb_1\rightarrow\tb_1+1}
\nn\\&\qquad
+\p_1\p_2
\Big(x_3 \bar{x}_3 Q_3\Big|_{\t_3\rightarrow\t_3+1,\,\tb_3\rightarrow\tb_3+1}
+x_4 \bar{x}_4 Q_4\Big|_{\t_4\rightarrow\t_4+1,\,\tb_4\rightarrow\tb_4+1}\Big)
\end{align}
where in the first line we used the fact that $\tb_1$ and $\tb_2$ enter $Q_0$ only in the combination $\tb_1+\tb_2$ and so the replacement $\tb_1\rightarrow \tb_1+1$ produces the same result as $\tb_2\rightarrow \tb_2+1$ allowing us to combine the two $Q_0$ terms.  We have in addition moved $Q_0$, $Q_3$ and $Q_4$ to the right (noting that all coefficients are independent of $\t_1$ and $\t_2$)  so as to be able to use the Euler equations for $\mathcal{I}_{\bs{\g}}$ to project to the physical hypersurface.   For this, we set  $\bar{x}_k\rightarrow 1$ and $\tb_k\rightarrow \t_k-\g_k$ inside all $Q_k$ coefficients giving 
\begin{align}\label{QWformula}
(W_{12}^{--})_{\mathrm{ph}} &= (\p_1+\p_2)Q_0\Big|_{\tb_k\rightarrow\t_k-\g_k+\delta_{k,1}}
\\&\quad
+\p_1\p_2
\Big(x_3  Q_3\Big|_{\t_3\rightarrow\t_3+1,\,\tb_k\rightarrow\t_k-\g_k+\delta_{k,3}}
+x_4 Q_4\Big|_{\t_4\rightarrow\t_4+1,\,\tb_k\rightarrow\t_k-\g_k+\delta_{k,4}}\Big).\nn
\end{align}
In all the replacements here, $\tb_k$ stands for any index $k=1,\ldots, 4$.
Evaluating this formula explicitly using the coefficients in \eqref{WQco}, we find
\begin{align}\label{4ptWv1}
&(W_{12}^{--})_{\mathrm{ph}} =
(\p_1+\p_2)(1-\g_0-\t_1-\t_2)\Big(2\t_3(\t_3-\g_3)+2\t_4(\t_4-\g_4) \\ &\qquad +
(1+2\t_3-\g_3+\t_1-\g_1+\t_2-\g_2)(1+2\t_4-\g_4+\t_1-\g_1+\t_2-\g_2)\Big)\nn\\&
\qquad
+\p_1\p_2\Big[x_3\Big((2+2\t_3-\g_3+2\t_4-\g_4)(2\t_4-\g_4)+(1+\t_3)(1+\t_3-\g_3)-\t_4(\t_4-\g_4)\Big)\nn\\&\qquad
+x_4\Big((2+2\t_3-\g_3+2\t_4-\g_4)(2\t_3-\g_3)+(1+\t_4)(1+\t_4-\g_4)-\t_3(\t_3-\g_3)\Big)\Big]\nn
\end{align}
where for the 4-point function $\g_0=d$ from \eqref{contparams}.

Alternatively, we can use the DWI \eqref{DWIcont}
projected to the physical hypersurface, 
\[
0=\Big(\frac{1}{2}(\g_0-\g_t)+\sum_{k=1}^4\t_k\Big)\mathcal{I}_{\bs{\g}},\qquad \g_t=\sum_{k=1}^4\g_k,
\]
to eliminate the factors of $\t_1+\t_2$ on the second line of \eqref{4ptWv1}.  After further moving all factors of  $\p_1$ and $\p_2$  to the right, this gives the equivalent form
\begin{align}\label{4ptWv2}
(W_{12}^{--})_{\mathrm{ph}} &=
-(\g_0+\t_1+\t_2)\Big((\t_3+\t_4)(\t_3+\t_4-\g_3-\g_4)
\\ &\qquad\qquad\qquad \qquad\quad 
+
\frac{1}{4}(2-\g_0-\g_t+2\g_3)(2-\g_0-\g_t+2\g_4 )\Big)(\p_1+\p_2)\nn\\&\hspace{-12mm}
+\Big[x_3\Big((2+2\t_3-\g_3+2\t_4-\g_4)(2\t_4-\g_4)+(1+\t_3)(1+\t_3-\g_3)-\t_4(\t_4-\g_4)\Big)\nn\\&\hspace{-12mm}
+x_4\Big((2+2\t_3-\g_3+2\t_4-\g_4)(2\t_3-\g_3)+(1+\t_4)(1+\t_4-\g_4)-\t_3(\t_3-\g_3)\Big)\Big]\p_1\p_2\nn
\end{align}
The action of $W_{12}^{--}$ is
\begin{align}\label{Weqn5}
W_{12}^{--}\mathcal{I}_{\bs{\g}} = b_W(\bs{\g})\Big|_{\g_0\rightarrow\g_0-1,\,\g_1\rightarrow\g_1-1}\mathcal{I}_{\bs{\g}}\Big|_{\g_1\rightarrow\g_1-1,\,\g_2\rightarrow\g_2-1}
\end{align}
where, once again, the shift on the $b$-function derives from the fact that in projecting from GKZ variables to the physical hypersurface we chose $W_{12}^{--}$ to act on the unshifted integral $\mathcal{I}_{\bs{\g}}$ requiring us to shift the $\bs{\g}$ parameters present in \eqref{Weqn2}.  Explicitly, this is
\begin{align}
 b_W(\bs{\g})\Big|_{\g_0\rightarrow\g_0-1,\,\g_1\rightarrow\g_1-1}
 &=\frac{1}{16}\Big(\g_3^2+\g_4^2-(\g_0+\g_1+\g_2-2)^2\Big)^2-\frac{1}{4}\g_3^2\g_4^2.
\end{align}
Acting on the holographic contact diagram, from \eqref{icontIg} we again have
\begin{align}\label{Woncontdi}
W_{12}^{--}i_{[d;\,\Delta_1,\Delta_2,\Delta_3,\Delta_4]} = \frac{1}{4(\g_1-1)(\g_2-1)} b_W(\bs{\g})\Big|_{\g_0\rightarrow\g_0-1,\,\g_1\rightarrow\g_1-1}i_{[d;\,\Delta_1-1,\Delta_2-1,\Delta_3,\Delta_4]}. 
\end{align}
To our knowledge,  this is the first time an operator that shifts the 4-point contact diagram in this fashion has been identified.  We emphasise that the 3-point operator \eqref{curlyWdef}, when applied to 4-point contact diagrams, generates shifted contact diagrams but with derivative vertices and hence does not satisfy this requirement  \cite{Bzowski:2022rlz}.

\paragraph{Examples:}

Contact diagrams for which the Bessel functions have half-integer indices can be evaluated directly.  This yields many simple examples for which the action of $W_{12}^{--}$ can be checked.  For instance, with  $(d,\Delta_1,\Delta_2,\Delta_3,\Delta_4)=(5,3,4,3,4)$, we find
\begin{align}
i_{[5;3,4,3,4]} &= \frac{1}{p_t^3}\Big(p_1^2+p_3^2+2(p_2^2+p_4^2)+3(p_1+p_3)(p_2+p_4)+2p_1p_3+6p_2p_4\Big)
\end{align}
where $p_t=\sum_{j=1}^4p_j$, while the shifted integral with $(d,\Delta_1,\Delta_2,\Delta_3,\Delta_4)=(5,2,3,3,4)$ is
\begin{align}
i_{[5;2,3,3,4]} &= -\frac{(p_t+2p_4)}{p_1 p_t^3}.
\end{align}
Evaluating the action of $W_{12}^{--}$ in \eqref{4ptWv2} using \eqref{badnotation}, we can verify \eqref{Woncontdi}, namely
\[
W_{12}^{--}i_{[5;3,4,3,4]} =-\frac{63}{2}i_{[5;2,3,3,4]}.
\]

\subsubsection{Combinations of operators}

To round up our discussion of shift operators for contact diagrams, we have identified operators mapping
\begin{align}
&\p_i:\quad \g_0\rightarrow\g_0+1,\quad \g_i\rightarrow\g_i-1\qquad
\bar{\p}_i:\quad \g_0\rightarrow\g_0+1,\quad \g_i\rightarrow\g_i+1\nn\\
&\mathcal{C}_i:\quad \g_0\rightarrow\g_0-1,\quad\g_i\rightarrow\g_i+1\qquad
\bar{\mathcal{C}}_i:\quad \g_0\rightarrow\g_0-1,\quad\g_i\rightarrow\g_i-1\nn\\
&W_{ij}^{\sigma_i\sigma_j}:\quad  \g_i\rightarrow\g_i+\sigma_i,\quad \g_j\rightarrow\g_j+\sigma_j,\qquad \{\sigma_i,\sigma_j\}\in\pm 1.
\end{align}
Combining these allows us to construct yet further shifts, for example:
\begin{align}
\mathcal{C}_i\bar{\mathcal{C}}_i:\quad \g_0\rightarrow \g_0 - 2,
\qquad
\mathcal{C}_i\bar{\partial}_i: \quad \g_i\rightarrow\g_i+2,\qquad
\bar{\mathcal{C}}_i\partial_i: \quad \g_i\rightarrow\g_i-2.
\end{align}
Acting on the 3-point function specifically,
\[
\mathcal{C}_1W_{23}^{++}:\quad \g_0\rightarrow\g_0-1,\quad \g_i\rightarrow\g_i+1\quad\forall\,\,i=1,2,3
\]
which is equivalent to shifting $d\rightarrow d-2$ while preserving all operator dimensions $\Delta_i$.

Finally, one might wonder why all these operators produce a shift of two units:  why, for example, can one not construct an operator shifting $\g_0\rightarrow \g_0+1$ only, or just $\g_1\rightarrow \g_1+1$?  The absence of such operators can be traced to the spacing of the singular hyperplanes of the contact diagram,
 specifically the term $-2m$ appearing in the singularity condition \eqref{contsing}.  As $m\in\mathbb{Z}^+$, this means that the singularities are effectively spaced by two units.  Any operator that produced a shift of a single unit would require a $b$-function containing an infinite number of factors, since
there are infinitely many finite integrals  that are only one unit away from a singular integral.  (Namely, those for which $m$ is half-integer.)  As the number of factors in the $b$-function corresponds to the order of the differential operator, there is thus no single-shift operator of finite order.
In contrast, for an operator shifting by two units, the number of finite integrals that can be mapped to singular integrals is finite, and hence the $b$-functions and shift operators are also of finite order.

\subsection{Exchange diagrams}
\label{sec:exch}

Having analysed contact diagrams, we now turn to the  $s$-channel exchange diagrams \eqref{iexch}.
Rather than  constructing an explicit GKZ representation,  here we simply note that shifts of the form
\[
 i_{[d;\,\Delta_1, \Delta_2; \,\Delta_3, \Delta_4; \,\Delta_x]}\rightarrow  i_{[d;\,\Delta_1+\sigma_1, \Delta_2+\sigma_2; \,\Delta_3, \Delta_4; \,\Delta_x]}
\]
for any $ \{\sigma_1,\sigma_2\}\in \pm 1$ can be obtained by combining the 3- and 4-point   $W_{12}^{--}$ operators given in \eqref{W3ptphys} and \eqref{4ptWv2} with the $s$-channel Casimir operator.
As with contact diagrams, it is sufficient to focus on the case $\sigma_1=\sigma_2=-1$, since all remaining operators follow by shadow conjugation according to \eqref{Wshadows}.
We emphasise however that both the original and the shifted exchange diagrams we consider have purely {\it non-derivative} vertices. Moreover, any operator and spacetime dimensions are permitted, provided we work in dimensional regularisation where necessary to avoid divergences.

For purposes of disambiguation, let us define the operator
  \begin{align}
\mathcal{W}^{--}_{12} 
&=(d+2\t_1+2\t_2)(\p_1+\p_2)-2s^2 \p_1\p_2
\end{align}
where $d$ is the boundary spacetime dimension and $\p_i=\p/\p x_i$ with $x_i=p_i^2$ as usual.
This is simply the 3-point operator $-2W_{12}^{--}$   in \eqref{curlyWdef}, but with $p_3^2$ replaced by the Mandelstam variable $s^2=(\bs{p}_1+\bs{p}_2)^2$ as appropriate for acting on $s$-channel exchange diagrams.  (The factor of $-2$ is included for consistency with the $\mathcal{W}^{--}_{12}$ defined in \cite{Baumann:2019oyu, Bzowski:2022rlz}.)
In the following, we will then use $W_{12}^{--}$ to refer exclusively to the {\it 4-point} $W_{12}^{--}$  operator given in \eqref{4ptWv2}.

As shown in \cite{Bzowski:2022rlz},  
the action of $\mathcal{W}^{--}_{12}$  on an $s$-channel exchange diagram is to produce a {\it linear combination} of a shifted exchange and a shifted contact diagram:
\begin{align}\label{Wonexch}
\mathcal{W}^{--}_{12}\,i_{[d;\,\Delta_1,\Delta_2;\Delta_3,\Delta_4;\,\Delta_x]} &= 
\mathcal{N}_{exch.}\,
i_{[d;\,\Delta_1-1,\Delta_2-1;\Delta_3,\Delta_4;\,\Delta_x]}
 \nn\\& \quad 
 +\mathcal{N}_{cont.}\, i_{[d;\,\Delta_1-1,\Delta_2-1,\Delta_3,\Delta_4]}
\end{align} 
where the coefficients\footnote{ Where the shifted exchange diagram has a pole (or double pole) in dimensional regularisation,   one (or both) of the factors on the right-hand side of \eqref{Nexch} vanish, see \cite{Bzowski:2022rlz}.  
}
\begin{align}\label{Nexch}
\mathcal{N}_{exch.} &=\Big(\frac{d}{2}-2+\g_1
+\g_2+\g_x\Big)\Big(\frac{d}{2}-2+\g_1
+\g_2-\g_x\Big)\mathcal{N}_{cont.} \\
\mathcal{N}_{cont.} &=-\frac{1}{8(\g_1-1)(\g_2-1)}
\label{Ncont}
\end{align}
where $\g_i=\Delta_i-d/2$ and  $\g_x = \Delta_x-d/2$.
Thus, in order to go from an exchange diagram to shifted exchange diagram only, the shifted contact contribution in \eqref{Wonexch} must be subtracted.  

This can be accomplished in two steps.  First, the {\it unshifted} contact diagram is obtained by acting on the original  exchange diagram with the reduced Casimir operator, 
\begin{align} 
\hat{\mathcal{C}}_{12}\,i_{[d;\,\Delta_1,\Delta_2;\Delta_3,\Delta_4;\,\Delta_x]}  = i_{[d;\,\Delta_1,\Delta_2,\Delta_3,\Delta_4]}, \label{Casonex} 
\end{align}
where
\begin{align}\label{redCas}
\hat{\mathcal{C}}_{12}&=
2s^2\Big((\t_1+1-\g_1)\p_1+(\t_2+1-\g_2)\p_2\Big)-\Big(2\t_1+2\t_2-\g_1-\g_2+\frac{d}{2}\Big)^2+\g_x^2
\end{align}
with $\t_i=x_i\p_i$.
The action of this operator  on an $s$-channel exchange  is equivalent to that of the Casimir operator plus the square of the exchanged mass \cite{Bzowski:2022rlz}.\footnote{Specifically, 
$\hat{\mathcal{C}}_{12} = \tilde{\mathcal{C}}_{12}+m_x^2$ with $\tilde{\mathcal{C}}_{12}$ as defined in (6.44) of \cite{Bzowski:2022rlz} and $m_x^2=\g_x^2-d^2/4$.}
If desired, $\hat{\mathcal{C}}_{12}$ can be shorted using the identity
\[
0=\Big((\t_1+1-\g_1)\p_1-(\t_2+1-\g_2)\p_2\Big)i_{[d;\,\Delta_1,\Delta_2,\Delta_3,\Delta_4\,x\,\Delta_x]} 
\]
which corresponds to the difference of the Bessel operators acting on legs 1 and 2, {\it i.e.,} $K_1-K_2$ where $K_i = \p_{p_i}^2 + (1-2\g_i)p_i^{-1}\p_{p_i}$.  However,  \eqref{redCas} is symmetric under $1\leftrightarrow 2$.

For the second step, we now construct the shifted contact diagram using the 4-point $W_{12}^{--}$ operator defined in \eqref{4ptWv2}.
From \eqref{Woncontdi}, this has the action 
\begin{align}
&W_{12}^{--}i_{[d;\,\Delta_1,\Delta_2,\Delta_3,\Delta_4]} =\mathcal{N}_{W} \,\mathcal{N}_{cont.}\,i_{[d;\,\Delta_1-1,\Delta_2-1,\Delta_3,\Delta_4]}
\end{align}
with $\mathcal{N}_{cont.}$ from \eqref{Ncont} and 
\begin{align}
\mathcal{N}_W=
- \frac{1}{8}
\Big[\left(\g_3^2+\g_4^2-(d+\g_1+\g_2-2)^2\right)^2-4\g_3^2\g_4^2\Big].
\end{align}

Putting everything together, we find the operator
\begin{align}
\Omega_{12}^{--} = 
\mathcal{N}_W\,\mathcal{W}_{12}^{--} - W_{12}^{--}\hat{\mathcal{C}}_{12}
\end{align}
whose action is
\begin{align}\label{OmegaAction}
\Omega^{--}_{12}\,i_{[d;\,\Delta_1,\Delta_2;\,\Delta_3,\Delta_4;\,\Delta_x]} &= \mathcal{N}_W
\mathcal{N}_{exch.}\,
i_{[d;\,\Delta_1-1,\Delta_2-1;\,\Delta_3,\Delta_4;\,\Delta_x]}.
\end{align} 
This is therefore the desired operator mapping an exchange to a shifted exchange diagram.

Written out explicitly, with $\g_t=\sum_{j=1}^4\g_j$, we have
\begin{align}\label{OmegaOp}
\Omega^{--}_{12}&= - \frac{1}{8}
\Big[\left(\g_3^2+\g_4^2-(d+\g_1+\g_2-2)^2\right)^2-4\g_3^2\g_4^2\Big]\Big((d+2\t_1+2\t_2)(\p_1+\p_2)-2s^2 \p_1\p_2\Big)\nn\\
&\quad 
-\Big[
-(d+\t_1+\t_2)\Big((\t_3+\t_4)(\t_3+\t_4-\g_3-\g_4)\nn
\\ &\qquad\qquad\qquad \qquad\qquad 
+
\frac{1}{4}(2-d-\g_t+2\g_3)(2-d-\g_t+2\g_4 )\Big)(\p_1+\p_2)\nn\\&\hspace{-2mm}
+x_3\Big((2+2\t_3-\g_3+2\t_4-\g_4)(2\t_4-\g_4)+(1+\t_3)(1+\t_3-\g_3)-\t_4(\t_4-\g_4)\Big)\p_1\p_2\nn\\&\hspace{-2mm}
+x_4\Big((2+2\t_3-\g_3+2\t_4-\g_4)(2\t_3-\g_3)+(1+\t_4)(1+\t_4-\g_4)-\t_3(\t_3-\g_3)\Big)\p_1\p_2\Big]\times\nn\\
&\hspace{-2mm}
\times \Big[2s^2\Big((\t_1+1-\g_1)\p_1+(\t_2+1-\g_2)\p_2\Big)-\Big(2\t_1+2\t_2-\g_1-\g_2+\frac{d}{2}\Big)^2+\g_x^2\Big].
\end{align}

\paragraph{Examples:}  All exchange diagrams involving fields of $\Delta=2,3$ in $d=3$ were computed recently in \cite{Bzowski:2022rlz} and are available in the associated Mathematica package {\tt HandbooK.wl}.  These results enable many tests of the operator $\Omega_{12}^{--}$ in \eqref{OmegaOp} and its shadow conjugates
\begin{align}\label{Omshadows}
\Omega_{12}^{+-}&= x_1^{1+\g_1} \Omega_{12}^{--}\,  x_1^{-\g_1},\\
\Omega_{12}^{-+}\,  &= x_2^{1+\g_2} \Omega_{12}^{--},  x_2^{-\g_2},\\
\Omega_{12}^{++}\,  &= x_1^{1+\g_1} x_2^{1+\g_2} \Omega_{12}^{--}\,  x_1^{-\g_1}x_2^{-\g_2}.
\end{align}
For this, we work in the dimensionally regulated theory with $d\rightarrow d+2\ep$ and $\Delta_i\rightarrow \Delta_i+\ep$ for all $i=1,2,3,4,x$.   This scheme has the virtue of preserving the half-integer values of all Bessel function indices  $\g_i=\Delta_i-d/2$.  The simplest such example is 
\begin{align}
i_{[3;22;22; 2]}&=-\frac{1}{2s}\mathcal{D}^{(+)},\\
i_{[3;33;22; 2]}&= \frac{1}{2}(p_3+p_4)\Gamma(2\ep) p_T^{-2\ep}+\frac{1}{4s}(p_1^2+p_2^2-s^2)\mathcal{D}^{(+)}
\nn\\&\quad 
+\frac{1}{2}(p_1+p_2)\Big[\log\Big(\frac{l_{34+}}{p_T}\Big)+1\Big]+\frac{7}{8}(p_3+p_4)+O(\ep)
\end{align}
where 
\[
p_T = \sum_{i=1}^4 p_i,\qquad l_{ij\pm} = p_i+p_j\pm s,
\]
and 
\[
\mathcal{D}^{(+)}=\Li_2\Big(\frac{l_{12-}}{p_T}\Big)+\Li_2\Big(\frac{l_{34-}}{p_T}\Big)+\log\Big(\frac{l_{12+}}{p_T}\Big)\log\Big(\frac{l_{34+}}{p_T}\Big)-\frac{\pi^2}{6}.
\]
By direct differentiation, one then finds
\[
\Omega_{12}^{--} i_{[3;33;22;2]}= (90+261\ep+O(\ep^2)) i_{[3;22;22;2]} + O(\ep)
\]
consistent with \eqref{OmegaAction}.  Note  $\mathcal{N}_W
\mathcal{N}_{exch.}$ on the right-hand side here is expanded to order $\ep$ since $i_{[3;22;22;2]}$ has an $\ep^{-1}$ pole.  
We have performed similar checks for all other values of the $\Delta_i$ and $\Delta_x$, and for the shadow conjugated operators.

This ability to shift exchange diagrams directly to other exchange diagrams means that, instead of computing all the  diagrams individually, we can compute the easiest diagram (namely, $i_{[3;22;22;2]}$) to sufficiently high order in the regulator $\ep$, and then obtain all others by acting with $\Omega_{12}^{\sigma_1\sigma_2}$ and $\Omega_{34}^{\sigma_3\sigma_4}$.

\section{Creation operators for Feynman diagrams}\label{sec:Feynman}

In this section, we analyse various Feynman integrals presenting their GKZ representations,  their singularities, and the associated creation operators.  Many of the examples we study have  appeared in the recent works \cite{de_la_Cruz_2019, Klausen:2019hrg, Feng_2020, Chestnov:2022alh}.  Here, our focus will be the construction of the creation operators and ways to automate this computation using standard Gr{\"o}bner basis and convex hulling algorithms.

In all cases, we start with an $L$-loop scalar integral in the momentum representation
\begin{align}\label{feynprop}
I=\Big(\prod_{j=1}^{L} \int \frac{\dd^d \bs{k}_j}{(2\pi)^d}\Big)\frac{1}{P_1^{\g_1}\ldots P_N^{\g_N}},
\end{align}
where the propagators $P_i$ for $i=1,\ldots, N$ are raised to  generalised powers $\g_i$.
As shown in appendix \ref{LPrepapp} (see also \cite{de_la_Cruz_2019, Weinzierl:2022eaz}), the corresponding GKZ integral is 
\begin{align}\label{feyngkz}
\mathcal{I}_{\bs{\g}}=  \Big(\prod_{i=1}^N \int_0^\infty \dd z_i \,z_i^{\g_i-1}\Big)\,\mathcal{D}^{-\g_0},\qquad \g_0=\frac{d}{2}
\end{align}
where the denominator $\mathcal{D}$ is formed from the Lee-Pomeransky denominator $\mathcal{G}=\mathcal{U}+\mathcal{F}$, the sum of first and second Symanzik polynomials, by replacing the coefficient of every term with an independent variable $x_k$.
The Feynman integral \eqref{feynprop} now corresponds to
\[\label{feyntogkz}
I=c_{\bs{\g}}\mathcal{I}_{\bs{\g}},\qquad c_{\bs{\g}} = \frac{(4\pi)^{-L\g_0}\Gamma\left(\g_0\right)}{\Gamma\left((L+1)\g_0-\g_t\right)\prod_{i=1}^N\Gamma(\g_i)},\qquad \g_t=\sum_{i=1}^N \g_i
\] 
with the $x_k$ restored to their physical (Lee-Pomeransky) values. 
Knowing the coefficient $c_{\bs{\g}}$ enables the action of a creation operator on the GKZ integral $\mathcal{I}_{\bs{\g}}$ to be related to its action on the  Feynman integral $I$.

\subsection{Bubble diagram}

First, we consider the 1-loop bubble integral with propagators of mass $m_1$ and $m_2$.
To warm-up, we begin with the single-mass case $(m_1,m_2)=(0,m)$ before turning to general masses.
The fully massless case $m_1=m_2=0$ is trivial (evaluating to a simple power of the momentum) and will be omitted.

\subsubsection{1-mass bubble}
\begin{figure}[t]
\centering
\begin{tikzpicture}
\draw [dashed, thick](0,1)circle (1);
\draw[thick,->](-1.5,1)--(-1.2,1);
\draw[thick] (-1.2,1)--(-1.0,1);
\draw[thick](1.0,1)--(1.5,1);
\draw[thick] (1,1) arc (0:-180:1);
\node[text width=0.5cm, text centered ] at (0,1.7) {$1$};
\node[text width=0.5cm, text centered ] at (0,-0.3) {$2$};
\node[text width=0.5cm, text centered ] at (-1.7,1.0) {$\bs{p}$};
\end{tikzpicture} 
\caption{The single-mass bubble integral \eqref{1massbub1}, with massless and massive propagators represented by dashed and undashed lines respectively.}
\label{onemass-bubble}
\end{figure}
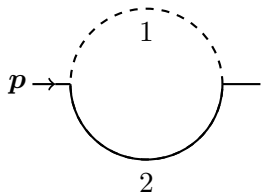
The single-mass bubble diagram
\begin{align}\label{1massbub1}
I= \int \frac{\dd^d \bs{k}}{(2\pi)^d}\frac{1}{\bs{k}^{2\g_1}\left((\bs{p}-\bs{k})^2+m^2\right)^{\g_2}},
\end{align}
corresponds via \eqref{feyntogkz} to the GKZ integral \cite{de_la_Cruz_2019, Klausen:2019hrg}
\begin{align}\label{feyn3gkz}
\mathcal{I}_{\bs{\g}}= \int_{\mathbb{R}_+^2} \dd z_1 \dd z_2\frac{z_1^{\g_1-1} z_2^{\g_2-1}}{(x_1z_1+x_2z_2+x_3 z_1 z_2+x_4 z_2^2)^{\g_0}}
\end{align}
evaluated on the physical hypersurface
\[\label{1bubphyshyp}
(x_1,x_2,x_3,x_4)=(1,1,p^2+m^2,m^2).
\]
In this simple case,  the GKZ integral can of course be evaluated directly,
\begin{align}\label{1massbubbleeval}
\mathcal{I}_{\bs{\g}}&=
 \frac{\Gamma(\g_1)\Gamma \left(\g_0 -\g_1\right)
 \Gamma \left(\g_1+\g_2-\g_0\right)\Gamma \left(2\g_0-\g_1-\g_2\right)}{\Gamma(\g_0)^2} \nn\\&\qquad \times m^{2(\g_0-\g_1-\g_2)}{}_2F_1\left(\g_1,\g_1+\g_2-\g_0;\g_0; -\frac{p^2}{m^2}\right),
\end{align}
enabling the action of all creation operators to be verified.
The $\mathcal{A}$-matrix  is
\begin{align}
\mathcal{A}=\left(
\begin{array}{cccc}
 1 & 1 & 1 & 1 \\
 1 & 0 & 1 & 0 \\
 0 & 1 & 1 & 2 \\
\end{array}
\right),
\label{pol-one-mass-bubble}
\end{align}
and from its kernel, we find a single toric equation 
\[
0=(\p_1\p_4-\p_2\p_3)\mathcal{I}_{\bs{\g}}.
\]
The Euler equations can be read off from the rows of the $\mathcal{A}$-matrix, 
\[\label{eul1mass}
0=(\g_0+\t_1+\t_2+\t_3+\t_4)\mathcal{I}_{\bs{\g}},\quad 0=(\g_1+\t_1+\t_3)\mathcal{I}_{\bs{\g}},\quad 0=(\g_2+\t_2+\t_3+2\t_4)\mathcal{I}_{\bs{\g}}.
\]
The (rescaled) Newton polytope derived from the column vectors of the $\mathcal{A}$-matrix is the parellelogram shown in figure \ref{fig:1-mass-bubble}. From \eqref{hypsings}, the GKZ integral is then singular for
\[\label{sing1mass}
2\g_0-\g_1-\g_2=-k_1,\quad \g_1+\g_2-\g_0=-k_2,\quad \g_0-\g_1=-k_3,\quad \g_1=-k_4,\quad k_i\in \mathbb{Z}^+
\]
consistent with the poles of the gamma functions in  \eqref{1massbubbleeval}.

\begin{figure}[t]
\centering

\begin{tikzpicture}[scale=1.8]
\draw[->,thick,gray] (-1/3,0) -- (2.4,0); 
\draw[->,thick,gray] (0,0) -- (0,2.4); 
\draw[thick,gray] (0,0) -- (0,-1/3); 
\draw[white] (0,-1/3) -- (0,-1/3-0.08);  

\draw[thick,black,fill=YellowGreen, fill opacity=0.2] (1,0)--(0,1) -- (0,2) -- (1,1) --cycle;

     \draw node[left] at (0,1) {($0,\gamma_0$)};   
   \draw node[below] at (1,0) {($\gamma_0,0$)};
   \draw node[left] at (0,2) {($0,2\gamma_0$)};
   \draw node[right] at (1,1) {($\gamma_0,\gamma_0$)};
    \draw node[right] at (2.4,0) {$\gamma_1$};
     \draw node[above] at (0,2.4) {$\gamma_2$};
   
   \end{tikzpicture}
\caption{The rescaled Newton polytope associated to the 1-mass bubble integral \eqref{feyn3gkz}.}
\label{fig:1-mass-bubble}
\end{figure}
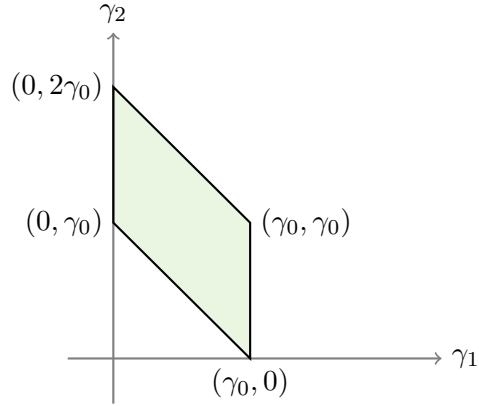
The annihilation operators $\partial_j$ send $\bs{\gamma}\rightarrow\bs{\gamma}'$ while the creation operators $\mathcal{C}_j$ send $\bs{\gamma}'\rightarrow\bs{\gamma}$
where, for each $j$, these parameters are related by  
\begin{align}\label{MlessbCshifts}
j&=1: & \gamma_0'&=\gamma_0+1,&\gamma_1'&=\gamma_1+1, & \gamma_2'&=\gamma_2, \nn\\
j&=2: & \gamma_0'&=\gamma_0+1, & \gamma_1'&=\gamma_1,&\gamma_2'&=\gamma_2+1,\nn\\
j&=3: & \gamma_0'&=\gamma_0+1, & \gamma_1'&=\gamma_1+1, & \gamma_2'&=\gamma_2+1, \nn\\
j&=4: & \gamma_0'&=\gamma_0+1,&\gamma_1'&=\gamma_1, & \gamma_2'&=\gamma_2+2.
\end{align}
Knowing the location of the singular  hyperplanes and the shifts generated by the creation operators, 
the $b$-functions can be  constructed according to \eqref{stdbfn},
\begin{align}
&b_1=\g_1(2\g_0-\g_1-\g_2),\nn\\
&b_2=(\g_0-\g_1)(2\g_0-\g_1-\g_2),\nn\\
&b_3=\g_1(\g_1+\g_2-\g_0),\nn\\
&b_4=(\g_0-\g_1)(\g_1+\g_2-\g_0).
\end{align}
Their zeros serve to cancel the singularities that arise whenever the action of a creation operator shifts us from a finite to a singular integral.
For example, $\mathcal{C}_4$ shifts $k_2\rightarrow k_2+1$ and $k_3\rightarrow k_3+1$ which, according to \eqref{sing1mass}, generates a singular integral when acting on finite integrals with either $k_2=-1$ or $k_3=-1$.  These singularities, however, are cancelled by the zeros of $b_4$.

Using the DWI and the Euler equations \eqref{eul1mass},  we can now re-write
\[\label{credef}
\mathcal{C}_j\p_j \mathcal{I}_{\bs{\g}} = b_j(\bs{\g}) \mathcal{I}_{\bs{\g}}=B_j(\t) \mathcal{I}_{\bs{\g}}
\]
where 
 \begin{align}
&B_1=(\t_1+\t_3)(\t_1+\t_2)=(\t_1+\t_2+\t_3)\t_1+\t_2\t_3,\nn\\
&B_2=(\t_2+\t_4)(\t_1+\t_2)=(\t_1+\t_2+\t_4)\t_2+\t_1\t_4,\nn\\
&B_3=(\t_1+\t_3)(\t_3+\t_4)=(\t_1+\t_3+\t_4)\t_3+\t_1\t_4,\nn\\
&B_4=(\t_2+\t_4)(\t_3+\t_4)=(\t_4+\t_2+\t_3)\t_4+\t_2\t_3.
\end{align}
By inspection, every term in $B_j$ either contains an explicit factor of $\p_j$ already  through $\t_j$, or else such a factor can be introduced using the toric equations.  In $B_1$ and $B_4$, for instance, we replace $\t_2\t_3=x_2x_3\p_2\p_3\rightarrow x_2x_3\p_1\p_4$.   This enables the $B_j$ to be factored (modulo the toric equations) in the form \eqref{credef} yielding the creation operators 
\begin{align}\label{bubcre}
\mathcal{C}_1 &=x_1(1+\t_1+\t_2+\t_3)+x_2x_3\p_4,\nn\\
\mathcal{C}_2 &=x_2(1+\t_1+\t_2+\t_4)+x_1x_4\p_3,\nn\\
\mathcal{C}_3 &=x_3(1+\t_1+\t_3+\t_4)+x_1x_4\p_2,\nn\\
\mathcal{C}_4 &=x_4(1+\t_2+\t_3+\t_4)+x_2x_3\p_1.
\end{align}
These creation operators act on the full GKZ integral \eqref{feyn3gkz}.
To obtain their counterparts acting on the  Feynman integral \eqref{1massbub1}, we must project to the physical hypersurface \eqref{1bubphyshyp}.  
Given the form of the operators \eqref{bubcre}, it is useful to first simplify using the DWI to
\begin{align}\label{bubcre2}
\mathcal{C}_1 &=x_1(1-\g_0-\t_4)+x_2x_3\p_4,\nn\\
\mathcal{C}_2 &=x_2(1-\g_0-\t_3)+x_1x_4\p_3,\nn\\
\mathcal{C}_3 &=x_3(1-\g_0-\t_2)+x_1x_4\p_2,\nn\\
\mathcal{C}_4 &=x_4(1-\g_0-\t_1)+x_2x_3\p_1.
\end{align}
Next, as all factors of $x_j$ are placed to the left of all derivatives, we set
\[\label{proj}
(x_1,x_2,x_3,x_4)\rightarrow (1,1,m^2+p^2,m^2)
\]
and replace all derivatives lying in directions off this hypersurface (namely $\p_1$ and $\p_2$)  with those lying along the hypersurface.  This can be accomplished using the Euler equations \eqref{eul1mass}  projected according to \eqref{proj}, namely
\[
\p_1\rightarrow -\g_1-(m^2+p^2)\p_3,\qquad
\p_2\rightarrow -\g_2-(m^2+p^2)\p_3-2m^2 \p_4.
\]
In addition, we use the chain rule 
with $p^2=x_3-x_4$ and $m^2=x_4$ 
to replace 
\[
\p_3  = \p_{p^2},\qquad \p_4 = -\p_{p^2}+\p_{m^2}.
\]
This yields
\begin{align}
\mathcal{C}_1^{\mathrm{ph}} &=1-\g_0+p^2\p_{m^2}-\t_{p^2},\nn\\
\mathcal{C}_2^{\mathrm{ph}} &=1-\g_0-\t_{p^2},\nn\\
\mathcal{C}_3^{\mathrm{ph}} &=(1-\g_0)m^2 + (1-\g_0+\g_2)p^2+(p^2-m^2)\t_{p^2}+2p^2 \t_{m^2}\nn\\
\mathcal{C}_4^{\mathrm{ph}} &=(1-\g_0)m^2+\g_1p^2 -(p^2+m^2)\t_{p^2}.
\end{align}
From \eqref{creationeq}, the  action on the projected GKZ integral is then 
\[
\mathcal{C}_1^{\mathrm{ph}}\mathcal{I}_{\gamma_0',\gamma_1',\gamma_2'}(p^2,m^2)=-\gamma_0^{-1}b_1\mathcal{I}_{\gamma_0,\gamma_1,\gamma_2}=-\gamma_0^{-1}\g_1(2\g_0-\g_1-\g_2)\mathcal{I}_{\gamma_0,\gamma_1,\gamma_2}
\]
and similarly for the other operators.  
When acting the original Feynman integral, there is an additional  factor of $c_{\bs{\g}'}/c_{\bs{\g}}$ from \eqref{feyntogkz} we must  take into account giving
\[
\mathcal{C}_1^{\mathrm{ph}}I_{\gamma_0',\gamma_1',\gamma_2'}(p^2,m^2)=-\frac{1}{4\pi}I_{\gamma_0,\gamma_1,\gamma_2}.
\]
All these results can be checked directly using \eqref{1massbubbleeval} and the standard shift operators for the ${}_2F_1$ (see {\it e.g.,} \cite{NIST:DLMF}).

\subsubsection{Massive bubble}
Next we consider the full bubble graph with general masses $m_1$ and $m_2$,
\begin{align}\label{massbub1}
I= \int \frac{\dd^d \bs{k}}{(2\pi)^d}\frac{1}{(\bs{k}^2+m_1^2)^{\g_1}\left((\bs{p}-\bs{k})^2+m_2^2\right)^{\g_2}}.
\end{align}
The corresponding GKZ integral is
\begin{align}\label{massbubgkz}
\mathcal{I}_{\bs{\g}}=\int_{\mathbb{R}_+^2} \dd z_1 \dd z_2\frac{z_1^{\g_1-1} z_2^{\g_2-1}}{(x_1z_1+x_2z_2+x_3 z_1^2+x_4 z_2^2+x_5 z_1 z_2)^{\g_0}},
\end{align}
where $\g_0=d/2$ and the physical hypersurface is 
\[\label{physhypfullbub}
(x_1,x_2,x_3,x_4,x_5)=(1,1,m_1^2,m_2^2,m_1^2+m_2^2+p^2).
\]
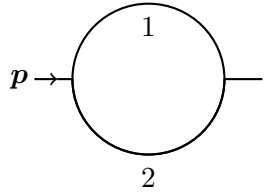
\begin{figure}
\centering
\begin{tikzpicture}
\draw [thick](0,1)circle (1);
\draw[thick,->](-1.5,1)--(-1.2,1);
\draw[thick] (-1.2,1)--(-1.0,1);
\draw[thick](1.0,1)--(1.5,1);
\draw[thick] (1,1) arc (0:-180:1);
\node[text width=0.5cm, text centered ] at (0,1.7) {$1$};
\node[text width=0.5cm, text centered ] at (0,-0.3) {$2$};
\node[text width=0.5cm, text centered ] at (-1.7,1.0) {$\bs{p}$};
\end{tikzpicture} 
\caption{The massive bubble integral \eqref{massbubgkz}.}
\label{mass-bubble}
\end{figure}
From the kernel of the $\mathcal{A}$-matrix 
\begin{align}
\mathcal{A}=\left(
\begin{array}{ccccc}
 1 & 1 & 1 & 1 & 1\\
 1 & 0 & 2 & 0 & 1\\
 0 & 1 & 0 & 2 & 1\\
\end{array}
\right)
\label{mass-bubble-mat}
\end{align}
we obtain the toric equations
\[
0=(\p_3\p_4-\p_5^2)\mathcal{I}_{\bs{\g}},\qquad 0=(\p_2\p_3-\p_1\p_5)\mathcal{I}_{\bs{\g}},\qquad 0=(\p_1\p_4-\p_2\p_5)\mathcal{I}_{\bs{\g}},
\]
while the DWI and the Euler equations can be read off from the rows:
\[
0=\Big(\g_0+\sum_{i=1}^5\t_i\Big)\mathcal{I}_{\bs{\g}},\qquad 0=(\g_1+\t_1+2\t_3+\t_5)\mathcal{I}_{\bs{\g}},\qquad 0=(\g_2+\t_2+2\t_4+\t_5)\mathcal{I}_{\bs{\g}}.
\]
The rescaled Newton polytope corresponding to this $\mathcal{A}$-matrix is the quadrilateral shown in figure \ref{fig:2-mass-bubble}. The singular hyperplanes lie parallel to and outside the facets of this polytope:
\[
\g_1=-k_1,\quad \g_2=-k_2,\quad 2\g_0-\g_1-\g_2=-k_3,\quad -\g_0+\g_1+\g_2=-k_4,\qquad k_i\in \mathbb{Z}^+.
\]

\begin{figure}[t]
\centering

\begin{tikzpicture}[scale=1.8]
\draw[->,thick,gray] (-1/3,0) -- (2.4,0); 
\draw[->,thick,gray] (0,-1/3) -- (0,2.4);

\draw[thick,black,fill=YellowGreen, fill opacity=0.2] (1,0)--(0,1) -- (0,2) -- (2,0) --cycle;

     \draw node[left] at (0,1) {($0,\gamma_0$)};   
   \draw node[below] at (1,0) {($\gamma_0,0$)};
   \draw node[left] at (0,2) {($0,2\gamma_0$)};
   \draw node[below] at (2,0) {($2\gamma_0,0$)};
    \draw node[right] at (2.4,0) {$\gamma_1$};
     \draw node[above] at (0,2.4) {$\gamma_2$};
   
   \end{tikzpicture}
\caption{The rescaled Newton polytope associated with the massive bubble GKZ integral \eqref{massbubgkz}.}
\label{fig:2-mass-bubble}
\end{figure}
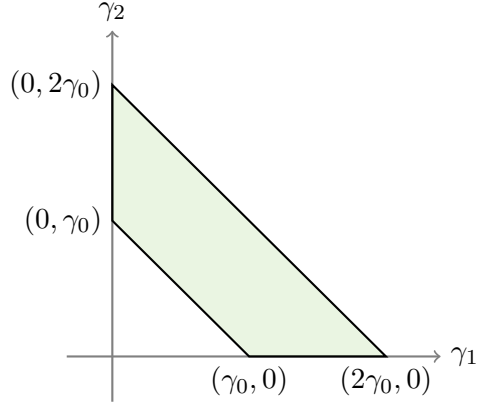

For illustration, let us now discuss the creation operator $\mathcal{C}_5$.  All others can be obtained by similar computations.  The annihilator $\partial_5$ sends $\bs{\gamma}\rightarrow\bs{\gamma}'$ where
\[
\p_5:\qquad \gamma_0'=\gamma_0+1,\qquad \gamma_1'=\gamma_1+1,\qquad \gamma_2'=\gamma_2+1,
\]
while the creation operator $\mathcal{C}_5$ acts  in the opposite direction sending $\bs{\gamma}'\rightarrow\bs{\gamma}$. 
Given this shift and the location of the singular hyperplanes, we identify  the $b$-function as
\[
b_5=\g_1\g_2(\g_0-\g_1-\g_2).
\]
Using DWI and Euler equations, this can be re-written  in terms of Euler operators as
\[
B_5=(\t_1+2\t_3+\t_5)(\t_2+2\t_4+\t_5)(\t_3+\t_4+\t_5).
\]
This expression can now be factorised as $\mathcal{C}_5\p_5$ by expanding out and using the  toric equations to replace any terms not involving $\p_5$ with equivalent terms containing this factor.
Stripping off the factor of $\p_5$ then yields  $\mathcal{C}_5$ in GKZ variables,
\begin{align}
\mathcal{C}_5&=x_5\big[2\t_3^2+2\t_4^2+8\t_3\t_4+3(\t_3+\t_4)(1+\t_5)+(1+\t_5)^2+\t_2(1+3\t_3+\t_4+\t_5)\nn\\& \qquad+\t_1(1+\t_2+\t_3+3\t_4+\t_5)
\big]
+x_2x_3\p_1(1+\t_1+2\t_3)
+x_1x_4\p_2(1+\t_2+2\t_4)\nn\\& \quad +2x_3x_4\p_5(4+\t_1+\t_2+2\t_3+2\t_4).
\end{align}
To project this operator to the physical hypersurface \eqref{physhypfullbub}, 
we first use the Euler equations to replace
\[\label{Eulrepl}
\t_1\rightarrow -\g_1-2\t_3-\t_5,\qquad
\t_2\rightarrow -\g_2-2\t_4-\t_5.
\]
The two occurrences of $\p_1$ and $\p_2$ can be dealt with similarly by writing $\p_i =(x_i)^{-1}\t_i$ for $i=1,2$ and using \eqref{Eulrepl}.  Then, setting 
$(x_1,x_2,x_3,x_4,x_5)\rightarrow (1,1,x_3,x_4,x_5)$, 
we obtain
\begin{align}
\mathcal{C}_5^{\mathrm{ph}}=&x_5\bigl[(1-\g_1)(1-\g_2)+
(1-\g_1-\g_2-\t_5)(\t_3+\t_4)
\bigr]
\nn\\&
+x_3(\g_1+2\t_3+\t_5)(\g_1-1+\t_5)
+x_4(\g_2+2\t_4+\t_5)(\g_2-1+\t_5)\nn\\&\quad
-2x_3x_4\p_5(\g_1+\g_2-4+2\t_5)
\end{align}
The remaining variables here are all physical since
\[
(x_3,x_4,x_5)=(m_1^2,m_2^2,m_1^2+m_2^2+p^2)
\]
and
\begin{align}
&\p_3=\p_{m_1^2}-\p_{p^2}, \qquad
\p_4=\p_{m_2^2}-\p_{p^2},\qquad
\p_5=\p_{p^2}.
\end{align}

\subsection{Massive triangle}\label{toricidealdisc}

Since the {\it massless} triangle integral is equivalent \cite{Bzowski:2013sza} to the 3-point contact Witten diagram studied in sections \ref{sec:3KC} and \ref{sec:3KW},  let us examine here the massive triangle integral
\begin{align}\label{masstri}
I= \int \frac{\dd^d \bs{k}}{(2\pi)^d}\frac{1}{(\bs{k}^2+m_3^2)^{\g_3}\left((\bs{k}-\bs{p}_1)^2+m_2^2\right)^{\g_2}\left((\bs{k}+\bs{p}_2)^2+m_1^2\right)^{\g_1}}.
\end{align}
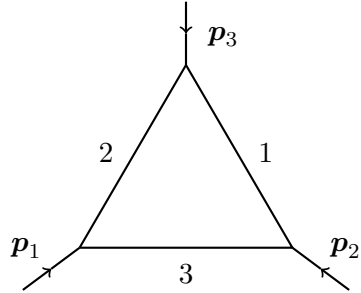
\begin{figure}[t]
\centering
\begin{tikzpicture}[scale=1.4]
   \draw[thick] (-1,0)--(1,0) -- (0,1.73); 
   \draw[,thick] (0,1.73) -- (0,1.73+0.3);
   \draw[<-,thick] (0,1.73+0.3)--   (0,1.73+0.6) ;
   \draw[thick] (0,1.73) -- (-1,0);
  \draw[thick] (-1,0)--(-1.2666,-0.2);
  \draw[<-,thick](-1.2666,-0.2)-- (-4/3-0.2,-0.4) ;
   \draw[thick] (1,0) -- (1.2666,-0.2);
   \draw[<-,thick] (1.2666,-0.2)--(4/3+0.2,-0.4) ;
   
   \node[text width=0.5cm, text centered ] at (-1.5,0) {$\bs{p}_1$};
   \node[text width=0.5cm, text centered ] at (1.5,0) {$\bs{p}_2$};
   \node[text width=0.5cm, text centered ] at (1/4+0.1,2) {$\bs{p}_3$};
   \node[text width=0.5cm, text centered ] at (-3/4,0.9) {$2$};
   \node[text width=0.5cm, text centered ] at (3/4,0.9) {$1$};
   \node[text width=0.5cm, text centered ] at (0,-1/4) {$3$};

\end{tikzpicture}
\caption{The massive triangle graph \eqref{masstri}.}
\label{fig:mass-tri}
\end{figure}
The corresponding GKZ integral according to \eqref{feyntogkz} is
\begin{align}\label{masstrigkz}
\mathcal{I}_{\bs{\g}}= \int_{\mathbb{R}_+^3} \dd z_1 \dd z_2 \dd z_3 \,z_1^{\g_1-1} z_2^{\g_2-1}z_3^{\g_3-1}\mathcal{D}^{-\g_0},
\end{align}
where 
\begin{align}\label{gkzden-masstri}
\mathcal{D}=&x_1z_1+x_2z_2+x_3z_3+x_4z_2z_3+x_5z_1z_3+x_6z_1z_2+x_7z_1^2+
x_8z_2^2+x_9z_3^2
\end{align}
with $\g_0=d/2$. The physical hypersurface is
\[\label{masstriphys}
\bs{x}= (1,1,1,p_1^2+m_2^2+m_3^2,p_2^2+m_1^2+m_3^2,
p_3^2+m_1^2+m_2^2,m_1^2,m_2^2,m_3^2)
\]
and the $\mathcal{A}$-matrix  reads
\begin{align}
\mathcal{A}=\left(
\begin{array}{ccccccccc}
 1 & 1 & 1 & 1 & 1 & 1 & 1 & 1 & 1\\
 1 & 0 & 0 & 0 & 1 & 1 & 2 & 0 & 0\\
 0 & 1 & 0 & 1 & 0 & 1 & 0 & 2 & 0\\
 0 & 0 & 1 & 1 & 1 & 0 & 0 & 0 & 2\\
\end{array}
\right).\label{masstri-mat}
\end{align}
For larger $\mathcal{A}$-matrices such as this one, it is useful to automate the calculation of creation operators using 
Gr{\"o}bner basis algorithms. 
To this end, 
in place of the five independent toric equations spanning the kernel of the $\mathcal{A}$-matrix, we will use instead the full set of 17 (non-independent) toric equations forming the toric ideal:\footnote{
These can be obtained using the Singular code \cite{singular}:
\vspace{-1mm}
\begin{alltt}
LIB "toric.lib";\\
ring r=0,(x1,x2,x3,x4,x5,x6,x7,x8,x9),dp;\\
intmat A[4][9]=1,1,1,1,1,1,1,1,1,1,0,0,0,1,1,2,0,0,0,1,0,1,0,1,0,2,0,0,0,1,1,1,0,0,0,2;\\
ideal I=toric{\_}ideal(A,"du");\\
I;
\end{alltt}
} 
\begin{align}\label{Itoric}
I_{toric}=\{&\p_2\p_6-\p_1\p_8,~\p_7\p_8-\p_6^2,~\p_1\p_6-\p_2\p_7,~\p_1\p_5-\p_3\p_7,~\p_3^2\p_9-\p_7,~\p_1\p_4-\p_3\p_6,\nn\\&~\p_4^2\p_9-\p_8,~\p_5\p_6-\p_4\p_7,~\p_4\p_6-\p_5\p_8,~\p_4\p_5\p_9-\p_6,~\p_3^2\p_8\p_9-\p_2^2,~\p_3^2\p_7\p_9-\p_1^2,\nn\\&~\p_3^2\p_6\p_9-\p_1\p_2,~\p_3\p_5\p_9-\p_1,~\p_3\p_4\p_9-\p_2,~\p_2\p_5-\p_3\p_6,~\p_2\p_4-\p_3\p_8\}.
\end{align}
Each entry here corresponds to a toric equation, for example the first is $0=(\p_2\p_6-\p_1\p_8)\mathcal{I}_{\bs{\g}}$ and similarly for the rest.
Since all the partial derivatives commute, these equations can be treated as a system of polynomial equations by mapping $\p_i$ to an ordinary commutative variable $y_i$.  As we will show below, this enables the factorisation step to be handled via ordinary commutative Gr{\"o}bner basis methods.  (For alternative constructions of creation operators  using {\it non-commutative} Gr{\"o}bner bases over the Weyl algebra, see  \cite{saito_sturmfels_takayama_1999}.)

The DWI and Euler equations for this $\mathcal{A}$-matrix are
\begin{align}
&0=\Big(\g_0+\sum_{i=1}^9\t_i\Big)\mathcal{I}_{\bs{\g}}, \nn\\ &0=(\g_1+\t_1+\t_5+\t_6+2\t_7)\mathcal{I}_{\bs{\g}},\nn\\&0=(\g_2+ \t_2+\t_4+\t_6+2\t_8)\mathcal{I}_{\bs{\g}}, \nn\\
 &0=(\g_3+\t_3+\t_4+\t_5+2\t_9)\mathcal{I}_{\bs{\g}}
\end{align}
and the corresponding Newton polytope  is depicted in figure \ref{fig:masstri}.
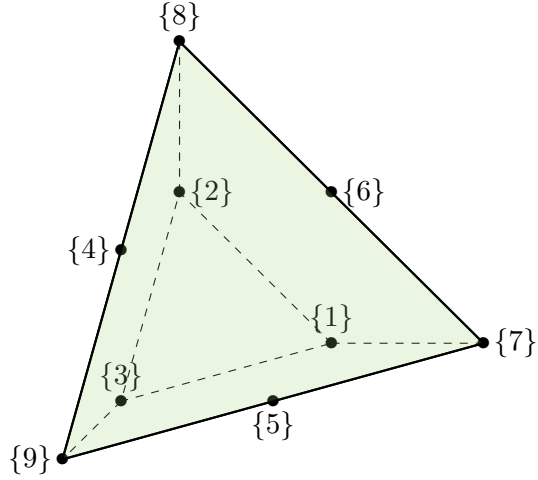
\begin{figure}[t]
\centering
\begin{tikzpicture}[scale=2.0]
   \draw[thick,black] (0,0,2)--(2,0,0) -- (0,2,0)  ; 
   \draw[thick,black] (0,2,0) -- (0,0,2) ;
    \draw[style=dashed, color=black] (0,1,0) --(0,2,0);
    \draw[style=dashed, color=black] (0,1,0) --(0,2,0);
    \draw[style=dashed, color=black] (1,0,0) --(0,1,0)-- (0,0,1)-- (1,0,0);
     \draw[style=dashed, color=black] (1,0,0) --(2,0,0);
     \draw[style=dashed, color=black] (0,0,1) --(0,0,2);
    \draw node[right, above] at (1,0,0) {\{1\}};      
     \draw[color=black] node at (1,0,0) {$\bullet$}; 
   \draw node[right] at (0,1,0) {\{2\}};
   \draw[color=black] node at (0,1,0) {$\bullet$}; 
   \draw node[above] at (0,0,1) {\{3\}};
   \draw[color=black] node at (0,0,1) {$\bullet$}; 
   \draw node[left] at (0,1,1) {\{4\}};
   \draw[color=black] node at (0,1,1) {$\bullet$}; 
   \draw node[below] at (1,0,1) {\{5\}};
   \draw[color=black] node at (1,0,1) {$\bullet$}; 
    \draw node[right] at (1,1,0) {\{6\}};
    \draw[color=black] node at (1,1,0) {$\bullet$}; 
    \draw node[right] at (2,0,0) {\{7\}};
    \draw[color=black] node at (2,0,0) {$\bullet$}; 
    \draw node[above] at (0,2,0) {\{8\}};
    \draw[color=black] node at (0,2,0) {$\bullet$}; 
    \draw node[left] at (0,0,2) {\{9\}};
    \draw[color=black] node at (0,0,2) {$\bullet$}; 

    \draw[thick,black,fill=YellowGreen, fill opacity=0.2] (0,0,2)--(2,0,0) -- (0,2,0) --cycle;
    
   \end{tikzpicture}
   \caption{The Newton polytope associated to the denominator of the massive triangle integral \eqref{gkzden-masstri}. The label \{i\} denotes the vector defined by the $i$th column of the $\mathcal{A}$-matrix.}
\label{fig:masstri}
\end{figure}
From its facets, we obtain the  singularity conditions
\begin{align}
\begin{array}{llll}
\{23489\}:& \g_1=-k_1, &\{1379\}:& \g_2=-k_2,\\
\{1278\}:& \g_3=-k_3, &\{123\}:& \g_0-\g_1-\g_2-\g_3=+k_4,\\
\{4789\}:& \g_1+\g_2+\g_3-2\g_0=+k_5, \,\,\,\,\,& & 
\end{array}
\end{align}
where all $k_i\in \mathbb{Z}^+$.  For the facet $\{123\}$, for example, we have the outward-pointing normal $\bs{n}=(-1,-1,-1)$ which leads via \eqref{Jshift} to the spacing  of singular hyperplanes $\delta^{(J)}=1$.

Let us now compute the creation operator $\mathcal{C}_4$ which acts on the GKZ integral to shift $\bs{\g}'\rightarrow \bs{\g}$, where
\[
\g_0'=\g_0+1,\qquad \g_1'=\g_1, \qquad \g_2'=\g_2+1,\qquad \g_3'=\g_3+1.
\]
From these parameter shifts and the location of the singular hyperplanes,  the corresponding $b$-function is
\[
b_4=\g_2\g_3(\g_0-\g_1-\g_2-\g_3).
\]
Using the DWI and Euler equations, this can be re-expressed as 
\[
B_4=\Big(\sum_{i=4}^9\t_i\Big)(\t_2+\t_4+\t_6+2\t_8)(\t_3+\t_4+\t_5+2\t_9).
\]
Our goal is now to  factorise $B_4$ as $\mathcal{C}_4\p_4$ using the toric equations.  
To achieve this in an automated fashion, we decompose $B_4$ over the Gr{\"o}bner basis formed from the toric ideal \eqref{Itoric} and $\p_4$.
Treating the partial derivatives as ordinary commutative variables and computing 
 this Gr{\"o}bner basis, we obtain 
\[
\bs{g}=\{\p_4,~\p_2,~\p_8,~\p_3^2\p_7\p_9-\p_1^2,~\p_6,~\p_1\p_5-\p_3\p_7,~\p_3\p_5\p_9-\p_1,~\p_5^2\p_9-\p_7\}.
\]
Expanding out $B_4$ and rewriting all terms in the form \eqref{Bterm} so that all partial derivatives $\p_i$ lie to the right of all $x_i$, we can now decompose each term of $B_4$ in this Gr{\"o}bner basis.
This yields 
\[
B_4=\bs{Q}\cdot \bs{g}=Q_1\p_4+\sum_{i=2}^{8}Q_ig_i.
\]
where the coefficients $Q_i$ are polynomials in the $x_j$ and $\t_j$ (with $j=1,..,9$) which can be computed automatically.\footnote{
In Mathematica, for example, after writing $B_4$  in the form \eqref{Bterm} with all derivatives to the right, we replace all  $\p_i$  (both in $B_4$ and in the toric ideal) by commutative variables $y[i]$.  The code
\vspace{-1mm}
\begin{alltt}
v = \lcb y[5], y[6], y[7], y[8], y[9], y[1], y[2], y[3], y[4]\rcb;\\ 
toric = \lcb y[2] y[6] - y[1] y[8], y[6]$^2$ - y[7] y[8], y[1] y[6] - y[2] y[7], ...\rcb;\\
g = GroebnerBasis[Append[toric, y[4]], v]\\
Q = PolynomialReduce[B4, g, v][[1]]
\end{alltt}
\vspace{-1mm}
then evaluates the $Q_i$ coefficients with all derivatives $y[i]$ placed to the right.  These can then be re-expressed in terms of Euler operators by rewriting $y_i^n = x_i^{-n}\t_i(\t_i-1)\ldots(\t_i-n+1)$ leading to \eqref{Qcotri}.
}
To extract the required overall factor of $\p_4$, we now re-express those $g_i$ ($i=2,..,18$) that are not already complete toric equations (and hence zero) in terms of $\p_4$.  For example, using the third from last toric equation in \eqref{Itoric}, we can replace
\[
g_2=\p_2~\rightarrow~\p_3\p_4\p_9.
\]
In this fashion, we can replace the basis $\bs{g}$ with the equivalent basis  (modulo the toric equations)
\[
\bs{\tilde{g}}=\{\p_4,~\p_3\p_4\p_9,~\p_4^2\p_9,~0,~\p_4\p_5\p_9,~0,~0\}.
\]
All surviving terms then have an explicit factor of $\p_4$ which can be removed to obtain the creation operator 
\begin{align}
\mathcal{C}_4&=Q_1+Q_2\p_3\p_9+Q_3\p_4\p_9+Q_5\p_5\p_9,
\end{align}
where the coefficients are
\begin{align}\label{Qcotri}
Q_1&=x_4 \big[ (1+\t_4)(\t_5+\t_7+\t_9)+\t_5(\t_6+\t_7+3\t_8+\t_9)+\t_6(\t_7+\t_8+3\t_9)\nn\\&\qquad\quad+(\t_8+\t_9)(1+2\t_7+\t_8+\t_9)+4\t_8\t_9+(1+\t_4+\t_5+\t_6+\t_8+\t_9)^2\nn\\&\qquad\quad +\t_3(1+\t_4+\t_5+2\t_6+\t_7+3\t_8+\t_9)\nn\\&\qquad\quad +\t_2(1+\t_3+\t_4+2\t_5+\t_6+\t_7+\t_8+3\t_9)
\big],\nn\\
Q_2&=x_2(\t_3+\t_5+2\t_9)(\t_5+\t_6+\t_7+\t_8+\t_9),\nn\\
Q_3&=x_8(\t_3+\t_5+2\t_9)\bigl(2\t_5+3\t_6+2(1+\t_7+\t_8+\t_9)\bigr),\nn\\ 
Q_5&=x_6(\t_3+\t_5+2\t_9)(1+\t_5+\t_6+\t_7+\t_9).
\end{align}
Finally, to project  to the physical hypersurface, we use the Euler equations  to eliminate the unphysical variables $\t_1,\t_2, \t_3$ and set $x_1=x_2=x_3=1$. This yields the physical creation operator
\begin{align}
\mathcal{C}_4^{\mathrm{ph}}=&x_4\bigl[(1-\g_2)(1-\g_3)+(\t_5+\t_6+\t_7+\t_8+\t_9)(1-\g_2-\g_3-\t_4)\bigl]\nn\\
&+(\g_3+\t_4)\bigl[\p_9(\t_5+\t_6+\t_7+\t_8+\t_9)(\g_3+\g_4+\g_5+2\t_9)\nn\\&\quad -x_6\p_5(1+\t_5+\t_6+\t_7+3\t_8+\t_9)-2x_8\p_4\p_9(1+\t_5+\t_7+\t_8+\t_9)\bigr]
\end{align}
where the $x_i$ are as given in \eqref{masstriphys} and
\begin{align}
\p_4&=\p_{p_1^2}, & \p_5&=\p_{p_2^2}, & \p_6&=\p_{p_3^2},\nn\\ \p_7&=\p_{m_1^2}-\p_{p_2^2}-\p_{p_3^2}, & \p_8&=\p_{m_2^2}-\p_{p_1^2}-\p_{p_3^2}, & \p_9&=\p_{m_3^2}-\p_{p_1^2}-\p_{p_2^2}.
\end{align}
The automated approach outlined here can be  applied similarly to other examples.

\subsection{Massless on-shell box}\label{sec:hulling}

Next we  consider the massless  box integral
\begin{align}\label{box1}
I &= \int \frac{\mathrm{d}^d\bs{q}}{(2\pi)^d}\frac{1}{|\bs{q}|^{2\g_1}|\bs{q}+\bs{P}_1|^{2\g_2}|\bs{q}+\bs{P}_2|^{2\g_3}|\bs{q}+\bs{P}_3|^{2\g_4}},
\end{align}
where 
\[\bs{P}_k=\sum_{j=1}^k\bs{p}_j,\quad \mathrm{for}\quad k=1,2,3, \qquad\sum_{i=1}^4 \bs{p}_i=0.
\] 
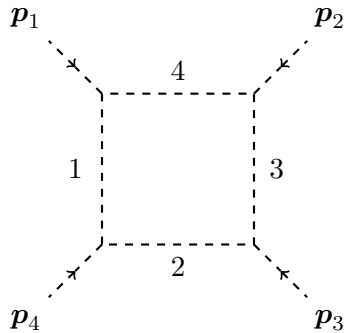
\begin{figure}[t]
\centering
\begin{tikzpicture}
\draw[dashed,thick](-1,-1)--(-1.35,-1.35);
\draw[<-,thick,dashed](-1.35,-1.35)--(-1.7,-1.7);
\draw[dashed,thick](1,-1)--(1.35,-1.35);
\draw[<-,thick,dashed](1.35,-1.35)--(1.7,-1.7);
\draw[dashed,thick](1,1)--(1.35,1.35);
\draw[<-,thick,dashed](1.35,1.35)--(1.7,1.7);
\draw[dashed,thick](-1,1)--(-1.35,1.35);
\draw[<-,thick,dashed](-1.35,1.35)--(-1.7,1.7);
\draw[dashed, thick](-1,-1)--(-1.0,1.0)--(1.0,1.0)--(1.0,-1.0)--cycle;
\node[text width=0.5cm, text centered ] at (-1.35,0) {$1$};
\node[text width=0.5cm, text centered ] at (0,-1.3) {$2$};
\node[text width=0.5cm, text centered ] at (1.3, 0) {$3$};
\node[text width=0.5cm, text centered ] at (0,1.3) {$4$};
\node[text width=0.5cm, text centered ] at (-2,2){$\bs{p}_1$};
\node[text width=0.5cm, text centered ] at (2,2){$\bs{p}_2$};
\node[text width=0.5cm, text centered ] at (2,-2){$\bs{p}_3$};
\node[text width=0.5cm, text centered ] at (-2,-2){$\bs{p}_4$};
\end{tikzpicture} 
\caption{The on-shell massless box integral \eqref{box1}.}
\label{os-massless-box}
\end{figure}
For simplicity, we will restrict to the on-shell case\footnote{ Creation operators for the off-shell box are also computable but the results are rather long.} 
where all $p_i^2=0$ for $i=1,..,4$.   
According to \eqref{feyntogkz}, the corresponding GKZ integral  is 
\begin{align}
\mathcal{I}_{\bs{\g}}=\prod_{i=1}^4\left(\int_0^\infty \dd z_i \:z_i^{\g_i-1}\right)  
(x_1z_1+x_2z_2+x_3z_3+x_4z_4 +x_5 z_1 z_3 + x_6 z_2 z_4)^{-\g_0} ,
 \label{integral-box}
\end{align}
where the physical hypersurface 
\[\label{boxphyshyp}
\bs{x}=(1,1,1,1,s^2,t^2)
\] 
with $s^2=(\bs{p}_1+\bs{p}_2)^2$ and $t^2=(\bs{p}_2+\bs{p}_3)^2$  the Mandelstam invariants. The integral can be evaluated  as a linear combination of the hypergeometric function $_3F_2$ \cite{Tarasov:2017jen}.

The $\mathcal{A}$-matrix 
\begin{align}\label{Amatboxon}
\mathcal{A}=\left(
\begin{array}{cccccc}
 1 & 1 & 1 & 1 & 1 & 1 \\
 1 & 0 & 0 & 0 & 1 & 0 \\
 0 & 1 & 0 & 0 & 0 & 1 \\
 0 & 0 & 1 & 0 & 1 & 0 \\
 0 & 0 & 0 & 1 & 0 & 1 \\
\end{array}
\right)
\end{align}
yields a single toric equation
\[\label{boxtoric}
0=(\p_1\p_3\p_6-\p_2\p_4\p_5)\mathcal{I}_{\bs{\g}},
\]
along with the DWI and Euler equations
\begin{align}\label{eulboxon}
&0=\Big(\g_0+\sum_{i=1}^6\t_i\Big)\mathcal{I}_{\bs{\g}},\quad 0=(\g_1+\t_1+\t_5)\mathcal{I}_{\bs{\g}},\quad 0=(\g_2+\t_2+\t_6)\mathcal{I}_{\bs{\g}},\nn\\& 0=(\g_3+\t_3+\t_5)\mathcal{I}_{\bs{\g}},\quad 0=(\g_4+\t_4+\t_6)\mathcal{I}_{\bs{\g}}.
\end{align}
To determine the singularities of the integral, we need to find the equations of the facets of the rescaled Newton polytope corresponding to the GKZ denominator in \eqref{integral-box}. 
As this polytope lives in four dimensions, 
it is convenient to use an automated hulling algorithm.  
Using the Mathematica package \cite{Nhull}, for example, we can enter the vertices $\bs{a}_j$ of the non-rescaled Newton polytope (where $\bs{a}_j$ is the $j$th column of the $\mathcal{A}$-matrix without the top row)  as row vectors:
\begin{verbatim} 
   verts = {{1,0,0,0},{0,1,0,0},{0,0,1,0},{0,0,0,1},{1,0,1,0},{0,1,0,1}};
\end{verbatim}
The command {\tt CHNQuickHull[verts]} then returns a list of the vertex vectors that make up the convex hull (labelled according to  the numbering specified in the input), followed by a list of the facets.  The latter are specified by the vertex vectors they contain.
Thus, in this example, we obtain
\begin{verbatim}
      {{1,2,3,4,5,6}, {{1,2,3,5}, {1,2,4,3}, {1,2,5,6}, {1,2,6,4},
            {1,3,4,5}, {1,4,6,5}, {2,3,5,6}, {2,3,6,4}, {3,4,5,6}}}
\end{verbatim}
where the first set
indicates that all six vertices belong to the convex hull, while the remainder  ($\{1,2,3,5\}$, $\{1,2,4,3\}$, {\it etc}.) list  the facets.  Here, each $\{ijkl\}$ is a co-dimension one facet containing the points $(\bs{a}_i,\bs{a}_j,\bs{a}_k,\bs{a}_l)$.

The equations for the facets of the {\it rescaled} Newton polytope with vertices $\g_0\bs{a}_i$ can now be computed through a determinant such as \eqref{fullgammaAdet}.  
For the facet $\{1,2,3,4\}$, for example, we have
\[
0=\bs{\g}\cdot\bs{N} = \det\,(\,\bs{\g}\,|\,\bs{\mathcal{A}}_1\,|\,\bs{\mathcal{A}}_2\,|\,\bs{\mathcal{A}}_3\,|\,\bs{\mathcal{A}}_4\,)=\g_0-\g_1-\g_2-\g_3-\g_4 
\]
and hence $\bs{N} = (n_0,\bs{n}) = (1,-1,-1,-1,-1)$.
The fact that $\bs{n}$ is outwards-pointing can be verified by showing $d_i^{(J)}=-\bs{\mathcal{A}}_i\cdot\bs{N}>0$ for any vertex $i=5,6$ not lying in the facet. 
 The spacing of the set of singular hyperplanes parallel to this facet is then $\delta^{(J)}=1$ using \eqref{Jshift} and \eqref{ddef},  with the singular hyperplanes themselves then following from \eqref{hypsings}.  
Automating this procedure and applying it to the other facets, the singularities for the GKZ integral \eqref{integral-box} are 
\begin{align}
&\g_i = -k_i, \quad i=1,2,3,4, & \g_1+\g_2+\g_3+\g_4-\g_0=-k_5,\nn \\
& \g_1+\g_2-\g_0=+k_6,   & \g_2+\g_3-\g_0=+k_7, \label{sing-box-un} \\ & \g_3+\g_4-\g_0=+k_8, & \g_4+\g_1-\g_0=+k_9,\nn
\end{align}
where all $k_i\in \mathbb{Z}^+$. 

We are now in a position to  compute the creation operators.  Let us choose $\mathcal{C}_1$, which acts on the GKZ integral to shift $\bs{\g}'\rightarrow \bs{\g}$ where
\[
\g_0'=\g_0+1,\qquad\qquad \g_1'=\g_1+1.
\]
From the singularities \eqref{sing-box-un}, the corresponding $b$-function is
\[
b_1=-\g_1(\g_2+\g_3-\g_0)(\g_3+\g_4-\g_0),
\]
which in terms of the Euler operators reads
\begin{align}
B_1&=(\t_1+\t_2)(\t_1+\t_4)(\t_1+\t_5)\nn\\& =
\big[ (\t_1 + \t_2) (\t_1 +\t_4) +  (\t_1 + \t_2 + \t_4) \t_5\big]\t_1+\t_2\t_4\t_5
\end{align}
Applying the toric equation \eqref{boxtoric} to the final term  now enables us to factorise $B_1$ as $\mathcal{C}_1\p_1$ giving the creation operator
\[
\mathcal{C}_1=x_1\left[(1+\t_1+\t_2)(1+\t_1+\t_4)+(1+\t_1+\t_2+\t_4)\t_5\right]+x_2x_4x_5\p_3\p_6,
\]
where we shifted the factor of $x_1$ to the left sending each $\t_1\rightarrow \t_1+1$.
Finally, to obtain the creation operator acting on the physical variables,  we use the Euler equations \eqref{eulboxon} to replace $\t_i$ for $i=1,2,3,4$ with $\t_5$ and $\t_6$ and project to  \eqref{boxphyshyp}.  This yields the operator 
\[
\mathcal{C}_1^{\mathrm{ph}}=(1-\g_1-\g_2-\t_{t^2})(1-\g_1-\g_4-\t_{t^2})+(\g_1-1)\t_{s^2}- s^2(\g_3+\t_{s^2})\p_{t^2}.
\]
Using the automated determination of the convex hull in this example,  and the  factorisation of the $b$-function via Gr{\"o}bner basis methods in the previous example,  the calculation of any creation operator can be fully automated.

\section{Discussion}
\label{sec:Discussion}

As we have seen, the GKZ formalism enables the construction of  non-trivial shift operators known as creation operators.  
The calculation is highly systematic.  First, a Feynman or Witten diagram is represented as a GKZ or $\mathcal{A}$-hypergeometric function. 
Second, the $b$-function
is identified by examining the parameter shifts produced by the creation operator in conjunction with the location of all singular hyperplanes of the integral.  
Physically, as the $b$-function is the function of parameters 
that multiplies the shifted integral,  its
zeros serve to cancel the singularities that would otherwise occur when the creation operator maps a finite to a singular integral.
Such singularities cannot arise under the action of a finite differential operator on a finite integral.
Next, using the Euler equations and DWI, the $b$-function is expressed as a function of Euler operators 
and factorised into a product of a creation and an annihilation operator with the aid of the toric equations.
The creation operator thus extracted is then re-expressed in terms of physical variables ({\it i.e.,} the momenta and masses) 
using once again the Euler equations and DWI.

This algorithm has a number of interesting features.  
First, the parametric singularities of the integral all lie on hyperplanes parallel to the facets of the Newton polytope associated with the integral's denominator.   We derived a precise formula for the spacing of these hyperplanes in \eqref{hypsings}.
  The $b$-function therefore has a geometrical character, as originally shown by Saito in \cite{Saito_param_shift}. 
Second, the algorithm makes heavy use of 
the higher-dimensional GKZ space obtained by promoting the coefficient of every term in the Lee-Pomeransky denominator to an independent variable.    
This systematises the set of PDEs obeyed by the integral into  two distinct classes: the Euler equations and DWI, and the toric equations.  
Using the former, we can uplift to GKZ space by exchanging all dependence on the parameters $\bs{\g}$ for dependence on the additional unphysical coordinates.  Conversely, we can project back to physical variables by using the Euler equations and DWI to exchange derivatives with respect to the unphysical variables for derivatives with respect to the physical variables and dependence on the parameters $\bs{\g}$.

This last step is however a potential weakness of the algorithm.  To project a creation operator from GKZ space back to the physical hypersurface, the total number of Euler equations (including the DWI) must be equal to, or greater than, the number of unphysical coordinates.  
This enables every derivative in unphysical variables to be  replaced by an equivalent expression in purely physical variables.
For higher-loop Feynman integrals, however, the number of terms in the Lee-Pomeransky denominator, and hence the dimension of the full GKZ space, typically grows more rapidly than the number of propagators and hence Euler equations.
Thus, while a full set of creation operators can be constructed in GKZ space, in general we lack sufficient Euler equations to project back to the physical hypersurface.  For this reason, we have focused here on 1-loop Feynman integrals. 

One possible workaround for this issue is to construct an alternative projectible GKZ system based on some representation other than the Lee-Pomeransky.
For example, for higher-loop massive sunset (aka melon or banana) diagrams, one can construct a GKZ representation based on their position-space formulation as a product of Bessel functions \cite{Klemm_2020,Zhang:2023fil}.  In this manner, these diagrams can be related to (analytic continuations of) the momentum-space contact Witten diagrams for which we have already constructed creation operators.   For more general classes of diagrams,  projectible GKZ representations can also be obtained from   Mellin-Barnes representations as shown in \cite{Feng_2020, Feng:2022ude, Zhang:2023fil}.  A further possibility might be to develop a GKZ representation starting from the Baikov representation.

Nevertheless, using the simplest formulation based on the Lee-Pomeransky representation, we have already identified 
a number of useful new shift operators.  In particular, for computations in AdS/CFT, we have found:
\begin{itemize}
\item The creation operators \eqref{3KC1} and \eqref{4KC1}, along with their permutations and shadow conjugates, connecting 3- and 4-point momentum-space contact Witten diagrams of different operator and spacetime dimensions.  These new operators are the inverse of the simple annihilators first identified in \cite{Bzowski:2013sza, Bzowski:2015yxv}.
The corresponding operators   can also be obtained in position space as detailed in Appendix \ref{sec:Dfn}.
\item The creation  operators \eqref{W3ptphys} and  \eqref{4ptWv2}, plus their permutations and shadow conjugates, relating 3- and 4-point  momentum-space contact Witten diagrams of different operator dimensions but the same spacetime dimension.  While the 3-point operator \eqref{W3ptphys}  is known  \cite{Karateev:2017jgd,Baumann:2019oyu}, the 4-point operator \eqref{4ptWv2}  is new.

\item  Using \eqref{4ptWv2}, we obtained a further new operator \eqref{OmegaOp} connecting exchange Witten diagrams of different external operator dimensions but the same spacetime dimension. 
Unlike any previous construction, this operator connects  exchange diagrams with purely {\it non-derivative vertices}.  Working in dimensional regularisation where necessary to avoid divergences \cite{Bzowski:2022rlz}, it also applies for arbitrary operators dimensions.  
\end{itemize}
There is ample scope for building on this first application of creation operators to Witten diagrams.   In particular, our results for exchange Witten diagrams were obtained from our analysis of  contact diagrams.  It may be preferable to develop a GKZ representation for the exchange diagram directly, both in momentum and in position space, potentially enabling a more compact set of shift operators to be found, as well as operators acting to shift the dimension of the exchanged leg.  Operators achieving this latter goal are at present known only for a very restricted set of  external operator dimensions \cite{Baumann:2019oyu, Bzowski:2022rlz}.
The application of the creation operator formalism to
cosmological correlators 
in de Sitter spacetime is also worthy of exploration.  
We hope to address some of these matters in future.

\end{chapter}\clearpage{}
  
  \part{Conclusions}
  \clearpage{}\begin{chapter}{\label{cha7}Conclusions}
In this thesis we presented studies of conformal field theory in momentum space, focusing on integral representations and shift operators. The mathematical and physical tools that we used were quite diverse, spanning from electrical circuits to multivariable hypergeometric systems. We believe this work shows the importance of studying an object from different perspectives. On one side, the recent formulation of conformal symmetry in momentum space enlarges the domain of possible physical applications of conformal symmetry -- borrowing Whitman's words, we could say that conformal symmetry is large, it contains multitudes. And we have shown that different representations of the same function reveal different properties.

In Chapter \ref{cha4} we took as input the general $n$-point solution of conformal Ward identities in the form of the simplex integral \eqref{simplex} and derived the new scalar parametrisations, \eqref{Schwrep}, \eqref{Cay2}, for this integral which allowed us to find the new shift operators $\mathcal{S}_{ij}$ and $\mathcal{S}$ for the $n$-point functions. To this aim, we first parametrised the integral in terms of the inverse Schwinger parameters $v_{ij}$ and obtained the associated Kirchhoff polynomials, expressible in terms of the Gram determinant or the Laplacian matrix. We then interpreted the parameters $v_{ij}$ as conductances of the corresponding simplicial electrical network and computed the effective resistances $s_{ij}$ between the nodes of the simplex. This led to a new reparametrisation of the simplex integral in terms of the Cayley-Menger matrix. This parametrisation features an exponential diagonal in the $s_{ij}$, and a product of powers of the determinant $|m|$ and first minors $|m^{(i,j)}|$ of the Cayley-Menger matrix. The structure of this integral then naturally reveals the shift operators. Multiplying the integrand by either the determinant or a first minor of the Cayley-Menger matrix corresponds to shifting their powers. The diagonal exponential allowed us to translate the polynomials in $s_{ij}$, defining $|m|$ and $|m^{(i,j)}|$, into differential operators with respect to $\bs{p}_i\cdot \bs{p}_j$. Besides finding new shift operators, we discussed other advantages of these parametrisations of the simplex. We noted that the number of integrals to be computed reduces from $(n-1)(n-2)d/2$ to $n(n-1)/2$, and the correspondence to the position-space solution simplifies evaluating the action of certain differential operators on the momentum-space integral. In particular, we showed that the special conformal Ward identity corresponds to a total derivative acting on the integrand, and we computed the action of the known weight-shifting operators $\mathcal{W}_{ij}^{\pm\pm}$.

Still, much remains to be explored about the $n$-point function. We have seen how the simplex perspective makes the recursive structure evident, and how these new parametrisations make the shift operators naturally arise. Perhaps by looking for other representations of the simplex, new properties would become manifest. For instance, we might ask whether a representation invariant under a shadow transform exists. Moreover, notice that we have only analysed scalar correlators. Having available a full set of shift operators which generalise the 3-point $\mathcal{L}_i$ operators, together with the other known spin- and weight-shifting operators \cite{Karateev:2017jgd, Arkani-Hamed:2018kmz, Baumann:2019oyu}, we now have the tools to build tensorial correlators that are also of interest in cosmology.

The study of integral representations led us to deepen our knowledge of the multivariable hypergeometric functions in the form of GKZ systems. In Chapter \ref{cha5} we discussed their features, giving a physical interpretation. We presented the formulation of GKZ integrals and showed how their properties are encoded in the $\mathcal{A}$-matrix. We discussed the spectral singularities and their geometrical interpretation via the Newton polytope associated with the $\mathcal{A}$-matrix. We then used this formulation to derive the creation operators of holographic contact Witten diagrams and some generalised Feynman integrals. The construction of these operators is very systematic and uses the $b$-function as input. The $b$-function is a polynomial in the parameters and the whole algorithm is based on the fact that it can be factorised in terms of annihilation and creation operators as $B(\theta)\sim \mathcal{C}\p$. We gave a physical interpretation of the form of the $b$-function. We derived it by requiring that the associated creation operator cannot send a finite integral to a divergent one. Moreover, we used an analogous algorithm to find operators that act on contact and exchange Witten diagrams to shift only the scaling dimensions and preserve the form of the functions. As discussed, this class of operators is new and applies to any values of the parameters. Our construction of creation operators for Witten diagrams is moreover valid at $n$ points. 

The GKZ formalism entered the physics literature only recently. While various studies focused on the series solutions of Feynman integrals, to our knowledge our work is the first application of creation operators to physics, therefore this is only the beginning. There are many paths to explore. Can we find a more optimal way for constructing the GKZ representations of multi-loop Feynman integrals? Can we find a GKZ representation of exchange Witten diagrams? Coming back to the simplex representation, in Chapter \ref{cha4}, we derived its Lee-Pomeransky representation from which a GKZ representation can be obtained. This would give us the spectral singularities of the conformal $n$-point functions. Moreover, conformal blocks in position space were found to have a description in terms of multivariable hypergeometric systems based on root systems \cite{SchomerusBig, ChenKyono}. An interesting question is whether we can construct a momentum-space description of conformal blocks, and also whether a connection between GKZ and root systems exists. A more immediate consequence of knowing various families of shift operators is their application to cosmological correlators. 

With these various open directions we conclude this thesis but certainly not the research ahead.

\end{chapter}
\clearpage{}
  
  \appendix
  
  \clearpage{}\begin{chapter}[{\texorpdfstring{The master integral $I_{1\{000\}}$}{The master integral I_{1\{000\}}}}]{\texorpdfstring{The master integral \boldmath{$I_{1\{000\}}$}}{The master integral I_{1\{000\}}}}\label{I1000comp}

In this appendix we evaluate the master integral $I_{1\{000\}}$. The strategy we follow is to start with the Feynman parametrisation of the 1-loop triangle diagram \eqref{triangle}, and solve the resulting integral by making suitable changes of variables and using partial fractions method. We will show that the $z$-variables defined in \eqref{z_var_def} arise naturally. 

For the master integral $d=4$ and $\Delta_j$ ($j=1,2,3$) and using \eqref{aijtobi}, $a_{jk}=-1$. Setting the parameters at these values and Feynman parametrising the triangle representation we have
\[
I_{1\{000\}}=\frac{1}{4}\int_{[0,1]^3}\text{d}X\frac{1}{p_1^2x_2x_3+p_2^2x_1x_3+p_3^2x_1x_2},
\]
where $\text{d}X=\text{d}x_1\text{d}x_2\text{d}x_3\d(x_1+x_2+x_3-1)$. Setting $u=p_1^2/p_3^2$ and $v=p_2^2/p_3^2$, the integral reads:
\[
4p_3^2I_{1\{000\}}=\int_0^1\text{d}x_3\int_0^{1-x_3}\text{d}x_2\frac{1}{x_1x_2+x_2x_3u+x_1x_3v},
\]
where $x_1=1-x_2-x_3$. Re-parametrising
\[
y_1=\frac{x_1}{x_3},~~~y_2=\frac{x_2}{x_3}~~\Rightarrow~~x_2=\frac{y_2}{1+y_1+y_2},~~~x_3=\frac{1}{1+y_1+y_2}
\]
and computing the Jacobian $|\p x/\p y|=(1+y_1+y_2)^{-3}$, we get
\[
4p_3^2I_{1\{000\}}=\int_0^\infty\text{d}y_1\int_0^\infty\text{d}y_2\frac{1}{(1+y_1+y_2)(y_1y_2+y_1v+y_2u)}.
\]
We then perform a first partial fraction
\begin{align}
\frac{1}{(1+y_1+y_2)(y_1y_2+y_1v+y_2u)}=&\frac{1}{y_1^2+(1+u-v)y_1+u}\nn\\&\times\left(\frac{u+y_1}{(u+y_1)y_2+vy_1}-\frac{1}{1+y_1+y_2}\right),
\end{align}
and, integrating over $y_2$, we find
\[\label{log&poly}
4p_3^2I_{1\{000\}}=\int_0^\infty\text{d}y_1\frac{1}{y_1^2+(1+u-v)y_1+u}\log\left(\frac{(1+y_1)(u+y_1)}{vy_1}\right).
\]
The denominator is quadratic in $y_1$, hence we can factorise it. This, indeed, leads to defining the $z$-variables as in \eqref{z_var_def} and the integral reads
\[\label{logterm}
4p_3^2I_{1\{000\}}=\int_0^\infty\text{d}y_1\frac{1}{(y_1+z)(y_1+\bar{z})}\log\left(\frac{(1+y_1)(y_1+z\bar{z})}{y_1(1-z)(1-\bar{z})}\right).
\] 
Then, we perform a second partial fraction
\[
\frac{1}{(y_1+z)(y_1+\bar{z})}=\frac{1}{\bar{z}-z}\left(\frac{1}{y_1+\bar{z}}-\frac{1}{y_1+z}\right),
\]
and, after some manipulations, the integral becomes
\[
4p_3^2(z-\bar{z})I_{1\{000\}}=\int_0^\infty\text{d}y_1\left(\frac{1}{y_1+z}-\frac{1}{y_1+\bar{z}}\right)\log\left(\frac{(1+y_1)^2}{(1-z)(1-\bar{z})}\right).
\]
We now can evaluate it in terms of the dilogarithm giving the expression in equation \ref{I1000z}, with $z_{1,2}\in \mathbb{C}-]-\infty,0]\cup [1,+\infty[$. To see this, note that there will be terms of the form
\[\label{dilog_int}
\int_0^\infty\text{d}y\frac{\log(1+y)}{y+z},
\]
which can be computed by applying the Cauchy integral formula taking into account that the discontinuity of Li$_2(z)$ across the branch cut $[1,\infty]$ is equal to $2\pi i\log|z|$. Setting $t=1+y$:
\[
\int_1^\infty\text{d}t\frac{\log t}{t-1+z}=\frac{1}{2\pi i}\int_1^\infty\text{d}t\frac{\text{disc}(\text{Li}_2(t))}{t-1+z}=\frac{1}{2\pi i}\oint_C\mathrm{d}t\frac{\text{Li}_2(t)}{t-1+z}=\text{Li}_2(1-z),
\]
where $C$ is the ``pac-man" closed contour. Then, using dilogarithm's properties one can express $\text{Li}_2(1-z)$ in terms of $\text{Li}_2(z)$ and find the explicit result anticipated in \eqref{I1000z}:
\[\label{I1000zbis}
I_{1\{000\}}=\frac{-1}{2p_3^2(z-\bar{z})}\left[\mathrm{Li}_2(\bar{z})-\mathrm{Li}_2(z)-\frac{1}{2}\ln(z\bar{z})\ln\left(\frac{1-z}{1-\bar{z}}\right)\right].
\]

\end{chapter}
\clearpage{}                    
 \clearpage{}\begin{chapter}{Appendix to chapter \ref{cha4}\label{app:simplex}}
\allowdisplaybreaks
\section{Derivation of graph polynomials}
\label{incidence_app}

In this appendix, we show that Schwinger representation of the simplex integral \eqref{simplex} is given by \eqref{Schwrep} with   graph polynomials given in \eqref{Kirchhoff1}.  Our discussion builds on that in  \cite{Itzykson:1980rh}.
Labelling the vertices of the simplex by $i=1,\ldots,n$, and the (directed) legs by $a=1,\ldots,N$ where $N=n(n-1)/2$, we introduce the incidence matrix 
\[
\ep_i^a = \begin{cases} +1\qquad &\mathrm{if\, leg}\, a\, \mathrm{is\, ingoing\, to\, vertex}\, i\\
-1\qquad& \mathrm{if\, leg}\, a\, \mathrm{is\, outgoing\, to\, vertex}\, i\\
0\qquad &\mathrm{otherwise}
\end{cases}
\]
where for clarity we will write the vertex index downstairs and the leg index upstairs. 
Thus, for example, if we choose $a=\{(12),(13),(14),(23),(24),(34)\}$ as the legs of the 4-point function, where the leg $(i,j)$ runs from vertex $i$ to vertex $j$, the incidence matrix is 
\[
\ep = \left(\begin{matrix}
-1 & -1 & -1 &0 &0 &0\\
1&0&0&-1&-1&0\\
0&1&0&1&0&-1\\
0&0&1&0&1&1
\end{matrix}\right).
\]
Momentum conservation at vertex $k$ of the simplex can now be re-expressed as 
\[
0=\bs{p}_k + \sum_{l\neq k}^n \bs{q}_{lk}=\bs{p}_k + \sum_{a}^N \ep_k^a\,\bs{q}_{a}
\]
where $\bs{q}_a$ is the internal momentum flowing along the directed leg $a$.  As always,  all sums are assumed to begin at one unless otherwise specified. 
The Laplacian matrix $\tilde{g}_{ij}$ defined in \eqref{tildeg} can now be written 
\[\label{Laplep}
\tilde{g}_{ij} = \sum_a^N v_a \ep_i^a \ep_j^a,
\]
which follows by noting that for $i\neq j$ only the leg for which $a$ runs between vertices $i$ and $j$ contributes giving $-v_{ij}$, while for $i=j$ all legs running into this vertex contribute giving $\sum_{k\neq i}v_{ik}$ as required. 

Turning to the simplex integral \eqref{simplex}, we first rewrite the delta functions of momentum conservation in Fourier form 
\[
\prod_{k}^n \,(2\pi)^d\delta\Big(\bs{p}_k + \sum_{l\neq k}^n \bs{q}_{lk}\Big) = \prod_k^n \int \mathrm{d}^d\bs{y}_k \, \exp\Big(-i\bs{y}_k\cdot\big(\bs{p}_k + \sum_{a}^N \ep_k^a\,\bs{q}_{a}\big)\Big).
\]
Next, exponentiating all propagators of internal momenta (labelled by their leg indices) using the Schwinger representation \eqref{Schwingerintegral}, we find\footnote{Note the argument of the arbitrary function $f$ changes from the momentum cross ratios $\hat{\bs{q}}$ in \eqref{conf_ratio_q}  to the Schwinger parameter cross ratios $\hat{\bs{v}}$ in \eqref{vcrossratio}.   This can be seen by temporarily representing the arbitrary function in Mellin-Barnes form ({\it i.e.,} (4.18) of \cite{Bzowski:2020kfw}) allowing all $q_{ij}$, including those from the cross ratios, to be exponentiated via the Schwinger parametrisation \eqref{Schwingerintegral}.  Performing the Mellin-Barnes integration then generates $f(\hat{\bs{v}})$, since the Schwinger parametrisation replaces powers of $q_{ij}$  by powers of $v_{ij}$.}
\begin{align}
&\< \O_1(\bs{p}_1) \ldots \O_n(\bs{p}_n) \>
=\Big(\prod_a^N \frac{1}{\Gamma(\alpha_a+d/2)}\int_0^\infty \dd v_a\,
v_a^{-d/2-\alpha_a-1}\Big)f(\hat{\bs{v}}) \nn\\&\qquad\times
\Big(\prod_{k}^n \int\dd^d\bs{y}_k \, \exp(-i\bs{y}_k\cdot\bs{p}_k)
\Big)\int\frac{\dd^d\bs{q}_a}{(2\pi)^d} \exp\Big(-\sum_a^N\Big(\frac{q_a^2}{v_a}+i\sum_l^n\ep_l^a\bs{y}_l\cdot \bs{q}_a\Big)\Big).
\end{align}
Evaluating the $\bs{q}_a$ integrals by completing the square and using \eqref{Laplep} now gives
\begin{align}
\< \O_1(\bs{p}_1) \ldots \O_n(\bs{p}_n) \>&
=\Big(\prod_a^N \frac{\pi^{d/2}}{\Gamma(\alpha_a+d/2)}\int_0^\infty \dd v_a\,
v_a^{-\alpha_a-1}\Big)f(\hat{\bs{v}}) \nn\\&\qquad\times
\Big(\prod_{k}^n \int\dd^d\bs{y}_k \Big)\, \exp\Big(-i\sum_k^n \bs{y}_k\cdot\bs{p}_k - \frac{1}{4}\sum_{k,l}^n \tilde{g}_{kl}\,\bs{y}_k\cdot\bs{y}_l\Big).
\end{align}
Since the Laplacian matrix has no inverse, to compute the $\bs{y}_k$ integrals we must first shift
\begin{align}
 \bs{y}_n = \bs{z}_n, \qquad \bs{y}_{k}=\bs{z}_k +\bs{z}_n, \qquad k=1,\ldots n-1.
\end{align}
This transformation has unit Jacobian, but moreover greatly simplifies the exponent.   Since all row and column sums of the Laplacian matrix vanish, 
\[
\sum_{l}^{n-1} \tilde{g}_{kl} = -\tilde{g}_{kn},\qquad\qquad  \sum_k^{n-1}\tilde{g}_{kn} = -\tilde{g}_{nn},
\]
and using these identities we then find
\begin{align}
&-i\sum_k^n \bs{y}_k\cdot\bs{p}_k - \frac{1}{4}\sum_{k,l}^n \tilde{g}_{kl}\,\bs{y}_k\cdot\bs{y}_l \nn\\&\qquad 
=-i\bs{z}_n\cdot\bs{p}_n -i\sum_k^{n-1} (\bs{z}_k+\bs{z}_n)\cdot\bs{p}_k \nn\\&\qquad\quad
-\frac{1}{4}\tilde{g}_{nn}z_n^2-\frac{1}{2}\sum_k^{n-1}\tilde{g}_{kn}(\bs{z}_k+\bs{z}_n)\cdot\bs{z}_n - \frac{1}{4}\sum_{k,l}^{n-1} \tilde{g}_{kl}\,(\bs{z}_k+\bs{z}_n)\cdot(\bs{z}_l+\bs{z}_n)
\nn\\&\qquad=
-i\bs{z}_n\cdot\Big(\sum_k^n\bs{p}_k\Big) -i\sum_k^{n-1}\bs{z}_k\cdot\bs{p}_k -\frac{1}{4}\sum_{k,l}^{n-1}g_{kl}\,\bs{z}_k\cdot\bs{z}_l.
\end{align}
In the final line here, all the $\bs{z}_k\cdot\bs{z}_n$ and $z_n^2$ terms cancel while the Laplacian matrix $\tilde{g}_{kl}$ reduces to $g_{kl}$ for $k,l=1,\ldots ,n$.   The $\bs{z}_n$ integral now gives the overall delta function of momentum conservation which we  strip off to obtain the reduced correlator \eqref{redcorr}.  The remaining $\bs{z}_k$ integrals can be evaluated by completing the square, given that the inverse $g_{kl}^{-1}$ exists. 
This yields our desired result,
\begin{align}
&\lla \O_1(\bs{p}_1) \ldots \O_n(\bs{p}_n) \rra = 
\mathcal{C}\prod_a^N \int_0^\infty\dd v_a\, v_a^{-\alpha_a-1} f(\hat{\bs{v}})\, |g|^{-d/2} \exp\Big(-\sum_{k,l}^{n-1}g_{kl}^{-1}\,\bs{p}_k\cdot\bs{p}_l\Big)
\end{align}
where the constant
\[
\mathcal{C} =(4\pi)^{(n-1)d/2} \prod_a^N \frac{\pi^{d/2}}{\Gamma(\alpha_a+d/2)}
\]
can simply be re-absorbed into the arbitrary function $f(\hat{\bs{v}})$.
Rewriting the product of legs $a$ as a product over vertices $i<j$ 
and replacing $\bs{p}_k\cdot\bs{p}_l$ with the Gram matrix $G_{kl}$, we recover precisely \eqref{Schwrep} with graph polynomials \eqref{Kirchhoff1}.

\section{Jacobian matrix} 

In this appendix, we compute the Jacobian matrix for the change of variables from $v_{ij}$ to $s_{ij}$.  In section \ref{Jac_app} we evaluate the Jacobian determinant, then in section \ref{sec:Jacobimatrixels}
we give expressions for its matrix elements enabling conversion between partial derivatives.

\subsection{Jacobian determinant}
\label{Jac_app}

Our first goal is to derive the relation \eqref{Jac0} for  the Jacobian determinant, namely 
\begin{align}\label{Jac1}
\left|\frac{\partial s}{\partial v}\right| = \left|\frac{\partial^2 \ln |g|}{\partial v\,\partial v}\right| \propto |g|^{-n},
\end{align}
where the constant of proportionality is not required since it can be re-absorbed into the arbitrary function $f(\hat{\bs{v}})$.
For small values of $n$ this result can be verified by direct calculation, and the exponent is simply fixed  by power counting, but our aim  is nevertheless to prove this relation for general $n$.

We start by noting
\begin{align}\label{matrixprod}
\frac{\partial^2 \ln |g|}{\partial v_{ij}\partial v_{kl}} = \frac{\partial g_{pq}}{\partial v_{ij}}\frac{\partial^2 \ln |g|}{\partial g_{pq}\partial g_{rs}} \frac{\partial g_{rs}}{\partial v_{kl}}
\end{align}
can be re-expressed as a product of three square matrices of dimension $n(n-1)/2$.  
Each of the index pairs $(p,q)$ and $(r,s)$ is replaced by a single index running over  the $n(n-1)/2$ independent entries of the  $(n-1)\times(n-1)$ symmetric matrix $g$, while $(i,j)$ and $(k,l)$ are each replaced by a single index running over the $n(n-1)/2$ edges of the simplex.  Noting the elements of $g$ are linear in the $v$, the matrix determinant $|\partial g/\partial v|$ evaluates to a nonzero constant.  On taking the determinant of \eqref{matrixprod}, we find 
\[
 \left|\frac{\partial^2 \ln |g|}{\partial v\,\partial v}\right| \propto  \left|\frac{\partial^2 \ln |g|}{\partial g \,\partial g }\right|
\]
hence it suffices to show that
\[\label{Hess0}
 \left|\frac{\partial^2 \ln |g|}{\partial g \,\partial g }\right| \propto |g|^{-n}.
\]
This relation in fact holds for any invertible symmetric square matrix  $g$ of dimension $n-1$.

To see this, from Jacobi's relation we have
\[\label{Hess1}
\frac{\partial^2 \ln |g|}{\partial g_{pq}\partial g_{rs}}
=\frac{\partial}{\partial g_{rs}}\Big(\frac{1}{|g|}(\mathrm{adj}\, g)_{pq}\Big)  = \frac{\partial (g^{-1})_{pq}}{\partial g_{rs}}=-2(g^{-1})_{p(r}(g^{-1})_{s)q}.
\]
Diagonalising $g$ via an orthogonal matrix $O$, 
\[
\Lambda = O g O^{-1},
\] 
since
$
g^{-1} = O^{-1}\Lambda^{-1}O
$
the chain rule gives 
\[
\frac{\partial g^{-1}}{\partial{g}} = \frac{\partial (O^{-1}\Lambda^{-1}O)}{\partial \Lambda^{-1}}\frac{\partial\Lambda^{-1}}{\partial \Lambda}\frac{\partial(OgO^{-1})}{\partial g}
\]
where the last factor is just $\partial \Lambda/\partial g$.
Regarding this as a matrix product, the first and last matrices depend only on $O$ and are inverses of each other.   On taking the determinant of the right-hand side, their contributions therefore cancel giving
\[
\left|\frac{\partial g^{-1}}{\partial{g}} \right| = \left|\frac{\partial\Lambda^{-1}}{\partial \Lambda}\right|.
\]
We thus only need to evaluate the latter determinant for the diagonal matrix $\Lambda$.

From \eqref{Hess1}, the Hessian 
is nonzero only when the index pairs are equal $(p,q)=(r,s)$, and  is thus diagonal when regarded as a square matrix of dimension $n(n-1)/2$:
\begin{align}
\frac{\partial^2\ln |\Lambda|}{\partial \Lambda_{pq}\partial \Lambda_{rs}} = \frac{\partial (\Lambda^{-1})_{pq}}{\partial \Lambda_{rs}}=\begin{cases} -\Lambda_{pp}^{-1}\Lambda_{qq}^{-1} \qquad &\mathrm{if}\quad (p,q) = (r,s) \\
0 \qquad& \mathrm{otherwise}\end{cases}
\end{align}
The determinant is now
\begin{align}
\left|\frac{\partial^2\ln |\Lambda|}{\partial \Lambda_{pq}\Lambda_{rs}}\right|\,
\propto \,\prod_{p=1}^{n-1} (\Lambda_{pp})^{-n} = |\Lambda|^{-n} = |g|^{-n}
\end{align}
since each eigenvalue $\Lambda_{pp}$ appears a total of $n$ times along the diagonal: for example, $\Lambda_{11}$ appears  quadratically in  the position $(1,1)$ and then linearly in each of the $(n-2)$ entries indexed by $(1,q)$ for $q=2,\ldots n-1$. 
We have thus established \eqref{Hess0}, and hence \eqref{Jac1}.

\subsection{Matrix elements}
\label{sec:Jacobimatrixels}

We now compute the elements of the Jacobian matrix required to establish the relation
\[\label{dvinds}
\partial_{v_{kl}} = -\frac{1}{4}\sum_{i<j}^n(s_{ki}-s_{li}-s_{kj}+s_{lj})^2\partial_{s_{ij}}
\]
which we used in \eqref{Wmmid}.
Starting with \eqref{srel1},
\begin{align}
\frac{\partial s_{ij}}{\partial v_{kl}} = \frac{\partial}{\partial v_{kl}}\Big((g^{-1})_{ab}\frac{\partial g_{ab}}{\partial v_{ij}}\Big) =\frac{\partial (g^{-1})_{ab}}{\partial v_{kl}}\frac{\partial g_{ab}}{\partial v_{ij}}
\end{align}
where since $g_{ab}$ is linear in the $v_{ij}$ its second derivative vanishes.  Using
\[
\frac{\partial (g^{-1})_{ab} }{\partial v_{kl}} = - (g^{-1})_{ae}(g^{-1})_{bf}\frac{\partial g_{ef} }{\partial v_{kl}} 
\]
then gives 
\begin{align}
\frac{\partial s_{ij}}{\partial v_{kl}} &= -\mathrm{tr}\Big(g^{-1}\cdot \frac{\partial g}{\partial v_{ij}}\cdot g^{-1}\cdot\frac{\partial g}{\partial v_{jk}}\Big) =-\sum_{a,b,e,f}^{n-1}(g^{-1})_{ae}\frac{\partial g_{ef}}{\partial v_{kl}}(g^{-1})_{fb}\frac{\partial g_{ba}}{\partial v_{ij}}. \label{dsdv1}
\end{align}
For $i,j,k,l \neq n$, we can evaluate this as 
\begin{align}
&\frac{\partial s_{ij}}{\partial v_{kl}} 
= -\sum_{a,b,e,f}^{n-1}(g^{-1})_{ae}(-2\delta_{k(e}\delta_{f)l}
+\delta_{ek}\delta_{fk}+\delta_{el}\delta_{fl}))
(g^{-1})_{fb}(-2\delta_{i(a}\delta_{b)j}+\delta_{ai}\delta_{bi}+\delta_{aj}\delta_{bj})\nn\\
&=-2(g^{-1})_{ik}(g^{-1})_{jl}-2(g^{-1})_{il}(g^{-1})_{jk}+2(g^{-1})_{ik}(g^{-1})_{jk}+2(g^{-1})_{il}(g^{-1})_{jl}\nn\\&
\quad +2 (g^{-1})_{ik}(g^{-1})_{il}+2(g^{-1})_{jk}(g^{-1})_{jl}
-((g^{-1})_{ik})^2-((g^{-1})_{il})^2-((g^{-1})_{jk})^2-((g^{-1})_{jl})^2\nn\\[1ex]
&= -\big((g^{-1})_{ik}-(g^{-1})_{il}-(g^{-1})_{jk}+(g^{-1})_{jl})\big)^2\nn\\&
=-\frac{1}{4}\big(s_{ik}-s_{il}-s_{jk}+s_{jl}\big)^2\label{srel2}
\end{align}
where we used the symmetry of the inverse matrix $g^{-1}_{ij}$, and in the last line we used \eqref{ginvs}.
For $j=n$ but $i,k,l\neq n$, 
\begin{align}
\frac{\partial s_{in}}{\partial v_{kl}} 
&= -\sum_{a,b,e,f}^{n-1}(g^{-1})_{ae}(-2\delta_{k(e}\delta_{f)l}
+\delta_{ek}\delta_{fk}+\delta_{el}\delta_{fl}))
(g^{-1})_{fb}(\delta_{ai}\delta_{bi})\nn\\
&
=-\big((g^{-1})_{ik}- (g^{-1})_{il}\big)^2
=-\frac{1}{4}\big(s_{ik}-s_{il}-s_{kn}+s_{ln}\big)^2
\end{align}
which is equivalent to \eqref{srel2} setting $j=n$.
The same also holds for $l=n$ but $i,j,k\neq n$ due to the symmetry of \eqref{dsdv1}.  Finally
\begin{align}
\frac{\partial s_{in}}{\partial v_{kn}} 
&= -\sum_{a,b,e,f}^{n-1}(g^{-1})_{ae}(\delta_{ek}\delta_{fk}))
(g^{-1})_{fb}(\delta_{ai}\delta_{bi})\nn\\&
=-((g^{-1})_{ik})^2=-\frac{1}{4}(s_{ik}-s_{in}-s_{kn})^2,
\end{align}
also equivalent to \eqref{srel2} since $s_{nn}=0$.
Thus \eqref{srel2} in fact holds for all values of the indices and we obtain \eqref{dvinds}.

For completeness, we can also calculate the inverse Jacobian by similar means:
\begin{align}
\frac{\partial v_{ij}}{\partial s_{kl}} &= \frac{\partial}{\partial s_{kl}}\Big( (m^{-1})_{ab}\frac{\partial m_{ab}}{\partial s_{ij}}\Big) = \frac{\partial (m^{-1})_{ab}}{\partial s_{kl}}\frac{\partial m_{ab}}{\partial s_{ij}}=-(m^{-1})_{ae}\frac{\partial m_{ef}}{\partial s_{kl}}(m^{-1})_{fb}\frac{\partial m_{ba}}{\partial s_{ij}}\nn\\
&=-(m^{-1})_{ae}(\delta_{ek}\delta_{fl}+\delta_{el}\delta_{fk})(m^{-1})_{fb}(\delta_{bi}\delta_{aj}+\delta_{bj}\delta_{ai})\nn\\&
=-2\Big((m^{-1})_{ik}(m^{-1})_{jl}+(m^{-1})_{li}(m^{-1})_{jk}\Big).
\label{dvds1}
\end{align}
Apart from the final $(n+1)^{\mathrm{th}}$ row and column, the inverse Cayley-Menger matrix is  minus one half the Laplacian matrix $\tilde{g}_{ij}$ as we showed in \eqref{minvisLa1} and \eqref{minvisLa2}.  This gives
\[
\frac{\partial v_{ij}}{\partial s_{kl}} =-\frac{1}{2}\Big(\tilde{g}_{ik}\tilde{g}_{jl}+\tilde{g}_{il}\tilde{g}_{jk}\Big),\label{dvds}
\]
where $\tilde{g}_{ij}=-v_{ij}$ for $i\neq j$ and $\tilde{g}_{ii}=\sum_{a=1}^n v_{ia}$. 
For $i,j,k,l$ all different, we therefore have 
\[
\frac{\partial v_{ij}}{\partial s_{kl}} =-\frac{1}{2}\Big(v_{ik}v_{jl}+v_{il}v_{jk}\Big), \qquad i\neq j\neq k\neq l
\]
while if $j=l$, 
\[
\frac{\partial v_{ij}}{\partial s_{kj}} =-\frac{1}{2}\Big(v_{ij}v_{jk}-v_{ik}\Big(\sum_{a=1}^n v_{ja}\Big)\Big)
\]
and if $i=k$ and $j=l$,
\[
\frac{\partial v_{ij}}{\partial s_{ij}} =-\frac{1}{2}\Big(v_{ij}^2+\Big(\sum_{a=1}^n v_{ia}\Big)\Big(\sum_{b=1}^n v_{jb}\Big)\Big).
\]

\section{Landau singularities}\label{Landau_app}

The Landau singularities of the simplex integral are best studied in the Lee-Pomeransky representation \eqref{LPrep}.  
They follow from solving simultaneously for all $v_{ij}$ the conditions
\[
0=\mathcal{U}+\mathcal{F},\qquad 0 = v_{ij}\frac{\partial}{\partial v_{ij}}(\mathcal{U}+\mathcal{F}).
\]
Here, the first Landau equation stipulates the vanishing of the Lee-Pomeransky denominator, while the second requires that this vanishing is either a double zero (for $v_{ij}\neq 0$), corresponding to a pinching of the $v_{ij}$ integration contour between two converging singularities of the integrand, or else a pinch of the integration contour between a singularity and the end-point of the integration ($v_{ij}=0$).  The second condition thus ensures the singularity generated by the vanishing denominator cannot be avoided by a deformation of the integration contour. 
Where the Landau conditions have more than one solution, the 
 solution with the greatest number of $v_{ij}\neq 0$ is  referred to as the {\it leading} singularity.

An important feature of the $\mathcal{U}$ polynomial  \eqref{Kirchhoff1} is that it is {\it multilinear} in the $v_{kl}$: from  the determinant structure  one sees that all the quadratic $v_{kl}^2$ terms cancel, and that no higher powers can appear since $v_{kl}$ enters only in the row/columns $(k,k)$, $(k,l)$, $(l,k)$ and $(l,l)$. 
Alternatively, this result follows from the matrix tree theorem where the Kirchhoff polynomial  $\mathcal{U}$ is the generator of spanning trees on the simplex.
Since $\mathcal{U}$ is also homogeneous of degree $(n-1)$, it follows that  
\[\label{Uid}
\sum_{k<l}\frac{\partial\mathcal{U}}{\partial v_{kl}}v_{kl} = (n-1)\mathcal{U}.
\]
We now find 
\[\label{Landau1eval}
\mathcal{U}+\mathcal{F} = \mathcal{U}+\sum_{k<l}\frac{\partial\mathcal{U}}{\partial v_{kl}}V_{kl} = \sum_{k<l}\frac{\partial\mathcal{U}}{\partial v_{kl}}\Big(\frac{v_{kl}}{n-1}+V_{kl}\Big)
\]
and so a solution of the first Landau condition for all $k<l$ is 
\[
v_{kl} = \lambda \,V_{kl} =- \lambda \,\bs{p}_k\cdot\bs{p}_l \qquad \mathrm{and}\qquad |G| = |\bs{p}_k\cdot\bs{p}_l|=0
\]
for some constant $\lambda$.  Evaluating 
the second Landau condition  on this solution $(\ast)$ of the first gives
\[
 \Big[v_{ij}\frac{\partial}{\partial v_{ij}}(\mathcal{U}+\mathcal{F})\Big]_{\ast}
=\Big[ v_{ij}\frac{\partial\mathcal{U}}{\partial v_{ij}}+v_{ij}\sum_{k<l}\frac{\partial^2\mathcal{U}}{\partial v_{ij}\partial v_{kl}}V_{kl}\Big]_{\ast}=
(\lambda +n-2)
\lambda^{n-2}V_{ij}\frac{\partial\mathcal{U}|_{v\rightarrow V
}}{\partial V_{ij}}
\]
using again the homogeneity of $\mathcal{U}$.  The second Landau condition is thus solved for all $i,j$ when $\lambda = 2-n$, and indeed this is the leading singularity since the $v_{kl}$ are generically nonzero.   Returning to \eqref{Landau1eval}, on the solution $(\ast)$ we have 
\[
(\mathcal{U}+\mathcal{F})_\ast= (2-n)^{n-2}|G|=0,
\] 
so to solve the first Landau condition we do indeed need the Gram determinant  to vanish.
Generally this requires analytic continuation to non-physical momentum configurations, since the only physical configurations (in Euclidean signature) for which the Gram determinant vanishes are collinear ones, and on physical grounds there are no collinear singularities.   There is no contradiction here since the Landau equations are necessary, but not sufficient, conditions for a singularity.

\section{Bernstein-Sato operators} 

\label{app:Bernstein}

In this appendix, we construct a Cayley-Menger analogue of the  classic  identity
\[\label{Cayley}
\mathrm{det(\partial)} (\mathrm{det}\,X)^a = a(a+1)\ldots (a+n-1) (\mathrm{det}X)^{a-1},
\]
where $X=(x_{ij})$ is an $n\times n$ matrix of independent variables and $\partial=(\partial/\partial x_{ij})$ is the corresponding matrix of partial derivatives.   For proofs and variants of this identity, traditionally attributed to Cayley, see,  {\it e.g.,} \cite{Caracciolo_2013, Fulmek}.  From a modern perspective,  \eqref{Cayley} is an example of a Bernstein-Sato operator, a  
differential operator whose action lowers the power $a$ to which some polynomial of interest is raised, generating in the process an auxiliary polynomial in $a$ known as the $b$-function \cite{Budur}.  Thus we have
\[
\mathcal{B}_f\, f(x_{ij})^a=b_f(a)f(x_{ij})^{a-1}
\] 
where for \eqref{Cayley}, $\mathcal{B}_f=\mathrm{det}(\partial)$, $f=\mathrm{det}\,X$ and $b_f(a) = a(a+1)\ldots (a+n-1)$.
In the following, we construct analogous operators for the Cayley-Menger determinant and other polynomials arising in our parametric representations \eqref{Cay2} and \eqref{Schwrep}.  Such relations are potentially a source of new weight-shifting operators, see {\it e.g.,} \cite{Tkachov:1996wh, Bitoun:2017nre}. 

Our starting point is the observation that
\[
\mathcal{B}_{|m|} = (|g|)\Big|_{v_{ij}\rightarrow \partial_{s_{ij}}}
\]
is a Bernstein-Sato operator for the Cayley-Menger determinant,
\[\label{Bernstein1}
\mathcal{B}_{|m|}\, |m|^a = b_{|m|}(a) |m|^{a-1}, \qquad b_{|m|}(a) =- \prod_{k=1}^{n-1}(1-k-2a).
\]
The operator $\mathcal{B}_{|m|}$ thus corresponds to evaluating the Kirchhoff polynomial $\mathcal{U}=|g|$ and replacing all $v_{ij}\rightarrow \partial_{s_{ij}}$ to generate a polynomial differential operator  in the $\partial_{s_{ij}}$.   We have verified  \eqref{Bernstein1} by direct calculation for matrices up to and including $n=5$.  Moreover, the leading behaviour at order $a^{n-1}$ follows by noting that such terms can only arise from all $n-1$ partial derivatives in $\mathcal{B}_{|m|}$ hitting a power of $|m|$ rather than a derivative of $|m|$.  Using \eqref{vdef} in the form  $\partial_{s_{ij}}|m|^a = a v_{ij} |m|^{a}$ along with \eqref{mginvreln}, then gives
\[\label{leadinga}
\mathcal{B}_{|m|}\,|m|^{a} = a^{n-1}|m|^{a}|g|+ O(a^{n-2})= (-1)^n \,2^{n-1}a^{n-1}|m|^{a-1}+ O(a^{n-2})
\]
in agreement with \eqref{Bernstein1}.\footnote{  
A full proof of \eqref{Bernstein1} 
likely follows via the methods of \cite{Caracciolo_2013},
though we will not pursue this here. }

Similarly, we find
\[
\mathcal{B}_{|g|} = (|m|)\Big|_{s_{ij}\rightarrow \partial_{v_{ij}}}
\]
({\it i.e.,} the Cayley-Menger determinant replacing each $s_{ij}\rightarrow \partial_{v_{ij}}$) 
is the Bernstein-Sato operator for the Kirchhoff polynomial $\mathcal{U}=|g|$,
\[\label{Bernstein2}
\mathcal{B}_{|g|} |g|^a = b_{|g|}(a) |g|^{a-1},\qquad b_{|g|}(a) =- \prod_{k=1}^{n-1}(1-k-2a).
\]
The $b$-function here is the same as that in \eqref{Bernstein1}, and the leading $a^{n-1}$ behaviour can be understood via the  analogous argument to that in \eqref{leadinga}.  
We note the result \eqref{Bernstein2} is equivalent to Theorem 2.15 of \cite{Caracciolo_2013}, since  
$|m| = |m^{(n+1,n+1)}|-|m^{(n+1,n+1)}+J|$ where $J$ is the $n\times n$ all-1s matrix, and $|m^{(n+1,n+1)}|$ is  the Cayley-Menger minor formed by deleting the final row and column consisting of 1s and 0s.
In addition, we find 
\[
\mathcal{B}_{|g|} (\partial_{v_{ij}}|g|)^a = 0.
\]

Some further results worth recording are the following. For the second minors of the Laplacian matrix, $|\tilde{g}^{(ij,ij)}|=\partial_{v_{ij}}|g|$, we find the operator
\[
\mathcal{B}_{\partial_{v_{ij}}|g|} = (\partial_{s_{ij}}|m|)\Big|_{s_{kl}\rightarrow \partial_{v_{kl}}}
\]
satisfies
\begin{align}\label{Bern3}
\mathcal{B}_{\partial_{v_{ij}}|g|}
\,(\partial_{v_{ij}}|g|)^a &= b(a)\, (\partial_{v_{ij}}|g|)^{a-1}, \qquad 
\mathcal{B}_{\partial_{v_{ij}}|g|} \,|g|^a =b(a)v_{ij} |g|^{a-1} 
\end{align}
where the $b$-function is proportional to that in \eqref{Bernstein1} but is missing the final factor, 
\[\label{bfn2}
b(a) = 2\prod_{k=1}^{n-2}(1-k-2a).
\]
Again, we have verified these identities for values up to and including $n=5$.  
This operator further annihilates all $(\partial_{v_{kl}}|g|)^a$
corresponding to other legs, {\it i.e.,} 
\[
\mathcal{B}_{\partial_{v_{ij}}|g|}(\partial_{v_{kl}}|g|)^a =0 \qquad \mathrm{for\,\, all}\quad (i,j)\neq (k,l).
\]
Similarly,
\[
(\partial_{v_{ij}}|g|)\Big|_{v_{kl}\rightarrow \partial_{s_{kl}}} |m|^a = b(a) s_{ij} |m|^{a-1}
\]
with the same $b$-function \eqref{bfn2}, 
but this operator does not appear to act simply (for $n>3$) on 
$(\partial_{s_{ij}}|m|)^a$, in contrast to \eqref{Bern3}.  Finding a Bernstein-Sato operator for  $(\partial_{s_{ij}}|m|)^a$ would be useful since by \eqref{dmdstominnors} this corresponds to the Cayley-Menger minors featuring in \eqref{Cay2}.

In principle, given a Bernstein-Sato relation such as \eqref{Bernstein1}, one might hope to apply it inside the parametric representation \eqref{Cay2} and integrate by parts to obtain an operator acting solely on the Schwinger exponential.  Since the exponential is diagonal in the representation \eqref{Cay2}, the result could then be translated to a differential operator in the external momenta.  This would then yield a new weight-shifting operator.

In practice, however, we must account for all the other powers of Cayley-Menger minors present in \eqref{Cay2}, as well as the arbitrary function.
Either we must find a modified Bernstein-Sato operator that acts appropriately on the {\it entire} non-exponential prefactor in \eqref{Cay2}, which seems hard to do, or else we must find some means of removing and then restoring these other factors.  The Cayley-Menger minors, for example, can be removed and then restored via a conjugation $\Omega\,\mathcal{B}_{|m|}\Omega^{-1}$ where $\Omega=\prod_{i<j}^n |m^{(i,j)}|^{-\alpha_{ij}-1}$.  After multiplying out, however, this conjugated operator is not in the Weyl algebra ({\it i.e.,} is non-polynomial in the $s_{ij}$ and their derivatives) and so does not trivially translate into an operator in the external momenta.  On the other hand, if we include additional powers of the $|m^{(i,j)}|$ on the left, so as to recover an operator in the Weyl algebra, besides lowering $\alpha$ in \eqref{Cay2} we also lower some of the $\alpha_{ij}$.  The operator then does not lower the spacetime dimension $d$.  
Thus we have not succeeded in finding new weight-shifting operators via this route, though with some variation the method might yet  be successful.

\end{chapter}
\clearpage{} 
  \clearpage{}\begin{chapter}{Appendix to chapter \ref{cha5}\label{app:gkz}}
\allowdisplaybreaks
\section{GKZ representation of Feynman integrals}
\label{LPrepapp}

In this appendix we relate a generic $L$-loop Feynman integral of the form \eqref{feynprop} to the corresponding GKZ integral \eqref{feyngkz}.  Related discussions can be found in, {\it e.g.,} \cite{de_la_Cruz_2019, Weinzierl:2022eaz}. 

After exponentiating the propagators and integrating out the loop momenta, \eqref{feynprop} has  the
Schwinger parametrisation 
\[
I = (4\pi)^{-\g_0 L}\Big(\prod_{i=1}^N \frac{1}{\Gamma(\g_i)}\int_0^\infty \dd t_i \, t_i^{\g_i-1}\Big)\, \mathcal{U}[t]^{-\g_0}\exp\Big(-\frac{\mathcal{F}[t]}{\mathcal{U}[t]}\Big),\qquad \g_0=\frac{d}{2},
\]
where $\mathcal{U}[t]$ and $\mathcal{F}[t]$ are the first and second Symanzik polynomials respectively, which are homogeneous  of weights $L$ and $L+1$ in the Schwinger parameters $t_i$.   
The prefactor of $(4\pi)^{-\g_0L}$ is simply  that in \eqref{feynprop} multiplied by $L$ factors of $\pi^{d/2}$ from integrating out the loop momenta.
The corresponding Feynman representation is obtained by reparametrising
\[
t_i = \sigma y_i, \qquad y_t = \sum_{i=1}^N y_i = 1
\]
and integrating out the variable $\sigma$. 
Using the Jacobian\footnote{See, {\it e.g.,} Appendix B of \cite{Bzowski:2020kfw}.}
\[
\prod_{i=1}^N \dd t_i = \sigma^{N-1}\dd\sigma \prod_{i=1}^N \dd y_i \,\delta(1-y_t),
\]
as well as the homogeneity of the Symanzik polynomials, we find
\begin{align}
I &= (4\pi)^{-\g_0 L}\Big(\prod_{i=1}^N\frac{1}{\Gamma(\g_i)} \int_0^1 \dd y_i \, y_i^{\g_i-1}\Big)\delta(1-y_t)\mathcal{U}[y]^{-\g_0}\int_0^\infty \dd \sigma\, \sigma^{\g_t-\g_0L-1}\exp\Big(-\sigma\frac{\mathcal{F}[y]}{\mathcal{U}[y]}\Big)\nn\\
&= (4\pi)^{-\g_0 L}\Gamma(\g_t-\g_0L)\Big(\prod_{i=1}^N\frac{1}{\Gamma(\g_i)} \int_0^1 \dd y_i \, y_i^{\g_i-1}\Big)\delta(1-y_t)\mathcal{U}[y]^{\g_t-\g_0(L+1)}\mathcal{F}[y]^{-\g_t+\g_0L}.
\end{align}
In special cases where $\g_t-\g_0(L+1)$ vanishes ({\it e.g.,} $d=2$ multi-loop sunsets with standard propagators) one can use the $\mathcal{F}$ polynomial alone to construct a GKZ representation \cite{Vanhove:2018mto}.  More generally, one can use the 
Lee-Pomeransky representation \cite{Lee:2013hzt, de_la_Cruz_2019} obtained by combining the two Symanzik polynomial factors using the Euler beta identity
\[
\mathcal{U}[y]^{-a}\mathcal{F}[y]^{a-b} = \frac{\Gamma(b)}{\Gamma(a)\Gamma(b-a)}\int_0^\infty\dd s\,s^{a-1}(\mathcal{F}[y]+s\,\mathcal{U}[y])^{-b}
\]
with $a=\g_0(L+1)-\g_t$ and $b=\g_0$ giving
\begin{align}
I&=c_{\bs{\g}}\,\Big(\prod_{i=1}^N\int_0^1 \dd y_i \, y_i^{\g_i-1}\Big)\delta(1-y_t)\int_0^\infty \dd s\, s^{\g_0(L+1)-\g_t-1}(\mathcal{F}[y]+s\,\mathcal{U}[y])^{-\g_0}
\end{align}
where
\[
c_{\bs{\g}}=\frac{(4\pi)^{-L\g_0}\Gamma\left(\g_0\right)}{\Gamma\left((L+1)\g_0-\g_t\right)\prod_{i=1}^N\Gamma(\g_i)}.
\]
Setting $y_i = s z_i$ and using once again the homogeneity of the Symanzik polynomials, we can eliminate the $s$ integral since
\[
\int_0^\infty\frac{\dd s}{s}\,\delta(1-s z_t) = \int_0^\infty \frac{\dd s}{sz_t}\delta(z_t^{-1}-s) = 1
\]
after which 
\begin{align}
I &= c_{\bs{\g}}\, \Big(\prod_{i=1}^N\int_0^\infty \dd z_i \, z_i^{\g_i-1}\Big)(\mathcal{F}[z]+\mathcal{U}[z])^{-\g_0}.
\end{align}
Finally, this Lee-Pomeransky representation is upgraded to the GKZ representation by replacing the coefficient of every term in the denominator $\mathcal{F}[z]+\mathcal{U}[z]$ with an independent variable $x_k$.  
For the massless triangle integral, for example, 
\[
\mathcal{U}[z] = z_1+z_2+z_3,\qquad \mathcal{F}[z]=p_1^2 z_2z_3+p_2^2 z_3z_1+p_3^2 z_1 z_2
\]
and so we replace the Lee-Pomeransky denominator
\[
\mathcal{G} = \mathcal{F}[z]+\mathcal{U}[z] = p_1^2 z_2z_3+p_2^2 z_3z_1+p_3^2 z_1 z_2+z_1+z_2+z_3
\]
with the GKZ denominator
\[
\mathcal{D} = x_1 z_2z_3+x_2 z_3z_1+x_3 z_1 z_2+x_4 z_1+ x_5 z_2+x_6 z_3.
\]
The GKZ integral 
\[
\mathcal{I}_{\bs{\g}} = \Big(\prod_{i=1}^N\int_0^\infty \dd z_i \, z_i^{\g_i-1}\Big)\mathcal{D}^{-\g_0}
\]
is then related to the massless triangle integral by
\[
I = c_{\bs{\g}}\mathcal{I}_{\bs{\g}} 
\]
evaluated on the physical hypersurface
\[
x_i = p_i^2, \qquad x_{i+3}=1, \qquad i=1,2,3.
\]

\section{Creation operators for the position-space contact Witten diagram}
\label{sec:Dfn}

In position space, the $n$-point AdS contact Witten diagram 
\[
I_n = \int_0^\infty\frac{\dd z}{z^{d+1}}\int\dd^d\bs{x}_0\prod_{i=1}^n C_{\Delta_i}\Big(\frac{z}{z^2+x_{i0}^2}\Big)^{\Delta_i},\qquad
C_{\Delta_i} = \frac{\Gamma(\Delta_i)}{\pi^{d/2}\Gamma(\Delta_i-\frac{d}{2})},
\]
has the parametric representation\footnote{See, {\it e.g.,} equations (5.46)--(5.51) and (B.1)--(B.11) of \cite{Bzowski:2020kfw}.}
\begin{align}\label{Sym_res1}
I_n &= C_n\,\Big(\prod_{i=1}^n \int_0^\infty \dd z_i\, z_i^{\Delta_i-1} \Big)\delta\big(1-\sum_{i=1}^n \kappa_i z_i\big) \Big(\sum_{i<j} z_i z_j x_{ij}^2\Big)^{-\Delta_t/2}.
\end{align}
where
\[
C_n = \frac{\pi^{d/2}}{2}\Gamma\Big(\frac{\Delta_t}{2}\Big)\Gamma\Big(\frac{\Delta_t-d}{2}\Big)\prod_{i=1}^n\frac{C_{\Delta_i}}{\Gamma(\Delta_i)}, \qquad \Delta_t = \sum_{i=1}^n\Delta_i,\qquad \bs{x}_{ij}=\bs{x}_i-\bs{x}_j.
\]
The parameters $\kappa_i\ge 0$ can be chosen arbitrarily provided they are not all zero.  For the 4-point function specifically, choosing $\kappa_i=\delta_{i4}$ and eliminating $y_4$ using the delta function leads to the GKZ representation
\begin{align}
I_4&=C_4 \mathcal{I}_{\bs{\g}},\qquad \mathcal{I}_{\bs{\g}}= \Big(\prod_{i=1}^3 \int_0^\infty \dd z_i \, z_i^{\g_i-1}\Big)\mathcal{D}^{-\g_0}
 \label{Dfunction}
\end{align}
where
\[
\mathcal{D} =  x_1 z_2 z_3 + x_2 z_1 z_3+x_3 z_1 z_2+x_4 z_1 + x_5 z_2 + x_6 z_3,
\]
the parameters
\[
\g_1=\Delta_1,\qquad \g_2 =\Delta_2,\qquad
\g_3=\Delta_3, \qquad \g_0 =\frac{1}{2}(\Delta_1+\Delta_2+\Delta_3+\Delta_4),
\]
and the GKZ variables are related to the physical coordinate separations by 
\[\label{Dfnphyshyp}
(x_1,x_2,x_3,x_4,x_5,x_6)=
(x_{23}^2,x_{13}^2,x_{12}^2,x_{14}^2,x_{24}^2,x_{34}^2).
\]
Comparing with \eqref{triD}, the position-space 4-point contact diagram, also known as the holographic $D$-function \cite{DHoker:1999kzh}, is thus equivalent to the massless triangle integral (see also \cite{Bzowski:2020kfw}).
As shown on page \pageref{tripleKex}, the massless triangle integral is itself equivalent to the triple-$K$ integral (or momentum-space 3-point contact diagram) under  affine reparametrisation of the GKZ integral. 
The creation operators for the position-space contact diagram are thus  those analysed in section \ref{sec:3KC} and \ref{sec:3KW}, except that no final projection to the physical hypersurface is required as all the GKZ variables in \eqref{Dfnphyshyp} are  physical.

Concretely, the $\mathcal{A}$-matrix \eqref{Atri} leads to the  Euler equations \eqref{triEulers} and DWI \eqref{triDWI}, and the toric equations \eqref{tritorics}. 
The Newton polytope corresponds to the right-hand panel in figure \ref{fig:tripolytope}.
From its facets we obtain the singularity conditions
\begin{align}\label{Dfun-sing}
& \g_i=-n_i,\qquad\qquad \g_0- \g_i=-m_i,\qquad\qquad i =1,2,3,\nn\\&
 \g_1+\g_2+\g_3-\g_0=-n,\qquad  2\g_0-\g_1-\g_2-\g_3=-m,
\end{align}
where  $n_i,m_i,n,m\in \mathbb{Z}^+$. 
The action of the annihilator $\p_1 = \p/\p x_{23}^2$ is to raise $\g_0$, $\g_2$ and $\g_3$ by one which corresponds to  raising $\Delta_2$ and $\Delta_3$ by one, and the 
 action of the creation operator $\mathcal{C}_1$ is the reverse of this.
The corresponding $b$-function 
\[
b_1=\g_2\g_3(\g_0-\g_1)(\g_1+\g_2+\g_3-\g_0),
\]
when re-expressed in terms of Euler operator is 
\[
B_{1}=(\t_1+\t_3+\t_5)(\t_1+\t_2+\t_6)(\t_1+\t_5+\t_6)(\t_1+\t_2+\t_3).
\]
As expected, this is simply  \eqref{3ptbigBex} under the mapping $\bar{\t}_i=\t_{i+3}$ 
since the affine reparametrisation from the $\mathcal{A}$-matrix \eqref{Atri} to \eqref{3KA} leaves the creation operators unchanged.
Expanding out and using the toric equations to factorise $B_1 = \mathcal{C}_1\p_1$, we recover the creation operator \eqref{3KC1gkz} in GKZ variables.  In our present variables \eqref{Dfnphyshyp}, this is 
\begin{align}
\mathcal{C}_1&
 = x_1(\t_1+1+u_2+u_3)\big((\t_1+1+u_2)(\t_1+1+u_3)+2(v_2+v_3)\big)\notag\\
&\quad +x_2x_5\p_4\bigl(
1+u_2+v_2-v_3+(u_2+u_3+2)u_3\bigr)\notag\\
&\quad +x_3x_6\p_4\bigl(1+u_3+v_3-v_2+(u_2+u_3+2)u_2\bigr)
\end{align}
where $u_i = \t_i+\t_{i+3}$ and $v_i = \t_i\t_{i+3}$.  
One likewise obtains the operator \eqref{W3ptgkz}, namely
\begin{align}
W_{12}^{--}&=(\t_4+\t_5+\t_6+\t_3)(x_4\p_2+x_5\p_1)+x_3x_6 \p_2\p_1. 
\end{align}
Both these operators can be rewritten in various equivalent forms using the DWI and Euler equations.
Their action on the position-space contact diagram follows from \eqref{affinetransfofg}, namely
\[
\mathcal{C}_1:\,\,\Delta_2\rightarrow\Delta_2-1,\quad \Delta_3\rightarrow\Delta_3-1,\qquad
W_{12}^{--}:\,\,\Delta_3\rightarrow \Delta_3+1,\quad\Delta_4\rightarrow\Delta_4-1.
\]

\section{Non-minimal b-functions}
\label{sec:nonnormal}

As we have seen, creation operators are constructed starting from a polynomial $b(\bs{\g})$ in the spectral parameters  known as the $b$-function.
In section \ref{sec:findingb}, we argued that $b(\bs{\g})$ must possess a certain {\it minimal} set of zeros, namely,  those required to cancel the singularities arising when a creation operator shifts us from a finite to a singular integral.  Notice however that this  argument does  not preclude the existence of {\it additional} zeros besides this minimal set.  
For all the Feynman and Witten diagram examples in the main text, the minimal $b$-functions were sufficient for the construction of all creation operators.  As these $b$-functions contain the fewest factors, the resulting creation operators were moreover of lowest possible order in derivatives.
Nevertheless, there are instances where the minimal $b$-function is not sufficient: a simple example, which we analyse in this appendix, is the GKZ integral \eqref{houseint}.
As we will show, additional factors must be appended to the minimal $b$-functions in order to be able to apply the toric equations and factorise 
into a product of creation and annihilation operators.  The zeros of these additional factors are all parallel to the facets of the rescaled Newton polytope, and in most (though not all) cases correspond to additional singular hyperplanes of the GKZ integral. 

Let us recall the necessary analysis of section \ref{sec:sings}.  The integral \eqref{houseint}, namely
\[\label{houseint2}
\mathcal{I}_{\bs{\gamma}} = \int_0^\infty\dd z_1\int_0^\infty\dd z_2 \,z_1^{\g_1-1}z_2^{\g_2-1} (x_1+x_2 z_2+x_3 z_1^2+x_4 z_1 z_2^2)^{-\gamma_0},
\]
corresponds to the $\mathcal{A}$-matrix 
\[\label{houseA2}
\mathcal{A} = \left(\begin{matrix} 1 & 1 & 1& 1\\ 0 & 0 & 2 & 1\\ 0 & 1 & 0 & 2\end{matrix}\right)
\]
with DWI and Euler equations
 \[\label{houseEul}
0=(\g_0+\t_1+\t_2+\t_3+\t_4)\mathcal{I}_{\bs{\gamma}},\quad 0=(\g_1+2\t_3+\t_4)\mathcal{I}_{\bs{\gamma}},\quad 0=(\g_2+\t_2+2\t_4)\mathcal{I}_{\bs{\gamma}},
\]
and a single toric equation
\[\label{housetor}
0=(\p_1^3\p_4^2-\p_2^4\p_3)\mathcal{I}_{\bs{\g}}.
\] 
The  singularities of this integral, derived in \eqref{housesing}, are 
\begin{align}\label{housesing2}
\gamma_1=-m_1,\quad \gamma_2= -m_2,\quad \gamma_0+\gamma_1-\gamma_2 = -m_3,\quad
4\gamma_0-2\gamma_1-\gamma_2=-3m_4,
\end{align}
for all $m_j\in\mathbb{Z}^+$.
The annihilation operators $\partial_j$ send $\bs{\gamma}\rightarrow\bs{\gamma}'$ while the creation operators $\mathcal{C}_j$ send $\bs{\gamma}'\rightarrow\bs{\gamma}$,
where for each $j$ these parameters are related by
\begin{align}
j&=1: & \gamma_0'&=\gamma_0+1,&\gamma_1'&=\gamma_1, & \gamma_2'&=\gamma_2 \nn\\
j&=2: & \gamma_0'&=\gamma_0+1, &\gamma_1'&= \gamma_1, & \gamma_2'&=\gamma_2+1\nn\\
j&=3: & \gamma_0'&=\gamma_0+1, & \gamma_1'&=\gamma_1+2,&\gamma_2'&=\gamma_2,\nn\\
j&=4: & \gamma_0'&=\gamma_0+1, & \gamma_1'&=\gamma_1+1, & \gamma_2'&=\gamma_2+2.
\end{align}
According to \eqref{stdbfn}, the minimal $b$-functions containing only the  zeros necessary to cancel the singularities produced by the action of the $\mathcal{C}_j$ are 
\begin{align}\label{housebmin}
b_1^{\mathrm{min}}&=(\g_0+\g_1-\g_2)\prod_{m_4=0}^1(4\gamma_0-2\g_1-\gamma_2+3m_4),\nn\\
b_2^{\mathrm{min}}&=\gamma_2(4\gamma_0-2\g_1-\gamma_2),\nn\\
b_3^{\mathrm{min}}&=\prod_{m_1=0}^1(\g_1+m_1)\prod_{m_3=0}^2(\g_0+\g_1-\g_2+m_3),\nn\\
b_4^{\mathrm{min}}&=\gamma_1\prod_{m_2=0}^1(\g_2+m_2).
\end{align}
For example, $\mathcal{C}_3$ shifts $m_1\rightarrow m_1+2$ and $m_3\rightarrow m_3+3$, and so the five singular integrals with $m_1=0, 1$ and $m_3=0,1,2$ in \eqref{housesing2} are accessible starting from finite integrals.  This means $b_3^{\mathrm{min}}$ has the five zeros shown, which act to cancel these singularities. 
The operator $\mathcal{C}_1$ is however a special case: this sends $m_3\rightarrow m_3+1$ and $m_4\rightarrow m_4 + 4/3$, corresponding to a non-integer $F_1^{(4)}=4/3$ in \eqref{Fdef}.  Only the singularities with $m_3=0$ and $m_4=0,1$ are then accessible starting from finite integrals for which all $m_j<0$.  (In other words, integrals for which the GKZ  representation \eqref{houseint2} converges without meromorphic continuation.)

Using the DWI and Euler equations to rewrite these  $b$-functions in terms of Euler operators, we then find
\begin{align}\label{Bjmin}
B_1^{\mathrm{min}}&=-(\t_1+3\t_3)\prod_{m_4=0}^1(4\t_1+3\t_2-3m_4),\nn\\
B_2^{\mathrm{min}}&=(\t_2+2\t_4)(4\t_1+3\t_2),\nn\\
B_3^{\mathrm{min}} &= -(2\t_3+\t_4)(2\t_3+\t_4-1)\prod_{m_3=0}^2(\t_1+3\t_3-m_3),\nn\\
B_4^{\mathrm{min}}&=-(2\t_3+\t_4)(\t_2+2\t_4)(\t_2+2\t_4-1).
\end{align}
At this point a problem appears:
to extract a creation operator requires factorising
\[\label{housefac}
B_j\mathcal{I}_{\bs{\g}} = \mathcal{C}_j\p_j\mathcal{I}_{\bs{\g}},
\]
however the only toric equation we have available for this purpose, \eqref{housetor}, is of {\it fifth} order in derivatives.  While $B_3^{\mathrm{min}}$ is indeed of fifth order, the remaining $B_j^{\mathrm{min}}$ are of at most third order.
Upon expanding out and ordering terms according to \eqref{Bterm}, we find
\begin{align}
B_1^{\mathrm{min}} &= (\ldots)\p_1 - 27 x_2^2 x_3\p_2^2\p_3,\nn\\
B_2^{\mathrm{min}} &= (\ldots)\p_2+8x_1x_4\p_1\p_4, \nn\\
B_3^{\mathrm{min}} &= (\ldots)\p_3-x_1^3x_4^2 \p_1^3\p_4^2,\nn\\
B_4^{\mathrm{min}} &= (\ldots)\p_4 -2x_2^2 x_3\p_2^2\p_3.
\end{align}
For $B_3^{\mathrm{min}}$, we obtain the necessary factorisation \eqref{housefac} upon using \eqref{housetor} allowing a successful construction of $\mathcal{C}_3$.  
For the others, the order in derivatives is too low  to apply \eqref{housetor}. 

To  find $\mathcal{C}_1$, $\mathcal{C}_2$ and $\mathcal{C}_4$, therefore, we look  for new (non-minimal) $B_j$ of the form:
\begin{align}
B_1 &= (\ldots)\p_1 +(\ldots)x_2^4 x_3\p_2^4\p_3,\nn\\
B_2 &= (\ldots)\p_2+(\ldots) x_1^3 x_4^2 \p_1^3\p_4^2, \nn\\
B_4 &= (\ldots)\p_4 +(\ldots)x_2^4 x_3\p_2^4\p_3.
\end{align}
By construction, these are all of fifth order and can be factorised into the desired form \eqref{housefac} using \eqref{housetor}.  
Since the $B_j$ must be functions of the Euler operators, and 
\[
x_i^n\p_i^n=\t_i(\t_i-1)\ldots(\t_i-n+1),
\] 
this is equivalent to seeking
\begin{align}\label{newBform}
B_1 &= (\ldots)\t_1 +(\ldots)\t_2(\t_2-1)(\t_2-2)(\t_2-3)\t_3,\nn\\
B_2 &= (\ldots)\t_2+(\ldots) \t_1(\t_1-1)(\t_1-2)\t_4(\t_4-1), 
\nn\\
B_4 &= (\ldots)\t_4 +(\ldots)\t_2(\t_2-1)(\t_2-2)(\t_2-3)\t_3.
\end{align}
For comparison, the singular hyperplanes in \eqref{housesing2}, when translated to Euler operators via \eqref{houseEul}, correspond to the zeros of
\begin{align}\label{houseEulsings}
(2\t_3+\t_4-m_1),\quad
(\t_2+2\t_4-m_2),\quad
(\t_1+3\t_3-m_3),\quad
(4\t_1+3\t_2-3m_4).
\end{align}
As the non-minimal $B_j$ in \eqref{newBform} must still contain the factors present in the minimal $B_j^{\mathrm{min}}$ in \eqref{Bjmin}, we see that for $B_1$,  and $B_4$ it suffices simply to append factors corresponding to additional singular hyperplanes:
\begin{align}\label{nonminBs}
B_1 &= -(\t_1+3\t_3)\prod_{m_4=0}^3(4\t_1+3\t_2-3m_4),\nn\\
B_4 &= -(2\t_3+\t_4)\prod_{m_2=0}^3 (\t_2+2\t_4-m_2).
\end{align}
Each of these non-minimal $B_j$ contain the factors already present in the minimal $B_j^{\mathrm{min}}$.  Moreover, they are of the form \eqref{newBform} since they correspond to performing a linear shift in $\t_j$ on each of the factors present in the second term of each $B_j$ in \eqref{newBform}.  (Equivalently, setting $\t_j$ to zero in each of the $B_j$ in \eqref{nonminBs} yields the second term of each $B_j$ in \eqref{newBform}.)  This also shows that they are of the smallest order in derivatives consistent with \eqref{newBform}. 

For $B_2$, the additional factors we must append to $B_2^{\mathrm{min}}$ are parallel to the singular hyperplanes in \eqref{houseEulsings} but have different spacing.  Explicitly, we require
\[\label{nonminB2}
B_2 = -\prod_{m_1=0}^1 (\t_2+2\t_4-2m_1)\prod_{m_2=0}^2(4\t_1+3\t_2-4m_2)
\]
so that, when expanded in $\t_2$, we obtain an expression of the form given in \eqref{newBform}.  Note this is not possible using the spacings in \eqref{houseEulsings}.\footnote{The zeros of \eqref{nonminB2}, and of the corresponding of $b_2$ in \eqref{finalbs}, {\it do} however coincide with the singular hyperplanes of the integral obtained by deleting the second column of the $\mathcal{A}$-matrix.  This removes a vertex from the Newton polytope changing the spacings of the singular hyperplanes; a procedure consistent with confining all $\t_2$ dependence in $B_2$ to the first factor in \eqref{newBform}.}

By construction,  the non-minimal $B_j$ in \eqref{nonminBs} and \eqref{nonminB2} all derive from corresponding non-minimal  $b$-functions which are polynomials in the spectral parameters,
\begin{align}\label{finalbs}
b_1&=(\g_0+\g_1-\g_2)\prod_{m_4=0}^3(4\g_0-2\g_1-\g_2+3m_4),\nn\\
b_2 &= \prod_{m_1=0}^1(\g_2+2m_1)\prod_{m_2=0}^2(4\g_0-2\g_1-\g_2+4m_2),\nn\\
b_4&=\g_1\prod_{m_2=0}^3(\g_2+m_2).
\end{align}and all lead to valid creation operators via \eqref{housefac}.  From $B_2$, for example, we find 
\begin{align}
\mathcal{C}_2&=x_2 \Bigl[12 \t_4^2 \Big(48 \t_1^2+12 \t_1 (3 \t_2-5)+9\t_2 (\t_2-2) +5\Big)\nn\\&\qquad +4 \t_4 \Big(64 \t_1^3+48 \t_1^2 (3 \t_2-4) +4 \t_1 \Big(9 \t_2 (3 \t_2-5)+32\Big)+3 \t_2 \Big(9 \t_2(\t_2-2) +5\Big)\Big)\nn\\&\qquad +(\t_2-1) (4 \t_1+3 \t_2-5) (4 \t_1+3 \t_2-1) (4 \t_1+3 \t_2+3)\Big]+256x_1^3x_4^2\p_2^3\p_3.
\end{align}

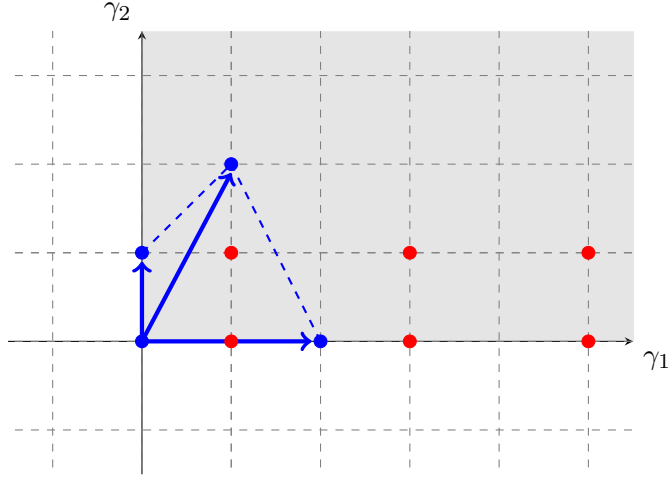
\begin{figure}[t]
\centering
\hspace{2cm}
\begin{tikzpicture}[scale=1.2][>=latex]
\begin{axis}[
  axis x line=middle,
  axis y line=middle,
  xtick,
  ytick,
  xlabel={$\gamma_1$},
  ylabel={$\gamma_2$},
  xlabel style={below right},
  ylabel style={above left},
  xmin=-1.5,
  xmax=5.5,
  ymin=-1.5,
  ymax=3.5,
  axis equal image]

\addplot [mark=none,domain=0:10,draw=none,name path=A] {3.5};
 \addplot [mark=none,domain=0:10,draw=none,name path=B] {0};
\addplot [gray, fill opacity=0.2] fill between[of=A and B,soft clip={domain=0.01:10}];

\foreach \k in {0,...,8} {
 
\addplot [mark=none,dashed,gray,domain=-2.5:10] {\k+1};
\addplot [mark=none,dashed,gray,domain=-2.5:10] (\k+1,x);
\addplot [mark=none,dashed,gray,domain=-2.5:10] {-\k+1};
\addplot [mark=none,dashed,gray,domain=-2.5:10] (-\k+1,x);

}
\foreach \k in {0,...,8} {
\addplot[holdot,red] coordinates{(2*\k+1,0)(2*\k+1,1)};
}
\addplot [holdot,blue] coordinates{(0,0)(0,1)(1,2)(2,0)};
\draw[->, blue,ultra thick](0,0)--(0,0.9);
\draw[->,blue,ultra thick](0,0)--(1,1.9);
\draw[->,blue,ultra thick](0,0)--(1.9,0);
\draw[blue,thick, dashed](0,1)--(1,2);
\draw[blue,thick, dashed](2,0)--(1,2);
\end{axis}
\end{tikzpicture}
\caption{For purposes of illustration, a two-dimensional example of a non-normal lattice can be obtained by projecting to the $(\gamma_1,\gamma_2)$ plane.  The points $(2k+1,0)$ and $(2k+1,1)$ for $k\in\mathbb{Z}^+$ (shown in red) lie within the positive cone  $\sum_j \mathbb{R}^+\bs{a}_j$ (shaded region) spanned by the vertex vectors $\bs{a}_j$ of the Newton polytope (shown in blue). These points also belong to $\sum_j\mathbb{Z}\bs{a}_j$ since $(2k+1,0)=k(2,0)+(1,2)-2(0,1)$ and $(2k+1,1)=k(2,0)+(1,2)-(0,1)$.  However, these points cannot be expressed in the form $\sum_j \mathbb{Z}^+\bs{a}_j$ and hence the lattice generated by the $\bs{a}_j$ is non-normal.
\label{fig:non-normal}}
\end{figure}

Having solved this example, let us note that the failure of the minimal $b$-functions in \eqref{housebmin} can also be understood geometrically.   For $B_1^{\mathrm{min}}$, $B_2^{\mathrm{min}}$ and $B_4^{\mathrm{min}}$ in \eqref{Bjmin} to be factorisable as $\mathcal{C}_j\p_j$, we would need each of 
$
\p_2^2\p_3\p_1^{-1}$, $\p_1\p_4\p_2^{-1}$ and $\p_2^2\p_3\p_4^{-1}$ to be expressible as $\prod_{k=1}^N \p_k^{c_k}$ for some set of $c_k\in\mathbb{Z}^+$.  (The inverses here are purely formal: we mean that $\p_2^2\p_3 = \p_1\prod_{k=1}^N \p_k^{c_k}$, {\it etc}.)
In terms of the shifts produced by these operators on the spectral parameters $\bs{\g}=(\g_0,\g_1,\g_2)$, this is equivalent to requiring that 
\begin{align}
2\bs{\mathcal{A}}_2+\bs{\mathcal{A}}_3-\bs{\mathcal{A}}_1&=\left(\begin{matrix}2\\2\\2\end{matrix}\right),  \quad \bs{\mathcal{A}}_1+\bs{\mathcal{A}}_4-\bs{\mathcal{A}}_2= \left(\begin{matrix}1\\1\\1\end{matrix}\right),\quad  2\bs{\mathcal{A}}_2+\bs{\mathcal{A}}_3-\bs{\mathcal{A}}_4= \left(\begin{matrix}2\\1\\0\end{matrix}\right)
\end{align}
are all expressible as  $\sum_{k=1}^N c_k\bs{\mathcal{A}}_k$ for some set of $c_k\in\mathbb{Z}^+$, where $\bs{\mathcal{A}}_k$ denotes the $k$th column of the $\mathcal{A}$-matrix including the top row of ones. 
Clearly this is not possible, although these vectors do all lie in the positive cone corresponding to solutions with $c_k\in\mathbb{R}^+$ since
\begin{align}
2\bs{\mathcal{A}}_2+\bs{\mathcal{A}}_3-\bs{\mathcal{A}}_1&=2(\bs{\mathcal{A}}_1+\bs{\mathcal{A}}_4-\bs{\mathcal{A}}_2) = \frac{2}{3}(\bs{\mathcal{A}}_2+\bs{\mathcal{A}}_3+\bs{\mathcal{A}}_4),\nn\\
2\bs{\mathcal{A}}_2+\bs{\mathcal{A}}_3-\bs{\mathcal{A}}_4&=\frac{1}{2}(3\bs{\mathcal{A}}_1+\bs{\mathcal{A}}_3).
\end{align}   
 Mathematically, this is precisely the condition that lattice generated by the $\bs{\mathcal{A}}_k$ (or equivalently, the toric ideal associated with the $\mathcal{A}$-matrix) is {\it non-normal}.  Conversely, when the normality condition
\[\label{normaltoric}
\Big(\sum_k \mathbb{R}^+\bs{\mathcal{A}}_k\Big)\cap\Big(\sum_k \mathbb{Z}\bs{\mathcal{A}}_k\Big) = \Big(\sum_k \mathbb{Z}^+\bs{\mathcal{A}}_k\Big)
\]
is satisfied, it can be shown that the minimal $b$-functions \eqref{stdbfn} generate valid creation operators \cite{Saito_param_shift, Saito_restrictions}.
The non-triviality of the normality condition is illustrated in figure \ref{fig:non-normal}.

\end{chapter}
\clearpage{}

\providecommand{\href}[2]{#2}\begingroup\raggedright\endgroup

          \bibliographystyle{JHEP}

\begin{thebibliography}{100}

\bibitem{WeilSimone}
S.~Weil, R.~Chenavier, A.~Devaux, M.~Chenavier, A.~Devaux and O.~Rey,
  \emph{{Oeuvres compl{\`e}tes 7 Correspondance 1 Correspondance familiale}},
  Gallimard, {Paris} (2012).

\bibitem{DiFrancesco}
P.~Di~Francesco, P.~Mathieu and D.~Senechal, \emph{{Conformal Field Theory}},
  Springer-Verlag, New York (1997).

\bibitem{Polchinski}
J.~Polchinski, \emph{{String theory. Vol. 1: An introduction to the bosonic
  string}}, Cambridge University Press (2007).

\bibitem{Maldacena97}
J.M.~Maldacena, \emph{{The Large N limit of superconformal field theories and
  supergravity}}, \href{https://doi.org/10.1023/A:1026654312961,
  10.4310/ATMP.1998.v2.n2.a1}{\emph{Int. J. Theor. Phys.} {\bfseries 38} (1999)
  1113} [\href{https://arxiv.org/abs/hep-th/9711200}{{\ttfamily
  hep-th/9711200}}].

\bibitem{Maldacena99}
O.~Aharony, S.S.~Gubser, J.M.~Maldacena, H.~Ooguri and Y.~Oz, \emph{{Large N
  field theories, string theory and gravity}},
  \href{https://doi.org/10.1016/S0370-1573(99)00083-6}{\emph{Phys. Rept.}
  {\bfseries 323} (2000) 183}
  [\href{https://arxiv.org/abs/hep-th/9905111}{{\ttfamily hep-th/9905111}}].

\bibitem{Elvang:2020lue}
H.~Elvang, \emph{{Bootstrap and amplitudes: a hike in the landscape of quantum
  field theory}}, \href{https://doi.org/10.1088/1361-6633/abf97e}{\emph{Rept.
  Prog. Phys.} {\bfseries 84} (2021) 074201}
  [\href{https://arxiv.org/abs/2007.08436}{{\ttfamily 2007.08436}}].

\bibitem{Polyakov:1970xd}
A.M.~Polyakov, \emph{{Conformal symmetry of critical fluctuations}},
  {\emph{JETP Lett.} {\bfseries 12} (1970) 381}.

\bibitem{Mack:1969rr}
G.~Mack and A.~Salam, \emph{{Finite component field representations of the
  conformal group}},
  \href{https://doi.org/10.1016/0003-4916(69)90278-4}{\emph{Annals Phys.}
  {\bfseries 53} (1969) 174}.

\bibitem{Parisi:1971zza}
G.~Parisi and L.~Peliti, \emph{{Calculation of critical indices}},
  \href{https://doi.org/https://doi.org/10.1007/BF02784709}{\emph{Lett. Nuovo
  Cimento} {\bfseries 2} (1971) 627}.

\bibitem{FERRARA197277}
S.~Ferrara, A.~Grillo, G.~Parisi and R.~Gatto, \emph{Covariant expansion of the
  conformal four -point function},
  \href{https://doi.org/https://doi.org/10.1016/0550-3213(72)90587-1}{\emph{Nuclear
  Physics B} {\bfseries 49} (1972) 77}.

\bibitem{Mack:1972kq}
G.~Mack and K.~Symanzik, \emph{{Currents, stress tensor and generalized
  unitarity in conformal invariant quantum field theory}},
  \href{https://doi.org/10.1007/BF01645514}{\emph{Commun. Math. Phys.}
  {\bfseries 27} (1972) 247}.

\bibitem{Fradkin:1994zf}
E.S.~Fradkin and M.Y.~Palchik, \emph{{Method of solving conformal models in
  D-dimensional space. 1}},
  \href{https://doi.org/10.1006/aphy.1996.0065}{\emph{Annals Phys.} {\bfseries
  249} (1996) 44}.

\bibitem{Osborn:1993cr}
H.~Osborn and A.C.~Petkou, \emph{{Implications of conformal invariance in field
  theories for general dimensions}},
  \href{https://doi.org/10.1006/aphy.1994.1045}{\emph{Annals Phys.} {\bfseries
  231} (1994) 311} [\href{https://arxiv.org/abs/hep-th/9307010}{{\ttfamily
  hep-th/9307010}}].

\bibitem{Giannotti:2008cv}
M.~Giannotti and E.~Mottola, \emph{{The Trace Anomaly and Massless Scalar
  Degrees of Freedom in Gravity}},
  \href{https://doi.org/10.1103/PhysRevD.79.045014}{\emph{Phys. Rev. D}
  {\bfseries 79} (2009) 045014}
  [\href{https://arxiv.org/abs/0812.0351}{{\ttfamily 0812.0351}}].

\bibitem{Armillis:2009pq}
R.~Armillis, C.~Coriano and L.~Delle~Rose, \emph{{Conformal Anomalies and the
  Gravitational Effective Action: The TJJ Correlator for a Dirac Fermion}},
  \href{https://doi.org/10.1103/PhysRevD.81.085001}{\emph{Phys. Rev. D}
  {\bfseries 81} (2010) 085001}
  [\href{https://arxiv.org/abs/0910.3381}{{\ttfamily 0910.3381}}].

\bibitem{Schwimmer:2023nzk}
A.~Schwimmer and S.~Theisen, \emph{{Comments on Trace Anomaly Matching}},
  \href{https://arxiv.org/abs/2307.14957}{{\ttfamily 2307.14957}}.

\bibitem{Witten:2001kn}
E.~Witten, \emph{{Quantum gravity in de Sitter space}},  in \emph{{Strings
  2001: International Conference}}, 6, 2001
  [\href{https://arxiv.org/abs/hep-th/0106109}{{\ttfamily hep-th/0106109}}].

\bibitem{Strominger:2001pn}
A.~Strominger, \emph{{The dS / CFT correspondence}},
  \href{https://doi.org/10.1088/1126-6708/2001/10/034}{\emph{JHEP} {\bfseries
  10} (2001) 034} [\href{https://arxiv.org/abs/hep-th/0106113}{{\ttfamily
  hep-th/0106113}}].

\bibitem{Antoniadis:2011ib}
I.~Antoniadis, P.O.~Mazur and E.~Mottola, \emph{{Conformal invariance, dark
  energy, and CMB non-Gaussianity}},
  \href{https://doi.org/10.1088/1475-7516/2012/09/024}{\emph{JCAP} {\bfseries
  09} (2012) 024} [\href{https://arxiv.org/abs/1103.4164}{{\ttfamily
  1103.4164}}].

\bibitem{Maldacena:2011nz}
J.M.~Maldacena and G.L.~Pimentel, \emph{{On graviton non-Gaussianities during
  inflation}}, \href{https://doi.org/10.1007/JHEP09(2011)045}{\emph{JHEP}
  {\bfseries 09} (2011) 045} [\href{https://arxiv.org/abs/1104.2846}{{\ttfamily
  1104.2846}}].

\bibitem{Creminelli:2011mw}
P.~Creminelli, \emph{{Conformal invariance of scalar perturbations in
  inflation}}, \href{https://doi.org/10.1103/PhysRevD.85.041302}{\emph{Phys.
  Rev. D} {\bfseries 85} (2012) 041302}
  [\href{https://arxiv.org/abs/1108.0874}{{\ttfamily 1108.0874}}].

\bibitem{Bzowski:2012ih}
A.~Bzowski, P.~McFadden and K.~Skenderis, \emph{{Holography for inflation using
  conformal perturbation theory}},
  \href{https://doi.org/10.1007/JHEP04(2013)047}{\emph{JHEP} {\bfseries 04}
  (2013) 047} [\href{https://arxiv.org/abs/1211.4550}{{\ttfamily 1211.4550}}].

\bibitem{Mata:2012bx}
I.~Mata, S.~Raju and S.~Trivedi, \emph{{CMB from CFT}},
  \href{https://doi.org/10.1007/JHEP07(2013)015}{\emph{JHEP} {\bfseries 07}
  (2013) 015} [\href{https://arxiv.org/abs/1211.5482}{{\ttfamily 1211.5482}}].

\bibitem{Kehagias:2012pd}
A.~Kehagias and A.~Riotto, \emph{{Operator product expansion of inflationary
  correlators and conformal symmetry of de Sitter}},
  \href{https://doi.org/10.1016/j.nuclphysb.2012.07.004}{\emph{Nucl.Phys.}
  {\bfseries B864} (2012) 492}
  [\href{https://arxiv.org/abs/1205.1523}{{\ttfamily 1205.1523}}].

\bibitem{McFadden:2013ria}
P.~McFadden, \emph{{On the power spectrum of inflationary cosmologies dual to a
  deformed CFT}}, \href{https://doi.org/10.1007/JHEP10(2013)071}{\emph{JHEP}
  {\bfseries 10} (2013) 071} [\href{https://arxiv.org/abs/1308.0331}{{\ttfamily
  1308.0331}}].

\bibitem{Ghosh:2014kba}
A.~Ghosh, N.~Kundu, S.~Raju and S.P.~Trivedi, \emph{{Conformal Invariance and
  the Four Point Scalar Correlator in Slow-Roll Inflation}},
  \href{https://doi.org/10.1007/JHEP07(2014)011}{\emph{JHEP} {\bfseries 07}
  (2014) 011} [\href{https://arxiv.org/abs/1401.1426}{{\ttfamily 1401.1426}}].

\bibitem{Anninos:2014lwa}
D.~Anninos, T.~Anous, D.Z.~Freedman and G.~Konstantinidis, \emph{{Late-time
  Structure of the Bunch-Davies De Sitter Wavefunction}},
  \href{https://doi.org/10.1088/1475-7516/2015/11/048}{\emph{JCAP} {\bfseries
  1511} (2015) 048} [\href{https://arxiv.org/abs/1406.5490}{{\ttfamily
  1406.5490}}].

\bibitem{Arkani-Hamed:2015bza}
N.~Arkani-Hamed and J.~Maldacena, \emph{{Cosmological Collider Physics}},
  \href{https://arxiv.org/abs/1503.08043}{{\ttfamily 1503.08043}}.

\bibitem{Arkani-Hamed:2018kmz}
N.~Arkani-Hamed, D.~Baumann, H.~Lee and G.L.~Pimentel, \emph{{The Cosmological
  Bootstrap: Inflationary Correlators from Symmetries and Singularities}},
  \href{https://doi.org/10.1007/JHEP04(2020)105}{\emph{JHEP} {\bfseries 04}
  (2020) 105} [\href{https://arxiv.org/abs/1811.00024}{{\ttfamily
  1811.00024}}].

\bibitem{Baumann:2019oyu}
D.~Baumann, C.~Duaso~Pueyo, A.~Joyce, H.~Lee and G.L.~Pimentel, \emph{{The
  cosmological bootstrap: weight-shifting operators and scalar seeds}},
  \href{https://doi.org/10.1007/JHEP12(2020)204}{\emph{JHEP} {\bfseries 12}
  (2020) 204} [\href{https://arxiv.org/abs/1910.14051}{{\ttfamily
  1910.14051}}].

\bibitem{Baumann:2020dch}
D.~Baumann, C.~Duaso~Pueyo, A.~Joyce, H.~Lee and G.L.~Pimentel, \emph{{The
  Cosmological Bootstrap: Spinning Correlators from Symmetries and
  Factorization}},
  \href{https://doi.org/10.21468/SciPostPhys.11.3.071}{\emph{SciPost Phys.}
  {\bfseries 11} (2021) 071}
  [\href{https://arxiv.org/abs/2005.04234}{{\ttfamily 2005.04234}}].

\bibitem{Sleight:2019hfp}
C.~Sleight and M.~Taronna, \emph{{Bootstrapping Inflationary Correlators in
  Mellin Space}}, \href{https://doi.org/10.1007/JHEP02(2020)098}{\emph{JHEP}
  {\bfseries 02} (2020) 098}
  [\href{https://arxiv.org/abs/1907.01143}{{\ttfamily 1907.01143}}].

\bibitem{apk2016}
A.~Bzowski, P.~McFadden and K.~Skenderis, \emph{{Scalar 3-point functions in
  CFT: renormalisation, beta functions and anomalies}},
  \href{https://doi.org/10.1007/JHEP03(2016)066}{\emph{JHEP} {\bfseries 03}
  (2016) 066} [\href{https://arxiv.org/abs/arXiv:1510.08442}{{\ttfamily
  arXiv:1510.08442}}].

\bibitem{Bzowski:2017poo}
A.~Bzowski, P.~McFadden and K.~Skenderis, \emph{{Renormalised 3-point functions
  of stress tensors and conserved currents in CFT}},
  \href{https://doi.org/10.1007/JHEP11(2018)153}{\emph{JHEP} {\bfseries 11}
  (2018) 153} [\href{https://arxiv.org/abs/1711.09105}{{\ttfamily
  1711.09105}}].

\bibitem{Bzowski:2018fql}
A.~Bzowski, P.~McFadden and K.~Skenderis, \emph{{Renormalised CFT 3-point
  functions of scalars, currents and stress tensors}},
  \href{https://doi.org/10.1007/JHEP11(2018)159}{\emph{JHEP} {\bfseries 11}
  (2018) 159} [\href{https://arxiv.org/abs/1805.12100}{{\ttfamily
  1805.12100}}].

\bibitem{Bzowski:2022rlz}
A.~Bzowski, P.~McFadden and K.~Skenderis, \emph{{A handbook of holographic
  4-point functions}},
  \href{https://doi.org/10.1007/JHEP12(2022)039}{\emph{JHEP} {\bfseries 12}
  (2022) 039} [\href{https://arxiv.org/abs/2207.02872}{{\ttfamily
  2207.02872}}].

\bibitem{Raju:2012zr}
S.~Raju, \emph{{New Recursion Relations and a Flat Space Limit for AdS/CFT
  Correlators}}, \href{https://doi.org/10.1103/PhysRevD.85.126009}{\emph{Phys.
  Rev. D} {\bfseries 85} (2012) 126009}
  [\href{https://arxiv.org/abs/1201.6449}{{\ttfamily 1201.6449}}].

\bibitem{Farrow:2018yni}
J.A.~Farrow, A.E.~Lipstein and P.~McFadden, \emph{{Double copy structure of CFT
  correlators}}, \href{https://doi.org/10.1007/JHEP02(2019)130}{\emph{JHEP}
  {\bfseries 02} (2019) 130}
  [\href{https://arxiv.org/abs/1812.11129}{{\ttfamily 1812.11129}}].

\bibitem{Lipstein:2019mpu}
A.E.~Lipstein and P.~McFadden, \emph{{Double copy structure and the flat space
  limit of conformal correlators in even dimensions}},
  \href{https://doi.org/10.1103/PhysRevD.101.125006}{\emph{Phys. Rev. D}
  {\bfseries 101} (2020) 125006}
  [\href{https://arxiv.org/abs/1912.10046}{{\ttfamily 1912.10046}}].

\bibitem{Armstrong:2020woi}
C.~Armstrong, A.E.~Lipstein and J.~Mei, \emph{{Color/kinematics duality in
  AdS$_{4}$}}, \href{https://doi.org/10.1007/JHEP02(2021)194}{\emph{JHEP}
  {\bfseries 02} (2021) 194}
  [\href{https://arxiv.org/abs/2012.02059}{{\ttfamily 2012.02059}}].

\bibitem{Albayrak:2020fyp}
S.~Albayrak, S.~Kharel and D.~Meltzer, \emph{{On duality of color and
  kinematics in (A)dS momentum space}},
  \href{https://doi.org/10.1007/JHEP03(2021)249}{\emph{JHEP} {\bfseries 03}
  (2021) 249} [\href{https://arxiv.org/abs/2012.10460}{{\ttfamily
  2012.10460}}].

\bibitem{Bzowski:2013sza}
A.~Bzowski, P.~McFadden and K.~Skenderis, \emph{{Implications of conformal
  invariance in momentum space}},
  \href{https://doi.org/10.1007/JHEP03(2014)111}{\emph{JHEP} {\bfseries 03}
  (2014) 111} [\href{https://arxiv.org/abs/1304.7760}{{\ttfamily 1304.7760}}].

\bibitem{Coriano:2013jba}
C.~Corian\`{o}, L.~Delle~Rose, E.~Mottola and M.~Serino, \emph{{Solving the
  Conformal Constraints for Scalar Operators in Momentum Space and the
  Evaluation of Feynman's Master Integrals}},
  \href{https://doi.org/10.1007/JHEP07(2013)011}{\emph{JHEP} {\bfseries 07}
  (2013) 011} [\href{https://arxiv.org/abs/1304.6944}{{\ttfamily 1304.6944}}].

\bibitem{evaluation}
A.~Bzowski, P.~McFadden and K.~Skenderis, \emph{{Evaluation of conformal
  integrals}}, \href{https://doi.org/10.1007/JHEP02(2016)068}{\emph{JHEP}
  {\bfseries 02} (2016) 068}
  [\href{https://arxiv.org/abs/[arXiv:1511.02357]}{{\ttfamily
  [arXiv:1511.02357]}}].

\bibitem{Bzowski:2015pba}
A.~Bzowski, P.~McFadden and K.~Skenderis, \emph{{Scalar 3-point functions in
  CFT: renormalisation, beta functions and anomalies}},
  \href{https://doi.org/10.1007/JHEP03(2016)066}{\emph{JHEP} {\bfseries 03}
  (2016) 066} [\href{https://arxiv.org/abs/1510.08442}{{\ttfamily
  1510.08442}}].

\bibitem{Bzowski:2020kfw}
A.~Bzowski, P.~McFadden and K.~Skenderis, \emph{{Conformal correlators as
  simplex integrals in momentum space}},
  \href{https://doi.org/10.1007/JHEP01(2021)192}{\emph{JHEP} {\bfseries 01}
  (2021) 192} [\href{https://arxiv.org/abs/2008.07543}{{\ttfamily
  2008.07543}}].

\bibitem{Bzowski:2019kwd}
A.~Bzowski, P.~McFadden and K.~Skenderis, \emph{{Conformal $n$-point functions
  in momentum space}},
  \href{https://doi.org/10.1103/PhysRevLett.124.131602}{\emph{Phys. Rev. Lett.}
  {\bfseries 124} (2020) 131602}
  [\href{https://arxiv.org/abs/1910.10162}{{\ttfamily 1910.10162}}].

\bibitem{Dirac1930-DIRTPO}
P.A.M.~Dirac, \emph{The Principles of Quantum Mechanics}, Clarendon Press,
  Oxford, (1930).

\bibitem{Edmonds:1955fi}
A.R.~Edmonds, \emph{{Angular momentum in quantum mechanics}},
  \href{https://doi.org/10.5170/CERN-1955-026}{\emph{CERN-55-26} (1955) }.

\bibitem{Sakurai:2011zz}
J.J.~Sakurai and J.~Napolitano, \emph{{Modern Quantum Mechanics}}, Quantum
  physics, quantum information and quantum computation, Cambridge University
  Press (10, 2020),
  \href{https://doi.org/10.1017/9781108587280}{10.1017/9781108587280}.

\bibitem{Dirac:1926vy}
P.A.M.~Dirac, \emph{{Quantum mechanics and a preliminary investigation of the
  hydrogen atom}}, \href{https://doi.org/10.1098/rspa.1926.0034}{\emph{Proc.
  Roy. Soc. Lond. A} {\bfseries 110} (1926) 561}.

\bibitem{Pauli:1926qpj}
W.~Pauli, \emph{{\"Uber das Wasserstoffspektrum vom Standpunkt der neuen
  Quantenmechanik}}, \href{https://doi.org/10.1007/BF01450175}{\emph{Z. Phys.}
  {\bfseries 36} (1926) 336}.

\bibitem{cohen}
C.~Cohen-Tannoudji and B.~Diu, \emph{Quantum mechanics}, John Wiley And Sons,
  New York (1977).

\bibitem{Pauli:1925nmn}
W.~Pauli, \emph{{\"Uber den Zusammenhang des Abschlusses der Elektronengruppen
  im Atom mit der Komplexstruktur der Spektren}},
  \href{https://doi.org/10.1007/BF02980631}{\emph{Z. Phys.} {\bfseries 31}
  (1925) 765}.

\bibitem{Peskin:1995ev}
M.E.~Peskin and D.V.~Schroeder, \emph{{An Introduction to quantum field
  theory}}, Addison-Wesley, Reading, USA (1995).

\bibitem{Delto:2023kqv}
M.~Delto, C.~Duhr, L.~Tancredi and Y.J.~Zhu, \emph{{Two-loop QED corrections to
  the scattering of four massive leptons}},
  \href{https://arxiv.org/abs/2311.06385}{{\ttfamily 2311.06385}}.

\bibitem{Tarasov:2022pwt}
O.V.~Tarasov, \emph{{Calculation of One-Loop Integrals for Four-Photon
  Amplitudes by Functional Reduction Method}},
  \href{https://doi.org/10.1134/S1547477123030676}{\emph{Phys. Part. Nucl.
  Lett.} {\bfseries 20} (2023) 287}
  [\href{https://arxiv.org/abs/2211.15535}{{\ttfamily 2211.15535}}].

\bibitem{Bottcher:2023wsr}
N.~B\"ottcher, N.~Schwanemann and S.~Weinzierl, \emph{{Box integrals with
  fermion bubbles for low-energy measurements of the weak mixing angle}},
  \href{https://arxiv.org/abs/2312.06773}{{\ttfamily 2312.06773}}.

\bibitem{Devoto:2023rpv}
F.~Devoto, K.~Melnikov, R.~R\"ontsch, C.~Signorile-Signorile and
  D.M.~Tagliabue, \emph{{A fresh look at the nested soft-collinear subtraction
  scheme: NNLO QCD corrections to $N$-gluon final states in $q\bar{q}$
  annihilation}},  \href{https://arxiv.org/abs/2310.17598}{{\ttfamily
  2310.17598}}.

\bibitem{Agarwal:2023suw}
B.~Agarwal, F.~Buccioni, F.~Devoto, G.~Gambuti, A.~von Manteuffel and
  L.~Tancredi, \emph{{Five-Parton Scattering in QCD at Two Loops}},
  \href{https://arxiv.org/abs/2311.09870}{{\ttfamily 2311.09870}}.

\bibitem{Gasparotto:2023roh}
F.~Gasparotto, S.~Weinzierl and X.~Xu, \emph{{Real time lattice correlation
  functions from differential equations}},
  \href{https://doi.org/10.1007/JHEP06(2023)128}{\emph{JHEP} {\bfseries 06}
  (2023) 128} [\href{https://arxiv.org/abs/2305.05447}{{\ttfamily
  2305.05447}}].

\bibitem{Heckelbacher:2022hbq}
T.~Heckelbacher, I.~Sachs, E.~Skvortsov and P.~Vanhove, \emph{{Analytical
  evaluation of cosmological correlation functions}},
  \href{https://doi.org/10.1007/JHEP08(2022)139}{\emph{JHEP} {\bfseries 08}
  (2022) 139} [\href{https://arxiv.org/abs/2204.07217}{{\ttfamily
  2204.07217}}].

\bibitem{Caloro:2022zuy}
F.~Caloro and P.~McFadden, \emph{{Shift operators from the simplex
  representation in momentum-space CFT}},
  \href{https://doi.org/10.1007/JHEP03(2023)106}{\emph{JHEP} {\bfseries 03}
  (2023) 106} [\href{https://arxiv.org/abs/2212.03887}{{\ttfamily
  2212.03887}}].

\bibitem{Arkani-Hamed:2023kig}
N.~Arkani-Hamed, D.~Baumann, A.~Hillman, A.~Joyce, H.~Lee and G.L.~Pimentel,
  \emph{{Differential Equations for Cosmological Correlators}},
  \href{https://arxiv.org/abs/2312.05303}{{\ttfamily 2312.05303}}.

\bibitem{Weinzierl:2022eaz}
S.~Weinzierl, \emph{{Feynman Integrals}},
  \href{https://arxiv.org/abs/2201.03593}{{\ttfamily 2201.03593}}.

\bibitem{Chetyrkin:1981qh}
K.G.~Chetyrkin and F.V.~Tkachov, \emph{{Integration by parts: The algorithm to
  calculate $\beta$-functions in 4 loops}},
  \href{https://doi.org/10.1016/0550-3213(81)90199-1}{\emph{Nucl. Phys. B}
  {\bfseries 192} (1981) 159}.

\bibitem{Smirnov:2010hn}
A.V.~Smirnov and A.V.~Petukhov, \emph{{The Number of Master Integrals is
  Finite}}, \href{https://doi.org/10.1007/s11005-010-0450-0}{\emph{Lett. Math.
  Phys.} {\bfseries 97} (2011) 37}
  [\href{https://arxiv.org/abs/1004.4199}{{\ttfamily 1004.4199}}].

\bibitem{Grozin:2011mt}
A.G.~Grozin, \emph{{Integration by parts: An Introduction}},
  \href{https://doi.org/10.1142/S0217751X11053687}{\emph{Int. J. Mod. Phys. A}
  {\bfseries 26} (2011) 2807}
  [\href{https://arxiv.org/abs/1104.3993}{{\ttfamily 1104.3993}}].

\bibitem{Bitoun:2017nre}
T.~Bitoun, C.~Bogner, R.P.~Klausen and E.~Panzer, \emph{{Feynman integral
  relations from parametric annihilators}},
  \href{https://doi.org/10.1007/s11005-018-1114-8}{\emph{Lett. Math. Phys.}
  {\bfseries 109} (2019) 497}
  [\href{https://arxiv.org/abs/1712.09215}{{\ttfamily 1712.09215}}].

\bibitem{Tarasov:1996br}
O.V.~Tarasov, \emph{{Connection between Feynman integrals having different
  values of the space-time dimension}},
  \href{https://doi.org/10.1103/PhysRevD.54.6479}{\emph{Phys. Rev. D}
  {\bfseries 54} (1996) 6479}
  [\href{https://arxiv.org/abs/hep-th/9606018}{{\ttfamily hep-th/9606018}}].

\bibitem{Tarasov:1997kx}
O.V.~Tarasov, \emph{{Generalized recurrence relations for two loop propagator
  integrals with arbitrary masses}},
  \href{https://doi.org/10.1016/S0550-3213(97)00376-3}{\emph{Nucl. Phys. B}
  {\bfseries 502} (1997) 455}
  [\href{https://arxiv.org/abs/hep-ph/9703319}{{\ttfamily hep-ph/9703319}}].

\bibitem{Lee:2010wea}
R.N.~Lee, \emph{{Calculating multiloop integrals using dimensional recurrence
  relation and $D$-analyticity}},
  \href{https://doi.org/10.1016/j.nuclphysbps.2010.08.032}{\emph{Nucl. Phys. B
  Proc. Suppl.} {\bfseries 205-206} (2010) 135}
  [\href{https://arxiv.org/abs/1007.2256}{{\ttfamily 1007.2256}}].

\bibitem{Dolan:2003hv}
F.A.~Dolan and H.~Osborn, \emph{{Conformal partial waves and the operator
  product expansion}},
  \href{https://doi.org/10.1016/j.nuclphysb.2003.11.016}{\emph{Nucl. Phys.}
  {\bfseries B678} (2004) 491}
  [\href{https://arxiv.org/abs/hep-th/0309180}{{\ttfamily hep-th/0309180}}].

\bibitem{Dolan:2011dv}
F.A.~Dolan and H.~Osborn, \emph{{Conformal Partial Waves: Further Mathematical
  Results}},  \href{https://arxiv.org/abs/1108.6194}{{\ttfamily 1108.6194}}.

\bibitem{El-Showk:2012cjh}
S.~El-Showk, M.F.~Paulos, D.~Poland, S.~Rychkov, D.~Simmons-Duffin and
  A.~Vichi, \emph{{Solving the 3D Ising Model with the Conformal Bootstrap}},
  \href{https://doi.org/10.1103/PhysRevD.86.025022}{\emph{Phys. Rev. D}
  {\bfseries 86} (2012) 025022}
  [\href{https://arxiv.org/abs/1203.6064}{{\ttfamily 1203.6064}}].

\bibitem{Li:2017lmh}
D.~Li, D.~Meltzer and D.~Poland, \emph{{Conformal Bootstrap in the Regge
  Limit}}, \href{https://doi.org/10.1007/JHEP12(2017)013}{\emph{JHEP}
  {\bfseries 12} (2017) 013}
  [\href{https://arxiv.org/abs/1705.03453}{{\ttfamily 1705.03453}}].

\bibitem{Kulaxizi:2017ixa}
M.~Kulaxizi, A.~Parnachev and A.~Zhiboedov, \emph{{Bulk Phase Shift, CFT Regge
  Limit and Einstein Gravity}},
  \href{https://doi.org/10.1007/JHEP06(2018)121}{\emph{JHEP} {\bfseries 06}
  (2018) 121} [\href{https://arxiv.org/abs/1705.02934}{{\ttfamily
  1705.02934}}].

\bibitem{Iliesiu:2015qra}
L.~Iliesiu, F.~Kos, D.~Poland, S.S.~Pufu, D.~Simmons-Duffin and R.~Yacoby,
  \emph{{Bootstrapping 3D Fermions}},
  \href{https://doi.org/10.1007/JHEP03(2016)120}{\emph{JHEP} {\bfseries 03}
  (2016) 120} [\href{https://arxiv.org/abs/1508.00012}{{\ttfamily
  1508.00012}}].

\bibitem{Dymarsky:2017xzb}
A.~Dymarsky, J.~Penedones, E.~Trevisani and A.~Vichi, \emph{{Charting the space
  of 3D CFTs with a continuous global symmetry}},
  \href{https://doi.org/10.1007/JHEP05(2019)098}{\emph{JHEP} {\bfseries 05}
  (2019) 098} [\href{https://arxiv.org/abs/1705.04278}{{\ttfamily
  1705.04278}}].

\bibitem{Dymarsky:2017yzx}
A.~Dymarsky, F.~Kos, P.~Kravchuk, D.~Poland and D.~Simmons-Duffin, \emph{{The
  3d Stress-Tensor Bootstrap}},
  \href{https://doi.org/10.1007/JHEP02(2018)164}{\emph{JHEP} {\bfseries 02}
  (2018) 164} [\href{https://arxiv.org/abs/1708.05718}{{\ttfamily
  1708.05718}}].

\bibitem{Polchinski:2016xgd}
J.~Polchinski and V.~Rosenhaus, \emph{{The Spectrum in the Sachdev-Ye-Kitaev
  Model}}, \href{https://doi.org/10.1007/JHEP04(2016)001}{\emph{JHEP}
  {\bfseries 04} (2016) 001}
  [\href{https://arxiv.org/abs/1601.06768}{{\ttfamily 1601.06768}}].

\bibitem{Murugan:2017eto}
J.~Murugan, D.~Stanford and E.~Witten, \emph{{More on Supersymmetric and 2d
  Analogs of the SYK Model}},
  \href{https://doi.org/10.1007/JHEP08(2017)146}{\emph{JHEP} {\bfseries 08}
  (2017) 146} [\href{https://arxiv.org/abs/1706.05362}{{\ttfamily
  1706.05362}}].

\bibitem{Karateev:2017jgd}
D.~Karateev, P.~Kravchuk and D.~Simmons-Duffin, \emph{{Weight Shifting
  Operators and Conformal Blocks}},
  \href{https://doi.org/10.1007/JHEP02(2018)081}{\emph{JHEP} {\bfseries 02}
  (2018) 081} [\href{https://arxiv.org/abs/1706.07813}{{\ttfamily
  1706.07813}}].

\bibitem{Pimentel2019}
D.~Baumann, C.~Duaso~Pueyo, A.~Joyce, H.~Lee and G.L.~Pimentel, \emph{{The
  Cosmological Bootstrap: Weight-Shifting Operators and Scalar Seeds}},
  \href{https://arxiv.org/abs/arXiv:1910.14051}{{\ttfamily arXiv:1910.14051}}.

\bibitem{Bzowski:2015yxv}
A.~Bzowski, P.~McFadden and K.~Skenderis, \emph{{Evaluation of conformal
  integrals}}, \href{https://doi.org/10.1007/JHEP02(2016)068}{\emph{JHEP}
  {\bfseries 02} (2016) 068}
  [\href{https://arxiv.org/abs/1511.02357}{{\ttfamily 1511.02357}}].

\bibitem{Caloro}
F.~Caloro and P.~McFadden, \emph{{$\mathcal{A}$-hypergeometric functions and
  creation operators for Feynman and Witten diagrams}}, {\emph{submitted to
  JHEP} } [\href{https://arxiv.org/abs/2309.15895}{{\ttfamily 2309.15895}}].

\bibitem{Rychkov}
S.~Rychkov, \emph{{EPFL Lectures on Conformal Field Theory in D\ensuremath{>}=
  3 Dimensions}}, SpringerBriefs in Physics, Springer (1, 2016),
  \href{https://doi.org/10.1007/978-3-319-43626-5}{10.1007/978-3-319-43626-5},
  [\href{https://arxiv.org/abs/1601.05000}{{\ttfamily 1601.05000}}].

\bibitem{DiFrancesco:1997nk}
P.~Di~Francesco, P.~Mathieu and D.~Senechal, \emph{{Conformal field theory}},
  Springer, New York (1997).

\bibitem{Gillioz:2022yze}
M.~Gillioz, \emph{{Conformal field theory for particle physicists}},
  SpringerBriefs in Physics, Springer (2023),
  \href{https://doi.org/10.1007/978-3-031-27086-4}{10.1007/978-3-031-27086-4},
  [\href{https://arxiv.org/abs/2207.09474}{{\ttfamily 2207.09474}}].

\bibitem{Wald}
R.M.~Wald, \emph{{General Relativity}}, Chicago Univ. Pr., Chicago, USA (1984),
  \href{https://doi.org/10.7208/chicago/9780226870373.001.0001}{10.7208/chicago/9780226870373.001.0001}.

\bibitem{Kravchuk:2016qvl}
P.~Kravchuk and D.~Simmons-Duffin, \emph{{Counting Conformal Correlators}},
  \href{https://doi.org/10.1007/JHEP02(2018)096}{\emph{JHEP} {\bfseries 02}
  (2018) 096} [\href{https://arxiv.org/abs/1612.08987}{{\ttfamily
  1612.08987}}].

\bibitem{Simmons-Duffin:2016gjk}
D.~Simmons-Duffin, \emph{{The Conformal Bootstrap}},  in \emph{{Proceedings,
  Theoretical Advanced Study Institute in Elementary Particle Physics: New
  Frontiers in Fields and Strings (TASI 2015): Boulder, CO, USA, June 1-26,
  2015}}, pp.~1--74, 2017,
  \href{https://doi.org/10.1142/9789813149441_0001}{DOI}
  [\href{https://arxiv.org/abs/1602.07982}{{\ttfamily 1602.07982}}].

\bibitem{Hogervorst:2013sma}
M.~Hogervorst and S.~Rychkov, \emph{{Radial Coordinates for Conformal Blocks}},
  \href{https://doi.org/10.1103/PhysRevD.87.106004}{\emph{Phys. Rev. D}
  {\bfseries 87} (2013) 106004}
  [\href{https://arxiv.org/abs/1303.1111}{{\ttfamily 1303.1111}}].

\bibitem{Coriano:2020ees}
C.~Corian\`o and M.M.~Maglio, \emph{{Conformal field theory in momentum space
  and anomaly actions in gravity: The analysis of three- and four-point
  function}}, \href{https://doi.org/10.1016/j.physrep.2021.11.005}{\emph{Phys.
  Rept.} {\bfseries 952} (2022) 1}
  [\href{https://arxiv.org/abs/2005.06873}{{\ttfamily 2005.06873}}].

\bibitem{Coriano:2019sth}
C.~Corian\`o and M.M.~Maglio, \emph{{On Some Hypergeometric Solutions of the
  Conformal Ward Identities of Scalar 4-point Functions in Momentum Space}},
  \href{https://doi.org/10.1007/JHEP09(2019)107}{\emph{JHEP} {\bfseries 09}
  (2019) 107} [\href{https://arxiv.org/abs/1903.05047}{{\ttfamily
  1903.05047}}].

\bibitem{abramowitz1972handbook}
M.~Abramowitz and I.A.~Stegun, \emph{Handbook of mathematical functions with
  formulas, graphs, and mathematical tables. national bureau of standards
  applied mathematics series 55. tenth printing.}, .

\bibitem{Davydychev1992}
A.I.~Davydychev, \emph{{Recursive algorithm of evaluating vertex type Feynman
  integrals}}, {\emph{J. Phys. A} {\bfseries 25} (1992) 5587}.

\bibitem{tHooft1978}
G.~'t~Hooft and M.~Veltman, \emph{{Scalar One Loop Integrals}},
  \href{https://doi.org/10.1016/0550-3213(79)90605-9}{\emph{Nucl. Phys. B}
  {\bfseries 153} (1979) 365}.

\bibitem{Zagier}
D.~Zagier, \emph{{The Dilogarithm Function}},  in \emph{{Proceedings, Les
  Houches School of Physics: Frontiers in Number Theory, Physics and Geometry
  II: On Conformal Field Theories, Discrete Groups and Renormalization: Les
  Houches, France, March 9-21, 2003}}, pp.~3--65, 2007,
  \href{https://doi.org/10.1007/978-3-540-30308-4_1}{DOI}.

\bibitem{apk2014}
A.~Bzowski, P.~McFadden and K.~Skenderis, \emph{Implications of conformal
  invariance in momentum space},
  \href{https://doi.org/10.1007/JHEP03(2014)111}{\emph{JHEP} {\bfseries 03}
  (2014) 111} [\href{https://arxiv.org/abs/arXiv:1304.7760}{{\ttfamily
  arXiv:1304.7760}}].

\bibitem{appell1926fonctions}
P.~Appell and J.~de~F{\'e}riet, \emph{Fonctions hyperg{\'e}om{\'e}triques et
  hypersph{\'e}riques: polynomes d'Hermite}, Gauthier-Villars (1926).

\bibitem{H_Exton_1995}
H.~Exton, \emph{On the system of partial differential equations associated with
  appell's function f4},
  \href{https://doi.org/10.1088/0305-4470/28/3/017}{\emph{Journal of Physics A:
  Mathematical and General} {\bfseries 28} (1995) 631}.

\bibitem{Prudnikov}
A.~Prudnikov, Y.~Brychkov and O.~Marichev, \emph{Integrals and Series. Vol. 2.
  Special Functions}, vol.~2, {Gordon and Breach Science Publishers} (01,
  1992).

\bibitem{Bzowski:2020lip}
A.~Bzowski, \emph{{TripleK: A Mathematica package for evaluating triple-K
  integrals and conformal correlation functions}},
  \href{https://doi.org/10.1016/j.cpc.2020.107538}{\emph{Comput. Phys. Commun.}
  {\bfseries 258} (2021) 107538}
  [\href{https://arxiv.org/abs/2005.10841}{{\ttfamily 2005.10841}}].

\bibitem{Raju:2012zs}
S.~Raju, \emph{{Four point functions of the stress tensor and conserved
  currents in AdS$_4$/CFT$_3$}},
  \href{https://doi.org/10.1103/PhysRevD.85.126008}{\emph{Phys.Rev.} {\bfseries
  D85} (2012) 126008} [\href{https://arxiv.org/abs/1201.6452}{{\ttfamily
  1201.6452}}].

\bibitem{Serino:2020pyu}
M.~Serino, \emph{{The four-point correlation function of the energy-momentum
  tensor in the free conformal field theory of a scalar field}},
  \href{https://doi.org/10.1140/epjc/s10052-020-8208-z}{\emph{Eur. Phys. J. C}
  {\bfseries 80} (2020) 686}
  [\href{https://arxiv.org/abs/2004.08668}{{\ttfamily 2004.08668}}].

\bibitem{Witten:1998qj}
E.~Witten, \emph{{Anti-de Sitter space and holography}},
  \href{https://doi.org/10.4310/ATMP.1998.v2.n2.a2}{\emph{Adv. Theor. Math.
  Phys.} {\bfseries 2} (1998) 253}
  [\href{https://arxiv.org/abs/hep-th/9802150}{{\ttfamily hep-th/9802150}}].

\bibitem{DHoker:1998ecp}
E.~D'Hoker and D.Z.~Freedman, \emph{{General scalar exchange in AdS(d+1)}},
  \href{https://doi.org/10.1016/S0550-3213(99)00169-8}{\emph{Nucl. Phys. B}
  {\bfseries 550} (1999) 261}
  [\href{https://arxiv.org/abs/hep-th/9811257}{{\ttfamily hep-th/9811257}}].

\bibitem{Gillioz:2018mto}
M.~Gillioz, \emph{{Momentum-space conformal blocks on the light cone}},
  \href{https://doi.org/10.1007/JHEP10(2018)125}{\emph{JHEP} {\bfseries 10}
  (2018) 125} [\href{https://arxiv.org/abs/1807.07003}{{\ttfamily
  1807.07003}}].

\bibitem{Gillioz:2019iye}
M.~Gillioz, X.~Lu, M.A.~Luty and G.~Mikaberidze, \emph{{Convergent
  Momentum-Space OPE and Bootstrap Equations in Conformal Field Theory}},
  \href{https://doi.org/10.1007/JHEP03(2020)102}{\emph{JHEP} {\bfseries 03}
  (2020) 102} [\href{https://arxiv.org/abs/1912.05550}{{\ttfamily
  1912.05550}}].

\bibitem{Gillioz:2020wgw}
M.~Gillioz, \emph{{Conformal partial waves in momentum space}},
  \href{https://doi.org/10.21468/SciPostPhys.10.4.081}{\emph{SciPost Phys.}
  {\bfseries 10} (2021) 081}
  [\href{https://arxiv.org/abs/2012.09825}{{\ttfamily 2012.09825}}].

\bibitem{Lee:2013hzt}
R.N.~Lee and A.A.~Pomeransky, \emph{{Critical points and number of master
  integrals}}, \href{https://doi.org/10.1007/JHEP11(2013)165}{\emph{JHEP}
  {\bfseries 11} (2013) 165} [\href{https://arxiv.org/abs/1308.6676}{{\ttfamily
  1308.6676}}].

\bibitem{kirchhoff1847}
G.~Kirchhoff, \emph{{\"U}ber die aufl{\"o}sung der gleichungen, auf welche man
  bei der untersuchung der linearen vertheilung galvanischer str{\"o}me
  gef{\"u}hrt wird}, {\emph{Annalen der Physik} {\bfseries 148} (1847) 497}.

\bibitem{kirchhoff1958}
G.~Kirchhoff, \emph{On the solution of the equations obtained from the
  investigation of the linear distribution of galvanic currents}, {\emph{IRE
  transactions on circuit theory} {\bfseries 5} (1958) 4}.

\bibitem{fiedler_2011}
M.~Fiedler, \emph{Simplex geometry},  in \emph{Matrices and Graphs in
  Geometry}, Encyclopedia of Mathematics and its Applications, Cambridge
  University Press (2011),
  \href{https://doi.org/10.1017/CBO9780511973611.003}{DOI}.

\bibitem{Devriendt_2022}
K.~Devriendt, \emph{Effective resistance is more than distance: Laplacians,
  simplices and the schur complement},
  \href{https://doi.org/10.1016/j.laa.2022.01.002}{\emph{Linear Algebra and its
  Applications} {\bfseries 639} (2022) 24}.

\bibitem{dorfler2012kron}
F.~Dorfler and F.~Bullo, \emph{Kron reduction of graphs with applications to
  electrical networks}, {\emph{IEEE Transactions on Circuits and Systems I:
  Regular Papers} {\bfseries 60} (2012) 150}
  [\href{https://arxiv.org/abs/arXiv:1102.2950}{{\ttfamily arXiv:1102.2950}}].

\bibitem{Loebbert:2020hxk}
F.~Loebbert, J.~Miczajka, D.~M\"uller and H.~M\"unkler, \emph{{Massive
  Conformal Symmetry and Integrability for Feynman Integrals}},
  \href{https://doi.org/10.1103/PhysRevLett.125.091602}{\emph{Phys. Rev. Lett.}
  {\bfseries 125} (2020) 091602}
  [\href{https://arxiv.org/abs/2005.01735}{{\ttfamily 2005.01735}}].

\bibitem{Rigatos:2022eos}
K.C.~Rigatos and X.~Zhou, \emph{{Yangian Symmetry in Holographic Correlators}},
  \href{https://doi.org/10.1103/PhysRevLett.129.101601}{\emph{Phys. Rev. Lett.}
  {\bfseries 129} (2022) 101601}
  [\href{https://arxiv.org/abs/2206.07924}{{\ttfamily 2206.07924}}].

\bibitem{Regge}
T.~Regge, \emph{{Algebraic topology methds in the theory of Feynman
  relativistic amplitudes}},  in \emph{{Battelle Recontres: Lectures in
  Mathematics and Physics}}, pp.~433--458, 1967.

\bibitem{Kashiwara:1977nf}
M.~Kashiwara and T.~Kawai, \emph{{Holonomic Systems of Linear Differential
  Equations and Feynman Integrals}},
  \href{https://doi.org/10.2977/prims/1195196602}{\emph{Publ. Res. Inst. Math.
  Sci. Kyoto} {\bfseries 12} (1977) 131}.

\bibitem{Vanhove:2018mto}
P.~Vanhove, \emph{{Feynman integrals, toric geometry and mirror symmetry}},  in
  \emph{{KMPB Conference}: {Elliptic Integrals, Elliptic Functions and Modular
  Forms in Quantum Field Theory}}, pp.~415--458, 2019
  [\href{https://arxiv.org/abs/1807.11466}{{\ttfamily 1807.11466}}].

\bibitem{de_la_Cruz_2019}
L.~de~la Cruz, \emph{{Feynman integrals as A-hypergeometric functions}},
  \href{https://doi.org/10.1007/JHEP12(2019)123}{\emph{JHEP} {\bfseries 12}
  (2019) 123} [\href{https://arxiv.org/abs/1907.00507}{{\ttfamily
  1907.00507}}].

\bibitem{Klausen:2019hrg}
R.P.~Klausen, \emph{{Hypergeometric Series Representations of Feynman Integrals
  by GKZ Hypergeometric Systems}},
  \href{https://doi.org/10.1007/JHEP04(2020)121}{\emph{JHEP} {\bfseries 04}
  (2020) 121} [\href{https://arxiv.org/abs/1910.08651}{{\ttfamily
  1910.08651}}].

\bibitem{Klausen:2021yrt}
R.P.~Klausen, \emph{{Kinematic singularities of Feynman integrals and principal
  A-determinants}}, \href{https://doi.org/10.1007/JHEP02(2022)004}{\emph{JHEP}
  {\bfseries 02} (2022) 004}
  [\href{https://arxiv.org/abs/2109.07584}{{\ttfamily 2109.07584}}].

\bibitem{Klausen:2023gui}
R.P.~Klausen, \emph{{Hypergeometric Feynman integrals}}, Ph.D. thesis, Mainz
  U., 2023.
\newblock \href{https://arxiv.org/abs/2302.13184}{{\ttfamily 2302.13184}}.

\bibitem{Klemm_2020}
A.~Klemm, C.~Nega and R.~Safari, \emph{{The l-loop banana amplitude from {GKZ}
  systems and relative Calabi-Yau periods}},
  \href{https://doi.org/10.1007/jhep04(2020)088}{\emph{Journal of High Energy
  Physics} {\bfseries 2020} (2020) }
  [\href{https://arxiv.org/abs/1912.06201}{{\ttfamily 1912.06201}}].

\bibitem{Feng_2020}
T.-F.~Feng, C.-H.~Chang, J.-B.~Chen and H.-B.~Zhang,
  \emph{{{GKZ}-hypergeometric systems for Feynman integrals}},
  \href{https://doi.org/10.1016/j.nuclphysb.2020.114952}{\emph{Nuclear Physics
  B} {\bfseries 953} (2020) 114952}
  [\href{https://arxiv.org/abs/arXiv:1912.01726}{{\ttfamily
  arXiv:1912.01726}}].

\bibitem{Chestnov:2022alh}
V.~Chestnov, F.~Gasparotto, M.K.~Mandal, P.~Mastrolia, S.J.~Matsubara-Heo,
  H.J.~Munch et~al., \emph{{Macaulay matrix for Feynman integrals: linear
  relations and intersection numbers}},
  \href{https://doi.org/10.1007/JHEP09(2022)187}{\emph{JHEP} {\bfseries 09}
  (2022) 187} [\href{https://arxiv.org/abs/2204.12983}{{\ttfamily
  2204.12983}}].

\bibitem{Ananthanarayan:2022ntm}
B.~Ananthanarayan, S.~Banik, S.~Bera and S.~Datta, \emph{{FeynGKZ: A
  Mathematica package for solving Feynman integrals using GKZ hypergeometric
  systems}}, \href{https://doi.org/10.1016/j.cpc.2023.108699}{\emph{Comput.
  Phys. Commun.} {\bfseries 287} (2023) 108699}
  [\href{https://arxiv.org/abs/2211.01285}{{\ttfamily 2211.01285}}].

\bibitem{Zhang:2023fil}
H.-B.~Zhang and T.-F.~Feng, \emph{{GKZ hypergeometric systems of the three-loop
  vacuum Feynman integrals}},
  \href{https://doi.org/10.1007/JHEP05(2023)075}{\emph{JHEP} {\bfseries 05}
  (2023) 075} [\href{https://arxiv.org/abs/2303.02795}{{\ttfamily
  2303.02795}}].

\bibitem{GKZ_1}
I.~Gel'fand, M.~Kapranov and A.~Zelevinsky, \emph{{Generalized Euler integrals
  and A-hypergeometric functions}},
  \href{https://doi.org/https://doi.org/10.1016/0001-8708(90)90048-R}{\emph{Advances
  in Mathematics} {\bfseries 84} (1990) 255}.

\bibitem{GKZ_2}
I.~Gel'fand, A.~Zelevinsky and M.~Kapranov, \emph{{Hypergeometric functions and
  toral manifolds}}, {\emph{Functional Analysis and its Applications}
  {\bfseries 23} (1989) 94}.

\bibitem{GKZ_3}
I.~Gel'fand, M.~Kapranov and A.~Zelevinsky, \emph{{Hypergeometric functions,
  toric varieties and Newton polyhedra}},  in \emph{{Special Functions, ICM-90
  Satellite Conference Proceedings}}, Springer, 1991.

\bibitem{GKZ_book}
I.~Gel'fand, M.~Kapranov and A.~Zelevinsky, \emph{Discriminants, Resultants,
  and Multidimensional Determinants}, Mathematics (Birkh{\"a}user), Springer
  (1994).

\bibitem{Stienstra:2005nr}
J.~Stienstra, \emph{{GKZ hypergeometric structures}},  in \emph{{Istanbul 2005:
  CIMPA Summer School on Arithmetic and Geometry Around Hypergeometric
  Functions}}, 11, 2005 [\href{https://arxiv.org/abs/math/0511351}{{\ttfamily
  math/0511351}}].

\bibitem{BeukersNotes}
F.~Beukers, \emph{{Notes on A-hypergeometric functions, in Arithmetic and
  Galois theories of differential equations}}, {\emph{S{\'e}minaires et
  Congr{\`e}s, Soci{\`e}t{\'e} math{\'e}matique de France} {\bfseries 23}
  (2011) 25}.

\bibitem{cattani2006three}
E.~Cattani, ``{Three lectures on hypergeometric functions}.''
  {\url{http://people.math.umass.edu/~cattani/hypergeom_lectures.pdf}}, 2006.

\bibitem{takayama2020hypergeometric}
N.~Takayama, \emph{{A-Hypergeometric Functions}},  in \emph{{Encyclopedia of
  Special Functions: The Askey-Bateman Project}}, vol.~2, Cambridge University
  Press, 2020.

\bibitem{saito2013grobner}
M.~Saito, B.~Sturmfels and N.~Takayama, \emph{Gr{\"o}bner deformations of
  hypergeometric differential equations}, vol.~6, Algorithms \& Computations in
  Mathematics, Springer (2013).

\bibitem{reichelt2021algebraic}
T.~Reichelt, M.~Schulze, C.~Sevenheck and U.~Walther, \emph{Algebraic aspects
  of hypergeometric differential equations},
  \href{https://arxiv.org/abs/2004.07262}{{\ttfamily 2004.07262}}.

\bibitem{Bytev:2009kb}
V.V.~Bytev, M.Y.~Kalmykov and B.A.~Kniehl, \emph{{Differential reduction of
  generalized hypergeometric functions from Feynman diagrams: One-variable
  case}}, \href{https://doi.org/10.1016/j.nuclphysb.2010.03.025}{\emph{Nucl.
  Phys. B} {\bfseries 836} (2010) 129}
  [\href{https://arxiv.org/abs/0904.0214}{{\ttfamily 0904.0214}}].

\bibitem{Bytev:2013gva}
V.V.~Bytev, M.Y.~Kalmykov and S.-O.~Moch, \emph{{HYPERgeometric functions
  DIfferential REduction (HYPERDIRE): MATHEMATICA based packages for
  differential reduction of generalized hypergeometric functions: $F_D$ and
  $F_S$ Horn-type hypergeometric functions of three variables}},
  \href{https://doi.org/10.1016/j.cpc.2014.07.014}{\emph{Comput. Phys. Commun.}
  {\bfseries 185} (2014) 3041}
  [\href{https://arxiv.org/abs/1312.5777}{{\ttfamily 1312.5777}}].

\bibitem{Dolan:2000ut}
F.A.~Dolan and H.~Osborn, \emph{{Conformal four point functions and the
  operator product expansion}},
  \href{https://doi.org/10.1016/S0550-3213(01)00013-X}{\emph{Nucl. Phys.}
  {\bfseries B599} (2001) 459}
  [\href{https://arxiv.org/abs/hep-th/0011040}{{\ttfamily hep-th/0011040}}].

\bibitem{Costa:2018mcg}
M.S.~Costa and T.~Hansen, \emph{{AdS Weight Shifting Operators}},
  \href{https://doi.org/10.1007/JHEP09(2018)040}{\emph{JHEP} {\bfseries 09}
  (2018) 040} [\href{https://arxiv.org/abs/1805.01492}{{\ttfamily
  1805.01492}}].

\bibitem{Saito_param_shift}
M.~Saito, \emph{{Parameter shift in normal generalized hypergeometric
  systems}}, {\emph{Tohoku Mathematical Journal} (1992) 523–534}.

\bibitem{Saito_restrictions}
M.~Saito and N.~Takayama, \emph{{Restrictions of A-hypergeometric systems and
  connection formulas on the $\Delta_1\times\Delta_{n-1}$-hypergeometric
  function}}, {\emph{International Journal of Mathematics} (1994) 537}.

\bibitem{saito_sturmfels_takayama_1999}
M.~Saito, B.~Sturmfels and N.~Takayama, \emph{{Hypergeometric Polynomials and
  Integer Programming}},
  \href{https://doi.org/10.1023/A:1000609524994}{\emph{Compositio Mathematica}
  {\bfseries 115} (1999) 231–240}.

\bibitem{nilsson2010mellin}
L.~Nilsson and M.~Passare, \emph{Mellin transforms of multivariate rational
  functions}, {\emph{Journal of Geometric Analysis} {\bfseries 23} (2013) 24}
  [\href{https://arxiv.org/abs/1010.5060}{{\ttfamily 1010.5060}}].

\bibitem{berkesch2013eulermellin}
C.~Berkesch, J.~Forsg{\aa}rd and M.~Passare, \emph{{Euler-Mellin integrals and
  A-hypergeometric functions}}, {\emph{Michigan Mathematical Journal}
  {\bfseries 63} (2014) 101} [\href{https://arxiv.org/abs/1103.6273}{{\ttfamily
  1103.6273}}].

\bibitem{singular}
W.~Decker, G.-M.~Greuel, G.~Pfister and H.~Sch\"onemann, ``{\sc Singular}
  {4-3-0} --- {A} computer algebra system for polynomial computations.''
  \url{http://www.singular.uni-kl.de/Manual/4-0-3/sing_1375.htm}, 2022.

\bibitem{Arkani-Hamed:2022cqe}
N.~Arkani-Hamed, A.~Hillman and S.~Mizera, \emph{{Feynman polytopes and the
  tropical geometry of UV and IR divergences}},
  \href{https://doi.org/10.1103/PhysRevD.105.125013}{\emph{Phys. Rev. D}
  {\bfseries 105} (2022) 125013}
  [\href{https://arxiv.org/abs/2202.12296}{{\ttfamily 2202.12296}}].

\bibitem{Matsubara-Heo:2023ylc}
S.-J.~Matsubara-Heo, S.~Mizera and S.~Telen, \emph{{Four Lectures on Euler
  Integrals}},  \href{https://arxiv.org/abs/2306.13578}{{\ttfamily
  2306.13578}}.

\bibitem{Kaneko:2009qx}
T.~Kaneko and T.~Ueda, \emph{{A geometric method of sector decomposition}},
  \href{https://doi.org/10.1016/j.cpc.2010.04.001}{\emph{Comput. Phys. Commun.}
  {\bfseries 181} (2010) 1352}
  [\href{https://arxiv.org/abs/0908.2897}{{\ttfamily 0908.2897}}].

\bibitem{Schultka:2018nrs}
K.~Schultka, \emph{{Toric geometry and regularization of Feynman integrals}},
  \href{https://arxiv.org/abs/1806.01086}{{\ttfamily 1806.01086}}.

\bibitem{Nhull}
L.~Petrich, ``Computational geometry.''
  \url{http://lpetrich.org/Science/#CHDV}.

\bibitem{MillerBook}
W.~Miller, \emph{Lie Theory and Special Functions}, vol.~43, Mathematics in
  Science and Engineering, Academic Press, New York (1968).

\bibitem{Miller}
E.~Kalanins, H.~Manocha and W.~Miller, \emph{{The Lie theory of two-variable
  hypergeometric functions}}, {\emph{Studies in Applied Mathematics} {\bfseries
  62} (1980) 143}.

\bibitem{smeets2000}
M.J.E.~Smeets, \emph{{Parameter shift in GKZ-hypergeometric systems}}, Ph.D.
  thesis, Eindhoven University of Technology, 2000.
\newblock \url{https://doi.org/10.6100/IR532761}.

\bibitem{NIST:DLMF}
``{NIST Digital Library of Mathematical Functions}.''
  \url{https://dlmf.nist.gov/}.

\bibitem{Tarasov:2017jen}
O.V.~Tarasov, \emph{{Massless on-shell box integral with arbitrary powers of
  propagators}}, \href{https://doi.org/10.1088/1751-8121/aac57f}{\emph{J. Phys.
  A} {\bfseries 51} (2018) 275401}
  [\href{https://arxiv.org/abs/1709.07526}{{\ttfamily 1709.07526}}].

\bibitem{Feng:2022ude}
T.-F.~Feng, H.-B.~Zhang, Y.-Q.~Dong and Y.~Zhou, \emph{{GKZ-system of the
  2-loop self energy with 4 propagators}},
  \href{https://doi.org/10.1140/epjc/s10052-023-11438-6}{\emph{Eur. Phys. J. C}
  {\bfseries 83} (2023) 314}
  [\href{https://arxiv.org/abs/2209.15194}{{\ttfamily 2209.15194}}].

\bibitem{SchomerusBig}
M.~Isachenkov and V.~Schomerus, \emph{{Integrability of conformal blocks. Part
  I. Calogero-Sutherland scattering theory}},
  \href{https://doi.org/10.1007/JHEP07(2018)180}{\emph{JHEP} {\bfseries 07}
  (2018) 180} [\href{https://arxiv.org/abs/arXiv:1711.06609}{{\ttfamily
  arXiv:1711.06609}}].

\bibitem{ChenKyono}
H.-Y.~Chen and H.~Kyono, \emph{{On conformal blocks, crossing kernels and
  multi-variable hypergeometric functions}},
  \href{https://doi.org/10.1007/JHEP10(2019)149}{\emph{JHEP} {\bfseries 10}
  (2019) 149} [\href{https://arxiv.org/abs/arXiv:1906.03135}{{\ttfamily
  arXiv:1906.03135}}].

\bibitem{Itzykson:1980rh}
C.~Itzykson and J.B.~Zuber, \emph{{Quantum Field Theory, pp 294-7}},
  International Series In Pure and Applied Physics, McGraw-Hill, New York
  (1980).

\bibitem{Caracciolo_2013}
S.~Caracciolo, A.D.~Sokal and A.~Sportiello, \emph{Algebraic/combinatorial
  proofs of cayley-type identities for derivatives of determinants and
  pfaffians}, \href{https://doi.org/10.1016/j.aam.2012.12.001}{\emph{Advances
  in Applied Mathematics} {\bfseries 50} (2013) 474}.

\bibitem{Fulmek}
M.~Fulmek, \emph{{A combinatorial proof for Cayley's identity}},
  \href{https://doi.org/https://doi.org/10.37236/3775}{\emph{{Electronic
  Journal of Combinatorics}} {\bfseries 21} (2014) }
  [\href{https://arxiv.org/abs/1309.6801}{{\ttfamily 1309.6801}}].

\bibitem{Budur}
N.~Budur, \emph{{Bernstein-Sato polynomials (Lecture Notes UPC, Barcelona)
  2015}},
  {\url{https://perswww.kuleuven.be/~u0089821/Barcelona/BarcelonaNotes.pdf}}.

\bibitem{Tkachov:1996wh}
F.V.~Tkachov, \emph{{Algebraic algorithms for multiloop calculations. The First
  15 years. What's next?}},
  \href{https://doi.org/10.1016/S0168-9002(97)00110-1}{\emph{Nucl. Instrum.
  Meth. A} {\bfseries 389} (1997) 309}
  [\href{https://arxiv.org/abs/hep-ph/9609429}{{\ttfamily hep-ph/9609429}}].

\bibitem{DHoker:1999kzh}
E.~D'Hoker, D.Z.~Freedman, S.D.~Mathur, A.~Matusis and L.~Rastelli,
  \emph{{Graviton exchange and complete four point functions in the AdS / CFT
  correspondence}},
  \href{https://doi.org/10.1016/S0550-3213(99)00525-8}{\emph{Nucl. Phys. B}
  {\bfseries 562} (1999) 353}
  [\href{https://arxiv.org/abs/hep-th/9903196}{{\ttfamily hep-th/9903196}}].

\end{thebibliography}

\end{document}